\documentclass[11pt,tightenlines,preprintnumbers,superscriptaddress,nofootinbib,fleqn]{revtex4}
\pdfoutput=1
\makeatletter
\def\l@subsubsection#1#2{}
\renewcommand{\p@subsection}{}
\renewcommand{\p@subsubsection}{}
\renewcommand{\p@paragraph}{}
\makeatother
\usepackage[pdftitle={Event Generators for High-Energy Physics Experiments},pdfauthor={},hidelinks]{hyperref}
\usepackage{amsmath,amssymb,xspace,graphicx,xcolor}
\usepackage{siunitx,scalefnt,listings,multirow}
\usepackage[draft]{listlbls}
\usepackage[caption=false]{subfig}

\newcommand{\eg}{\emph{e.g.}\xspace}
\newcommand{\ie}{\emph{i.e.}\xspace}

\newcommand{\fortran}{\textsc{Fortran}\xspace}
\newcommand{\python}{\textsc{Python}\xspace}
\newcommand{\cxx}{\textsc{C++}\xspace}
\newcommand{\cpp}{\textsc{C++}\xspace}

\newcommand{\myscalefont}[1]{{\scalefont{0.8}#1}}
\newcommand{\nlox}{N\myscalefont{LO}X\xspace}
\newcommand{\tred}{TR\myscalefont{ED}\xspace}
\newcommand{\Matrix}{M\myscalefont{ATRIX}\xspace}
\newcommand{\Munich}{M\myscalefont{UNICH}\xspace}
\newcommand{\Python}{P\myscalefont{YTHON}\xspace}
\newcommand{\qgraf}{QG\myscalefont{RAF}\xspace}
\newcommand{\form}{F\myscalefont{ORM}\xspace}

\newcommand{\NLOCT}{NLOCT\xspace}
\newcommand{\MadLoop}{M\myscalefont{AD}L\myscalefont{OOP}\xspace}
\newcommand{\CutTools}{C\myscalefont{UT}T\myscalefont{OOLS}\xspace}
\newcommand{\Ninja}{N\myscalefont{INJA}\xspace}
\newcommand{\Golem}{G\myscalefont{OLEM}95\xspace}
\newcommand{\MGaMC}{M\myscalefont{AD}G\myscalefont{RAPH5}\_aM\myscalefont{C}@N\myscalefont{LO}\xspace}
\newcommand{\RadISH}{R\myscalefont{adISH}\xspace}
\newcommand{\GoSam}{G\myscalefont{O}S\myscalefont{AM}\xspace}
\newcommand{\OpenLoops}{O\myscalefont{PEN}L\myscalefont{OOPS}\xspace}
\newcommand{\collier}{C\myscalefont{OLLIER}\xspace}\newcommand{\Collier}{\collier}
\newcommand{\aNLO}{M\myscalefont{AD}G\myscalefont{RAPH}5\_aM\myscalefont{C}@N\myscalefont{LO}}
\newcommand{\aNLOs}{\aNLOs}\newcommand{\madgraphee}{\aNLO}\newcommand{\mgamc}{\aNLO}
\newcommand{\madgraph}{M\myscalefont{AD}G\myscalefont{RAPH}\xspace}

\newcommand{\mcmule}{M\myscalefont{C}M\myscalefont{ULE}\xspace}
\newcommand{\dire}{D\myscalefont{IRE}\xspace}
\newcommand{\lhe}{LHE\xspace}
\newcommand{\hepmc}{H\myscalefont{EP}MC\xspace}
\newcommand{\rivet}{R\myscalefont{IVET}\xspace}

\newcommand{\hztool}{HZT\myscalefont{OOL}\xspace}
\newcommand{\lhapdf}{LHAPDF\xspace}
\newcommand{\contur}{C\myscalefont{ONTUR}\xspace}
\newcommand{\professor}{P\myscalefont{ROFESSOR}\xspace}
\newcommand{\feynrules}{F\myscalefont{EYN}R\myscalefont{ULES}\xspace}
\newcommand{\sherpa}{S\myscalefont{HERPA}\xspace}
\newcommand{\amegic}{A\myscalefont{MEGIC}\xspace}
\newcommand{\comix}{C\myscalefont{OMIX}\xspace}
\newcommand{\achilles}{A\myscalefont{CHILLES}\xspace}

\newcommand{\herwig}{H\myscalefont{ERWIG}\xspace}
\newcommand{\evtgen}{E\myscalefont{VT}G\myscalefont{EN}\xspace}

\newcommand{\ariadne}{A\myscalefont{RIADNE}\xspace}
\newcommand{\pythia}{P\myscalefont{YTHIA}\xspace}
\newcommand{\vincia}{V\myscalefont{INCIA}\xspace}
\newcommand{\photos}{P\myscalefont{HOTOS}\xspace}
\newcommand{\horace}{H\myscalefont{ORACE}\xspace}
\newcommand{\babayaga}{B\myscalefont{ABA}Y\myscalefont{AGA}\xspace}
\newcommand{\bhlumi}{BHL\myscalefont{UMI}\xspace}
\newcommand{\bhwide}{BHW\myscalefont{IDE}\xspace}
\newcommand{\genie}{G\myscalefont{ENIE}\xspace}
\newcommand{\nuisance}{N\myscalefont{UISANCE}\xspace}
\newcommand{\neut}{N\myscalefont{EUT}\xspace}
\newcommand{\cernlib}{C\myscalefont{ERNLIB}\xspace}
\newcommand{\gibuu}{GiBUU\xspace}
\newcommand{\nuwro}{NuWro\xspace}

\newcommand{\recola}{R\myscalefont{ECOLA}\xspace}
\newcommand{\stripper}{S\myscalefont{TRIPPER}\xspace}
\newcommand{\whizard}{W\myscalefont{HIZARD}\xspace}
\newcommand{\omegaMEG}{O\myscalefont{MEGA}\xspace}
\newcommand{\circe}{C\myscalefont{IRCE}\xspace}
\newcommand{\circetwo}{C\myscalefont{IRCE2}\xspace}
\newcommand{\guineapig}{G\myscalefont{UINEA}-P\myscalefont{IG}\xspace}
\newcommand{\cain}{C\myscalefont{AIN}\xspace}
\newcommand{\vegas}{V\myscalefont{EGAS}\xspace}
\newcommand{\vamp}{VAMP\xspace}
\newcommand{\kkmc}{KKMC\xspace}
\newcommand{\kkmcee}{KKMC\myscalefont{EE}\xspace}
\newcommand{\koralw}{K\myscalefont{ORAL}W\xspace}
\newcommand{\foam}{F\myscalefont{OAM}\xspace}
\newcommand{\bbmc}{BBMC\xspace}
\newcommand{\mocanlo}{M\myscalefont{O}C\myscalefont{A}NLO\xspace}
\newcommand{\powhegbox}{P\myscalefont{OWHEG}-B\myscalefont{OX}\xspace}
\newcommand{\matchbox}{M\myscalefont{ATCH}B\myscalefont{OX}\xspace}
\newcommand{\mcatnlo}{M\myscalefont{C}@N\myscalefont{LO}\xspace}
\newcommand{\powheg}{P\myscalefont{OWHEG}\xspace} 

\newcommand{\KrkNLO}{K\myscalefont{RK}NLO\xspace}
\newcommand{\NNLOJET}{NNLOJ\myscalefont{ET}\xspace}
\newcommand{\HEJ}{H\myscalefont{EJ}\xspace}
\newcommand{\tomte}{T\myscalefont{OMTE}\xspace}
\newcommand{\caesar}{C\myscalefont{AESAR}\xspace}
\newcommand{\ares}{A\myscalefont{RES}\xspace}
\newcommand{\geneva}{G\myscalefont{ENEVA}\xspace}
\newcommand{\minlo}{M\myscalefont{iNLO}$^{\prime}$}
\newcommand{\minnlo}{M\myscalefont{iNNLO}$_{\rm PS}$}
\newcommand{\nnlops}{N\myscalefont{NLOPS}}

\newcommand{\POWHEG}{P\myscalefont{OWHEG}}
\newcommand{\GENIE}{G\myscalefont{ENIE}\xspace}
\newcommand{\tauola}{T\myscalefont{AUOLA}\xspace}
\newcommand{\geant}{G\myscalefont{EANT}\xspace}
\newcommand{\panscales}{P\myscalefont{AN}S\myscalefont{CALES}\xspace}

\newcommand{\di}{\mathrm{d}}
\newcommand{\modulo}[1]{\left |#1\right |}
\newcommand{\sqmodulo}[1]{{\modulo{#1}}^{2}}
\newcommand{\M}{{\cal{M}}}
\newcommand\as{\alpha_{s}}
\newcommand\aem{\alpha}

\newcommand{\alphaS}{\ensuremath{\alpha_\text{s}}\xspace}

\newcommand{\qT}{\ensuremath{q_\mathrm{T}}\xspace}

\newcommand{\Tr}{Tr}
\newcommand\sss{\scriptscriptstyle}

\newcommand\gE{\gamma_{\sss\rm E}}
\newcommand{\epem}{e^+e^-}
\newcommand{\lp}{e^+}
\newcommand{\lm}{e^-}

\newcommand{\zp}{z_+}
\newcommand{\zm}{z_-}

\newcommand{\yp}{y_+}
\newcommand{\ym}{y_-}
\newcommand\hsig{\hat{\sigma}}

\newcommand\MSb{\overline{\rm MS}}
\newcommand\msbar{\overline{\rm MS}}

\newcommand\mZ{m_Z}

\def\iket#1{|{#1}\rangle} 
\def\ibra#1{\langle{#1}|}
  
\newcommand{\order}[1]{${\cal O}(#1)$}
\newcommand{\Mmf}{\mathfrak{M}}

\def\hbeta{\hat{\beta}}
\newcommand{\Meu}{{\cal M}}
\newcommand{\sfac}{\mathfrak{s}}

\newcommand\Tau{\mathcal{T}}
\newcommand{\df}{\mathrm{d}}

\newcommand \cut{\rm cut}

\newcommand{\FF}{${\rm F}$}
\newcommand{\FJ}{${\rm FJ}$}

\newcommand{\PhiB}{\Phi_{\scriptscriptstyle \rm F}}
\newcommand{\PhiBJ}{\Phi_{\scriptscriptstyle \rm FJ}}

\newcommand{\mathd}{\mathrm{d}}

\newcommand{\tmop}[1]{\ensuremath{\operatorname{#1}}}
\newcommand{\ptlocal}{\ensuremath{p_{\text{\scalefont{0.77}T}}}}
\newcommand{\muF}{{\mu_{\text{\scalefont{0.77}F}}}}
\newcommand{\muR}{{\mu_{\text{\scalefont{0.77}R}}}}

\newcommand{\abarmu}[1]{\frac{\as(#1)}{2\pi}}
\newcommand{\ptvec}{{\vec{p}_T}}

\def\PDF#1#2{\Gamma_{\!#1/#2}}

\newcommand{\mrm}[1]{{\mathrm{#1}}}

\newcommand{\phantomsubfloat}[1]{
    {
        \captionsetup[subfigure]{labelformat=empty}
        \subfloat[][]{#1}
    }%
}

\newcommand{\affAmsterdamU}{Department of Physics and Astronomy, Vrije Universiteit, 1081 HV Amsterdam, The Netherlands}
\newcommand{\affANL}{Argonne National Laboratory, Lemont, IL, 60439}
\newcommand{\affBNL}{Brookhaven National Laboratory, Upton, NY 11973, USA}
\newcommand{\affBonnU}{Bethe Center for Theoretical Physics \& Physikalisches Institut, Universit\"at Bonn, 53113 Bonn, Germany}
\newcommand{\affBuffalo}{University at Buffalo, The State University of New York, Buffalo
14260, USA}
\newcommand{\affCambridge}{University of Cambridge, Cambridge CB3 0HE, United Kingdom}
\newcommand{\affCERN}{Theoretical Physics Department, CERN, 1211 Geneva 23, Switzerland}
\newcommand{\affCERNIT}{Information Technology Department, CERN, 1211 Geneva 23, Switzerland}
\newcommand{\affCincinnatiU}{Department of Physics, University of Cincinnati, Cincinnati, OH 45221, USA}
\newcommand{\affCISMLouvain}{Centre de Calcul Intensif et stockage de Masse, Universit\'e Catholique de Louvain, Louvain-la-Neuve, Belgium}
\newcommand{\affCornellU}{Department of Physics, Cornell University, Ithaca, NY 14850, USA}

\newcommand{\affDelawareU}{Bartol Research Institute, Department of Physics and Astronomy, University of Delaware, Newark, DE 19716, USA}
\newcommand{\affDESYHamburg}{Deutsches Elektronen-Synchrotron DESY, Notkestr. 85, 22607 Hamburg, Germany}
\newcommand{\affDESYZeuthen}{Deutsches Elektronen-Synchrotron DESY, Platanenallee 6, 15738 Zeuthen, Germany}
\newcommand{\affDresdenU}{Institute of Nuclear and Particle Physics, Technische Universit{\"a}t Dresden, 01062 Dresden, Germany}
\newcommand{\affEdinburghU}{Higgs Centre for Theoretical Physics, The University of Edinburgh, Edinburgh EH9 3FD, Scotland, United Kingdom}
\newcommand{\affETHZurich}{Institute for Theoretical Physics, ETH, CH-8093 Z\"urich, Switzerland}
\newcommand{\affFermilab}{Fermi National Accelerator Laboratory, Batavia, IL 60510, USA}
\newcommand{\affFloridaU}{Department of Physics, University of Florida, Gainesville, FL 32611, USA}
\newcommand{\affFreiburgU}{Physikalisches Institut, Albert-Ludwigs-Universit\"at Freiburg, 79104 Freiburg, Germany}
\newcommand{\affFSU}{Department of Physics, Florida State University, Tallahassee, FL 32306-4350, USA }
\newcommand{\affUPadovaandINFN}{Dipartimento di Fisica e Astronomia, Universit\`a degli Studi di Padova\\ and INFN, Sezione di Padova, I-35131 Padova, Italy}
\newcommand{\affCUNY}{City University of New York, New York, NY 10016, USA}
\newcommand{\affUNapoliandINFN}{Dipartimento di Fisica, Universit\`a di Napoli and INFN, sezione di Napoli, 80126 Napoli, Italy}
\newcommand{\affUBologna}{Dipartimento di Fisica e Astronomia, Universit\`a di Bologna, I-40126 Bologna, Italy}
\newcommand{\affUBolognaandINFN}{Dipartimento di Fisica e Astronomia, Universit\`a di Bologna and INFN, Sezione di Bologna, I-40126 Bologna, Italy}
\newcommand{\affGiessenU}{Institut f{\"u}r Theoretische Physik, Universit{\"a}t Giessen, 35390 Giessen, Germany}
\newcommand{\affGlasgowU}{School of Physics \& Astronomy, University of Glasgow, Glasgow G12~8QQ, Scotland, United Kingdom}
\newcommand{\affGoettingenU}{Institut f\"ur Theoretische Physik, Georg-August-Universit\"at G\"ottingen, 37073 G\"ottingen, Germany}
\newcommand{\affGrazU}{NAWI Graz, University of Graz, 8010 Graz, Austria}
\newcommand{\affHeidelbergU}{Institut f\"ur theoretische Physik, Universit\"at Heidelberg, 69117 Heidelberg, Germany}
\newcommand{\affIAPKarlsruhe}{Institute for Astroparticle Physics, Karlsruhe Institute of Technology, 76344 Eggenstein-Leopoldshafen, Germany}
\newcommand{\affIDSDurham}{Institute for Data Science, Durham University, DH1 3LE Durham, United Kingdom}
\newcommand{\affINFNGenova}{INFN Sezione di Genova, Via Dodecaneso 33, Genova 16146, Italy}
\newcommand{\affIPPPDurham}{Institute for Particle Physics Phenomenology, Durham University, Durham DH1 3LE, United Kingdom}

\newcommand{\affITPCAS}{Institute of Theoretical Physics, Chinese Academy of Sciences, Beijing 100190, China}
\newcommand{\affJyvaskyla}{University of Jyvaskyla, Department of Physics, P.O. Box 35, FI-40014 University of Jyvaskyla, Finland}
\newcommand{\affMilanBicoccaAndINFN}{Universit{\`a} degli Studi di Milano-Bicocca \& INFN, Milano 20126, Italy}
\newcommand{\affMilanAndINFN}{TIF Lab, Dipartimento di Fisica, Universit{\`a} degli Studi di Milano and INFN Sezione di Milano}
\newcommand{\affOxford}{Rudolf Peierls Centre for Theoretical Physics, University of Oxford, Oxford OX1 3PU, United Kingdom}
\newcommand{\affOxfordAllSouls}{All Souls College, University of Oxford, Oxford OX1 4AL, United Kingdom}
\newcommand{\affTelAvivU}{Tel Aviv University, Tel Aviv 6997801, Israel}
\newcommand{\affMIT}{The Massachusetts Institute of Technology, Department of Physics, Cambridge, MA 02139, USA}
\newcommand{\affUCL}{Department of Physics and Astronomy, University College London, London, WC1E 6BT, United Kingdom}
\newcommand{\affKITKarlsruhe}{Institute for Theoretical Physics, Karlsruhe Institute of Technology, 76131 Karlsruhe, Germany}
\newcommand{\affINFNPavia}{INFN, Sezione di Pavia, via Bassi 6, 27100 Pavia, Italy}
\newcommand{\affPaviaU}{Dipartimento di Fisica, Universit\`a di Pavia, via Bassi 6, 27100 Pavia, Italy}
\newcommand{\affWuerzburgU}{Institut f\"ur Theoretische Physik und Astrophysik, Universit\"at W\"urzburg, 97074~W\"urzburg, Germany}
\newcommand{\affJLAB}{Thomas Jefferson National Accelerator Facility, Newport News, VA 23606, USA}
\newcommand{\affPittsburghU}{Department of Physics and Astronomy, University of Pittsburgh, Pittsburgh, PA 15260, USA}
\newcommand{\affPSIVilligen}{Paul Scherrer Institut, CH-5232 Villigen PSI, Switzerland}
\newcommand{\affZurichU}{Physik-Institut, Universit\"at Z\"urich, CH-8057 Z\"urich, Switzerland}
\newcommand{\affTuftsU}{Department of Physics and Astronomy, Tufts University, Medford, MA 02155, USA}
\newcommand{\affLMUMunich}{Arnold Sommerfeld Center for Theoretical Physics, Ludwig-Maximilians-Universit\"at M\"unchen, D-80333 M\"unchen, Germany}
\newcommand{\affMPIMunich}{Max-Planck-Institut f\"ur Physik, F\"ohringer Ring 6, 80805 M\"unchen, Germany}
\newcommand{\affMSU}{Department of Physics and Astronomy, Michigan State University, MI 48824, USA}
\newcommand{\affINPKrakow}{Institute of Nuclear Physics, Physics Polish Academy of Sciences, 31-342 Krak\'ow, Poland}
\newcommand{\affSiegenU}{Department of Physics, University of Siegen, Walter-Flex-Strasse 3, 57068 Siegen, Germany}
\newcommand{\affSLAC}{SLAC National Accelerator Laboratory, Menlo Park, CA 94025, USA}
\newcommand{\affSMU}{Department of Physics, Southern Methodist University, Dallas, TX 75275, USA}

\newcommand{\affRutgers}{NHETC, Department of Physics and Astronomy, Rutgers University, Piscataway, NJ 08854, USA}
\newcommand{\affKoreaU}{Department of Physics, Korea University, Seoul 136-713, Korea}
\newcommand{\affUCLouvain}{Centre for Cosmology, Particle Physics and Phenomenology, Universit\'e Catholique de Louvain, 1348 Louvain-la-Neuve, Belgium}
\newcommand{\affKennesawU}{Department of Physics, Kennesaw State University, Kennesaw, GA 30144, USA}
\newcommand{\affSussexU}{School of Mathematical and Physical Sciences, University of Sussex, Brighton BN1 9RH, United Kingdom}
\newcommand{\affSussexPA}{Department of Physics and Astronomy, University of Sussex, Brighton BN1 9QH, United Kingdom}
\newcommand{\affRoyalHolloway}{Royal Holloway, University of London, Egham TW20 0EX, United Kingdom}
\newcommand{\affWayneState}{Department of Physics and Astronomy, Wayne State University, Detroit, Michigan 48201, USA}
\newcommand{\affLosAlamos}{Theoretical Division, Los Alamos National Laboratory, Los Alamos, NM 87545, USA}
\newcommand{\affLIPLisbon}{Laboratory of Instrumentation and Experimental Particle Physics, 1649-003 Lisbon, Portugal}
\newcommand{\affLisbonU}{Faculdade de Ci\^encias, Universidade de Lisboa, 1749-016 Lisboa, Portugal}
\newcommand{\affViennaU}{Particle Physics, University of Vienna, 1010 Vienna, Austria}
\newcommand{\affLundU}{Department of Astronomy and Theoretical Physics, Lund University, SE-223 62 Lund, Sweden}
\newcommand{\affLAPThGrenoble}{LAPTh, Universit\'e Grenoble Alpes, USMB, CNRS, 74940 Annecy, France}
\newcommand{\affNikhef}{Nikhef Theory Group, Science Park 105, 1098 XG Amsterdam, The Netherlands}
\newcommand{\affMainzU}{Johannes Gutenberg University, 55122 Mainz, Germany}
\newcommand{\affJagiellonianU}{Jagiellonian University, 31-007 Krak\'ow, Poland}
\newcommand{\affMonashU}{School of Physics and Astronomy, Monash University, Wellington Rd, VIC-3800, Australia}
\newcommand{\affOregonU}{Department of Physics, University of Oregon, Eugene, OR 97401, USA}
\newcommand{\affTorinoU}{Universit\`a di Torino, Dipartimento di Fisica, 10125 Torino, Italy}
\newcommand{\affOldDominionU}{Physics Department, Old Dominion University, Norfolk, VA 23529, USA}
\newcommand{\affNTU}{National Taiwan University, Taipei 10617, Taiwan}
\newcommand{\affKCL}{King's College London, London, WC2R 2LS, United Kingdom}

\newcommand{\affRochesterU}{University of Rochester, Rochester, NY 14627, USA}
\newcommand{\affLBNL}{Physics Division, Lawrence Berkeley National Laboratory, Berkeley, CA 94720, USA}

\newcommand{\affLiverpoolU}{University of Liverpool, Liverpool L69 7ZE, United Kingdom}
\newcommand{\affRAL}{Science and Technology Facilities Council, Rutherford Appleton Laboratory, Oxfordshire OX11 0QX, United Kingdom}
\newcommand{\affIIT}{Department of Physics, Illinois Institute of Technology, Chicago, IL 60616, USA}
\newcommand{\affWro}{University of Wroc\l aw, Institute of Theoretical Physics, 50-204 Wroc\l aw, Poland}
\newcommand{\affIPhT}{Universit\'e Paris-Saclay, CNRS, CEA, Institut de physique 
th\'eorique, 91191, Gif-sur-Yvette, France}
\newcommand{\affGent}{Department of Physics and Astronomy, Ghent University, 9000 Gent, Belgium}

\newcommand{\affManchesterU}{Department of Physics and Astronomy, University of Manchester, Manchester M13 9PL, United Kingdom}
\newcommand{\affCAMK}{Astrocent, Nicolaus Copernicus Astronomical Center Polish Academy of Sciences, ul. Rektorska 4, 00-614, Warsaw, Poland}
\newcommand{\affNCBJ}{National Centre for Nuclear Research, Pasteura 7, 02-093 Warsaw, Poland}
\newcommand{\affICL}{Imperial College London, London, SW7 2BW, United Kingdom}

\newcommand{\affNWU}{Department of Physics and Astronomy, Northwestern University, Evanston, IL, 60208, USA}
\newcommand{\affJYFL}{University of Jyvaskyla, Department of Physics, P.O. Box 35, FI-40014 University of Jyvaskyla, Finland}
\newcommand{\affHIP}{Helsinki Institute of Physics, P.O. Box 64, FI-00014 University of Helsinki, Finland}
\newcommand{\affCPHT}{CPHT, CNRS, Ecole polytechnique, IP Paris, F-91128 Palaiseau, France}

\newcommand\snowmass{\begin{center}\rule[-0.2in]{\hsize}{0.01in}\\\rule{\hsize}{0.01in}\\
\vskip 0.1in Submitted to the US Community Study\\ 
on the Future of Particle Physics (Snowmass 2021)
\rule{\hsize}{0.01in}\\\rule[+0.2in]{\hsize}{0.01in} \end{center}}

\begin{document}
\pagestyle{empty}

\snowmass
\centerline{\large\bf Event Generators for High-Energy Physics Experiments}\bigskip
We provide an overview of the status of Monte-Carlo event generators for high-energy particle physics. Guided by the experimental needs and requirements, we highlight areas of active development, and opportunities for future improvements. Particular emphasis is given to physics models and algorithms that are employed across a variety of experiments. These common themes in event generator development lead to a more comprehensive understanding of physics at the highest energies and intensities, and allow models to be tested against a wealth of data that have been accumulated over the past decades. A cohesive approach to event generator development will allow these models to be further improved and systematic uncertainties to be reduced, directly contributing to future experimental success.
Event generators are part of a much larger ecosystem of computational tools. They typically involve a number of unknown model parameters that must be tuned to experimental data, while maintaining the integrity of the underlying physics models. Making both these data, and the analyses with which they have been obtained accessible to future users is an essential aspect of open science and data preservation. It ensures the consistency of physics models across a variety of experiments.

\bigskip\bigskip\bigskip
\centerline{\includegraphics[width=0.33\textwidth]{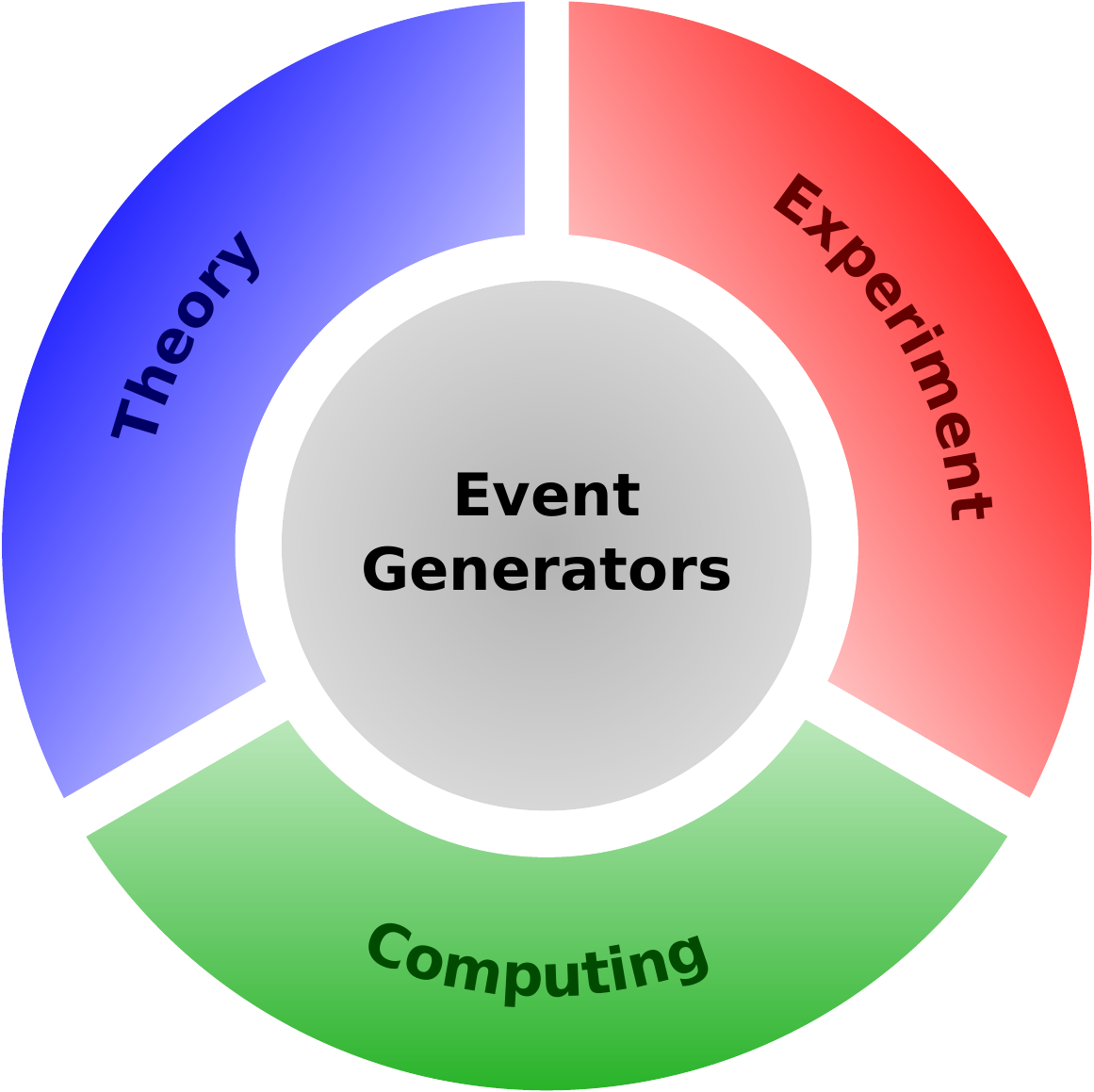}}
\vfill\noindent
CP3-22-12\hfill DESY-22-042\hfill FERMILAB-PUB-22-116-SCD-T\hfill IPPP/21/51\\
JLAB-PHY-22-3576\hfill KA-TP-04-2022 \hfill LA-UR-22-22126 \hfill LU-TP-22-12 \\
MCNET-22-04\hfill OUTP-22-03P\hfill P3H-22-024\hfill PITT-PACC 2207 \hfill UCI-TR-2022-02
\clearpage

\author{J.~M.~Campbell}\affiliation{\affFermilab}
\author{M.~Diefenthaler}\affiliation{\affJLAB}
\author{T.~J.~Hobbs}\affiliation{\affFermilab}\affiliation{\affIIT}
\author{S.~H{\"o}che}\affiliation{\affFermilab}
\author{J.~Isaacson}\affiliation{\affFermilab}
\author{F.~Kling}\affiliation{\affDESYHamburg}
\author{S.~Mrenna}\affiliation{\affFermilab}
\author{J.~Reuter$^{\;4}$ (editors)}\noaffiliation
\author{S.~Alioli}\affiliation{\affMilanBicoccaAndINFN}
\author{J.~R.~Andersen}\affiliation{\affIPPPDurham}
% ...

\maketitle
\clearpage

\centerline{\large\bf EXECUTIVE SUMMARY}\bigskip
The continued success of the high energy physics (HEP) experimental programs critically relies on advancements in physics modeling 
and computational techniques, driven by a close dialogue between large experimental collaborations and small teams of event generator authors.
The development, validation and long-term support of event generators can only be successfully achieved
with the help of a vibrant research program at the interface of theory, experiment, and computing -- one that
is intimately connected to all three areas.
Given the commonalities outlined in this white paper, there is an opportunity to bring together
the diverse communities of event generator developers to share
resources and ideas. A successful concept for this was established
in the context of the LHC with the European based MCnet collaboration~\cite{MCnetWeb}.
A similar collaboration addressing all experiments of interest to US particle physics, 
with a focus on cross-cutting aspects, would be highly beneficial in meeting the goals of the HEP program. 
This white paper, where domain experts from various sub-fields came together for the first time, is the initial step towards developing such a collaboration. We propose to continue the effort through the creation of a joint theoretical-experimental working group addressing common physics and computational challenges.

To support the above ideas, we survey existing physics models and their implementations, as well as computational challenges,
highlighting improvements that are needed to reach the precision goals of experiments in the next decade.
In the absence of obvious new physics signatures, such as new resonances, there are no alternative strategies 
to theory-driven simulations. Event generators also play an essential role in the planning and design of future experiments.

Uncertainties on the results of experimental analyses are often dominated by effects
associated with event generators.
They arise from the underlying physics models and theory, the truncation of perturbative
expansions, the PDFs and their implementation, the modeling of non-perturbative effects, the tuning of model parameters, 
and the extraction of fundamental parameters of the theory.
The need to address these uncertainties for various facilities and experiments is 
the driving force behind the efforts of the community of Monte-Carlo event generator developers. 
We discuss these questions in the context of the larger facilities that continue to drive 
the development in the near term.

\begin{itemize}
\item {\bf High Energy Colliders}:
In the coming decade experiments at the LHC will make precise measurements of Standard Model
parameters, such as the $W$ mass and the Higgs boson couplings. Both the extraction of these parameters and 
their interpretation will be limited primarily by the precision of perturbative Quantum Chromodynamics (QCD)
and electroweak (EW) calculations, both fixed order and resummed. 
Other major uncertainties arise from the strong coupling, and from the PDFs. The results of some analyses 
will also be limited by the number of Monte Carlo events that can be generated. 
Future highest-energy colliders, including a potential muon collider, may/will require electroweak effects 
to be treated on the same footing as QCD and electromagnetic effects.

\item {\bf Neutrino Experiments}:
The next generation neutrino experiments DUNE and HyperK will make precise measurements of the 
CP violating phase, mixing angles and the mass hierarchy. The SBN program will focus on precise measurements
of neutrino cross sections and searches for new physics. None of these experiments will be limited
by statistical uncertainties. Since all running and planned experiments use nuclear targets, one of the leading systematic uncertainties to the measurements arises 
from the modeling of neutrino-nucleus interactions. This requires the use of state-of-the-art nuclear-structure and -reaction theory calculations. The implementation of physics models with complete error budgets
will be required to reach the precision goals of these experiments.

\item {\bf Electron-Ion Collider}:
The EIC  will investigate the structure of nucleons and nuclei at an unprecedented level. This will be accomplished by performing precise measurements of DIS and other processes over the complete relevant kinematic range including the transition region from perturbative to non-perturbative QCD. Highly polarized beams and high luminosity will allow probes of the spatial and spin structure of nucleons and nuclei, leading to high-precision determinations of PDFs and other quantum correlation functions. These investigations will advance our understanding of hadronization as well as QCD factorization and evolution, and will require the development, validation and support of novel physics models.

\item {\bf Forward Physics Facility}:
The Forward Physics Facility at the LHC will leverage the intense beam of 
neutrinos, and possibly undiscovered particles, in the far-forward direction to 
search for new physics and calibrate forward particle production. 
These measurements will require an improved description of
forward heavy flavor -- particularly charm -- production,
neutrino scattering in the TeV range, 
and hadronization inside nuclear matter, including uncertainty quantification.

\item {\bf Lepton Colliders}:
Future lepton colliders would provide per mill level measurements of Higgs boson couplings and $W$ and top-quark masses. The unprecedented experimental precision will require event generators to
cover a much wider range of processes than at previous facilities, both in the Standard Model and beyond. In addition,
predictions for the signal processes must be made with extreme precision.
Some of the methodology is available from the LEP era, while other components will
need to be developed.
\end{itemize}

Event generation for the above facilities contains many common physics components, such as higher-order QCD and electroweak perturbative corrections, factorization theorems and parton evolution equations, resummation of QCD and Quantum Electrodynamic (QED) effects,  hadronization, and final-state modeling. 
Various experiments also require the understanding of heavy-ion collisions and nuclear dynamics at high energies as well as heavy-flavor effects.
A common aspect to most experiments is the search for new physics effects, which must be appropriately simulated.
In addition to the physics components, there are similar computational 
ingredients, such as interfaces to external tools for analysis, handling of tuning and systematics, and the need for improved computing efficiency.
Many of these aspects may profit from developments in artificial intelligence and machine learning.

\bigskip
\centerline{\includegraphics[width=0.55\textwidth]{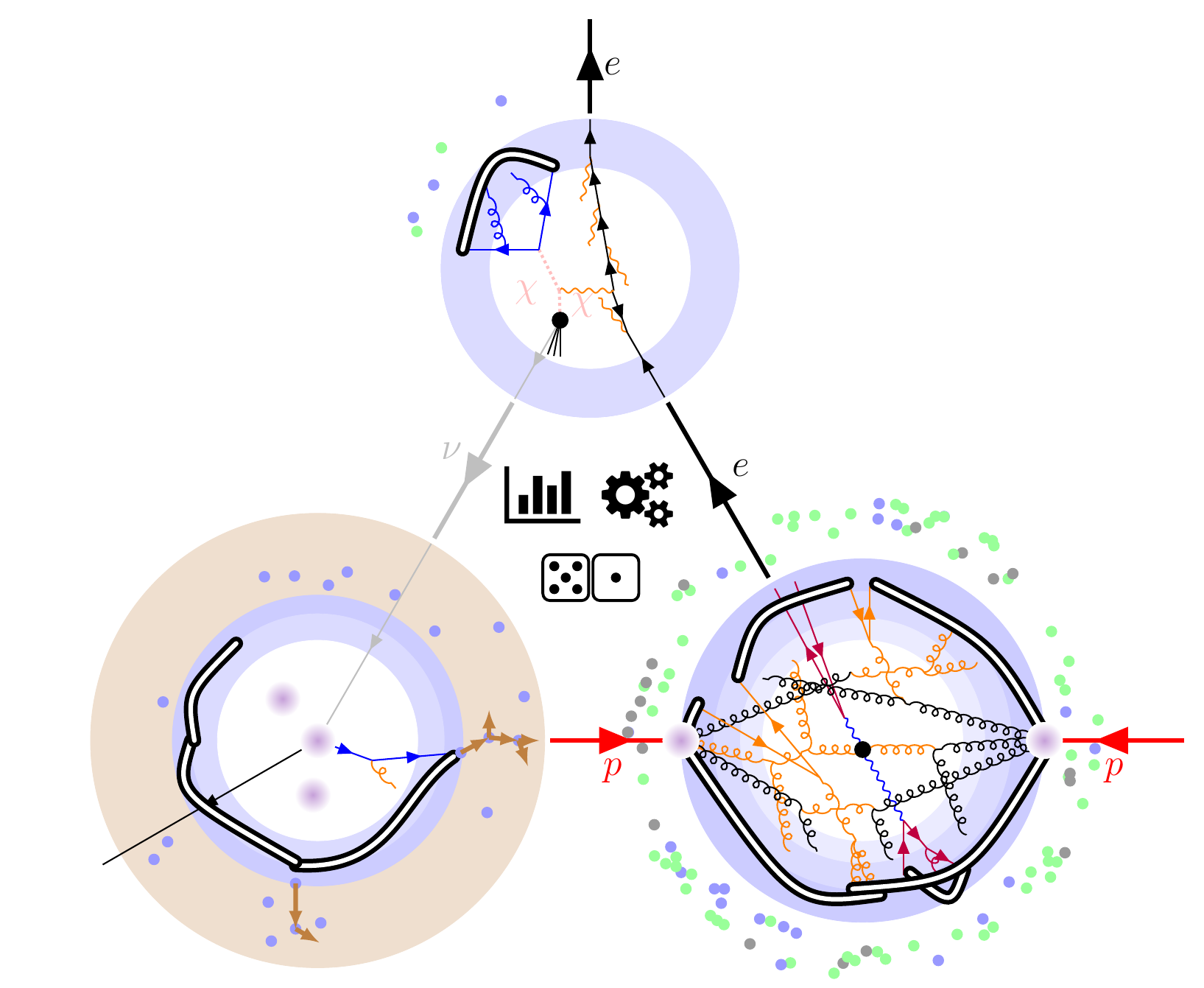}}
\clearpage
\tableofcontents

\newpage
\addcontentsline{toc}{part}{{\large\bf Part I: Cross-Cutting Aspects}}
\section{Introduction}
\pagestyle{myheadings}
\setcounter{page}{1}
\label{sec:intro}
Modern high-energy and nuclear physics experiments employ a multitude of collider and 
detector technologies that allow the details of collision events to be reconstructed with an 
unprecedented level of accuracy.
The full potential of the data obtained from such measurements can only be realized
if theoretical predictions, used to test for an underlying physics model, are provided 
at the same level of statistical accuracy and systematic precision. 
The quantum nature of the interactions and the often very large final-state particle multiplicity required
of calculations
make this a computationally hard problem. In most cases, a sensible description can be provided
only by using approximations 
and Monte-Carlo methods. The benefit of Monte-Carlo implementations of phase-space integrals 
and solutions to (integro-)differential equations is that one obtains theoretical predictions in the form of 
``events'' that can be treated like actual detector events in an experiment. 
These Monte-Carlo events are then subjected to additional simulations of detector response 
and acceptance effects, before being analyzed in the same fashion as data.
Eventually, simulated and experimental results are compared on a statistical basis.

Over the past decades, this paradigm has provided the foundations for the enormous success of high-energy physics: 
the establishment of the Standard Model of particle physics as the preeminent theory,
one that describes a multitude of experimental observations. Event generators have been essential in discoveries, precision measurements, 
the planning of future experiments and design studies for next-generation high-energy physics facilities.
Nevertheless, the Standard Model of particle physics must be considered incomplete, and future
experiments will need to establish which extensions of it are necessary in order to incorporate
dark matter as well as other anomalies, such as those found in the neutrino sector, $B$ flavor physics, 
and the anomalous magnetic moment of the muon.
Moreover, the quest to explore the Standard Model itself is heavily dependent on the output of event generators as well --- for instance, in understanding QCD to higher precision, the mass hierarchy and mixing of neutrinos, searching for novel strong-interaction dynamics like the onset of gluon saturation at low $x$, precision EW parameter measurements, or probing the properties of dense nuclear matter.
The development of event generators will continue to play an essential role in all these endeavors.

One of the most important tasks of both high energy particle theory and nuclear theory is
to extend the predictive power of event generators to higher formal precision and improve computational performance.
The construction, validation and maintenance of event generators has traditionally been 
carried out by theorists and phenomenologists, who themselves have developed the
underlying physics models that are implemented.
The presence of such a model also sets traditional event generators apart from 
surrogate based Monte-Carlo methods, such as generative neural networks for event generation. 
While the information content of such surrogate models 
is limited by the fitted input data, conventional event generators can be used to simulate
data sets at arbitrary statistical accuracy, and are therefore the most reliable means 
for hypothesis tests with controlled systematics, as required for precision experiments.

In this contribution to the Snowmass 2021 community study, we summarize the status of
current event generators for various branches of high energy physics in Secs.~\ref{sec:he_colliders},~\ref{sec:neutrino_experiments},~\ref{sec:eic},~\ref{sec:forward_facilities}, and~\ref{sec:lepton_colliders}. We list areas
of active development, highlight cross-cutting efforts that span several fields in Secs.~\ref{sec:cross_cuts} and~\ref{sec:cross_cuts_technical}, 
and comment on new efforts required for experimental success. 
We begin with a brief overview of the main areas covered in this document, 
and the associated physics goals.

\subsection{High Energy Colliders}
\label{sec:physics_goals_hadron_colliders}
With an integrated luminosity of 3~ab$^{-1}$ for each of the experiments ATLAS and CMS,
the high-luminosity Large Hadron Collider (HL-LHC) will reach a precision on Higgs boson
couplings in the 2-4\% range, the $W$-boson mass will be measured to about 6~MeV precision,
the top-quark mass will be determined with about 1~GeV accuracy, and the uncertainty on
$\sin\theta_W$ is expected to be reduced by $\sim 10-25\%$ through global fits~\cite{Azzi:2019yne}.
The production rate of two (three) electroweak gauge bosons is expected to be measured at between
6\% (11\%) and 10-40\% (36\%) for the channels with largest and smallest rates, respectively.
While LHCb is expected to collect a smaller data sample than ATLAS and CMS of 300~fb$^{-1}$ during the HL-LHC, the forward LHCb acceptance will provide a number of complementary precision measurements including the $W$-boson mass, $\sin\theta_W$, and PDF constraints~\cite{Azzi:2019yne}.
With all those measurements combined, the Higgs sector of the Standard Model and the mechanism
of electroweak symmetry breaking will be tested at an unprecedented accuracy.
Most of the measurements rely on the precise modeling of complete particle-level events
as well as their backgrounds with the aid of general-purpose event generators~\cite{Buckley:2011ms}.
In various processes, Monte-Carlo modeling is the dominant uncertainty, despite calculations 
being performed at NLO precision in QCD~\cite{ATLAS:2017fak}.
Compared with hadron colliders, lepton colliders have the advantage of providing a clean 
experimental environment for both precision measurements and for new particle discovery. 
Recently, the breakthrough of muon beam cooling technology~\cite{Delahaye:2019omf} 
paved a potential way to a multi-(tens of) TeV muon collider~\cite{Bartosik:2020xwr,Schulte:2020xvf,Long:2020wfp}. 
The experiment-specific developments needed for high-energy collider event generators 
in view of the HL-LHC and possible future experiments will be discussed in Sec.~\ref{sec:he_colliders}.
    
\subsection{Neutrino Experiments}
\label{sec:physics_goals_neutrino_experiments}
Current and future accelerator neutrino experiments will measure neutrino oscillation parameters with unprecedented accuracy.
For example, the DUNE experiment is designed to measure the neutrino generation mixing angle $\sin^22\theta_{23}$ and mass splitting $\Delta m^2_{31}$ with resolutions better than 0.5\%, measure the $CP$ violation phase at not worse than $18^\circ$ precision, and determine the neutrino mass ordering, i.e., the question whether $\Delta m^2_{31}$ is positive or negative, with a significance well over 5$\sigma$~\cite{DUNE:2020jqi}.
Similar considerations apply to the future Hyper-Kamiokande (HK) experiment~\cite{Hyper-Kamiokande:2018ofw}.
Moreover, accelerator neutrino experiments provide excellent opportunities to search for beyond Standard Model physics due to intense beams and large detectors.
The Short-Baseline Neutrino program~\cite{MicroBooNE:2015bmn} has in its core physics program the search for eV-scale sterile neutrinos and several new physics scenarios ranging from searches for dark matter produced in the beam to studies of novel neutrino interactions and beyond~\cite{Machado:2019oxb}.
Besides accelerator experiments, neutrino observatories like IceCube~\cite{IceCube:2016zyt} and ANTARES~\cite{ANTARES:2011hfw} detect atmospheric and astrophysical-source neutrinos with energies ranging from several GeV up to the PeV scale, in which deep inelastic scattering in the nucleus dominates the cross section.
The nature of the interaction depends on neutrino flavor, to some extent, which can shed light on the production mechanism of cosmogenic neutrinos.
High-precision event generators will be crucial for predicting both signal and background rates. According to the DUNE Conceptual Design Report~\cite{DUNE:2015lol}, ``uncertainties exceeding 1\% for signal and 5\% for backgrounds may result in substantial degradation of the sensitivity to CP violation and mass hierarchy.'' In addition, searches for new physics could be greatly improved by using exclusive channels in which backgrounds are mitigated.  However, simulating exclusive quantities, such as the number of outgoing protons or pions, presents a significant challenge due to a strong dependence on final-state interaction models and nonperturbative QCD effects. A detailed discussion on the needed improvements to obtain this sensitivity can be found in Sec.~\ref{sec:neutrino_experiments}.

\subsection{Electron-Ion Collider}
\label{sec:physics_goals_eic}
In the further exploration of the Standard Model, a vital research area is the study of the strong interaction and understanding how the properties of nucleons and nuclei emerge from their constituents, quarks and gluons, and their dynamics. The U.S.-based Electron-Ion Collider (EIC) will support the exploration of nuclear matter over a wide range of center-of-mass energies, $\sqrt{s}\!=\!20-140$ GeV, and ion species, using highly-polarized electrons to probe highly-polarized light ions and unpolarized heavy ions. The high instantaneous luminosity of up to $\mathcal{L}=$ \(10^{34}\,\mathrm{cm}^{-2}\mathrm{s}^{-1}\)and the highly-polarized beams at this frontier particle accelerator facility will allow for the precision study of the nucleon and the nucleus at the scale of the sea quarks and gluons, over a broad kinematic range adapted to explore the relevant parton-level dynamics. The success of the EIC science program will depend on accurate and precise simulations of electron-ion collisions. Developing event generators for the EIC requires a close dialog between the diverse communities studying perturbative and non-perturbative QCD phenomena at various energies, to combine advances in general-purpose event generation driven by hadron colliders with advances in non-perturbative QCD. Details will be discussed in Sec.~\ref{sec:eic}.
\subsection{Forward Physics Facility}
\label{sec:physics_goals_fpf}
High energy collisions at the LHC produce an intense beam of highly energetic neutrinos, and possibly other undiscovered weakly interacting neutral particles, in the far forward direction. The FASER and SND@LHC experiments will take advantage of this unexplored flux of particles during the upcoming Run~3 of the LHC, while the FPF is a proposal to continue and expand this program during the HL-LHC era, for more details on the experiments see Sec.~3 of the FPF white paper~\cite{Feng:2022inv}. The physics goals of these experiments include i) the search for decay or scattering signatures of new light particles as predicted by many models of BSM physics, ii) the study of neutrino interactions for all three flavors in the TeV energy range, and iii) the use of the neutrino fluxes as a probe of forward particle production with many implications for QCD and astro-particle physics. For this program to be successful,
it is crucial to not only make accurate predictions of forward particle production and neutrino interactions at TeV energies, but also to quantify associated uncertainties. Many components of existing simulations do not allow uncertainty quantification and their improvement is therefore a crucial ingredient to achieve the physics goals. A detailed discussion on the  generator related requirements can be found in Sec.~\ref{sec:forward_facilities}.
\subsection{Lepton Colliders}
\label{sec:physics_goals_lepton_colliders}
While the LHC, and its potential hadronic successors, may provide a deeper understanding of the nature of the Higgs boson, a future lepton-lepton collider or ``Higgs-Factory'' could provide unprecedented measurements of the electroweak nature of the SM.  Indeed, the European Strategy update from 2020 advocated for the development
and construction of a new lepton collider~\cite{EuropeanStrategyGroup:2020pow}.
There are two types of lepton colliders being considered: a circular collider, which has a large luminosity capability but is limited to sub-TeV energies, or a linear collider, which can reach TeV energies but has limited luminosity.
The physics goals include unprecedented tests of the Standard Model, in particular the model-independent measurement of Higgs couplings to mostly per-mil level and the measurement of the $W$ and top masses to 1-2 MeV and 20-70 MeV precision, respectively~\cite{Baer:2013cma,Behnke:2013lya,ilc_snowmass,Linssen:2012hp,Abramowicz:2016zbo,Blondel:2018mad,FCC:2018evy,Heinemeyer:2021rgq}.
In addition, there is a rich physics program at B factories~\cite{BaBar:2014omp,Belle-II:2018jsg}.
While there is a wealth of theoretical experience from the LEP era,
there are also challenges that are unique to these colliders, and even the specific facilities that are under consideration. Details will be discussed in Sec.~\ref{sec:lepton_colliders}.

\section{Physics Components}
\label{sec:cross_cuts}
The types of experiments discussed in Sec.~\ref{sec:intro} span a wide range of energies,
beam particles, targets (collider vs.\ fixed target), temperature
and chemical potential.
Not surprisingly, an event generator employed for a given experimental configuration
may require some dedicated physics models, while other parts can be similar or even identical
to those used for other configurations.
A particular strength of event generators derives from the factorization or assumed factorization 
of physics at different energy scales.   This principle allows some physics models to be universal
and often enables
the modular assembly of (parts of) a generator from existing tools when targeting a new
experiment.   In this manner, previously gained knowledge and experience can be transferred, and a more
comprehensive understanding of the physics models is made possible
by allowing them to be tested against a wealth of data gathered in past and current experiments.
These cross-cutting topics in event generation are highlighted in this section. 

A shared feature of all experiments is the hard interaction, which probes the colliding beams
or fixed targets at the shortest distance scales. This component of any reaction is 
the most interesting, since it is most susceptible to new physics effects. 
It is described in simulations using full quantum mechanical calculations, including 
interference, and typically at the highest order in perturbation theory that is practicable.
Since measurements happen at much larger distance scales, the particles 
produced in hard interactions can radiate a substantial number of additional quanta before being detected.
This radiation is implemented in quasi-classical cascade models, which are matched to the quantum 
mechanical calculation of the hard process to increase precision. Finally, if the active degrees of freedom
at the hard scale are the asymptotically free quarks and gluons of the QCD Lagrangian,
the transition to color-neutral hadrons must be accounted for, typically through phenomenological models inspired by the string effect
or local parton hadron duality. Particles produced in this simulation chain may also undergo transport
through nuclear matter. They may also still be unstable and decay on timescales that can 
in some cases be resolved by detectors.

In the following, we will discuss the above topics in more detail.
In some cases, different solutions are needed for individual types of experimental facilities 
and research programs, and we will refer to the various dedicated sections of this report as needed.

\subsection{Fixed-order perturbation theory}
Perturbative calculations in the Standard Model and beyond are employed 
in event generators to model interactions at the highest energy scales.
The computation of tree level cross sections can be performed fully automatically using a number 
of programs~\cite{Kanaki:2000ey,Krauss:2001iv,Kilian:2007gr,Gleisberg:2008fv,Alwall:2011uj}.
By fully automatic or automated, we mean an end-to-end calculation that does not require any intermediate manipulations by the user.
A similar automation of next-to-leading order (NLO) calculations was enabled 
by the development of generic methods for the subtraction of infrared
singularities~\cite{Frixione:1995ms,Catani:1996vz,Catani:2002hc}, which allow the 
differential cross section to be written as a combination of two types of terms: (1) real-emission corrections, subtracted by local counterterms that cancel all infrared
singularities in the matrix element and make the cross section finite in four dimensions, and (2) the combination of the Born result, virtual corrections and the integrated counterpart
of the real subtraction terms, the sum of which is again finite in four dimensions. 
A third necessary component, the automated calculation of one-loop amplitudes, is now available using techniques based on generalized 
unitarity~\cite{Bern:1994cg,Ossola:2006us,Ossola:2008xq,Berger:2008sj,Giele:2008ve,Ellis:2008ir}.
Some of the most challenging calculations at one-loop order in QCD~\cite{
 Ita:2011wn,Bern:2013gka,Badger:2013yda,Cullen:2013saa,FebresCordero:2015kfc,
 Hoche:2016elu} have been performed with the help of generic frameworks for leading order
calculations~\cite{Kanaki:2000ey,Krauss:2001iv,Kilian:2007gr,Gleisberg:2008fv,Alwall:2011uj},
making use of the fact that the infrared subtraction terms are tree-like, and that
one-loop amplitudes can be computed independently. The separation of tree-like
and one-loop components has also increased cross-talk and collaboration between
different communities~\cite{Binoth:2010xt,Alioli:2013nda}.
Numerical stability as well as computational efficiency have become focus areas
of development, in particular because the calculation of next-to-next-to-leading corrections 
requires the evaluation of the same matrix elements at very low resolution scales.
Recursive techniques for one-loop calculations~\cite{Cascioli:2011va,Actis:2012qn} 
and on-the-fly reduction methods~\cite{Buccioni:2017yxi}
have alleviated the efficiency problem, and numerical stability
has been achieved through a multitude of developments~\cite{Denner:2005nn,Binoth:2005ff,Denner:2016kdg}.
For many important physics applications, analytic results for one-loop amplitudes
still play an important role, due to reduced evaluation time
and increased numerical stability~\cite{Campbell:2021vlt}.
Most higher-order calculations at hadron colliders have focused on the large QCD corrections.
However, precision tests require 
fully-differential, fixed-order calculations at NLO in the electroweak theory. 
Such calculations are now fully automated as well~\cite{
Kallweit:2015dum,Chiesa:2015mya,Frederix:2016ost,Schonherr:2017qcj}.
For additional details see Sec.~\ref{sec:hadron_collider_hard_process}.

Over the last decade, fully-differential calculations at next-to-next-to-leading order (NNLO) in perturbative QCD have become more common~\cite{Amoroso:2020lgh,Heinrich:2020ybq,Cordero:2022gsh,Huss:2022ful}.  This has been driven by the development of a number of approaches for handling infrared singularities at this order, ranging from local subtraction procedures~\cite{Heinrich:2002rc,Boughezal:2011jf,Czakon:2014oma,Gehrmann-DeRidder:2005btv,Currie:2013vh,Caola:2017dug} to non-local subtraction or slicing methods~\cite{Catani:2007vq,Gaunt:2015pea,Boughezal:2015dva,Boughezal:2016wmq,Billis:2019vxg} and other more specialized techniques~\cite{Cacciari:2015jma,Dreyer:2016oyx}.
Computations for a wide range of color-singlet processes are available publicly~\cite{Grazzini:2017mhc,Campbell:2022gdq}, see subsections \ref{sec:MATRIX} and \ref{sec:MCFM}.  
However, more complex processes need dedicated calculations for which a freely-available code may not exist.  The current state-of-the-art for NNLO corrections in QCD, primarily for LHC applications, is summarized in Table~\ref{table:nnlosummary}.
Much work remains to be done to reach the NNLO level of precision for more complicated processes and to meet future experimental needs. 
This includes the calculation of some two-loop amplitudes that are currently not available, as well as reaching the level of automation, in terms of overall assembly and ease of use, that has already been achieved at NLO.
For a few select processes, N$^3$LO precision has been achieved~\cite{Chen:2021isd,Chen:2021vtu,Chen:2022cgv};  an appraisal of the prospects for reaching this level of precision more broadly is given in Ref.~\cite{Caola:2022ayt}.

The application of this technology to high-energy colliders will be discussed in Sec.~\ref{sec:he_colliders}.  This includes a discussion of the state-of-the-art tools available for providing fully-fledged differential predictions at NNLO and beyond in Sec.~\ref{sec:hadron_collider_hard_process}. This section also covers tools that are more generally applicable, ones that can provide one- and two-loop matrix elements for any scattering process, since many of these have been developed as a direct response to LHC needs. 
The matching between fixed-order calculations and particle-level simulations
poses a separate challenge, which will be discussed in detail in Sec.~\ref{sec:hadron_collider_matching}.

\begin{table}[t]
\begin{center}
\begin{tabular}{|l}
\hline
Higgs  \\
\hline
$H$~\cite{Anastasiou:2004xq,Anastasiou:2005qj,Catani:2007vq,Gaunt:2015pea} \\
\hphantom{$H$}~\cite{Mistlberger:2018etf,Cieri:2018oms,Chen:2021isd,Boughezal:2016wmq} \\
$W^\pm H$~\cite{Ferrera:2013yga,Caola:2017xuq,Campbell:2016jau} \\
$ZH$~\cite{Ferrera:2014lca,Campbell:2016jau} \\
H (VBF)~\cite{Cacciari:2015jma,Buckley:2021gfw} \\
HH~\cite{Grazzini:2018bsd} \\
HHH~\cite{deFlorian:2019app}\\
$H+$jet~\cite{Boughezal:2013uia,Chen:2014gva,Boughezal:2015dra,Boughezal:2015aha,Caola:2015wna,Chen:2016vqn,Campbell:2019gmd,Mondini:2021nck}\\
$W^{\pm}H$+jet~\cite{Gauld:2020ced} \\
$ZH$+jet~\cite{Gauld:2021ule} \\
\hline
\end{tabular}
\hspace*{-.8em}
\begin{tabular}{|l}
\hline
SM candles  \\
\hline
$W^\pm$~\cite{Melnikov:2006di,Catani:2009sm,Grazzini:2017mhc,Boughezal:2016wmq} \\
$Z$~\cite{Melnikov:2006kv,Catani:2009sm,Gaunt:2015pea,Grazzini:2017mhc,Boughezal:2016wmq} \\
$\gamma \gamma$~\cite{Catani:2011qz,Grazzini:2017mhc,Catani:2018krb,Alioli:2020qrd,Campbell:2016yrh}  \\
$W^\pm\gamma$~\cite{Grazzini:2016hai,Grazzini:2017mhc,Cridge:2021hfr,Campbell:2021mlr} \\
$Z\gamma$~\cite{Grazzini:2017mhc,Campbell:2017aul,Campbell:2017aul} \\
$W^+W^-$~\cite{Gehrmann:2014fva,Caola:2015rqy,Grazzini:2016ctr,Grazzini:2017mhc,Grazzini:2020stb,Poncelet:2021jmj}  \\
$WZ$~\cite{Grazzini:2016swo,Grazzini:2017ckn} \\
$ZZ$~\cite{Cascioli:2014yka,Grazzini:2015hta,Caola:2015psa,Heinrich:2017bvg,Grazzini:2017mhc,Kallweit:2018nyv,Grazzini:2018owa} \\
$\gamma \gamma \gamma$~\cite{Chawdhry:2019bji,Kallweit:2020gcp} \\
\\
\hline
\end{tabular}
\hspace*{-.8em}
\begin{tabular}{|l}
\hline
Jets  \\
\hline
dijets~\cite{Currie:2014upa,Currie:2016bfm,Gehrmann-DeRidder:2019ibf,Czakon:2019tmo}  \\
3 jets~\cite{Czakon:2021mjy} \\
$W^{\pm} +$jet~\cite{Boughezal:2015dva,Boughezal:2016dtm,Gehrmann-DeRidder:2017mvr,Pellen:2021vpi} \\
$Z+$jet~\cite{Gehrmann-DeRidder:2015wbt,Gehrmann-DeRidder:2016cdi,Boughezal:2015ded}\\
$\gamma+$jet~\cite{Chen:2019zmr,Campbell:2016lzl} \\
$Z+b$~\cite{Gauld:2020deh} \\
$W^{\pm}c$~\cite{Czakon:2020coa} \\
$\gamma \gamma$+jet~\cite{Chawdhry:2021hkp} \\
\\
\\
\hline
\end{tabular}
\hspace*{-.8em}
\begin{tabular}{|l|}
\hline
Other  \\
\hline
single top~\cite{Brucherseifer:2014ama,Campbell:2020fhf} \\
$t \bar{t}$~\cite{Czakon:2015owf,Czakon:2016ckf,Abelof:2015lna,Catani:2019hip,Catani:2019iny,Czakon:2020qbd,Catani:2020tko} \\
$b\bar{b}$~\cite{Catani:2020kkl} \\
$H \to b \bar{b}$~\cite{Anastasiou:2011qx,DelDuca:2015zqa,Mondini:2019gid} \\
$t$ decay~\cite{Gao:2012ja,Brucherseifer:2013iv,Campbell:2020fhf} \\
\(\mathrm{e}^+\mathrm{e}^-\to 3j\)~\cite{Gehrmann-DeRidder:2014hxk,Gehrmann:2017xfb} \\
DIS (di-)jets~\cite{Currie:2017tpe,Niehues:2018was} \\
\\
\\
\\
\hline
\end{tabular}
\end{center}
\caption{Calculations that can be performed differentially at NNLO in QCD and, in some cases, beyond NNLO  (for a $pp$ initial state, unless explicitly indicated). 
\label{table:nnlosummary} }
\end{table}

At high luminosity lepton colliders with \order{10^{-5}} precision targets,
such as the TeraZ option at FCC-ee, it will be necessary to compute Standard Model predictions for electroweak precision observables to high orders in perturbation theory. The electroweak  
corrections would need to be included at least up to second order. Additionally, QED corrections will also be sizable,
and would be required up to fourth order~\cite{Jadach:2019bye}.
For such experiments it is therefore important to have a consistent 
and practically feasible scheme for combining electroweak 
and QED corrections calculated to different orders, without violating 
gauge invariance and infrared cancellations.
Such a scheme is introduced in Sec.~\ref{sec:lepton_colliders_ceex}.
It should also be noted that, in contrast to most QCD calculations, QED calculations require finite fermion masses. Since QED with massive fermions has only soft singularities, this leads to a tremendous simplification of the infrared structure.
However, the presence of fermion masses introduces additional scales that complicate the calculation of virtual amplitudes, so that many two-loop results cannot be obtained analytically. Moreover, the presence of small but non-vanishing masses also presents numerical challenges in evaluating higher-order corrections.
These issues, together with tools for the computation of hard processes at lepton colliders, are discussed in Sec.~\ref{sec:lepton_colliders}.

\subsection{QCD factorization and parton evolution}\label{sec:factorization}
Calculations in the Standard Model depend crucially on factorization theorems. 
In general terms, these theorems allow cross sections or other observables to be separated 
into a short-distance component, such as parton-level scattering matrix elements that can be calculated in
perturbation theory, and long-distance components, such as the parton distribution functions (PDFs) or fragmentation functions.
In the case of hadronic collisions, these factorization theorems  assume the schematic form%
~\cite{Collins:1989gx,Ellis:1996mzs,Collins:2011zzd}
\begin{equation}
\label{eq:factorization_main}
\sigma[J] \approx \sum_{a,b}\int\!\mathrm{d}x_a \int\!\mathrm{d}x_b\
f_{a/A}(x_a,\mu_J^2)\, f_{b/B}(x_b,\mu_J^2)\
\hat \sigma[J]
\;,
\end{equation}
wherein $J$ is the observable of interest, and $\hat \sigma[J]$ is the partonic cross section, 
which can be evaluated to a given level of theoretical accuracy in fixed-order perturbative theory.
Similar formulae can be written for lepton-hadron or lepton-lepton collisions.
The quantities $f_{a/A}(x_a,\mu_J^2)$ and $f_{b/B}(x_b,\mu_J^2)$ in Eq.~(\ref{eq:factorization_main}) 
are the {\it collinear unpolarized} PDFs, which depend upon the fractional momentum, $x_a$ and $x_b$ of hadrons $A$ and $B$, 
carried by partons of flavor $a$ and $b$, respectively. They quantify long-distance correlations between 
partonic currents in the interacting hadron by providing the probability distribution for a parton of
given flavor with the specified fractional momentum.
Similar expressions for PDFs can be provided for different hadrons or the photon,
and can be defined depending on the spin of the struck hadron, or for tensor
distributions. Beyond these collinear quantities, various unintegrated distributions can also be
defined, including transverse-momentum dependent (TMD) distributions~\cite{Collins:1981uk,Collins:1981uw,Collins:1984kg}.
In general, the formal structure of factorization theorems like Eq.~(\ref{eq:factorization_main}) becomes
more complicated as one introduces transverse-momentum dependence or moves toward less-inclusive
observables. This topic will be discussed in more detail in Sec.~\ref{sec:eic}.
At the same time, it can be challenging
to understand the kinematic boundaries within which factorization can be considered reliable
for more inclusive quantities typified by
Eq.~(\ref{eq:factorization_main}) or, {\it e.g.}, similar expressions for Deep Inelastic Scattering (DIS) structure functions.
Dissecting these factorization limits and regions of validity remains an important topic of research
in QCD, along with efforts to establish factorization for novel observables involving the strong
interaction.   This nontrivial interplay between formal QCD and potential
measurements at hadron, lepton, and DIS colliders must be bridged by 
event generators.  

\subsubsection{Parton evolution}
\label{sec:cross_cuts_parton_evolution}
In the framework of collinear factorization, the PDFs obey evolution equations that determine their
dependence on the scale $\mu_J$ at fixed momentum fraction $x$~\cite{Gribov:1972ri,Lipatov:1974qm,          
  Dokshitzer:1977sg,Altarelli:1977zs}:
\begin{equation}\label{eq:pdf_evolution}
  \frac{{\rm d}\,xf_{a/A}(x,\mu_J^2)}{{\rm d}\ln \mu_J^2}=
  \sum_{b=q,g}\int_0^1{\rm d}\tau\int_0^1{\rm d} z\,\frac{\alpha_s}{2\pi}
  \big[zP_{ab}(z)\big]_+\,\tau f_{b/A}(\tau,\mu_J^2)\,\delta(x-\tau z)\;.
\end{equation}
In this expression, $P_{ab}(z)$ are unregularized splitting functions for the parton transition $a\to b$, 
which depend on the momentum fraction $z$, and potentially on other variables in the process.
The evolution of the PDFs is constrained by momentum sum rules, which is reflected by the 
plus distribution indicated by $[\ldots]_+$. Evolution equations structurally similar 
to Eq.~\eqref{eq:pdf_evolution} are solved by Monte-Carlo event generators using appropriate
algorithms that are known as parton showers~\cite{Marchesini:1983bm,Sjostrand:1985xi,Bengtsson:1986hr,Marchesini:1987cf}
and dipole showers~\cite{Lonnblad:1992tz,Kharraziha:1997dn}. They provide 
an event by event interpretation of the evolution in terms of final-state particles.
Their precise implementation is often adapted to the physics problem at hand, and we will therefore
discuss focus areas of current and future development separately in Secs.~\ref{sec:hadron_collider_evolution}
and~\ref{sec:eic}.

The PDFs that appear in Eqs.~\eqref{eq:factorization_main} and Eq.~\eqref{eq:pdf_evolution} 
are non-perturbative inputs that are extracted phenomenologically
from global analyses of high-energy data~\cite{PDFWP}.
These extractions are typically carried out at a specific
fixed order in $\alpha_s$ and with a variety of nontrivial choices,
such as the scheme to handle heavy-quark flavors, mass thresholds, and the inclusion or neglect of
higher-order electroweak corrections. As such, the use of a particular family of PDFs in a given generator framework entails a number of subtleties in both the
formal and statistical consistency of various theoretical, model-based, or methodological choices from which the PDFs were derived~\cite{PDFWP}.
Similar objects to the PDFs are the fragmentation functions needed for precision predictions of processes involving
final-state hadrons. The implementation of fragmentation in event generators is discussed
in Sec.~\ref{sec:fragmentation}.

This latter issue relates to a general aspect of modern event generators: achieving maximum 
consistency among parallel nonperturbative physics inputs.
The challenge of realizing such consistency may be illustrated through attempts to describe
{\it e.g.}, single-inclusive hadroproduction in lepton-hadron or hadron-hadron collisions ---
processes that generally require simultaneous knowledge of both the PDFs of the colliding hadron(s)
as well as the fragmentation functions associated with the final-state hadrons. While it is in
principle possible to simultaneously extract both quantities in a combined analysis \cite{PDFWP}, in 
practice single-inclusive hadroproduction data might be used to constrain nucleon PDFs in the presence
of an assumed, {\it frozen} set of fragmentation functions. Such evaluations may potentially result
in systematic uncertainties or biases in the resulting PDF extraction that can be difficult to control,
and place a significant emphasis on comprehensive uncertainty quantification across nonperturbative
quantities appearing in generators.
Similar considerations apply to neutrino generators, which consist of a complicated interplay
of theoretical models dependent on the neutrino energy $E_\nu$.
In this context, contributions
from DIS become sizable in the few-GeV region and ultimately dominate at higher energies. At present,
a detailed understanding of possible correlations among models used to describe DIS contributions
to the $\nu A$ cross section and, {\it e.g.}, resonance or hadronization models, is something that
must be developed further for higher precision in neutrino event generation. These issues are discussed
in greater detail in Secs.~\ref{sec:nu:RES},~\ref{sec:nu:DIS}, and~\ref{sec:nu:RES-DIS}.
In addition, we point out that improvements to the PDFs will be a central
feature of the EIC program discussed in Sec.~\ref{sec:eic}.

\subsubsection{Photoproduction}
\label{sec:xcuts_photoproduction}
An effect that is particularly important for lepton-lepton and lepton-hadron collisions is
the exchange of a low-virtuality photon that fluctuates into a hadronic state with partonic content.
Such photons are called resolved.
This partonic structure of photons can be described with PDFs that are derived in a similar fashion as for any other hadron.
Most of the available data come from LEP experiments, where the structure of low-virtuality photons that emerge from a lepton beam is probed with a high-virtuality photon from the other lepton beam.
Additional data are available from the lepton-hadron collisions at HERA that provide further constraints on the gluon content of a photon \cite{Nisius:2009xx}.
An important difference between resolved photons and ordinary hadrons is that a low-virtuality photon may split into a quark-antiquark pair perturbatively, which adds an extra term to the DGLAP equation beyond the hadron-like, non-perturbative, input.
Therefore the photon structure can be separated into a hadron-like part, often modeled with the vector-meson-dominance (VMD) model~\cite{Sakurai:1960ju, Gell-Mann:1961jim, Stodolsky:1966am}, and a point-like (anomalous) part that can be calculated perturbatively.
In the general-purpose event generators, the flux of incoming photons is usually described with the equivalent photon approximation (EPA), but the treatment of the different components varies.
In \pythia~8 \cite{Helenius:2019gbd} the point-like structure is included only after hard-process generation by ISR, which may collapse the photon back to an unresolved state, whereas in \pythia~6~\cite{Sjostrand:2006za} the different components were separately generated from the beginning~\cite{Schuler:1993td,Schuler:1995fk}.
\sherpa also has an implementation of photoproduction \cite{Archibald:2008zzb}.
In addition to photoproduction in lepton-hadron colliders, these frameworks can be applied to ultra-peripheral collisions of protons and heavy nuclei, which has been a recent area of interest for different LHC experiments \cite{Angerami:2017kot, CMS:2020rfg, ATLAS:2021jhn}.
The current Monte-Carlo implementations are all based on parton-\-level matrix elements evaluated at Born level.
While this framework has been sufficient to adequately model events at LEP and HERA, the anticipated increased accuracy of future lepton colliders or the planned EIC necessitates significant improvements.
At a minimum, it will be important to implement the well-proven higher--\-order matching and multi-\-jet--\-merging algorithms and include them for photo--\-production events.
In addition, it will become mandatory to better understand the transition from electro- to photo--\-production, often trivially connected to different sizes of the photon virtuality (with electro--\-production associated with the regime of large $Q^2$, while photo--\-production is connected to quasi--\-real photons, with $Q^2\approx 0$).
It will be interesting to differentiate these two regimes in a more robust way, with the help of relevant observable quantities.
Another future avenue is to further develop the non-perturbative side of event generation using an MPI model, discussed in the following section, and to allow realistic simulations of photon-nucleus interactions relevant for the EIC (see Sec.~\ref{sec:eic}) and ultra-peripheral heavy-ion collisions at the LHC (see Sec.~\ref{sec:ion_colliders}). 

\subsubsection{Factorization breaking effects}
\label{sec:xcuts_mpi}
Using factorization to compute hard scattering cross sections for composite beam particles 
like protons and resolved photons does not account for the possibility that the beam remnants
may also interact. These interactions can occur at scales that are not too dissimilar 
from the ones in the primary, hard scattering. Secondary interactions are commonly modeled
under the assumption that the non-diffractive cross section is saturated by the hard,
perturbative cross section, thus leading to multiple, independent hard scattering events~\cite{Sjostrand:1987su}. 

Semi-hard multiple scatterings are modeled as basic partonic QCD processes with a small minimum transverse momentum $p_\perp^{\rm min}$, that are evolved with parton showers quite similar to the hard process, albeit with modified parton distribution functions that keep track of the extraction of previous partons from the incoming hadrons.  The average number of scatterings depends on $p_\perp^{\rm min}$ and the spatial distribution of partons inside the proton that is characterized by a typical transverse radius of the partons.  In \pythia~\cite{Sjostrand:1987su} and \sherpa~\cite{Alekhin:2005dx}, the semi-hard scatterings are damped towards smaller $p_\perp$ by a saturation of \alphaS towards low scales.  In \herwig~\cite{Bahr:2008dy,Bellm:2019icn}, scatterings with transverse momentum below $p_\perp^{\rm min}$ are modeled via additional \emph{soft} partonic interactions that follow the principles of phenomenological strong particle production, and additional partons are distributed uniformly in rapidity and with a strongly limited transverse momentum resulting in additional soft activity in the final state~\cite{Bahr:2009ek}.

For the specific case of two hard scatterings, or double parton scattering, a Monte Carlo simulation called {\tt dShower} has recently been developed \cite{Cabouat:2019gtm} that is based on the first principles QCD description of this phenomenon developed in \cite{Diehl:2017kgu} (see also \cite{Blok:2011bu, Manohar:2012jr, Ryskin:2011kk, Gaunt:2012dd}). This code can incorporate the effects of non-perturbative correlations in the parton pairs initiating the DPS process, as well as appropriately accounting for the phenomenon where a pair of partons arises as a result of a single parton splitting perturbatively into two (`$1 \to 2$ splitting'). To achieve the latter requires careful account of the transverse separation $y$ between the partons. Non-perturbative correlations are fed into the simulation via double parton distributions (DPDs) that are the two-parton analogue of the standard PDFs. In principle the code can accommodate any DPDs supplied by the user (provided that they satisfy some basic theoretical constraints), but a default set is provided with the simulation that is a slight modification of that developed in \cite{Diehl:2017kgu}. This set encodes in an approximate way correlations due to momentum and valence number constraints \cite{Gaunt:2009re, Diehl:2018kgr}. At present only correlations between the $x$ fractions of the two partons and $y$ can be accounted for in the simulation, but in future it should be possible to incorporate the effects of colour and spin correlations between the partons.

Very briefly, the code simulates an event as follows. First, the $x$ fractions of the protons initiating the hard process and transverse separation between the two hard interactions $y$ is selected according to the DPS cross section formula from \cite{Diehl:2017kgu}, which contains a convolution of DPDs with parton-level cross sections. Then, QCD emissions are generated going backwards from the hard process - the algorithm here is very similar to that used for a single parton shower, except that there are two legs that can potentially emit in each proton (with the leg that emits being chosen using the competing veto algorithm \cite{Kleiss:2016esx}), and the PDF ratio used in the emission probability factor involves the appropriate ratio of DPDs. The emissions continue until the code reaches a scale of $\mu_y \sim 1/y$, at which point the partons start to overlap in transverse space and can merge together (see figure 4 of \cite{Cabouat:2019gtm}). At this point a merging can occur with a probability that is given by the fraction of the DPD that can be attributed to $1 \to 2$ splitting. If a merging occurs, the code proceeds with a usual single parton shower, otherwise it proceeds with a double parton shower, but using DPDs where the splitting component has been removed.

The DPS process in which a $1 \to 2$ splitting occurs in both protons has overlap with loop corrections to the single scattering (SPS) mechanism, and one needs to avoid double counting between the two. At large $y$ one should have a DPS description, whilst at small $y$ the SPS description is appropriate, and one should have a smooth transition between the two. For the total DPS cross section this was achieved in \cite{Diehl:2017kgu}. In \cite{Cabouat:2020ssr}, a fully-differential version of the procedure in \cite{Diehl:2017kgu} was formulated that could be used in a parton shower. This formulation bears some similarities to the \mcatnlo method of matching NLO calculations to parton showers \cite{Frixione:2002ik, Frixione:2010wd, Frederix:2012ps, Frederix:2020trv}. This mechanism has been implemented and validated in the {\tt dShower} simulation in the context of diboson production \cite{Cabouat:2020ssr}. Note that in some cases, the overlap issue may be in practice numerically irrelevant; there is a parameter in {\tt dShower} denoted by $\nu$, and by varying this one may gauge if the double counting issue is numerically important or not (see \cite{Cabouat:2019gtm, Diehl:2017kgu} for more details).

It has been seen via {\tt dShower} simulations that the number and momentum valence constraints and/or the effect of $1\to 2$ splittings can have an appreciable effect for certain processes and distributions~\cite{Cabouat:2019gtm, Andersen:2023hzm}.

\subsection{QED radiative corrections}\label{sec:qed_evolution}
Radiative corrections play an important role in calculations involving charged leptons (particularly electrons)
as well as other light, charged particles. These corrections can be estimated to all
orders in the electromagnetic coupling using the soft photon approximation.
Generic frameworks to resum soft photon effects and to match them to higher-order QED calculations
are of great importance for precision measurements in most high-energy physics experiments.
In addition, at high luminosity lepton colliders, the energy of the scattering leptons is dispersed
due to collective effects in the interaction of the colliding bunches, as well as the distortion 
of bunch shapes during acceleration and beam transport. The latter issue and those
related to QED initial-state radiation will be discussed in Sec.~\ref{sec:beamstrahlung}.
QED initial-state radiation is also addressed in a separate Snowmass white paper~\cite{Frixione:2022ofv}.
Here, the focus is on QED radiation in particle decays,
which is implemented in event generators in a fairly uniform fashion.
\subsubsection{QED showers}
\label{sec:cross_cuts_qed_shower}
The simplest way to include QED radiative corrections is to implement a parton shower.
The defining approximation of the all-orders decay rate ${\rm d}\Gamma$ 
in terms of a given leading order decay rate ${\rm d}\Gamma_0$ is given as
a solution of the DGLAP equation, Eq.~\eqref{eq:pdf_evolution} as
\begin{equation}\label{eq:methods:photosmaster}
  \begin{split}
    {\rm d}\Gamma^\text{QEDPS}
    =\;&{\rm d}\Gamma_0\bigg\{1+
        \sum\limits_{c=1}^{n_\text{ch}}\sum\limits_{n_\gamma}
        \frac{\left(\alpha\,L_c\right)^{n_\gamma}}{n_\gamma!}
        \bigg[\,\prod\limits_{i=1}^{n_\gamma}{\rm d} x_c^i\,\bigg]
        P_{\epsilon_{\text{cut}}}(x_c^1)\!\otimes\!.\;\!\!.\;\!\!.\!\otimes\! P_{\epsilon_{\text{cut}}}(x_c^{n_\gamma})
      \bigg\}\hspace*{-20pt}
  \end{split}
\end{equation}
where the radiative part is summed over all $n_\text{ch}$ charged particles. 
$L_c$ is the logarithm of the ratio of the decaying particle's mass 
over the mass of the charged particle $c$, and $x_c=\prod x_c^i$ 
is its retained energy fraction after the radiation of $n_\gamma$ photons. 
The phase space distribution of these photons is described by the 
Altarelli-Parisi splitting functions, $P_{\epsilon_{\text{cut}}}$, (cf.\ Eq.~\eqref{eq:pdf_evolution})
in the presence of an infrared cut-off $\epsilon_{\text{cut}}$, modified by suitable 
weights to recover the correct soft-photon limit and implement 
exact higher-order corrections, and iterated over all $n_\gamma$ 
emitted photons. 
QED showers have been implemented in \herwig~\cite{Masouminia:2021kne}, \photos~\cite{
  Barberio:1990ms,Barberio:1993qi,Golonka:2005pn,Davidson:2010ew},
\sherpa~\cite{Hoeche:2009xc} and \vincia~\cite{
  Kleiss:2020rcg,Skands:2020lkd}.
The \vincia implementation has the unique feature that it also correctly captures 
the multipole structure of soft photon emissions. Further independent implementations
of QED showers have been obtained in \horace~\cite{CarloniCalame:2003ux,
  CarloniCalame:2005vc,CarloniCalame:2006zq,CarloniCalame:2007cd}
and \babayaga~\cite{CarloniCalame:2000pz,CarloniCalame:2001ny,CarloniCalame:2003yt}.

\subsubsection{Soft photon resummation}
\label{sec:cross_cuts_soft_photons}
A robust method to resum the effects of real and virtual photons in the soft limit
and to all orders was developed by Yennie, Frautschi, and Suura (YFS)~\cite{Yennie:1961ad}.
A key feature of the technique is that it can be systematically
improved by including  exact, fixed-order expressions.

The total cross section for the production of a configuration of charged particles 
with momenta $q_\text{out}$ in the final state, ${\rm d}\Phi_Q$, 
including all higher-order corrections, is given by a summation over 
all real and virtual photons, $n_\gamma$ and $n_\gamma^V$, with respect to its leading 
order configuration, 
\begin{equation}
    \label{yfs:xs}
    {\rm d}\sigma =  \sum_{n_\gamma=0}^{\infty} \frac{1}{n_\gamma!}
    {\rm d}{\Phi_Q}
    \bigg[\prod_{i=1}^{n_\gamma}{\rm d}{\Phi_i^\gamma}\bigg]
    \left(2\pi\right)^4
    \delta^4\bigg(\sum_\text{in}q_\text{in}
                  -\sum_\text{out} q_\text{out}
                  -\sum_{i=1}^{n_\gamma} k_i\bigg)
    \bigg| \sum_{n_\gamma^V=0}^{\infty} 
    \mathcal{M}^{n_\gamma^V+\frac{1}{2}n_\gamma}_{n_\gamma}\bigg|^2.
\end{equation}
Here, ${\rm d}\Phi_i^\gamma$ is the phase space element of photon $i$ with momentum $k_i$. 
In this notation for the matrix element $\mathcal{M}$,
the Born level contribution is defined as $\mathcal{M}_{0}^{0}$, while
the matrix element $\mathcal{M}_{n_\gamma}^{p}$ refers to the Born
process plus $n_\gamma$ real photons evaluated at an overall power $p$
in the electromagnetic coupling $\alpha$. In principle, this expression
includes photon emissions to all orders, but in practice only the first
few terms in the perturbative series can be computed.
Following the YFS approach, the total cross section can be reformulated 
in such a way that the infrared divergences are resummed to all orders.
This leads to the expression
\begin{equation}\label{eq:masterYFS}
    {\rm d}\sigma = \sum_{n_\gamma=0}^\infty
        \frac{e^{Y(\Omega)}}{n_\gamma!}\,
        {\rm d}{\Phi_Q} 
        \bigg[\prod_{i=1}^{n_\gamma}{\rm d}{\Phi_i^\gamma}\,\tilde{S}\left(k_{i}\right)\bigg]
        \Bigg(\tilde{\beta}_0
        + \sum_{j=1}^{n_\gamma}\frac{\tilde{\beta}_{1}(k_j)}{\tilde{S}\left(k_{j}\right)}
        +\sum_{\substack{j,k=1\\j< k}}^{n_\gamma}
        \frac{\tilde{\beta}_{2}(k_j,k_k)}{\tilde{S}\left(k_{j}\right)\tilde{S}\left(k_{k}\right)}
        + \cdots
        \Bigg)\;,
\end{equation}
where $Y(\Omega)$ is the YFS form factor. The form factor contains contributions from real 
and virtual photons inside the soft domain $\Omega$, summed to infinite order and 
can be computed using a dipole decomposition.
$\tilde{S}\left(k\right)$ is the soft YFS eikonal associated with the emission of real 
photons and can also be decomposed into dipoles.
The remaining terms inside the brackets are the infrared finite residuals 
$\tilde{\beta}_{n_\gamma}$, comprising infrared subtracted squared matrix elements 
with $n_\gamma$ real photons. They can be systematically calculated order-by-order
in perturbation theory.

The YFS approach to QED radiation was championed by 
the LEP era Monte Carlo (MC) tools such as \kkmc~\cite{Arbuzov:2020coe},
which focused on $e^+e^-\rightarrow f\bar{f}$, and 
\koralw/YFSWW~\cite{Jadach:2001mp}, used to generate 
$e^+e^- \rightarrow W^+W^-$ (see Sec.~\ref{sec:lepton_colliders_ceex}). 
For hadron colliders, an implementation of the YFS formalism for the 
decay of massive resonances decaying into leptons has been implemented in the 
purpose-built \textsc{Winhac/Zinhac}~\cite{Placzek:2003zg} and KKMC-hh~\cite{Jadach:2016zsp} 
generators as well as the multi-purpose event generators \herwig~\cite{Hamilton:2006xz} 
and \sherpa~\cite{Schonherr:2008av,Krauss:2018djz}.

\subsection{Hadronization models}
\label{sec:fragmentation}
One of the biggest strengths of event generators is that they model 
the transition from partons to observable hadrons, a dynamic process 
at the boundary between perturbative and non-perturbative QCD.
Currently, this process cannot be described in any other approach.
Due to a lack of first-principles calculations, event generators employ models
inspired by preconfinement~\cite{Amati:1979fg,Marchesini:1980cr} 
and local parton hadron duality~\cite{Azimov:1984np,:2007cp}, 
the string effect~\cite{Andersson:1983jt,Andersson:1983ia}, 
or KNO scaling~\cite{Koba:1972ng}.
While these are not \textit{ab initio} calculations,
the string~\cite{Bengtsson:1987kr,Andersson:1997xwk}
and cluster~\cite{Gottschalk:1982yt,Gottschalk:1983fm,Webber:1983if} 
models encapsulate much of the dynamics of QCD, and yield reliable predictions
over a wide range of energies.
The extension and refinement of hadronization models is a topic of highest practical relevance.
It has a direct bearing for example on the determination of the strong coupling from event shapes at
lepton colliders, and it impacts the modeling of energy flow and flavor composition
of hadronic final states in experiments, which in turn determines the detector response.
For a discussion on low energy models important to neutrino physics see Sec.~\ref{sec:nu:had}.

\subsubsection{The Lund string model}
\label{sec:lund-string-model}
The basic assumption in the string approach is that a linear
confinement potential between a color charge and its anticharge
can be approximated by a color flux tube, or string,  stretched from one to the other
\cite{Andersson:1983ia}. In the aftermath of a
collision, if a string is connecting a receding quark $q_0$
and antiquark $\overline{q}_0$ pair, the initial
kinetic energy is converted into potential energy as the string is
stretched out. The stored energy can be used to produce new
quark--antiquark pairs $q_i\overline{q}_i$ that break the string
by screening the endpoint colors. This can lead to a sequence
$q_0\overline{q}_1 - q_1\overline{q}_2 - \ldots - q_n\overline{q}_0$
of mesons. The breaks occur with (wide) fluctuations around a hyperbola
of constant invariant time. Since the breaks are causally disconnected,
the fragmentation process can be considered in any order, and most
conveniently from the ends inwards. At each step, a fraction of
remaining lightcone momentum is removed from the string and transferred to the hadron according
to a probabilistic fragmentation function. This predicts a flat
rapidity distribution of produced particles, except near the
endpoints.

The evolution described so far produces hadrons only along the string axis.   A quantum mechanical tunneling process is required to produce quarks
with a non-vanishing transverse momentum, which leads to a
suppression of heavier flavors and larger transverse momenta.
Baryon production can be introduced in a straightforward fashion as a string break where
an antidiquark--diquark pair is produced instead of a $q\overline{q}$
one.  In a more sophisticated variant, referred to as the ``popcorn'' model, a
$q\overline{q}$ pair carrying a different color than the endpoints results in
a net attraction of quark to quark and antiquark to antiquark,
such that a field remains between them (recall the simple rule
that red + green = antiblue).   In this field, one or more $q\overline{q}$
pairs can be produced, such that the baryon and antibaryon do not have
to be nearest neighbors, but can be separated by a meson. The latter
approach provided a sufficient description of LEP data, but has required extensions
to describe LHC data.

The central addition of the Lund string model is to introduce gluons as kinks on a string
stretched between its two endpoints, as comes naturally in a
leading-color picture of QCD \cite{tHooft:1973alw}. The kinked
string can break along its length, and notably produce a leading
gluon-jet hadron that contains the gluon and pieces of string on
both sides of it (This is quite unlike the simpler handling of
gluons in cluster models).  While several parameters are needed
to describe the fragmentation of the simple $q\overline{q}$ string,
no further ones need to be introduced to include gluons.

The string picture lends itself to several other phenomenological implications.
The multiparton interactions (MPI) mechanism of hadronic collisions
\cite{Sjostrand:1987su} can lead to complicated string topologies,
where strings overlap in space--time, given that they have
hadron-scale transverse sizes. The overlap can lead to a number of
collective effects. Color reconnection is a broad topic (see Sec.~\ref{subsubsection:color-reconnection}), but at a
minimum it implies that the field between the many color charges
is rearranged so as to reduce the total string length being drawn out. 

One possibility that goes further is the formation of color
ropes, where several strings merge into one field of a higher color
representation \cite{Bierlich:2014xba}. The larger string tension in
ropes reduces the tunneling suppression, and thereby enhances strange
baryon production. This mechanism offers a good description of the
ALICE observation of an enhanced strange baryon production in
high-multiplicity events \cite{ALICE:2017jyt}, i.e.\ where rope
formation is more likely. An increased string tension could also be
motivated by close-packing of separate strings \cite{Fischer:2016zzs}.

Another possibility is junction formation. A junction topology
can be represented by a Y-shaped string configuration, where each
string piece stretches out to a quark, but the baryon number is carried
by the central junction in a topological sense. Two long parallel
strings can collapse to one in the central region (red + green =
antiblue again), with a junction and an anti-junction near each common
end \cite{Christiansen:2015yqa}. This increases the rate of strange
baryon production. Interestingly, it also explains the unexpectedly
large $\Lambda_c/D$ ratio observed by ALICE \cite{ALICE:2021dhb},
notably at small transverse momenta and at high multiplicities.

The close-packing of strings could also lead to a repulsive force
between them known as a shove \cite{Bierlich:2016vgw}. This can give rise to
collective flow, notably in heavy-ion collisions (see Sec.~\ref{sec:hi-soft-collective}).

While string fragmentation usually is formulated in energy--momentum
space, the linear nature of the string potential allows a translation
into a space--time picture, where notably the primary hadron production
vertices are defined \cite{Ferreres-Sole:2018vgo}. This can then be used
as the starting point for modeling hadronic rescattering
\cite{Sjostrand:2020gyg}, another mechanism for collective flow.

The multiple scattering framework (see Sec.~\ref{sec:xcuts_mpi}) 
implies that several strings can be pulled out to the
beam remnants of a hadronic collision. Color reconnection can partly
reduce this number. In the simple example of a gluon taken out of a
proton by a collision, the color octet remnant can be split into a
diquark and a quark, each with a string stretched towards the central
part of the event. It is then assumed that the diquark takes most of
the total remnant energy. However, comparing with LHCf data
\cite{LHCf:2015nel,LHCf:2015rcj}, the full framework
does not give hard enough neutrons, while pions are too hard.
Modifications have been implemented within the FPF studies
\cite{Anchordoqui:2021ghd,Feng:2022inv}, where the popcorn mechanism is not
allowed to act on a remnant diquark, and this shifts results in the
right direction.
 
Charm and bottom quarks are expected only to appear at the ends of
strings, since they are too heavy for any appreciable tunneling rate.
More of their energy is taken to form the leading heavy hadron than in
a light-quark jet. Interestingly, and often overlooked, the hadron can
even have a larger energy than the original quark. A typical example
where this may happen is when the heavy quark is attached via a string
to a beam remnant ahead of it, and so is pulled forwards by the string
rather than backwards.
This was observed in the 1990s in fixed-target
experiments, with large flavor/antiflavor asymmetries, a behavior that
is consistent with the string model \cite{Norrbin:1998bw}.

The string framework is intended to be applicable without modification over a wide
energy range, with a nontrivial energy dependence coming out of the
preceding perturbative processes, such as MPIs and parton showers. It
cannot be used at very low energies, where exclusive processes become
important.
However, at least an approximate description from almost threshold
energies up to FCC ones has been developed for a range of hadron--hadron
collisions, for use in the description of hadronic
cascades such as from cosmic ray interactions in the atmosphere \cite{Sjostrand:2021dal}.

There are also on-going hadronization studies related to heavy ion
collisions \cite{Bierlich:2018xfw}, photon-induced physics \cite{Helenius:2017aqz},
the formation of molecular states \cite{Ilten:2021jcb}, and the EIC
program (see Sec.~\ref{sec:eic}). 
  
Taking a broad view, it can be argued the flavour-composition picture
currently contains too many epicycles, and that a new starting point
is needed. This could involve more dependence on hadron masses and
less on quark/diquark ones. Another ingredient could be an enhanced
role for higher hadron multiplets. Part of the modelling problem there
is that the multiplet structure often is unclear, and decay tables
incomplete. Angular decay distributions presumably should be aligned
along the string axis, to some degree. Some hadrons in higher multiplets
have been observed at LEP with non-negligible rates, but generally
tunes do not improve if such multiplets are included, presumably for
the reasons already listed. The flavor picture becomes even more
complicated when moving from $e^+e^-$ to $pp$ and beyond, as already
noted.

The related energy--momentum and space--time pictures have been more
successful, especially in the clean $e^+e^-$ environment, but are
so far rooted in a leading-color interference-free picture of the
hadronization process, which can only be an approximation to the
full process. Bose--Einstein effects are only approximately
included as a non-default option, and Fermi--Dirac ones not at all.
In the more busy low-$p_{\perp}$ $pp$ environment, not to mention
the $AA$ one, the issue is what happens when strings become
close-packed. Although some possibilities have been mentioned, it is
not clear whether they will survive more detailed comparisons with
data. More work is likely to be necessary.

In summary, while the basic ingredients of the string framework in
\pythia \cite{Sjostrand:2006za,Bierlich:2022pfr} are 40+ years old,
several new extensions to it have been introduced in recent years.
A high priority in the short run is to provide a synthesis of them,
e.g.\ in terms of a new global tune. But, most likely, further ideas
and coding efforts will prove necessary to match the requirements of
future experiments.

\subsubsection{The Cluster model}
While string hadronization models in general and the Lund model in
particular build on linear confinement and the idea of practically
one-dimensional QCD flux tubes, local parton-hadron duality
(LPHD)~\cite{Azimov:1984np} and, in particular,
preconfinement~\cite{Amati:1979fg} have been the guiding principles
underpinning the development of cluster hadronization models 
in~\cite{Field:1982dg,Gottschalk:1982yt,Gottschalk:1983fm,Gottschalk:1986bv} and in~\cite{Marchesini:1983bm,Webber:1983if}.  

The ideas and algorithms of the latter have been implemented and
refined in the \herwig event
generator~\cite{Corcella:2000bw,Bellm:2015jjp}, and, later, in a modified
version, in \sherpa~\cite{Winter:2003tt}.
The underlying idea is that hadronization progresses along the lines 
of the formation and subsequent decay of clusters, i.e.\ color-neutral states 
that can be interpreted as resonances of hadrons with a continuous mass
spectrum.
These clusters are formed from pairs of quarks and anti-\-quarks in 
color-singlet states, and, similar to string models, 
baryonic quantum numbers are carried by diquarks replacing quarks.  
In the following, quarks and diquarks will be collectively denoted as
``flavors''. 

In a first step of cluster hadronization, all gluons at the end
of the parton shower are
split into flavor--\-anti-\-flavor pairs; 
the underlying large--$N_c$ formulation of the parton showers guarantee 
that color-singlets can be unambiguously defined and formed.
The \herwig and \sherpa realizations differ in this step: 
\herwig assumes gluons to have a (non-perturbative) mass of 
around 1\,GeV to facilitate their splitting into flavor pairs, 
which in turn implies a restriction on the flavors that can 
be produced due to their associated constituent masses, 
usually constrained to light quark pairs ($u$, $d$, and $s$).  
In contrast, \sherpa keeps gluons massless and ensures 
momentum conservation in the gluon splitting process by including  
the recoil of one of the color-connected spectators.
The invariant mass of the gluon-spectator pair, in turn, places  
dynamic limits on the produced masses and resulting 
flavors; as a consequence, baryonic quantum numbers, 
{\it i.e.}\ diquarks, can be produced already in the cluster formation step.
To some degree, this could be interpreted as the localized cluster 
hadronization version of the popcorn mechanism in string models.

The resulting clusters populate a continuous mass spectrum that 
depends mainly on the infrared cut-off of the parton shower.  
A flavor-dependent mass threshold, defined by the sum of the 
heaviest hadrons or hadron pairs available for the flavors 
constituting the cluster, determines whether clusters further decay into 
secondary clusters or if they transit or decay into hadrons.
If clusters are heavier than this threshold, they decay into 
clusters by producing a new flavor--\-anti-\-flavor pair 
and arranging the four flavors into two clusters, 
$[f_0\bar f_1]\to [f_0\bar f_2] + [f_2\bar f_1]$.
In \herwig, the kinematics and properties of the new clusters 
are determined by selecting their masses according to a probability 
distribution and fixing their orientation.
In \sherpa, the cluster fission is handled by fixing the kinematics 
of the four flavors with some ``cluster fragmentation functions''
and reconstructing the secondary cluster masses from it.

If clusters are light, that is, if their mass is below the 
flavor-dependent threshold, they will transit directly to hadrons with
the same flavor numbers or they will decay into pairs of hadrons.  
In \sherpa, both options -- transit to single hadrons with assigning 
momentum transfer to other clusters or hadrons and decays into hadron 
pairs -- are enabled.  In \herwig, the decay into hadrons pairs is the 
preferred route, while light clusters only transit to single hadrons in exceptional cases.
In both models, a cluster with flavors $f_0\bar f_1$ splits into two hadrons by
``popping'' a new flavor pair $f_2\bar f_2$; the hadron flavors are
selected probabilistically, taking
into account kinematics, the flavor wave functions of the hadrons
$(f_0\bar f_2)$ and $(f_2\bar f_1)$, and the popping probabilities of
$f_2$.
In \herwig, diquarks appear at this stage. In original versions of
cluster hadronization models and in the current \herwig model, 
the clusters are assumed to decay isotropically into hadrons.  
In \sherpa, this assumption has been abandoned in favor of a more anisotropic treatment.  

The clusters that emerge from MPIs are in general predicted to be too heavy,
since they can originate
from color-connected partons not necessarily close in momentum space. 
In this case, color reconnection models can be used to restore
a color preconfined state, resulting in smaller invariant cluster
masses. In \herwig, a model for color reconnection was introduced in
\cite{Gieseke:2012ft} and extended to include 
more complex cluster topologies in \cite{Gieseke:2017clv}.
The models for color reconnection have been tuned to underlying event (UE) and minimum bias (MB) data and
offer a good description of the observed enhancement of strange-baryon
production in \cite{ALICE:2017jyt}. The cluster hadronization
model of \herwig has been further extended to include a space-time
description of color reconnection in \cite{Bellm:2019wrh}.

Many aspects of cluster models deserve further scrutiny including
a detailed comparison between its two main implementations in \herwig and
\sherpa.
In view of upcoming experiments, such as the neutrino experiments, EIC and the forward physics facilities, it will also be important to revisit questions related to
the kinematics of beam fragmentation in the very forward direction.  
In addition, it will be interesting to see how ideas of color
reconnections, Bose-Einstein correlations, or analogies to the ``rope''
effects described in the previous section can be incorporated into
cluster models.

\subsubsection{Future improvements}
Despite their success in describing the bulk of high-energy collision data at hadron level, at the few-per cent accuracy level, there are a number of known short-comings of both the string and the cluster models that need to be addressed in view of future experiments:

Most importantly, while both models encapsulate in a quantitative way qualitatively well-known and well-tested features of the strong interaction, they rely on a multitude of parameters.  
These govern kinematic aspects, such as the break-up dynamics of the strings or clusters, the flavour composition of the break-ups, manifest through relative ``popping'' probabilities of new quark- or diquark-pairs during the break-ups, the relative yields of hadrons and their mulitplets, or the explicit suppression or enhancement of specific hadron types, such as the $\eta$ and $\eta'$ mesons.  
In total, and dependent on the model this results in typically about 15-20 parameters, which need to be tuned to data.  
Owing to their experimental accuracy,  detail, and statistics, the prime data source are results from the LEP experiments, usually supported by data from other $e^-e^+$ colliders.  
With hadronic events in such settings are mainly driven by quark production, the fragmentation of gluons is less well studied, and one may safely assume that this will lead to relatively larger uncertainties and differences between the models, when compared to the case of, say, quark jets. 
This clearly limits our ability to distinguish jets from gluons vs.\ light quarks at the current LHC experiments.
In a similar vein, the fragmentation of the (proton) rumps in the forward direction are less well studied, and data from e.g.\ HERA have typically not systematically been included into tuning efforts, with correspondingly increased uncertainties in modelling the (far) forward region of hadroproduction.  
This region, however, is relevant not only for LHC, but, even more so, for a future EIC.

In a somewhat different kinematic regime, both string and cluster models are, by construction, limited in the accurate description of low-multiplicity hadronic final states.  
These states emerge either as relatively rare events at high-energy colliders, where individual quarks fragment into very few hadrons, for example one single pion or kaon.  
Similarly, the correct description of the hadronization of low-mass colour-singlet systems -- with masses below 5 GeV or so -- is extremely challenging for both string and cluster models.
One could speculate about ways to improve the models there, for example by trying to systematically embed {\em exclusive} fragmentation functions into them; however, such an endeavour will certainly have to be driven by more and better data combined with a more detailed understanding of soft strong interactions.
In particular the low-mass regime is of considerable interest for the neutrino-experiments, which will hopefully lead to dedicated analysis strategies to support the necessary model building in this region.

Another aspect that is usually not covered in hadronization models concerns spin.
It is relatively straightforward to trace parton-level polarisations through the perturbative phase of event generation, a technology that has been systematically implemented for decay chains of both heavy elementary particles, such as the top, the gauge bosons, or the $\tau$ lepton, but also for parton showers. One of the reasons for not considering spin correlations in hadronization is because both strings and clusters are extended objects which may possess orbital angular momentum in addition to the spins of their constituents, similar to very high-mass hadron resonances. However, for low-mass systems, or systems containing heavy quarks, this picture changes, and it becomes interesting to embed spins into the models for these cases. Some progress in this direction has recently been made~\cite{Kerbizi:2021pzn,Kerbizi:2023cde}.

\subsection{Final-state interactions}
\label{sec:mpi_cr_rescattering}
In the standard factorization approach to computing hard scattering cross sections for composite beam particles,
like protons and resolved photons, the possibility that the beam remnants
may also interact and that final-state particles may re-interact is not considered.
In this section we discuss, some aspects of simulating these effects.

\subsubsection{Color Reconnection}
\label{subsubsection:color-reconnection}

The color structure of the partons that enter the hadronization process is only
partially determined within a single scattering.  In the parton shower, the color structure is usually approximated in the improved leading color limit.  The same is true for semi-hard scattering events. The color in soft scatterings is distributed based on phenomenological models. Improving the color structure of parton showers beyond the leading color approximation is an area of active research (cf.\ Sec.~\ref{sec:spin-color}).
However, the \emph{overall} color structure of a hadronic scattering process is poorly understood.  One may approach this from different directions.  On the perturbative side it is possible to consider at least double parton scattering to tie together two individual hard scatterings.  It is also possible to formulate an evolution of a partonic final state in color space, based on soft gluons~\cite{Gieseke:2018gff} and make a phenomenological connection to a full partonic event.  

The most widely used approach to correlate the whole partonic final state is based on the idea of color pre-confinement~\cite{Amati:1979fg} and modeled entirely phenomenologically as \emph{color reconnection}~\cite{Sjostrand:1993hi,Sjostrand:2004pf,Gieseke:2012ft,Argyropoulos:2014zoa,Christiansen:2015yca}.  Color pre-confinement means that all partons that are close to each other in momentum space will most likely also be close in color space, i.e.\ the color lines spanned mostly between pairs of quarks and anti-quarks should have a somewhat minimal length.  Hence, the basis for a color reconnection model is some measure of color length that can be a distance in momentum space~\cite{Gieseke:2012ft} or even in space-time~\cite{Bellm:2019wrh}.  A color reconnection model then interchanges color lines such that this color length is minimized or at least reduced substantially.
The shortening of color lines results in less hadrons from the hadronization that share the same initial transverse energy from the partons, resulting in fewer and harder hadrons being produced.
This effect plays an important role in particle production.
Indeed, the first application to hadronic collisions was to explain
the rising trend of $\langle p_{\perp}\rangle(n_{\mathrm{charged}})$
\cite{Sjostrand:1987su}, and this remains one
of the most compelling pieces of evidence.
The LEP collaborations also observed color reconnection in
$W^+W^- \to q_1\overline{q}_2 q_3\overline{q}_4$ events
\cite{ALEPH:2013dgf}, consistent with predictions \cite{Sjostrand:1993hi}.
More recently, enhanced baryon production by junction reconnection
could explain the large observed charm baryon fraction,
see Sec.~\ref{sec:lund-string-model}.

\subsubsection{Hadronic transport models}
\label{sec:hadronic_transport}

The description of hadronic interactions at lower energies has a long history with a large variety of models and Monte Carlo programs \cite{Wolter:2022rqc}. These models describe low energy interactions via the excitation of resonances. Some programs implement the excitation of strings for the description of interactions above the few GeV region with custom models (e.g.~UrQMD \cite{Bleicher:1999xi}) or rely entirely on \pythia (e.g.~GiBUU \cite{Buss:2011mx}). Recently, it has been shown that this description can model even neutrino interactions on nuclei in the TeV region for the
FPF \cite{Mosel:2022tqc} (see Sec.~\ref{sec:fpf:neutrino} for details). All of these prescriptions separate the final hadrons into two classes, `leading' and `non-leading'.  Traditionally, the distinction of the hadrons is mostly done on a phenomenological level by just selecting the hadrons with largest momenta. Recently, a procedure was presented to extract the information about the string fragmentation from \pythia directly from the event simulation~\cite{Gallmeister:2005ad}. Here, the `leading' hadrons are those that have a parton connected to the hard interaction, while the non-leading ones stem from the string breakings.

Comparisons with data suggest a reduction of the interaction cross section of the non-leading hadrons.
This can be accomplished by introducing a `formation time' for all hadrons, and also a non-vanishing `production time' for the non-leading hadrons.
The interaction cross section of hadrons is often set to zero until the production time elapses.
However, there exist different assumptions about the time evolution between the production time and the formation time. After the formation time, hadrons interact with their full hadronic vacuum cross section with the surrounding medium. In~\cite{Gallmeister:2005ad}, these times were extracted from the string decay for every single particle in every single event, contrary to the usual choice of setting these as some external average parameters.

Data from the EMC experiments at CERN and the HERMES experiment at DESY \cite{Airapetian:2007vu} together tested this model under different assumptions for the time dependence of the interaction cross section.
The result is that 
only a linear increase of the interaction cross section of hadrons during the hadronization is compatible with the data \cite{Gallmeister:2007an,gallmeister:2022sjq}, in full accordance with quantum diffusion models~\cite{Dokshitzer:1991wu,Farrar:1988me}.
In contrast, the usual prescription of setting the cross section to zero is inconsistent with data.
Describing the hadronic interactions in a multi-coupled-channel transport model, such as GiBUU, includes the absorption and the production of different species during the evolution of the interaction.
While the description of multi-differential experimental data from the HERMES and the CLAS collaboration are achieved without any further assumptions, models just based on absorption or on partonic energy loss fail to describe this data.

While experiments such as CLAS with 5 GeV or 12 GeV beam energies show less sensitivity to these kind of issues~\cite{Gallmeister:2007an} (the corresponding Lorentz factor is smaller and thus most of the hadronization happens inside the nucleus), the EIC will probe the very early phase of the time evolution of the interaction.

\subsubsection{Hadronic rescattering in \texorpdfstring{\pythia}{Pythia}}
After hadrons are produced in the hadronization process (see Sec.~\ref{sec:fragmentation}), it is possible for them to undergo secondary
collisions.  
A detailed discussion of how this is implemented in \pythia in the context of $\mathrm{pp}$ \cite{Sjostrand:2020gyg}
and heavy ion collisions \cite{Bierlich:2021poz} can be found elsewhere.
Here, we provide a brief overview of the current phenomenological approach.
The key to modeling such phenomenon is to understand the space-time structure of
the hadronization process.  This can be studied naturally in the context of the Lund string 
model \cite{Ferreres-Sole:2018vgo}.
In the \pythia implementation,
it is assumed that the hadron-hadron interaction probability depends
on the impact parameter $b$ and the total cross section $\sigma_{\mathrm{tot}}$, which, in turn,  depends on the collision center-of-mass energy $\sqrt{s}$ and the specific hadron species.  
The functional form is taken to be either a Gaussian or a disk (theta-function) model.
The effects of rescattering are implemented by first cataloging which hadron pairs could interact.
These pairs are then time-ordered and the earliest interaction in the list is considered. A collision is
simulated, with probabilities proportional to the partial cross sections for each process.
The newly produced hadrons are then tested for interactions and placed in the time-ordered list.   The process continues until there are no more 
potential rescatterings.

Short-lived hadrons can also decay during the rescattering phase. To model
this, the decay times of those hadrons are recorded in the list together with 
rescattering interaction times, and the decay occurs when it is chosen if it has not already rescattered.

Including rescattering has a few consequences for the properties of events.
First, the charged particle multiplicity increases, since
only processes with two incoming hadrons are allowed,
but inelastic processes can produce more than two outgoing ones.
Similarly, hadron composition will change. Baryon number in particular is
reduced in rescattering through annihilation processes. For example, 
$pp \to \pi\pi\pi$ is possible, but not the reverse process.  Also, resonant processes such as $\pi K \to K^*$ are possible.  
A particular consequence of the increased multiplicity is that hadrons
will on average have lower $p_T$.  On the other hand, protons
move slower than pions with similar $p_T$ and are pushed from behind.

Obviously, models like this are fairly simplistic, but the effects they describe
are expected to be present and cannot currently be explored in any other way.
Finally, what is discussed here for hadrons can also be applied to partonic interactions~\cite{Sjostrand:2004ef}.

\subsection{Heavy Ions}
\label{sec:ion_colliders}

Ultrarelativistic heavy ion (HI) collisions at RHIC and the LHC provide an
extreme laboratory to study multiple parton interactions.
This subsection provides a brief review of Monte Carlo event generators within this field, with a focus on the
overlap areas between high energy HI and high energy $\mathrm{pp}$ collisions.
Traditionally, the HI collider physics community has not
developed event generators coherently with the HEP community.
One reason for this is that there are several differences between HI
generators and generators for $\mathrm{pp}$ collisions, some of which
are technical and other are due to different foci in the physics
questions being asked, and the nature of the corresponding
observables being studied. 

The nature of the observables being studied with HI collisions differs from $\mathrm{pp}$ in several ways. The main assumption is the creation of a deconfined Quark--Gluon Plasma (QGP), and the main purpose of measurements is to probe this state of matter, to understand how it is formed, how it evolves and how it affects the measurements. 

Technically, the main difference lies in how the generation of events is 
organized. For small collision systems, generators typically start with a
high-scale subprocess, and dress up the event by secondary processes
with ever decreasing scales. In contrast HI generators typically start
with an initial-state model, often to build the QGP,
and then proceed in a time-ordered fashion, applying one or several different
models or calculations in succession. The models used often depend on the observable
in question, with different approaches for cold nuclear effects 
(nuclear shadowing, total multiplicities, scaling relations), soft collective effects 
(flow, femtoscopy, strangeness abundances, soft heavy flavor) and jet quenching effects (nuclear modification
factors, jet broadening, jet softening), see the reviews in~\cite{STAR:2005gfr,PHENIX:2004vcz,PHOBOS:2004zne,BRAHMS:2004adc} for a summary from RHIC. For LHC experiments, see~\cite{Armesto:2015ioy}, as reviews from experimental collaborations are still work in progress.
In many cases, the theoretical predictions are only available analytically or in special purpose simulation programs that cannot be directly compared to the measured particles in the detector. Therefore, the HI experiments normally do not unfold their data to particle level as is usually done for $\rm{pp}$ collisions. 
The remainder of this subsection is organized according to the aforementioned time-evolution picture,
where various aspects of initial-state modeling is discussed in Sec.~\ref{sec:hi-initial-state},
soft collective effects in Sec.~\ref{sec:hi-soft-collective} and jet quenching in Sec.~\ref{sec:hi-jet-quenching}. Finally in Sec.~\ref{sec:HImonolithicMCEG}, the differences
in event generation approaches in the heavy ion community are discussed.

\subsubsection{Initial-state modeling}
\label{sec:hi-initial-state}

A simple initial-state model is that of Glauber \cite{Glauber:1955qq} (see also the review \cite{Miller:2007ri}), which in spite of
its simplicity, does well in establishing basic scaling relations and
provides a rough geometrical picture of heavy ion collisions on an event--by--event basis. In its simplest form, a Glauber model is a pure geometrical picture, where the nuclear charge density is parameterized from data. In the optical limit, this allows the calculation of total cross sections and other geometrical quantities, such as the average number of participating nucleons and binary sub-collision, with the nucleon--nucleon cross section as input. In most applications, event--by--event Monte Carlo simulation is allowed by the identification of the nucleon radius with the quantity $\sqrt{\sigma^\text{inel}_\mrm{NN}/\pi}$, giving the direct physical interpretation that two nucleons overlapping in transverse space, will interact inelastically. 

At the LHC, data from $\mathrm{p}A$ collisions revealed \cite{ATLAS:2015hkr}
that color fluctuations of target and projectile states must be included in the model
in
order to describe in particular forward ($\mrm{p}$-going direction)
particle production. Several Glauber models and implementations exist
\cite{Alvioli:2009ab,Moreland:2014oya,Loizides:2014vua,Bozek:2019wyr}
that allow for nucleon fluctuations, manifesting in cross section
fluctuation.  However, a crucial insight was provided by the ``Glauber--Gribov'' model \cite{Alvioli:2014sba}, stating that once
a projectile (color) state is selected by interacting with one target
nucleon, it must remain frozen in that state while traversing the
nucleus. This effect can potentially be drastic at the EIC~\cite{Bierlich:2019wld}, and must therefore be further understood in
current data (see Sec.~\ref{sec:eic} for details).

A Monte Carlo Glauber initial state is often interfaced with a model for particle production, to either allow extraction of average quantities in
simplified models (such as ``wounded nucleon models'' \cite{Bialas:2004su} or experimental fits for centrality estimation \cite{ALICE:2013hur}) or as initial states for full event generators such as \textsc{HIJING} \cite{Wang:1991hta,Deng:2010mv}, \textsc{AMPT} \cite{Lin:2004en} or \pythia8/Angantyr \cite{Bierlich:2018xfw}. All three examples rely on \pythia to generate nucleon--nucleon interactions, which are then combined into a full event. 
The first two generators introduce a shadowing contribution, where not all possible sub-collisions will actually materialize, and the latter by including coherence effects when one (projectile) nucleon strikes several (target) nucleons.
The initial-state approach offered by the EPOS Monte Carlo event generator \cite{Pierog:2013ria}
starts from coherence considerations, instead of using the geometrical Glauber model. 
The complete $AA$ collision is a coherent sum of cut Pomeron exchanges, treated using Regge theory with added partonic degrees of freedom \cite{Drescher:2000ha}.

The Color--Glass Condensate (CGC) \cite{Gelis:2010nm} effective field theory framework
offers a somewhat different approach to high energy heavy ion initial states.
This is a type of \textit{ab initio} calculation, which describes the initial-state dynamics with gluon fields, with time evolution based on solutions of the classical Yang-Mills equations \cite{Schenke:2012wb}. The resulting energy-momentum tensor has traditionally been used as input for event--by--event hydrodynamics calculations \cite{Schenke:2020mbo}, but recently also to serve as an initial state fragmented either with Lund strings \cite{Schenke:2016lrs} or \herwig clusters \cite{Greif:2020rhi}. While the latter approaches are still much in their infancy, this serves as a good example of a common work-flow in high energy heavy ion Monte Carlo studies, where models are often mixed and matched.
We will address these issues later in Sec.~\ref{sec:HImonolithicMCEG}.

\subsubsection{Soft collective effects}
\label{sec:hi-soft-collective}

To obtain final-state observables from the initial state of a heavy ion collision, the dynamics of the intermediate phase must be understood.
In general, this phenomenon is treated using \textit{transport theory}, which then comes in various flavors.  Transport theories can roughly
be divided into macroscopic ones, based on relativistic hydrodynamics, and microscopic ones, based
on free quarks and/or gluons in the medium as the fundamental degrees of freedom.

In the first class of models, an initial state is cast in 
terms of an energy--momentum tensor ($T^{\mu\nu}$), which is time-evolved, based on local conservation of energy $\partial_\mu T^{\mu\nu}(X) = 0$,
up to a freeze--out time where final-state hadrons are produced. The system dynamics enter through an equation of state, which relates the \emph{local}
energy density with a \emph{local} pressure. This would in principle assume local thermal equilibrium, but has been shown also to work well even at multiplicities too low to make such an assumption \cite{Schenke:2021mxx}. To allow the description
of complete events, a hydrodynamic calculation cannot stand alone, but is supplemented by an initial-state description, including a translation to
an energy--momentum tensor, as well as a model for freeze-out, possibly followed by a hadronic cascade.

State--of--the--art hydrodynamic models implemented in event generators, are currently represented by 3+1 dimensional relativistic second--order
viscous hydrodynamics, which complements the above by equations of motion for the shear stress tensor and the bulk pressure \cite{Denicol:2012cn,Molnar:2013lta}.
This is realized in the MUSIC code \cite{Schenke:2010nt}, which gives good quantitative descriptions of flow observables in \rm{AA} collisions
at RHIC and LHC energies \cite{Schenke:2010rr,ALICE:2019zfl}.
A good theoretical candidate for a microscopic kinetic theory is offered by the so--called AMY framework \cite{Arnold:2002zm}. Monte Carlo event generators
based on kinetic theory are, however, not as well developed, with existing codes \cite{Lin:2004en,Fochler:2011en} being more \textit{ad hoc} adaptations
of the idea, rather than a rigorous realization.

With the observation of soft collective effects in \rm{pp} collisions,
such as flow \cite{CMS:2010ifv} and continuous enhancement of strangeness with multiplicity \cite{ALICE:2016fzo},
the general purpose Monte Carlo event generators for \rm{pp} physics have started the process of including models to describe these phenomena as well.
Notably, these approaches are \textit{not} based on the assumption of a deconfined QGP, but instead (in the case of flow) on Lund string interactions \cite{Bierlich:2017vhg}, and
in the case of hadrochemistry either on decays of overlapping strings as a ``rope'' \cite{Bierlich:2014xba}, or on advanced models of cluster reconnections \cite{Gieseke:2017clv,Duncan:2018gfk}. At the time of writing,
such models do not yet offer a competitive picture in \rm{AA} collisions \cite{Bierlich:2020naj}.
In the case of \rm{pp} collisions, they are seen to offer at least as good a description as QGP based approaches in several sensitive observables \cite{Bierlich:2018lbp,Bierlich:2019ixq,ALICE:2021nvv,ALICE:2021nir}.

Core--corona models \cite{Werner:2007bf}
offer a ``middle-ground'' approach between the two extremes presented above.
Here a collision event is separated into a core part, treated with hydrodynamics and HI freeze--out models, and a corona part, given a normal \rm{pp} treatment, \eg using Lund strings. The intended physical picture is that
of QGP ``droplets'' forming in the collision when the density gets high enough. The EPOS Monte Carlo event generator \cite{Pierog:2013ria} implements such a picture, while implementations based on \pythia 8/Angantyr also exist \cite{Kanakubo:2021qcw}. While the approach delivers a very competitive description of data, it comes
at the price of introducing an \textit{ad hoc} definition of what constitutes the ``core'' versus the  ``corona.''

\subsubsection{Jet quenching}
\label{sec:hi-jet-quenching}
The term \emph{jet quenching} refers to a suppression of the jet cross section and a
characteristic modifications of the internal structure of jets 
in heavy ion collisions compared to naively scaling predictions from $\mathrm{pp}$ collisions.
This phenomenon is attributed to interactions between the hard partons that will eventually end up forming a jet
and the dense background they are traversing. 

The hard partons that will evolve into jets are produced in hard
partonic scattering events that factorize from the rest of the event
due to their high scale and are -- by uncertainty principle -- the
first processes to take place in a heavy ion collision. The only
modification to the production processes compared to $\mathrm{pp}$
collisions thus enters through (moderately sized) nuclear
modifications of the parton distribution functions. The hard partons
then propagate for some time through the QGP and interact with it
before hadronization sets in. While the production and at least the
high scale part of the scale evolution are arguably perturbatively
hard processes, this does not necessarily apply to the interactions
between the hard partons evolving into a jet and the background. In
fact, it is an open question in the field whether these interactions
are weak (perturbative) or strong (non-perturbative). The relevant scale here is the transverse momentum transfer between the hard parton and the background (and not the transverse momentum of the hard parton as for the scale evolution). One can, however, argue on general grounds that the hard part of the scale evolution (manifesting itself in additional QCD radiation simulated with parton showers)  proceeds as in $\mathrm{pp}$ collisions. In this early phase of the evolution, the hard partonic sub-system stays coherent and its structure cannot be resolved by the interactions with the background.
Two jet generators ~\cite{Zapp:2012ak,Caucal:2019uvr} include modifications of the scale evolution,
while the others assume that the entire scale evolution factorizes from the interactions with the background. Only one model~\cite{Casalderrey-Solana:2014bpa} assumes a strong coupling between hard partons and background, while the others implement different weak coupling scenarios.

Interactions of hard partons in the background thus lead to modifications of jets, but the passage of hard partons also affects the evolution of the background. Both effects are relevant for jet observables, particularly those sensitive to the internal structure of jets.
An increasing number of jet and event generators account for the background's response to the energy and momentum deposited by hard partons in different approaches. The most common is a simultaneous evolution of hard partons and the background where the energy and momentum deposited shows up as a source term in the hydrodynamic evolution of the background.

Despite significant progress in the last decade, a number of open questions remain. A comprehensive discussion is beyond the scope of this white paper, but we would still like to highlight a few open questions: Firstly, the question remains how jets can be used to extract properties such as transport coefficients of the QGP. The determination of the transport coefficient $\hat q$ (essentially the squared transverse momentum transfer from the background medium to a hard parton per mean free path), for instance, has so far been based on single-inclusive hadron observables~\cite{JETSCAPE:2021ehl,JET:2013cls,Casalderrey-Solana:2014bpa,Chien:2015vja,Andres:2016iys,Noronha-Hostler:2016eow,Bianchi:2017wpt,Zigic:2018ovr,Xie:2019oxg,Xie:2020zdb} while an extraction from multi-particle final states, e.g.~by Bayesian inference as for other parameters~\cite{Novak:2013bqa,Bernhard:2016tnd,JETSCAPE:2020shq,Pratt:2015zsa} has not been performed yet. Secondly, it has been argued repeatedly that jet quenching is intimately related to thermalization~\cite{Kurkela:2014tla,Kurkela:2014tla}, even to the extent that a theory of jet quenching can be formulated as a limiting case of a more general kinetic theory capable of describing rapid thermalization in QCD~\cite{Kurkela:2014tea}. It would certainly be interesting to verify (or falsify) and exploit such a direct connection between the two phenomena. Thirdly, there are a number of interesting theoretical developments~\cite{Sievert:2019cwq,Mehtar-Tani:2019ygg,Andres:2020vxs,Feal:2019xfl,Blaizot:2014bha,Wu:2014nca,Liou:2013qya,Iancu:2014kga,Casalderrey-Solana:2012evi,Fister:2014zxa,Arnold:2020uzm,Arnold:2021pin} that have not found their way into Monte Carlo event generators yet. Finally it should be mentioned that jet quenching has not yet been observed in small systems ($\mathrm{pp}$ or $\mathrm{p}$A), and its existence is an open question of high importance for the field \cite{Adolfsson:2020dhm}.

\subsubsection{Hybrid models vs. monolithic approaches}
\label{sec:HImonolithicMCEG}
As discussed in the previous sections, heavy ion collisions are a complex, multi-scale problem with a large variety of theoretical concepts, ideas and calculational frameworks invoked to describe certain aspects of the collision.
This has lead to a sizable number of specialized Monte Carlo event generators, like the jet generators discussed in section~\ref{sec:hi-jet-quenching}, that encode one particular theoretical picture.
These can only describe a sub-set of observables, and, even for those, a quantitative comparison to data is often impeded by the incomplete modeling.
This has lead some authors to combine their codes with others to provide an adequate modeling of the entire collision. The degree to which the different components are integrated and intertwined varies considerably, and ranges from a loose tool chain, to integration into a common simulation framework. JETSCAPE~\cite{Putschke:2019yrg} has taken this approach even further by making the components exchangeable in a plug--and--play framework.
The benefit of of this approach is that the framework contains -- usually quite advanced -- models for all relevant aspects of the collision.   The disadvantages are that the resulting model is not clearly linked to a particular theoretical idea any more, and that it becomes difficult to control the theoretical uncertainties on the predictions.
Combining different models typically introduces arbitrary separation scales that can lead to large uncertainties. Even more difficult to quantify are uncertainties due to explicit or implicit model assumptions and inconsistencies in the modeling. This problem is greatly reduced in the monolithic approaches that aim for a coherent physics model of the entire collision. While this is appealing from a theoretical perspective,
the number of physics descriptions encompassing a large fraction of the dynamics of a heavy ion collision is unfortunately severely limited.

It should be noted that not all event generators discussed in this subsection are publicly available and open source. That they were  would be highly desirable, since it increases transparency and credibility.
Furthermore, public tools often spark new developments in the field. 

\subsection{Heavy-Flavor Physics}
\label{sec:hv_physics}

Heavy-flavor physics is relevant not only to current electron-positron
experiments, such as Belle II, but also to LHC experiments, including
LHCb, ATLAS, CMS, and the FPF.
Consequently, a broad spectrum of heavy-flavor production processes
must be available in event generators. The core of heavy-flavor
physics, however, requires the accurate modeling of heavy-flavor
hadron decays,
primarily those of $B$ and $D$ mesons. These heavy-flavor states can have
thousands of decay channels,
and their decays occur at energy scales that often require
phenomenological models for a sensible description.
Additionally, with the emergence of the LHC as a factory for competitive heavy-flavor measurements, 
local modeling around the decay of the heavy-flavor hadron has become more important. Given the data that will be collected by currently running heavy-flavor experiments, large-scale Monte Carlo samples will be necessary in the near future, and current inefficiencies in heavy-flavor generation models will make generating such samples impractical.

\subsubsection{Particle Decays}

Most experiments that have a program of heavy-flavor physics rely
heavily on \evtgen~\cite{Lange:2001uf}, a dedicated heavy-flavor decay
package. In turn, \evtgen utilizes \pythia and its decay models for a
large number of more inclusive heavy-flavor decays,
\tauola~\cite{Jadach:1990mz,Jadach:1993hs} or \pythia for dedicated
$\tau$ decays, and
\photos~\cite{Barberio:1990ms,Barberio:1993qi,Golonka:2005pn,Davidson:2010ew} for
final-state radiation in decays. Modern experimental software
frameworks are migrating to multi-threaded environments where the
\fortran-based \tauola and \photos packages are no longer
available. Consequently, significant work needs to be done to ensure
that the physics of these packages, primarily \photos, is available through \evtgen in a multi-threaded environment.

Reliance on a single package such as \evtgen hinders implementation cross-checks. Already, significant work on heavy flavor decays has been included in some multipurpose event generators. However, more comprehensive coverage of the physics is necessary for these generators to provide viable alternatives to \evtgen. One unique feature of \evtgen is the ability to simply specify an exclusive decay channel for a signal decay, while leaving the decays of the same particle species unaffected outside the signal decay. In multipurpose event generators, configuring such exclusive decays is significantly more difficult, if even possible. Calculating the weights of such exclusive signal decays is also not currently handled by event generators, requiring \textit{ad hoc} implementations by each experiment.

A larger issue is how branching fractions for decay channels are
handled between event generators, the PDG, and experimental
collaborations. Currently, there is no consistent methodology due to
the complexities of the problem. Some branching fractions are set from
experimental measurements, while others are determined from theory,
phenomenological models, or simple symmetry arguments. In many cases,
intermediate states are included, while in others, only the final
state is reported.
Even a simple case such as the decay $\tau^- \to \pi^- \pi^0 \nu_\tau$
can be problematic. This decay is typically modeled through an
intermediate $\rho^-$, but with additional contributions from a
$\rho^-(1450)$ and $\rho^-(1700)$. In the PDG, only the final-state
branching fraction is reported, while the intermediate resonances are
included in \herwig and \pythia.  However, explicit entries in the
event record are not included in \pythia.

For the decay of the $B^0$, \pythia includes over $800$ decay
channels, while the PDG lists approximately $500$ observed channels.
Over $900$ decay channels are included in \evtgen, with approximately
$20$ inclusive channels provided externally via \pythia.
Attempts have been made to ensure the consistency of \pythia and
\evtgen decay channels, but such efforts are challenging.
The decay channels for the $B^0$ in \evtgen were last updated in
$2010$.
Subsequent attempts within LHCb to update the $B^0$ decay channels
with more recent PDG data have resulted in inconsistent behavior for
inclusive decays. Clearly, maintaining such decay tables is well
beyond the scope of a single event generator group, or even a single experimental collaboration. A consistent approach needs to be developed between the PDG, various event generator groups, and experimental collaborations.

\subsubsection{Rare Production from Hadronization}

Whether at Belle II or the LHC, the standard method to generate the
production of heavy-flavor hadrons is through hadronization models.
Unlike the generation through a hard process, where a specific final state can be selected, hadronization is a Markov-chain process where the final-state particles are indeterminate, although dependent upon the initial state and the hadronization model parameters. The initial state is defined not only by the partons produced in the hard process, but also subsequent partons produced in the parton shower, and any color reconnection that is applied to the final partonic state. The hadronization model parameters typically dictate the relative formation of hadronic states in the Markov chain, with some states significantly suppressed or in some cases not even present.

This is problematic when attempting to generate hadrons that are
produced only rarely, such as the $B_s^0$. Most hadronization chains
will not generate a $B_s^0$, and so either $\mathcal{O}(10^4)$ events
at the LHC need to be generated to produce a single event with a
$B_s^0$, or repeated hadronization of a single event needs to be
performed until a $B_s^0$ is produced. This latter method introduces
weighting biases, and just as importantly, cannot guarantee the
production of a $B_s^0$, \ie if there are no $b$ partons in the
initial state. Work is ongoing to develop hadronization methods where
a final-state signal hadron can be selected, and an efficient
generation is performed with an appropriate
weight~\cite{LHCb:2011dpk,Muller:2018vny}.  This work is critical for the LHC heavy-flavor experiments, where even repeated hadronization can be costly due to color reconnection models.

\subsubsection{Hard Production}

Most heavy-flavor hadrons are produced via hadronization models, but
in some cases, the hard production
process is of interest. The production of quarkonia in most event generators is modeled via an NRQCD hard process~\cite{Baier:1983va,Gastmans:1986qv, Cho:1995ce, Yuan:1998gr, Humpert:1983yj,Qiao:2002rh}, rather than through hadronization, although in some cases quarkonia can still be produced through hadronization but at a significantly lower rate. In \pythia, an extensible framework for producing quarkonia via NRQCD hard processes is available, but with a notable lack of polarization effects. However, a number of other heavy-flavor hadrons are often modeled through hard processes, including $B_c$ states and double heavy baryons, \ie $\Xi_{cc}^+$. These states are notably missing from most hadronization models and are not available from hard processes in general purpose event generators.

Two specialized event generators, \textsc{BcVegPy}~\cite{Chang:2015qea} and \textsc{GenXicc}~\cite{Wang:2012vj}, are oftentimes used to generate $B_c$ and double heavy baryon hard processes, respectively. These hard processes are then subsequently passed through multipurpose event generators to produce complete events. However, both these packages are written in \fortran and cannot be included in modern multi-threaded environments. Additionally, a more flexible framework is necessary to allow the generation of an extended sector for these hadrons. Efforts are ongoing to understand the production of many of these states in both parton showers and hadronization, but further efforts need to be made to ensure that these hard processes are also available in multipurpose generators.

\subsubsection{Particle Transport}

The non-negligible lifetimes of heavy-flavor hadrons can lead to
ambiguities when interfacing event generation with detector
simulation.
Typically, the event generator is responsible for the decays of
heavy-flavor hadrons.
However, in principle, the heavy-flavor hadron can interact with the
detector material, and so the detector simulation, \ie
\geant~\cite{GEANT4:2002zbu,Allison:2016lfl}, must also be able to
model the interaction of these hadrons with material.
In many cases, these models are either missing, or not sufficiently
developed. Recent developments in low energy hadronic interactions in
the context of hadronic rescattering~\cite{Sjostrand:2021dal} can be
used to provide such models,
although significant work is needed to interface these event generator models to detector simulation packages.

An additional complication is the technical interface between handing
a particle after detector simulation back to an event generator for
decay. At the LHC, this is accomplished on a somewhat \textit{ad hoc}
basis within each experiments' software framework. Developing a standardized method for this technical interface would benefit the community at large and avoid duplication of work. Additionally, this hand-off between detector simulation and event generator is critically important not only for heavy-flavor physics but also for the simulation and decay of long-lived particles. This standardized method would also benefit the simulation of long-lived new physics particles at neutrino experiments.

\subsection{New-physics models}\label{sec:new_physics}
Automated matrix element (ME) generators, such as \amegic~\cite{Krauss:2001iv}, \comix~\cite{Gleisberg:2008fv},
HELAC~\cite{Kanaki:2000ey}, \madgraph~\cite{Alwall:2011uj}, and \whizard~\cite{Kilian:2007gr},
and semi-automated matrix-element generators such as \herwig~\cite{Bellm:2015jjp},
are a central component of the particle-level simulation frameworks in
most experiments.
These generators provide a description of the hard scattering process
that is at the core of particle-level predictions made by the general
purpose event generators.
The implementation of Beyond-the-Standard-Model (BSM) physics into these programs is a well-defined task
and has therefore been accomplished in a completely generic form, with the help of an automated
toolchain. The first component in this framework is the \feynrules package~\cite{Christensen:2008py, Alloul:2013bka},
which allows the Feynman rules from nearly arbitrary
Lagrangians to be extracted and stored in the
UFO format~\cite{Degrande:2011ua}, accessible by most of the current matrix element generators.
The flexibility of the UFO format plays a major role in the exploration of model and parameter space
by the experiments and phenomenologists at collider experiments, and has recently been extended
to neutrino-nucleus scattering~\cite{Isaacson:2021xty}. 
It is important to note that certain types of models require not only a modification 
of the Feynman rules used in automated perturbative calculations, but also a change to QCD evolution
and hadronization. These changes still require domain knowledge in order to be implemented in the
various particle-level Monte-Carlo simulations, which can be a limiting factor in new physics searches.

\begin{table}[t]
    \centering
    \scalebox{0.945}{
    \begin{tabular}{c|c|c|c|c|c|c}
        Generator & \multicolumn{6}{c}{Representation}\\
            & \parbox{1.2cm}{\centering singlet} & \parbox{1.2cm}{\centering triplet} & \parbox{1.2cm}{\centering octet} & \parbox{1.2cm}{\centering $\epsilon^{ijk}$} & \parbox{1.2cm}{\centering 6} & \parbox{1.2cm}{\centering 10} \\
            \hline
          MG5aMC & \checkmark & \checkmark & \checkmark & \checkmark & \checkmark & \\
          \sherpa & \checkmark & \checkmark & \checkmark &  &  & \\
          \whizard& \checkmark & \checkmark & \checkmark & & & \\
    \end{tabular}}\\[3mm]
    \scalebox{0.945}{
    \begin{tabular}{c|c|c|c|c|c|c|c|c|c|c}
        Generator & \multicolumn{3}{c|}{Representations} & \multicolumn{3}{c|}{Lorentz structures} & \multicolumn{4}{c}{Other aspects}\\
            & SM & Spin $\frac32$ & Spin 2 & Custom & Majorana & 4-Fermi & Propagator & Running & Form factor & Unitarity\\
            \hline
          MG5aMC & \checkmark & \checkmark & \checkmark & \checkmark & \checkmark & (\checkmark) &   \checkmark &  EFT & \checkmark &\\
          \sherpa & \checkmark & & (\checkmark) & \checkmark & \checkmark & (\checkmark) & & & & \\
          \whizard& \checkmark & \checkmark & \checkmark & \checkmark & \checkmark & \checkmark & \checkmark &  & \checkmark & \checkmark  \\
    \end{tabular}}
    \caption{Top: Color structures currently supported by commonly used matrix-element generators.
    Bottom: Representations of the Lorentz group and interaction vertex types currently supported by commonly used matrix-element generators. ``Custom'' indicates support of expressions of Lorentz structure beyond those predefined by the code. ``4-Fermi'' indicates the support of arbitrary four fermion interactions (codes with parentheses can handle four fermion interactions only in absence of Majorana particle/flow violation). ``Propagator'' means the handling of non standard propagators. ``Running'' means the handling of the running of non QCD type (EFT here means restricted to effective field theory type of running). "Form factor" is the possibility to link to an external library to compute form-factor and/or other means to include such type of factor. ``Unitarity'' means the possibility to re-unitarize the amplitude on the fly.
    \label{tab:lorentz_restriction}}
\end{table}    
Given a UFO model input, each matrix element generator has to convert the information on the particle
content and interaction structure of the theory into actual code that can be used to evaluate the 
matrix-elements. This means that even though the creation of the UFO model is now fully automated,
there are still restrictions on the type of UFO model supported by any given matrix element generator.
Table~\ref{tab:lorentz_restriction} lists the support for the 
color and Lorentz structures supported by a variety of commonly used programs. 
The conversions are mostly based on dedicated implementations of a UFO parser that becomes part of
the matrix-element generator, for example for \sherpa~\cite{Hoche:2014kca}.
In some cases, they are provided as more generic stand-alone programs~\cite{deAquino:2011ub}.

Current experimental results constrain  any new physics to be either 
very weakly coupled and/or to contain states that are very massive.
Effective field theories (EFTs) provide a means to probe the second scenario, where new physics 
is much heavier than the center of mass energy of the collision.
The impact on experimental measurements can then be approximated through an effective Lagrangian 
with dimension 6 operators, which are invariant under the Standard Model symmetries. 
This provides a single, unified framework to parameterize constraints for a large range 
of BSM scenarios~\cite{
Gupta:2012mi,Grojean:2013kd,Chen:2013kfa,Henning:2014wua,deBlas:2017xtg,
Brivio:2017vri,deBlas:2019rxi,Dawson:2020oco,Trott:2021vqa}.

In the absence of any new physics signals, the EFT provides a framework to quantify the level
of agreement of a measurement with the Standard Model theory by
constraining the coefficients of the operators most likely responsible for any deviations.
Many efforts have been made over the years to improve the accuracy and
extend the range of possible operators in EFTs, for example by including dimension 8 operators~\cite{Murphy:2020rsh,Li:2020gnx}.

The ability of matrix element generators to simulate the hard
scattering process for arbitrary EFTs
will be important in interpreting any new physics scenarios, if/when a deviation
from the Standard Model is observed in the experiments.

\section{Algorithms and Computing}
\label{sec:cross_cuts_technical}
In addition to the physics topics discussed in
Sec.~\ref{sec:cross_cuts}, there are a number of cross-cutting
technical aspects.
First, we discuss interfaces, which are standard data structures for handling inputs and outputs to
event generators.   
Second, we discuss the tools that are used to preserve and easily access data and those used to facilitate the comparison with event generator predictions.
This constitutes an important aspect of data preservation and open science.
Thirdly, and intimately tied to the treatment of data and the development of tools, we
treat the process of constraining or optimizing the parameters that appear in the event generator
models.   
While many aspects of event generators are based on first principle calculations, 
event generators also allow to interface these calculations with phenomenological models, 
which necessarily include tune-able parameters. A full exploration of the parameter and model space 
is a difficult problem, and semi-automated tuning tools are an active area of research.  
Fourthly, we address the computing performance of event generators and related tools.
The ability to provide predictions in a timely manner impacts the ability of experiments
to finalize analyses and react to potential signals of new physics.
Perturbative matrix element calculations, the unweighting of events, and the emergence 
of negative weights at higher orders in perturbation theory can lead to computational
bottlenecks and are addressed through algorithmic improvements and improved computing strategies.
Finally, we consider how the growing field of machine learning, in particular, could provide solutions to
some of the problems that arise in event generation.

In the following, we will discuss the above topics in more detail.
In some cases, different solutions are needed for different types of experimental facilities 
and research programs, and we will refer to the various dedicated sections of this report as needed.

\subsection{Standardized interfaces}
\label{sec:xcuts_interfaces}
The development of some of the components of the event generators, such as the PDFs and matrix element calculations, constitute research projects on their own with small, specialized research teams.   To communicate with this community and exploit their expertise, it is prudent to develop a common language and standards for transferring information.
Standardization of interfaces, both in terms of code APIs and data-interchange formats, are key to ensuring that effort is not unnecessarily duplicated between the limited resources of each generator development team.  Examples, largely but not universally supported, include the \lhe and \hepmc event formats and their supporting libraries~\cite{Alwall:2006yp,Dobbs:2001ck,Buckley:2019xhk}, specialist ``afterburner'' event generators like \evtgen~\cite{Lange:2001uf}, the \lhapdf interface and archive of parton-density functions~\cite{Buckley:2014ana}, and TMDlib~\cite{Abdulov:2021ivr}.

\lhe and \hepmc are built on community-agreed standards, some developed \emph{ad hoc}, others more formally, such as particle ID codes (PIDs)~\cite{ParticleDataGroup:2020ssz}, particle (and vertex) status codes, and an evolving standard for structured event-weight naming~\cite{Proceedings:2018het}. These standards, particularly the PID one, are used from event generators to MC analysis tools, from detector simulation to PDF querying, and their support and adherence is crucial. A missing factor with event formats is a suite of standard tools, both programmatic and executable, for manipulating, visualizing, and converting files. The presence of a single maintained set of such tools would help to reduce further duplication of effort and opportunities for error, and should be considered in the finalization and public adoption of new, efficient event formats for parallel-I/O HPC generation~\cite{Hoche:2019flt}. Truth-level analysis routines in \hepmc, acting between generator programs, needs to be actively coordinated to avoid incompatible forks of the format. The success of the \hepmc event record for the LHC community should be a strong incentive for other frontiers to adopt it. In the neutrino community, there is currently no standardized output format. The leading proposal is to adopt the \hepmc format based on the successful use at the LHC (see Sec~\ref{sec:nu:output} for additional details).

\lhapdf is the community standard resource for access to parton density fits, and, in its current incarnation (v6, since 2013), contains more than 1150 PDF sets encoded in a uniform data format and interpolated with standard algorithms. While these have generally met or exceeded the precision required for MC calculations, the expense of PDF interpolation is a non-trivial aspect of NLO calculations, and for calculations at N$^3$LO order the default local-bicubic interpolation has been found insufficiently stable~\cite{Dulat:2017prg,Nagar:2019gij,Diehl:2021gvs}.
The computing time for interpolation can also be appreciable, even for LO PDFs in general purpose event generators. This is because PDF evaluations occur in the treatment of the hard process, the parton shower, multiparton interactions, and the beam remnant.
Work in 2020-21 succeeded in both reducing the CPU cost (intrinsic to \lhapdf, as well as via optimized generator PDF-call strategies) and developing improved interpolators for high-precision calculations. The latter, as well as support for GPU workflows (cf.~Python-oriented tools like \textsc{PdfFlow}~\cite{Carrazza:2020qwu}) and more general error-set combination rules, will shortly appear in new \lhapdf releases. Longer-term requirements on PDFs, from precision hadron-collider studies, $e^+e^-$ collider prospects, and $ep$ physics at EIC, will also require extension of the current nucleon-specific \lhapdf machinery and interface to support also resolved-photon and transverse-momentum dependent (TMD) PDFs, similarly to the implementation in TMDlib. This generalization may also be a useful opportunity to agree upon community-standard interfaces for PDF querying, to allow better inter-operation of \lhapdf\,6 with PDF-fitting toolkits such as \textsc{ApfelGrid}~\cite{Bertone:2013vaa,Bertone:2017gds} and \textsc{xFitter}~\cite{Bertone:2017tig}.

\subsection{Physics analysis tools}
\label{sec:xcuts_analysis}
Even though not components of event generators \textit{per se}, generator level analysis tools are important for comparing predictions to data and greatly streamline the task of tuning model parameters. The status is rather distinct between neutrino-event analysis and other, mostly collider-physics, analysis; we hence split their summaries into two sections. This is followed up with a discussion on data preservation.

\subsubsection{Hadron and charged-lepton analysis}
The main MC event analysis tool for shower+hadronization generators is \rivet\cite{Bierlich:2019rhm,Buckley:2010ar}, written as a \texttt{C++} successor to the \hztool\cite{Bromley:1995np,Waugh:2006ip} application from the HERA experiments. \rivet contains a standard event-loop system, a wide set of observable calculators (called ``projections''), and user-oriented physics-objects types wrapping the \hepmc event, particle, and vertex objects. \rivet analysis routines are compiled into ``plugins'', loadable at runtime without modifying the core library. A total of more than 1000 analyses are currently bundled with \rivet releases, and maintained/validated as its interfaces and physics tools gradually evolve. \rivet has strong connections with the LHC experiments, and growing ones with \eg~RHIC and EIC collaborations, to streamline direct provision of official analysis codes from the experiments. \rivet also serves as a platform for further tool-building both in experiments (\eg~CMS calls \rivet to obtain truth-particle property definitions such as promptness, and ATLAS uses it in computing truth-level observable responses for EFT analysis), and in phenomenology (among others, the LHC MC tuning projects and the \contur \cite{Butterworth:2016sqg,Buckley:2021neu} BSM-interpretation tool).

The main development in the current v3 series of \rivet releases is automatic support for MC systematic modeling variations, as expressed by event weights: the \rivet system hides the details of weight dispatching, so analyses can be written in a simple form unaware of event weights, and histogramming operations are automatically distributed over multiple copies of each data object, including automatic synchronization of NLO counter-event groups. \rivet\,3 also provides a projection-based approach to detector-response emulation, suitable for reconstruction-level analyses, and a suite of higher-level tools for computing centrality-quantile calibrations and cross-event correlations, as required by many inclusive-QCD and heavy-ion analyses. Histograms are written out in the custom YODA format, with \rivet\,3 supporting merging and re-finalizing of histogram files written by parallel \rivet runs on either equivalent or distinct process types.

This latter feature preempts many speed issues in \rivet via trivial parallelization, but it is worth noting that the main performance bottlenecks in \rivet stem from the speed of copying and graph-walking operations in \hepmc, and intrinsically combinatoric or recursive truth-property checks on event graphs; projections themselves automatically cache results, so the exact same computation is never performed twice on the same event. Especially with the now-typical presence of hundreds of event-weight streams in generator runs, histogramming itself has become an expensive operation both in terms of fill-multiplexing and the resulting size of output histogram data: both aspects are in need of supported effort, where the core developer team are able to only make slow progress due to time constraints.

Three areas of clear future growth can be identified for \rivet: 1) heavy-ion and cosmic-ray oriented analyses, with the increasing work towards EIC (including the incorporation of legacy HERA analyses) and strong engagement from the HI/MC community; 2) uptake of \rivet for BSM physics, including the existing particle-level measurements but also detector-level search analysis preservation, using and extending existing features for \eg~custom smearing and efficiency functions encoded as data histograms; 3) use for future-detector physics prototyping, including EIC and FCC detector proposals, again via the detector-emulation features. These extra use-cases necessitate involvement from the associated communities not just in using \rivet, but in supporting coherent core-library feature development to meet their requirements. 

\subsubsection{Neutrino-event analysis}
Generator-level analysis tools are at a much earlier stage of development in the neutrino experimental community. Common standards for such tools, associated experimental data releases, and even a unified MC event format officially supported by multiple generators are not yet established. There is growing recognition of the urgency to address these basic technical needs~\cite{Barrow:2020gzb} and years of LHC experience underscore their importance in a different physics context. At present, a software package called
\nuisance~\cite{Stowell:2016jfr} provides the main capability for automated neutrino data/MC comparisons and serves as a \textit{de facto} community standard. \nuisance\ provides an unofficial unified interface to multiple neutrino event generator output formats (\neut, \genie, \nuwro, \gibuu) and the capability for automated comparison and tuning to a growing library of over 300 measured neutrino cross-section distributions.  There is no built-in assumption of neutrino-induced events within \nuisance, and limited studies of \genie and \neut electron-induced events have also been performed.

Much like \rivet, \nuisance facilitates: performing Monte Carlo truth studies, quantitatively comparing generator predictions against world-data, tuning of free parameters, sharing of experiment's \emph{in-house} generator tunes, and the dissemination and preservation of data releases. It also provides a framework for experimental collaborations to develop and test their own data releases as a consumer and find and correct trivial errors that damage the longevity or utility of their work. By being open source and actively seeking collaborations with publishing experimental collaborations, \nuisance encourages the robust usage of modern data releases.

The encapsulation of both historical and recent measurements in \nuisance is of broad importance to the field at a time when neutrino interaction uncertainties are becoming discovery-limiting. \nuisance contributors have digitized many historical measurements (for which there was no data release) and have extracted numerous signal definitions from published manuscripts (this process not infrequently requires contact with the authors of the original measurements as descriptions in published manuscripts were ambiguous). During the development of \nuisance, the authors have provided corrections to a number of published data releases. This is all work that is done by, for, and shared freely with, the neutrino-scattering community in the hope that it is of use to simulation-tuners and measurement-makers alike.

What follows is a non-exhaustive list of future developments that are important to \nuisance's long-term utility.
First, incorporating comparisons of measurements made with non-neutrino probes (\emph{e.g.} electrons, protons, or charged pions), where relevant for \nuisance's overall goal of reducing global uncertainties in neutrino--nucleus interactions.
Second, interfacing more closely with HepData~\cite{Maguire:2017ypu} to develop formats that allow the storage and retrieval of current and future neutrino measurements implemented in \nuisance from the HepData archive. This work will be of great importance for the longevity and utility of neutrino-scattering data beyond \nuisance.
Third, continuing to work with experimental collaborations to develop community standards for data releases and to ensure that public measurements are correctly implemented in \nuisance alongside, or soon-after, publication.
Fourth, investigating the development of modern and future data publication techniques, such as \emph{forward folding}, where smearing and efficiency functions are published alongside measurements made in reconstructed or observable projections. Such techniques have the potential to dramatically reduce model-dependence in published neutrino-scattering data.
Fifth, further developing interfaces with experiment-specific tools to smooth cross-collaboration use of neutrino interaction tools and model tunes.
Sixth, developing tools to support the brute force tuning of un-reweightable parameters. However, this must go hand-in-hand with the development of tools to use the outputs of such tunings, which are currently difficult to disseminate to analyzers. The difficulty is that the application of uncertainties of un-reweightable parameters requires a `regeneration' of the simulation, for the simulation and reconstruction tool chains currently used by neutrino-scattering experiments, which is completely infeasible. Tools for the effective reweighting of \emph{un-reweightable} parameters are developed and used by experimental collaborations internally, but standardizing and incorporating such tools into \nuisance would be of wide community interest.

Currently, the only other software tool with comparable functionality for any neutrino generator is the proprietary \textit{Comparisons} product maintained by the \genie\ Collaboration~\cite{GENIE:2021npt}.

\subsubsection{Data Preservation}
Another challenge, for all analysis-preservation systems, is developing the
ability to reliably preserve analyses reliant on machine-learning. This is a long-identified issue, with no current clear solution: initial work has been done to allow preservation of BDT-based analyses without need for an associated ML framework, but the situation is less advanced for neural- and graph-network analysis methods. This issue is deserving of top-down structural attention in HEP as it profoundly affects the reproducibility and long-term scientific impact of analysis results from the LHC and other large experimental investments, both current and upcoming.

An example of the power of particle-level analysis-preservation tools is the \contur program mentioned above, via which new physics models can be rapidly confronted with the huge range of collider final states probed by the measurements archived in \rivet. This model-neutral/model-independent approach to constraining new physics at the TeV scale has a potentially huge impact on Standard-Model-like final states in the high-luminosity LHC era, freeing search efforts to focus on truly exotic final states (dark showers, long-lived particles and the like) where model-independent particle-level measurements are more difficult or perhaps impossible.

\subsection{Tuning and systematic uncertainties}
\label{sec:xcuts_tuning}

MC event generators contain several classes of adjustable parameters. These include choices (such as scales or matching criteria) made in a first-principles calculation to some controlled order in the SM, as well as the choice of approximation itself. There are also phenomenological models for non-perturbative effects such as hadronization that can be freely adjusted to fit data. The choice of phenomenological model itself (e.g. string or cluster hadronization) may also be viewed as a ``parameter.'' In this section we discuss tools for systematically tuning especially parameters within a phenomenological model - although there will generally be some interdependency between all the above, discussed at the end of the section.

\subsubsection{Hadron- and lepton-collider tuning}
MC generator tuning was systematized in the early years of LHC operation, to operate via statistical metrics akin to PDF fits, albeit with more complex models and a wider variety of data. This means that while a single PDF can typically describe the desired (relatively small) set of data, there may well be no single MC model that is valid for all the observables that a MC generator is capable of simulating.

 The most commonly used statistical tuning framework is \professor~\cite{Buckley:2009bj}, associated with, but not dependent on, the \rivet analysis toolkit~\cite{Bierlich:2019rhm}. This implements a $\chi^2$-type fit metric, with scope to mask and weight reference distributions and include bin-to-bin correlations, minimized with respect to model parameters via a set of polynomial bin-response parameterizations.

An alternative approach is available in the MCNNTUNES~\cite{Lazzarin:2020uvv} framework, which models the generator response to changes in the input parameters using supervised machine learning algorithms, based on neural networks. The best tune is then obtained either via minimizing a $\chi^2$-type loss function or by directly reconstructing the best generator parameters using a feed-forward neural network with the bin values as input layer and with the generator parameters as output layer.

The construction of a surrogate model, based on individual MC evaluations,  
is the key feature of these approaches.
The surrogate model can be based on standard regression or machine learning methods.
Other optimizations include the automatic identification of data subsets that are most (or least) sensitive to
the parameters being tuned.
Methods are well established for generating compact representations of
tuning uncertainties from the fit-function, such as the eigentunes
provided by the ATLAS \pythia{8} A14 tune set~\cite{TheATLAScollaboration:2014rfk}.

The core machinery for any MC tuning is therefore available to the community: The tools include the \cxx-core \professor\,2 code, or its Python+numpy evolution known as \textsc{Apprentice}~\cite{Krishnamoorthy:2021nwv}. The latter code also adds rational approximants for the bin parameterization and optional portfolio optimization, and is likely to be the platform for future tuning-tool development.

\subsubsection{Tuning neutrino-nucleus interaction models}
The generator tuning landscape is considerably less well-developed for neutrino experiments and requires further investment of effort going forward. The experience and techniques developed for the LHC will be helpful in solving the needed technical challenges. In addition, there are open questions regarding the necessity of tuning  parameters that should be fixed by theory~\cite{Mosel:2019vhx}.
While some out-of-the-box support for event reweighting presently exists in multiple neutrino event generators, those capabilities are typically extended and applied to model tuning by individual experiments on an ad hoc basis~\cite{NOvATune,MINERvAPionTune,uBooNEGENIETune}.

The main software tool currently used for model parameter tuning against neutrino data, \nuisance\, primarily employs a simple event reweighting scheme (mostly delegated to the generators themselves) to calculate the effects of neutrino interaction model variations. Reliance upon this strategy has enabled useful results but also presents important limitations. Parameter changes that, for instance, open up previously inaccessible regions of phase space cannot be fully treated in this way: missing events in the base simulation results cannot be weighted into existence. Additionally, many of the weight calculators currently provided by neutrino generators are highly model-specific, creating a significant maintenance burden and preventing their reuse across multiple generators (or, often, for distinct models available within a single generator).

The \GENIE collaboration provides
several tunes to quantify realistic parametric and model uncertainties
of the underlying nuclear physics. The tunes are obtained via analyses of global neutrino and electron scattering data.
The \GENIE global analysis was made possible through the continued development of curated data archives, and their successful interface to the \professor tool~\cite{Buckley:2009bj}.
This interface enabled efficient multi-parameter scans and removed limitations from event reweighting procedures.
This concept goes beyond the existing reweighting scheme since the tuning of parameters that are not normally reweightable is now possible.
Initial results were for $\nu_\mu H$ tuning to hydrogen/deuterium data~\cite{GENIENucleonTune} and tuning of the AGKY hadronization model~\cite{GENIEHadronizationTune}. 
While promising as a more robust tuning strategy for neutrino experiments going forward, the \professor\ interface is currently \GENIE-specific and proprietary to that collaboration.

\subsubsection{Systematic uncertainties from tuning}
Conceptually, the main challenges in automated tuning are development of an uncertainty scheme that can cover both experimental-measurement uncertainties and systematic limitations in generator models, and an objective approach to deciding observable weights (it is common to have certain important physical measurements with only a few bins, which risk being drowned out by many less important, but accurate and highly differential, ones.)

The uncertainties associated with MC event generators, including those derived from the tuning procedures described above, are important whenever an event simulation is used to interpolate or extrapolate into unmeasured regions or predict previously unmeasured variables. In cases where the MC event generator provides the state-of-the-art calculation of a given process, they form an important part of the interpretation of data either in terms of validating the SM or probing physics beyond it. They may be classified as:
\begin{enumerate}
    \item Uncertainties in the underlying first-principles SM calculation itself    
    \item Uncertainties resulting from the allowed range of the parameters of a given phenomenological model, ideally constrained by and derived from a tune to data.
    \item Uncertainties resulting from the choice of phenomenological model
\end{enumerate}
Before any claim of a discrepancy between data and the SM can be made with confidence, all these have to be taken into account. The first two can be systematically controlled to some extent via perturbative techniques and tuning; the third class of uncertainty is harder to quantify and typically requires some form of judgment as to what constitutes a reasonable range of choices of model, if indeed there is even a range available. Note also that movement of a parameter of the first type within its allowed range can affect the second two types of uncertainty.

Problems can emerge when independent tuning is done on parameters that are linked by physics constraints, for example pion production and absorption, which are linked by time-reversal invariance. If the processes are obtained from very different theories (e.g.\ production from Rein-Sehgal, absorption from a MC cascade) and both ingredients have been tuned separately, this can lead to inconsistencies~\cite{Mosel:2019vhx}.

\subsection{Computing performance and portability}
\label{sec:xcuts_computing}
The computing performance of event generators and related tools has come to the forefront of discussion 
in the context of LHC experiments due to the increased precision demands in the high-luminosity
phase~\cite{HSFPhysicsEventGeneratorWG:2020gxw,snowmass_hsf_refs}.
The Physics Event Generator Working Group of the HEP Software
Foundation (HSF) has coordinated numerous activities~\cite{HSFPhysicsEventGeneratorWG:2021xti}
to address potential bottlenecks, which are described in more detail in 
Sec.~\ref{sec:performance_hadron_colliders}. The major aspects and common problems 
relevant to all types of experiments and event generators are briefly discussed here.

Because of the wide range of complex physics phenomena that must be simulated, the landscape of generator software
is extremely varied, even within a particular class of
generators, like the ones constructed for hadron collider physics.
Independent of the particular implementation, the most significant computational challenges
arise from the perturbative matrix element calculations, the unweighting 
of events, and the emergence of negative weights at higher orders 
in perturbation theory and through interference effects.

The existence of negative weights is a source of (sometimes large)
inefficiency, as larger event samples must be generated and passed
through the experiment simulation, reconstruction and analysis
software, increasing the compute and storage requirements. For a
fraction $r$ of $N$ events with weight $-1$,
the number of events needed to match a sample of $N$ events with only
weight $1$ increases by a factor $1/(1-2r)^2$.

The negative-weight fraction can be reduced through 
resampling techniques~\cite{Andersen:2020sjs,Andersen:2021mvw,Nachman:2021opi}.
These techniques eliminate negative weights by redistributing them between other events.
The resampling does not avoid the need to generate the original event sample.
A computational gain can therefore only be achieved when the events are subjected
to multiple processing steps, such as NLO matching or analysis, in which case the
gain stems from reduced fluctuations in the subsequent steps.
A new interesting approach determines cells based on a distance metric in a manner 
in which each cell is the minimum region of phase space with 
accumulated non-negative weight. This exploits the fact that a physically sound
Monte Carlo prediction with sufficient statistics must not contain
phase space regions with total negative weight that are large enough
to be resolved in a real-world experiment. The practical performance of the cell resampling procedure
was demonstrated on $W$ boson plus two jet production at next-to-leading order~\cite{Andersen:2021mvw}. 
This method can be easily extended beyond LHC applications.

Matrix element calculations are challenging because they are
inherently quantum mechanical and include all interference effects;
the computing time necessary to evaluate them typically
scales exponentially with the number of external particles.
Parallelization, vectorization and the use of accelerators have been
investigated to speed up these calculations.
The opportunity to use heterogeneous computing architectures at 
leadership class computing facilities has facilitated these
investigations and promises to provide solutions for future computing platforms. 
Currently, there is no full-fledged, portable event generator, but 
several concepts have been presented that may lead to solutions in the near term,
and various alpha release versions are currently in preparation~\cite{Valassi:2021ljk,
Bothmann:2021nch,Valassi:2022dkc,Bothmann:2022itv,Bothmann:2023siu,Valassi:2023yud}.
The emergence of cross-platform abstraction layers
is expected to simplify the construction of architecture-agnostic event generators.

While computing matrix elements is one bottleneck, another is performing 
the integral itself using Monte Carlo methods, especially when the goal is to generate unweighted events.
The construction of efficient phase-space sampling techniques is an important research topic
and is in many cases highly problem specific. While generic solutions to phase-space integration are known
and have been used for several decades, big improvements can often be made when the integrand is understood
in detail for a certain problem at hand, and the Monte-Carlo integrator is adapted accordingly.

Uncertainty estimates such as PDF variations, renormalization and factorization scale variations and physics 
model parameter variations can be handled through computing alternative weights for the same physical event.
However, some modeling uncertainties cannot be calculated through weights, such as for parton shower starting scale and underlying event tune variations. This causes an extra burden for the MC production systems of the experiments because additional samples with different parameters need to be generated. 
Efficient ways to include these variations are a key aspect of computing performance improvements.

Other computational bottlenecks in the generators themselves are worth
mentioning.   Models of Bose-Einstein correlations, hadron
rescattering, color reconnections \textit{etc.} sometimes require
looping over all possible pairs  of particles and partons in an event
and often increase the runtime substantially.   Such effects arise
when simulating collider, heavy ion, and neutrino interactions.
Also, the calculation of alternative event weights for the parton
showers, for example, -- though quite useful for estimating
uncertainties and more efficient than performing multiple event runs
-- rely on an inefficiency of the algorithm to sample a large number
of trial showers.   Finally,
some matching/merging schemes require the construction of ordered
event histories, and the factorial growth problem of parton level topologies reemerges.

\subsection{Machine learning techniques}
One successful application of machine learning to particle physics phenomenology was the development of the NNPDF parton densities~\cite{NNPDF:2014otw,NNPDF:2021uiq,Ball:2021leu}. It is expected that modern machine learning techniques can improve LHC simulations in other aspects as well~\cite{Butter:2020tvl}. For a more comprehensive overview of the topic, see~\cite{MLWP}.

Within the established event generation chain, modern neural networks can first be applied to phase-space integration~\cite{Bendavid:2017zhk,Klimek:2018mza,Chen:2020nfb} and phase-space sampling~\cite{Bothmann:2020ywa,Gao:2020vdv,Gao:2020zvv,Verheyen:2020bjw}. Current sampling methods, optimizing either the variance of an integral or weight distribution, already work by constructing surrogate models and encoding them, for instance, in the distribution of evaluation points~\cite{Lepage:1977sw}. Their main weakness is in dealing with complex correlations, which is, at the same time, the main strength of NN-based surrogates for the integrand. In addition, neural networks are known for their excellent fitting or interpolation properties in high-dimensional spaces. An open question related to NN-samplers is how to combine them optimally with improved multi-channel methods. Similarly, perturbative precision calculations already use interpolation methods to reduce the evaluation time for expensive loop amplitudes, defining a task where appropriately designed neural networks can be expected to outperform standard methods~\cite{Bishara:2019iwh,Badger:2020uow,Aylett-Bullock:2021hmo,Maitre:2021uaa,Danziger:2021eeg}. The challenge in NN-based surrogate models for integrands and amplitudes is to control that all relevant features are indeed encoded in the network and a reliable uncertainty treatment of the network training. Here, Bayesian networks can play a key role~\cite{bnn_early3,deep_errors,Bollweg:2019skg,Kasieczka:2020vlh}. Still at the amplitude level, machine learning can speed up the calculation of loop integrals by optimizing the Feynman integral calculation~\cite{Winterhalder:2021ngy}.
Following the hierarchy of scales in classic event generation, neural networks can efficiently combine different contributions, including subtraction terms~\cite{Butter:2019eyo}, improve the efficiency of event unweighting~\cite{Nachman:2020fff,Verheyen:2020bjw,Backes:2020vka,Danziger:2021eeg}, and approximate parton showers~\cite{shower,locationGAN,monkshower,juniprshower,Dohi:2020eda}.

Parton shower Monte Carlo generators have many free parameters that must be tuned to data and machine learning may provide the most effective way to simultaneously tune many parameters to data while also using all of the available information.  Preliminary studies with machine learning methods are promising and motivate additional exploration~\cite{Ilten:2016csi,Andreassen:2019nnm,Andreassen:2020gtw}.  Reference~\cite{Andreassen:2019nnm} in particular combines a surrogate model with gradient-based optimization -- an early example of using \textit{differentiable programming} to steer simulations for optimization.  See also Ref.~\cite{Heinrich:2022xfa} for an exploration of differentiable programming in the context of matrix element generators.

Aside from standard regression and classification networks and their applications, the most promising network architectures for event generation are generative networks. 
Besides the above-mentioned uncertainty treatment, generative networks are where particle physics applications can set standards for a wider ML community.
One use case for such generative networks is fully NN-based event generators~\cite{dutch,gan_datasets,DijetGAN2,Butter:2019cae,Alanazi:2020klf, Howard:2021pos}.
Structurally and technically, these generators are similar to ML-based detector simulations~\cite{calogan1,calogan2,Musella:2018rdi,aachen_wgan1,aachen_wgan2,ATLASsimGAN,Belayneh:2019vyx,Buhmann:2020pmy,Buhmann:2021lxj,Chen:2021gdz,Krause:2021ilc,Krause:2021wez}, which are already used in the LHC experiments (see e.g., Ref.~\cite{ATLAS:2021pzo}). In addition to driving progress in treating phase-space generation through deep networks, fully NN-based generators are required to be trained on any combination of simulated and/or measured events. Therefore, they are not a replacement to existing generators when it comes to simulating the underlying physics in an interpretative manner. They also provide an efficient way to define and ship standardized event samples, including a modest numerical advantage over a statistics-limited training dataset~\cite{Butter:2020qhk,Bieringer:2022cbs}, and can be used for post-processing of standard simulations. In these functions, it is again crucial to understand and control the precision of these generative networks at the level needed for precision measurements~\cite{Bellagente:2021yyh,Butter:2021csz}.

Finally, we should always look for simulation-related questions and problems where neural networks might lead to transformative progress. Examples include simulation-based inference~\cite{Brehmer:2019xox,Bieringer:2020tnw}, symbolic regression~\cite{Bieringer:2020tnw}, sample compression~\cite{Carrazza:2021hny}, or detection of symmetries~\cite{Barenboim:2021vzh,Desai:2021wbb}. An especially promising application is inverting the simulation chain or unfolding, using reweighting networks~\cite{Andreassen:2019cjw} or conditional generative networks~\cite{Datta:2018mwd,fcgan,Bellagente:2020piv}. ML-based methods are being studied by LHC experiments (see also Ref.~\cite{H1:2021wkz}) and a comparison of two leading ML-methodologies can be found in Ref.~\cite{Arratia:2021otl}.

\phantomsection
\addcontentsline{toc}{part}{{\large\bf Part II: Individual Facilities and Experiments}}
\newpage
\section{High Energy Colliders}
\label{sec:he_colliders}
In this section, we focus on aspects of theory and phenomenology that are important for event simulation at high-energy colliders, particularly the currently operational Large Hadron Collider.
The experimental program at these facilities is largely based on the
study of rare events characterized by a large momentum transfer scale $\mu_H$. For instance, the production of a new particle that is not part of the Standard Model entails momentum transfers on the order of the mass of the new particle, and Standard Model background events that must be distinguished from new physics events that involve similar scales.
Predictions of cross sections at large $\mu_H$ can often be made with high precision, because they rest on a firm theoretical understanding of collinear factorization (see Sec.~\ref{sec:factorization}) and of the size of power corrections~\cite{FerrarioRavasio:2020guj,Caola:2021kzt}.
For observables that are infrared safe at the scale $\mu_J$, parton
splittings at scales smaller than $\mu_J$ in a perturbative calculation 
have a negligible effect.
For example, 
$\sigma[J]$ in Eq.~\eqref{eq:factorization_main} could be a cross section to produce suitably defined jets,
which has a reduced dependence on the QCD dynamics at the hadronization scale.
In the collinear factorization approach, the inclusive cross sections at the parton level, $\hat \sigma[J]$, can 
then be reliably computed using an expansion in powers of the strong and electroweak coupling. Such perturbative 
expansions are useful when $\mu_H$ is large and $\mu_J \sim
\mu_H$, because the strong coupling $\as(\mu_J^2) \ll 1$ for large $\mu_J^2$.
We will discuss the corresponding techniques in Sec.~\ref{sec:hadron_collider_hard_process}.

If measurements involve multiple scales, or if $\mu_J\ll\mu_H$, the QCD evolution encoded in Eq.~\eqref{eq:pdf_evolution}
will induce large logarithmic corrections proportional to
$\alphaS(\mu_J^2) \log\left(\mu_H^2/\mu_J^2\right)$ that also depend on the color
flow in the hard process. Similar corrections will emerge in the electroweak sector, and these corrections often take on
a more complicated form because of the the more intricate flavor
structure.
Parton showers provide solutions to the corresponding evolution equations and
will be discussed in Secs.~\ref{sec:hadron_collider_evolution} and~\ref{sec:hadron_collider_ewps}.
The matching of QCD parton showers to fixed-order perturbative calculations will be addressed in Sec.~\ref{sec:hadron_collider_matching}.
Analytic techniques to approach the problem that have been implemented in event generators 
are discussed in Sec.~\ref{sec:hadron_collider_resummation}.

\subsection{Higher-order QCD and electroweak computations}
\label{sec:hadron_collider_hard_process}
Testing the validity of the Standard Model at the shortest distances
requires accurate and precise predictions in perturbation theory.
Several specialized programs exist for this task, each focused on a
particular calculational technique.
They all provide predictions at either next-to-leading-order (NLO) or
next-to-next-to-leading-order (NNLO) for key Standard Model processes.
These predictions are important inputs to event generators, because
the hard process determines features of the full event.   Having
multiple tools using different techniques is important for
cross-validation of complicated numerical calculations, for exposing
uncertainties related to the methods, and for driving innovation.
In this section, we describe their current status, and prospects for future developments.
We list projects in alphabetical order.

\subsubsection{GoSam}

The program package \GoSam~\cite{Cullen:2011ac,Cullen:2014yla} provides numerical results for one-loop amplitudes 
within and beyond the Standard Model.\footnote{\GoSam is publicly available at {\tt https://github.com/gudrunhe/gosam}.}
Inputs for beyond the Standard Model calculations can be provided using the UFO format~\cite{Degrande:2011ua}.
In addition to QCD, electroweak corrections can be calculated~\cite{Chiesa:2015mya}\footnote{The version
  containing the automation of EW corrections is available at {\tt https://github.com/tramonta/gosam-1l.git}}.
\GoSam can be used with complex masses as well as with a non-diagonal CKM matrix.

Using the BLHA interface~\cite{Binoth:2010xt,Alioli:2013nda},  \GoSam can be combined with NLO-capable Monte Carlo programs. It has interfaces to \sherpa~\cite{Gleisberg:2008ta,Hoeche:2012yf}, \POWHEG~\cite{Alioli:2010xd,Luisoni:2013kna} and \MGaMC~\cite{Alwall:2014hca}, and is included in \herwig7~\cite{Bellm:2015jjp} and \whizard~\cite{Kilian:2007gr,Weiss:2015npa}.

Based on the input card where the user specifies the process, the model and other optional settings, 
\GoSam first generates the expressions for one-loop  amplitudes starting with Feynman diagrams, generated
with \qgraf~\cite{Nogueira:1991ex}, and then writes the expressions as $D$-dimensional integrands over the loop momentum,
making use of \form~\cite{Kuipers:2012rf,Ruijl:2017dtg} and {\sc Spinney}~\cite{Cullen:2010jv} 
to generate the algebraic expressions. For non-standard applications, \python filter functions 
can be defined at input card level to exclude certain topologies.
For the  reduction it can either use  integrand reduction via Laurent expansion~\cite{Mastrolia:2012bu,vanDeurzen:2013saa} 
based on the library \Ninja~\cite{Peraro:2014cba}, or 
tensor integral reduction as implemented in \Golem~\cite{Binoth:2008uq,Heinrich:2010ax,Cullen:2011kv,Guillet:2013msa}.
The scalar one-loop integrals are computed with the {\sc OneLoop}~\cite{vanHameren:2010cp} integral library.
Alternatively
one can choose other reduction strategies such as the OPP reduction method~\cite{Ossola:2006us,Mastrolia:2008jb,Ossola:2008xq} 
which is implemented in  {\sc Samurai}~\cite{Mastrolia:2010nb}, or other scalar integral libraries such as {\sc QCDloop}~\cite{Ellis:2007qk}.

The code contains several options to improve the numerical stability in kinematic regions where the amplitude approaches
soft or collinear configurations, or spurious poles due to the reduction procedure.
Since version 2.1, \GoSam also has an option to switch automatically  to quadruple precision, 
thus improving the stability without too much cost in runtime, as the fraction of such ``unstable points''
usually does not exceed the few percent range.

While other automated NLO generators use recursive relations to generate the amplitudes, and therefore
lead to relatively compact generated codes, {\sc GoSam} with its diagram based approach
allows to select and study individual parts contributing to the amplitudes, from single diagrams to groups,
and for every helicity and/or color structure. Therefore, apart from the usage as a full amplitude generator, 
{\sc GoSam} is also a tool suitable for special purposes.

\paragraph{Examples of processes calculated with {\sc GoSam}}
Various applications of  \GoSam both within and beyond the Standard Model, as well as in combination
with different Monte Carlo programs, have shown that \GoSam is a very versatile tool providing one-loop amplitudes
for a broad range of use cases.

\GoSam in combination with Sherpa has been used to calculate NLO QCD corrections to Higgs boson production 
in association with three jets at NLO in the heavy top limit~\cite{Cullen:2013saa,Greiner:2015jha}, 
Higgs production in association with a top quark pair and up to one jet~\cite{Hoeche:2013mua,vanDeurzen:2013xla}, 
$W^+W^++$2\,jets production,  top quark pair production matched to a parton shower and $W^{+} W^{-}b\bar{b}$ 
production with leptonic decays~\cite{Heinrich:2013qaa,Heinrich:2017bqp}.

\GoSam in combination with the \powhegbox has been used to calculate the NLO QCD corrections to 
$pp\to HW/HZ+0,1$\,jet~\cite{Luisoni:2013kna} and $pp\to W b\bar{b}+1$\,jet~\cite{Luisoni:2015mpa}.
NLO QCD results for $t\bar{t}H$ and $t\bar{t}\gamma\gamma$ production have been calculated in combination with \MGaMC~\cite{vanDeurzen:2015cga}.

\GoSam also has been used to calculate the real-virtual contribution to $Z$-boson pair production
at NNLO in Ref.~\cite{Heinrich:2017bvg}.
NLO EW corrections have been obtained for $Z+0,1,2,3$\,jets~\cite{Chiesa:2013yma}, $W+2$\,jets~\cite{Chiesa:2015mya} 
and diphoton + jets~\cite{Chiesa:2017gqx}, using MadDipole~\cite{Frederix:2008hu,Frederix:2010cj} for the real radiation.

\GoSam in combination with \herwig7 has been used to calculate  $W^+W^-$ production including
dimension-8 operators mediating anomalous gluon-gauge boson couplings~\cite{Bellm:2016cks}.
Further examples of the calculation of processes beyond the Standard Model with \GoSam are
SUSY-QCD corrections to neutralino pair production in association with a jet~\cite{Cullen:2012eh},
NLO QCD corrections to diphoton plus jet production through graviton exchange~\cite{Greiner:2013gca},
the production of a top-philic resonance at the LHC~\cite{Greiner:2014qna} or interference contributions
to gluon initiated heavy Higgs production in the Two-Higgs-Doublet Model~\cite{Greiner:2015ixr}.

Recently, \GoSam has been employed to generate the one-loop five-point amplitudes entering the
NLO real radiation part of the loop-induced processes  $pp\to  HH$~\cite{Borowka:2016ehy}, 
$pp\to  HJ$~\cite{Jones:2018hbb} and $gg\to HZ$~\cite{Chen:2022rua}.
In combination with the \powhegbox, NLO QCD predictions for  $pp \to HH$ including anomalous
couplings within  HEFT~\cite{Heinrich:2019bkc,Heinrich:2022idm} and SMEFT~\cite{Heinrich:2022idm}
are available as public Monte Carlo tools.

\subsubsection{MadLoop}

\MadLoop~\cite{Hirschi:2011pa,Alwall:2014hca}
provides the virtual contribution (one-loop matrix elements) to NLO simulations performed by \MGaMC~\cite{Alwall:2014hca}, but it can also easily be used in stand-alone mode
to generate a library that can be interfaced to other codes (e.g.\ \sherpa~\cite{Campbell:2021vlt} and \powheg~\cite{Nason:2020lxx}).
\MadLoop uses a combination of reduction algorithms, mainly \Ninja~\cite{Peraro:2014cba,Hirschi:2016mdz}, \collier~\cite{Denner:2014gla,Denner:2016kdg} and \CutTools~\cite{Ossola:2007ax}, together with an in-house implementation of the original \OpenLoops~\cite{Cascioli:2011va} approach for computing numerator tensor coefficients. It also features robust on-the-fly numerical stability assessment, with automated fallback onto alternative reduction techniques and quadruple precision.
While \MadLoop still offers competitive run-time
speed~\cite{Campbell:2021vlt,Degrande:2018neu}, it is often not the
bottleneck in
NLO calculations.  The program is designed primarily for generality
and flexibility, which stems mainly from the fact that \emph{all} model-specific information required by \MadLoop is imported from a single stand-alone \Python module following an extension of the UFO~\cite{Degrande:2011ua} standards. This includes the ultraviolet (UV) renormalization counterterms and rational $R2$ contributions~\cite{Draggiotis:2009yb,Page:2013xla,Pozzorini:2020hkx} automatically generated by \NLOCT~\cite{Degrande:2014vpa} (a module of the \feynrules~\cite{Degrande:2011ua} package).
We now briefly detail particular aspects of the flexibility offered by \MadLoop and hint at future developments.

\paragraph{Flexible process definition} 
The construction of the one-loop matrix elements in \MadLoop allows
the user to select any gauge invariant subset of the contributing
diagrams. This selection is facilitated by the full support of
\madgraph process-definition syntax, including e.g. required
s-channels, forbidden particles, multiparticle labels and coupling
order specifications, both at the amplitude and squared amplitude
level. The user can even implement a custom loop-diagram filter to
select even more specific sets of diagrams; so long as the selection is gauge invariant, \MadLoop will generate the appropriate set of $R_2$ and UV counterterms.
This flexibility proves to be especially important to phenomenologists
interested in understanding the exact impact of particular sets of
contributions (in the Standard Model but also beyond) to collider signatures. \MadLoop also offers full support for loop-induced processes (i.e. processes receiving no contribution at tree-level), both for the matrix-element generation and their efficient, parallel integration in \MGaMC~\cite{Hirschi:2015iia}.

\paragraph{Wide support for Beyond-the-Standard-Model (BSM) Lagrangians}
The strong suit of \madgraph has always been its extensive support of
new physics models, and \MadLoop follows this trademark. The support
for Majorana particles, multi-point interactions, and  exotic color and
Lorentz structures  allows \MadLoop to be applied to a wide variety of
models.  A non-exhaustive list includes
the full MSSM~\cite{Degrande:2015vaa}, 2HDM~\cite{Degrande:2016hyf}, leptoquarks~\cite{Borschensky:2021jyk} and spin-two particles~\cite{Das:2016pbk}.
The clean separation of model-specific input provided via UFO models
allowed the community to independently perform their own BSM one-loop
computations (see the repository of NLO-ready UFO models available at \url{https://feynrules.irmp.ucl.ac.be/wiki/NLOModels}).
More recently, NLO computations in the six-dimensional Standard Model
effective theory (SMEFT) were automated~\cite{Degrande:2020evl}.
This is especially important as SMEFT is becoming a prime
UV-theory-agnostic framework for quantifying
the agreement between Standard Model predictions and data.
One particular challenge in the years to come will be to combine this effort with the systematic inclusion of the complete renormalization group flow dictating the running of all Wilson coefficients and other parameters of the theory.

\paragraph{Support for mixed QCD and Electro-Weak (EW) corrections}
From the perspective of one-loop computations alone, the support of
mixed QCD and EW corrections comes naturally as a by-product of the
support for arbitrary BSM models. However, in the mixed case, loop
matrix element evaluation time can be more of a limiting factor than
for pure QCD corrections.   Depending on the kinematic regime being probed, it is possible to build efficient and fast approximants leveraging the dominance of so-called Sudakov EW logarithms~\cite{Pagani:2021vyk}. 
One challenge in computing one-loop EW matrix elements is the support for the complex-mass scheme that was completed in~\cite{Frederix:2018nkq} together with the generalization in \MGaMC of FKS subtraction~\cite{Frederix:2009yq} for infrared singularities in mixed QCD and EW perturbative computations.
The fixed-order predictions hence obtained will be the basis for
future work seeking to support matching to parton showers including QED corrections as well as application to lepton colliders~\cite{Frixione:2021zdp}.

\paragraph{Beyond NLO corrections}
While \MadLoop focuses on one-loop calculations, \MGaMC still seeks to
natively access multi-loop amplitudes.
Over the last decade, a large number of two-loop amplitudes have been
computed, but these are often not readily available to the rest of the
community.  The UFO standard can provide a bridge 
between the Monte-Carlo and multi-loop provider community.
Multi-loop amplitudes can be projected onto a specific color and
Lorentz basis so as to be encoded as an effective vertex in a UFO
model. The form factors of this effective vertex are very complicated
functions of its kinematics that are computed by separate libraries
provided along with the UFO model.
This strategy was successfully applied to inclusive Higgs
production to include the two-loop matrix elements necessary for the computation of the mixed QCD+EW NLO contribution ~\cite{Becchetti:2020wof}.
One key advantage of such an approach is that the traditional
tree-level generator of Monte-Carlo software is then responsible for
constructing the helicity amplitude containing the multi-loop effective
vertex and interfering it with other amplitudes.
Also, arbitrary decay structures can be attached in this way.
By employing this strategy systematically, a collection of effective
vertices encoding all the multi-loop amplitudes computed
(semi-)analytically thus far by the theoretical community could be made readily available to all MC generators.

\subsubsection{MATRIX}
\label{sec:MATRIX}
The computational framework \Matrix
~\cite{Grazzini:2017mhc}
features a parton-level Monte Carlo generator
capable of computing fiducial cross sections and distributions for Higgs boson, 
vector-boson and vector-boson pair production processes up to NNLO accuracy in QCD.
Calculations include all relevant leptonic decay channels of the vector bosons with
spin correlations and off-shell effects from all resonant
and non-resonant diagrams.
This allows the user to apply arbitrary fiducial cuts
directly on the phase space of the respective leptonic final state.
Dedicated studies have been performed within early versions of the \Matrix framework for $Z\gamma$~\cite{Grazzini:2013bna,Grazzini:2015nwa}, $W\gamma$~\cite{Grazzini:2015nwa}, $ZZ$~\cite{Cascioli:2014yka,Grazzini:2015hta,Kallweit:2018nyv}, $W^+W^-$~\cite{Gehrmann:2014fva,Grazzini:2016ctr} and $W^\pm Z$~\cite{Grazzini:2016swo,Grazzini:2017ckn} production.
In its recent second public release, the functionality of \Matrix was extended
in several directions. Besides multiple technical improvements, NLO EW corrections are now included
for almost all of the aforementioned processes, and an approximation for the leading
effects of missing mixed QCD--EW corrections is provided by a sophisticated combination
of higher-order QCD and EW effects~\cite{Kallweit:2019zez}.
Moreover, in terms of corrections beyond NNLO in QCD, for massive diboson processes
the NLO corrections to the loop-induced gluon fusion channel is expected to be the
dominant component, and it has been added for all cases with an overall
neutral final state~\cite{Grazzini:2018owa,Grazzini:2020stb,Grazzini:2021iae}.

\paragraph{Features of the public \Matrix framework}
\Matrix achieves NNLO accuracy by using a process-independent implementation of the
\qT-subtraction formalism~\cite{Catani:2007vq} within the Monte Carlo program \Munich{},
from which it inherits a fully automated implementation of the dipole subtraction method
at NLO QCD~\cite{Catani:1996jh,Catani:1996vz} and NLO EW~\cite{Dittmaier:1999mb,Dittmaier:2008md,Gehrmann:2010ry,Kallweit:2017khh,Schonherr:2017qcj},
a highly efficient multi-channel integration algorithm, a process-independent amplitude interface
as well as an automated bookkeeping of the relevant partonic subprocesses.
The required tree-level and one-loop amplitudes, including spin and color correlations for the subtraction of
infrared divergences, are calculated with \OpenLoops~\cite{Cascioli:2011va,Buccioni:2017yxi,Buccioni:2019sur} using scalar integrals from \Collier{}~\cite{Denner:2014gla,Denner:2016kdg}.
For the required two-loop amplitudes, \Matrix employs its own implementations for associated
production of a $W/Z$ boson with a photon~\cite{Gehrmann:2011ab} and
$\gamma\gamma$~\cite{Anastasiou:2002zn} production. External codes are used for on-shell 
$ZZ$~\cite{Cascioli:2014yka} and $W^+W^-$~\cite{Gehrmann:2014fva} production
and the publicly available \textsc{VVamp} package~\cite{hepforge:VVamp} for off-shell production of massive vector-boson pairs~\cite{Gehrmann:2015ora,vonManteuffel:2015msa}.

The cancellation of IR divergences at NNLO is accomplished using a process-independent
implementation of the \qT-subtraction formalism \cite{Catani:2007vq}.
 This relies on the universality of the small transverse momentum
 $\qT$ behavior for the production of a colorless system.   This
 behavior is known to NNLO from transverse-momentum resummation~\cite{Collins:1984kg,Bozzi:2005wk}.
The remaining $\qT$-independent term from the hard-collinear coefficient
${\cal H}^{\mathrm{F}}_{\mathrm{NNLO}}$ is constructed from suitably subtracted two-loop amplitudes
in a universal manner~\cite{Catani:2011kr,Catani:2012qa,Catani:2013tia}, such that the method
can be directly applied to the hadroproduction of an arbitrary set of colorless particles
once the corresponding two-loop amplitudes are available.
Since the \qT subtraction method 
is non-local, a slicing cut-off $r_{\mathrm{cut}}$ is introduced on the dimensionless quantity $r=\qT/Q$,
where $Q$ is the invariant mass of the colorless system.
Below this cut-off, the difference between the fixed-order (N)LO prediction  
and the subtraction counterterm 
is neglected, which is correct up to power-suppressed
contributions that vanish in the limit $r_{\mathrm{cut}}\to0$ and can be controlled by monitoring
the $r_{\mathrm{cut}}$ dependence of the cross section.
\Matrix simultaneously computes the cross section at several $r_{\mathrm{cut}}$ values and uses
this information to perform an extrapolation to approach the limit $r_{\mathrm{cut}}\to0$ and to assign an extrapolation error estimate to the result. More details on the procedure are
given in \cite{Grazzini:2017mhc}.

\paragraph{New developments within the \Matrix framework}
Beyond the public release, there are multiple ongoing developments both to further improve
the performance and to extend the functionality of the \Matrix framework in several directions.
Given the chosen slicing approach to achieve NNLO accuracy, a good control of the remaining cutoff
dependence arising in terms of power corrections after subtracting the logarithmic divergences
is a primary task. While massive-boson production processes are known to exhibit at most quadratic
power corrections, there are several sources of (possibly logarithmically enhanced) linear power 
corrections like photon-isolation requirements, but also phase space restrictions like symmetric
or asymmetric cuts in (effective) two-particle final states. To eliminate a possible bias from
choosing a finite $r_{\mathrm{cut}}$ value, \Matrix applies a numerical extrapolation to
approach the limit $r_{\mathrm{cut}}\to0$ on the level of inclusive cross sections.
This feature has been extended towards a bin-wise extrapolation for differential distributions
and will be made available in an upcoming release of \Matrix.
Nevertheless, the need to perform calculations down to very low $r_{\mathrm{cut}}$ values renders
the integration more challenging and computationally expensive, such that theoretical approaches
to predict at least the linear power corrections are a very valuable achievement since they
allow a reliable $r_{\mathrm{cut}}\to0$ extrapolation to be performed based on significantly larger
$r_{\mathrm{cut}}$ values. To this end, following the considerations of Ref.~\cite{Ebert:2020dfc}
those linear power corrections that originate from symmetric or asymmetric phase space cut
configurations can be implemented by taking into account the recoil of the colorless
two-particle system in the counterterm~\cite{Buonocore:2021tke,Camarda:2021jsw}, as in the resummed calculation of Ref.~\cite{Catani:2015vma}.
The corresponding correction term that is particularly useful where it describes the only source
of linear power corrections has been implemented in the \Matrix framework~\cite{Buonocore:2021tke} and will also be added in an upcoming public release.

\paragraph{Color-singlet production processes}
Given the universality of our implementation, the inclusion of further color-singlet
production processes into the \Matrix framework does not involve any conceptual issues as soon as
the required two-loop amplitudes are known.
Processes of that class that have been calculated with \Matrix and will be added in
future releases are Higgs-boson pair production~\cite{deFlorian:2013jea,deFlorian:2016uhr} and
the neutral and charged Higgsstrahlung processes~\cite{Alioli:2019qzz}.
Most notably, a first $2\to3$ calculation at NNLO QCD has been performed in \Matrix for triphoton production~\cite{Kallweit:2020gcp}, which proved that in spite of being implemented as a slicing
method, \qT subtraction performs sufficiently well to produce per mill level precision results at
this level of complexity. Preliminary studies of the cut-off dependence for other triboson production
processes, which do not require the knowledge of the respective two-loop amplitudes, but test the
numerically most challenging part of the calculation, confirm that the whole class of triboson processes involving photons and/or massive weak bosons and their leptonic decay products can be addressed
in the \Matrix framework.

Beyond fixed-order calculations, the close relationship between \qT subtraction and
resummation enabled also the small-\qT resummation of logarithmically enhanced terms at NNLL
accuracy in the \Matrix framework, as discussed for $ZZ$ and $WW$ production in~\cite{Grazzini:2015wpa}, which will be added to the public framework in the future.
Another path towards resummed predictions not only in \qT but also in the transverse momentum of the leading jet as well as their joint resummation inside the \Matrix framework has been taken by establishing an interface to the \RadISH program~\cite{Monni:2016ktx,Bizon:2017rah,Monni:2019yyr}. The resulting \Matrix+\RadISH code is publicly available~\cite{Kallweit:2020gva,Wiesemann:2020gbm,MATRIXRadisH}.

\paragraph{Heavy-quark production processes}
With appropriate modifications, the \qT-subtraction method for the hadroproduction of colorless
systems can be extended to deal also with massive quarks in the final state.
The extension to top-quark pair production has been discussed in~\cite{Bonciani:2015sha}. The
construction of the NNLO cross section is analogous to the light flavor case with $F=t\bar t$, and the
IR-subtraction counterterm is obtained from the NNLO perturbative expansion
(see e.g.\ Refs.\cite{Bonciani:2015sha,Bozzi:2005wk,Bozzi:2007pn}) of the resummation formula of
the logarithmically enhanced contributions to the \qT distribution of the $t\bar t$
pair~\cite{Zhu:2012ts,Li:2013mia,Catani:2014qha}. The additional soft contributions
that enter ${\cal H}^{\mathrm{t\bar t}}_{\mathrm{NNLO}}$~\cite{Angeles-Martinez:2018mqh,ttbarsoftCDGM},
provide the necessary final ingredients to handle heavy-quark pair production. Using the two-loop amplitudes of%
~\cite{Barnreuther:2013qvf}, the corresponding \Matrix implementation has been applied
to top-quark~\cite{Catani:2019iny,Catani:2019hip} and bottom-quark~\cite{Catani:2020kkl} pair
production.
The extension to associated heavy-quark pair production does not
introduce any new QCD structures, but
is complicated by the evaluation of the required soft function for the
given kinematics.
As a proof of concept, the NNLO contributions in the non-diagonal partonic channels
for $t{\bar t} H$ production were calculated in this approach within the \Matrix
framework~\cite{Catani:2021cbl}.
The evaluation of the soft function for this process will pave the way for treating the production of
heavy-quark pairs in association with arbitrary colorless particles, given that two-loop
amplitudes for the respective processes or at least suitable approximations can be obtained.

\paragraph{Mixed NNLO QCD--EW corrections}
For reference processes measured with very high precision, like Drell--Yan~(DY) lepton pair production, fixed-order predictions at NNLO and beyond in QCD combined with NLO EW corrections are expected to be insufficient to match the experimental accuracy,
and theoretical calculations of mixed NNLO QCD--EW, possibly even NNLO EW corrections are required.
In those cases, the cancellation of IR singularities can be achieved by using a formulation of the $q_T$ subtraction formalism derived from the NNLO QCD computation of heavy-quark production through a suitable Abelianization procedure~\cite{deFlorian:2018wcj,Buonocore:2019puv}. To regularize collinear final-state singularities, charged particles need to be treated with non-vanishing
masses. There is an analog of the \qT-subtraction scheme for corrections of relative $\mathcal{O}(\alpha_s^m\alpha^n)$,
which involves mixed QCD/EW subtraction terms if $m=1$ and $n=1$. Those have been calculated in the \Matrix framework for the charged Drell--Yan process in a reweighted pole approximation~\cite{Buonocore:2021rxx}, and for the neutral DY process in a first complete calculation of $\mathcal{O}(\alpha_s\alpha)$ corrections~\cite{Bonciani:2021zzf}, using the amplitudes of \cite{Armadillo:2022bgm}. Along the same lines, the \qT subtraction formalism can be extended towards NNLO EW corrections (i.e.\ $m=0$, $n=2$); the corresponding two-loop amplitudes are still challenging, but may be considered to be in reach from recent developments.

\paragraph{Outlook}
A current limitation to \Matrix is the ability to handle 
final states with light jets at the Born level. The recently proposed
$k_T^{\rm ness}$ variable~\cite{Buonocore:2022mle} can be considered as an extension
of \qT towards processes involving jets at Born level and may provide a promising candidate to
eventually overcome this limitation.
With all these features at hand, the \Matrix framework is well placed
to face the
challenges for precision calculations posed by the high-luminosity phase at the LHC and future high-energy colliders.

\subsubsection{MCFM}
\label{sec:MCFM}
MCFM is
a code that calculates
parton-level 
cross-sections and differential distributions for a multitude of
scattering processes in the Standard Model. It 
provides fixed-order predictions at NLO in 
QCD~\cite{Campbell:1999ah,Campbell:2011bn} and meanwhile includes 
many NNLO predictions, as described below. More recently, all color singlet processes can 
include the effect of transverse momentum resummation \cite{Becher:2020ugp}. 
MCFM features automatic scale uncertainties, 
PDF uncertainties for multiple PDF sets simultaneously and differential slicing cutoff 
extrapolation. It is designed with OpenMP, MPI and Coarray support for numerically fast and high 
precision evaluations of the most complicated scattering processes \cite{Campbell:2019dru} and 
scales with near-linear speedup to many thousand cores on computer clusters. All one- and two-loop 
amplitudes are implemented as efficient analytic expressions.
All NNLO processes were originally computed using the jettiness
slicing method~\cite{Boughezal:2015dva,Gaunt:2015pea}. Since version 9.0, subleading jettiness 
power corrections \cite{Moult:2016fqy,Boughezal:2016zws,Ebert:2018lzn} are included that allow for 
the calculation of most NNLO processes to per mill 
precision on a desktop computer within a few hours \cite{Campbell:2019dru}. The latest
release of the code~\cite{Campbell:2022gdq} will also allow the calculation
to be performed using $q_T$ slicing~\cite{Catani:2007vq}. 

\paragraph{Fixed-order NNLO calculations}
MCFM is capable of computing a range of color-singlet Drell-Yan-like
processes at NNLO~\cite{Boughezal:2016wmq}, with phenomenological applications focusing on Standard 
Model measurements. In addition, the capability of the code to compute
diboson production processes has steadily expanded from
$\gamma\gamma$~\cite{Campbell:2016yrh} and $Z\gamma$~\cite{Campbell:2017aul}
to also include $W\gamma$~\cite{Campbell:2021mlr} and all remaining
($WW$, $ZZ$ and $WZ$) processes~\cite{Campbell:2022gdq}.

MCFM places special emphasis on Higgs production, including NNLO corrections for Higgs 
production in the infinite top-mass limit \cite{Boughezal:2016wmq}.
Finite top-quark mass effects in H+jet production can be included at NLO 
\cite{Neumann:2016dny,Neumann:2018bsx}. Together with the included transverse momentum resummation 
(see below) \cite{Becher:2020ugp}, state-of-the-art theory predictions for Higgs observables can be 
computed. Lastly, Higgs boson + jet predictions in gluon fusion at NNLO are included 
~\cite{Campbell:2019gmd} as well as $b\bar{b}$ induced contributions~\cite{Mondini:2021nck} at 
NNLO, 
but both are not in the public version yet.

The
public version of the code also includes the capability to calculate $t$-channel single
top production, and decay, at NNLO~\cite{Campbell:2020fhf,Campbell:2021qgd}. The implementation 
includes $b$-quark tagging and double-deep-inelastic-scattering scales \cite{Campbell:2021qgd} that 
allow for constraining PDFs. Predictions in the 
SMEFT have been included 
for off-shell t-channel single-top-quark production at NLO \cite{Neumann:2019kvk}. MCFM therefore 
provides state-of-the-art predictions for this process.

By using a 1-jettiness factorization theorem,
MCFM can also been used to compute a range of processes
that also contain a final-state jet: $W$~+ jet~\cite{Boughezal:2015dva}, 
$Z$~+jet~\cite{Boughezal:2015ded},
$\gamma$~+jet~\cite{Campbell:2016lzl}, and Higgs boson + jet~\cite{Campbell:2019gmd}.
These processes have not been made available in the public version yet~\cite{Campbell3:2022}.

\paragraph{Resummation of large logarithms: CuTe-MCFM}
\label{sec:cute-mcfm}
With the release of version $10.0$, MCFM includes transverse momentum ($q_T)$ resummation for the NNLO boson and 
diboson 
processes under the name CuTe-MCFM \cite{Becher:2020ugp}. CuTe-MCFM implements
a $q_T$-resummation 
formalism~\cite{Becher:2010tm,Becher:2011xn,Becher:2012yn}  where large logarithms
of argument $q_T/Q$ are resummed through RG evolution of hard function and beam functions, and 
rapidity logarithms are directly exponentiated through the collinear-anomaly formalism.
CuTe-MCFM accounts for linear power corrections in the resummation with a Lorentz-boost recoil
prescription. The boundary conditions of the RG evolution are set
directly in $q_T$ space, thereby avoiding any Landau pole
singularities.
The matching to fixed-order is performed using a 
flexible transition function that allows for a transparent matching and estimation of matching 
uncertainties.

Overall, the resummation in CuTe-MCFM is performed at order $\alpha_s^2$ in RG-improved 
perturbation theory (N$^3$LL+NNLO) for the processes $H,W^\pm, Z,
\gamma\gamma,Z\gamma,ZH, W^\pm H$, with
$ZZ, WZ, WW$ to be included soon.

For $\gamma\gamma$ production further three loop virtual corrections and three loop beam functions 
have been included \cite{Neumann:2021zkb} for predictions at N$^3$LL'+NNLO. Such three loop virtual 
corrections are also available for $W^\pm,Z$ and $H$ production and will lead to resummed 
predictions at N$^3$LL'+N$^3$LO and to resummation improved fixed-order prediction at N$^3$LO in 
the future~\cite{Campbell3:2022}.

\paragraph{Beyond QCD \& MCFM as a library}
Although MCFM emphasizes higher-order calculations in QCD, at this level of precision it is
often desirable -- or mandatory -- to also include the effect of electroweak 
contributions at NLO.  For a small number of processes, these effects may also
be computed~\cite{Campbell:2016dks,Campbell:2021mlr}. 

Efficient implementations of tree and one-loop amplitudes in MCFM are also available as a 
library interface \cite{Campbell:2021vlt}.   The use of this library
can decrease the execution time of parton 
shower codes 
relying on those 
amplitudes, sometimes by orders of magnitude compared to automated amplitude generators. The extension of
the MCFM library of processes using numerical reconstruction methods is an area
of current development~\cite{Campbell2:2022}.

\paragraph{Outlook}
Future developments for MCFM include providing fully differential N$^3$LO predictions for 
the standard candle processes of $W^\pm,Z$ and $H$ production and
increasing the 
precision of electroweak contributions, non-fixed-order and non-perturbative effects, as well as improvements 
that go beyond current approximations like mass effects and non-factorizable contributions. The 
authors are 
furthermore working on interfacing MCFM with a parton shower, providing access to hadronization and 
increased precision in kinematic regimes that benefit from the resummation of soft gluon radiation.

\subsubsection{NNLOJET}
\NNLOJET is a parton-level event generator to perform fully differential fixed-order calculations at next-to-next-to-leading order (NNLO) using the antenna subtraction method \cite{Gehrmann-DeRidder:2005btv,Currie:2013vh}.
It can compute arbitrary infrared safe observables, incorporates a
semi-automated tool chain to set up the subtraction for different
processes,
and includes a sophisticated testing suite to validate the construction and implementation.

A unique feature of the antenna method is the ability to exploit universal factorization properties of \emph{color-ordered amplitudes} that are particularly economical in setting up the subtraction due to the various singular limits that are captured within a single factorized object, called the antenna function.
While a construction of subtraction terms thus requires an appropriate decomposition of squared amplitudes in terms of different color structures, the resulting hierarchy of gauge-invariant color levels \(1\), \(N_f/N_c\), \(1/N_c^2\), \(N_f^2/N_c^2\), \(N_f/N_c^3\), \(1/N_c^4\), \(\ldots\) readily lends itself for multi-channel adaption techniques to optimize their sampling and thus significantly improve the efficiency of the calculation.
This is particularly relevant for NNLO computations where performance is increasingly becoming a major point of concern.
Although the use of a color decomposition may suggest that the method is solely applicable to QCD corrections, an extension to electroweak (EW) corrections can be realized in a straightforward manner.
Photonic real-emission corrections are reminiscent of \(U(1)\) gluons that arise from decoupling identities and which naturally appear in the sub-leading color contributions of QCD amplitudes.
As such, the subtraction  for EW corrections proceeds in an analogous manner as the sub-leading color corrections and are in fact significantly simpler owing to the absence of genuine color interference terms that can complicate the construction in the case of QCD corrections \cite{Currie:2013dwa}.

The \NNLOJET framework has been successfully applied to a large number of important processes in \(\mathrm{e}^+\mathrm{e}^-\), deep-inelastic scattering (DIS), and hadron-hadron collisions:
\(\mathrm{e}^+\mathrm{e}^-\to 3j\) \cite{Gehrmann-DeRidder:2014hxk,Gehrmann:2017xfb},
(di-)jets in DIS \cite{Currie:2017tpe,Niehues:2018was},
\(pp\to \text{(di)-jets}\) \cite{Currie:2016bfm,Currie:2017eqf},
\(pp\to \gamma\gamma\) \cite{Gehrmann:2020oec},
\(pp\to \gamma+j/X\) \cite{Chen:2019zmr},
\(pp\to V+j\) \cite{Gehrmann-DeRidder:2015wbt,Gehrmann-DeRidder:2016cdi,Gehrmann-DeRidder:2017mvr},
\(pp\to H+j\) \cite{Chen:2016zka},
\(pp\to VH(+\mathrm{jet})\) \cite{Gauld:2019yng,Gauld:2020ced,Gauld:2021ule},
and Higgs production in VBF \cite{Cruz-Martinez:2018rod}.
A generic blueprint for the subtraction is in place for any hadron-collider process up to two resolved jets and any color-less final state.
These calculations have demonstrated repeatedly the importance of NNLO corrections to match the precision of the experiments and in some cases even resolve some tensions observed at lower orders.
With the goal of consolidating our description of nature and increasing the sensitivity to subtle deviations, it becomes crucial to extend calculations at this order to a wider set of processes, in particular, tackling the frontier of \(2\to3\) scattering reactions.
This will require an extended level of automation of the antenna formalism to eliminate the remaining manual steps in the assembly of subtraction terms.

Most of the precise predictions in the Standard Model that are
considered standard candles are also crucial ingredients for constraining the parton distribution function (PDF) of the proton; the associated errors often being among the dominant sources of theory uncertainties.
The use of full NNLO calculations in the extraction of PDFs and other
Standard Model parameters, however, is still limited due to the
computational cost of repeated calculations.
The use of fast interpolation grids addresses this bottleneck and has been implemented in the APPLgrid \cite{Carli:2005ji,Carli:2010rw} and fastNLO \cite{Kluge:2006xs,Britzger:2012bs} packages.
The extension of this approach to NNLO predictions has been achieved for the case of DIS in \cite{Britzger:2019kkb} and successfully applied to the extraction of \(\alpha_\mathrm{s}\) and PDFs using HERA data \cite{H1:2017bml,Britzger:2019kkb,ZEUS:2021sqd}.
Further details on the associated APPLfast interface are given in Sec.~10 of~\cite{PDFWP}.

\paragraph{Beyond on-shell states and challenges with identified states}
While tremendous progress has been made in computing NNLO predictions, essentially completing all relevant \(2\to1\) and \(2\to2\) scattering processes, in some cases there remain residual approximations as well as a mismatch between theory and experiment that can potentially limit the interpretation of the data.

\paragraph*{Beyond on-shell states} The Higgs boson is a prime example where an
intermediate state can be treated as a stable particle because of its narrow width.
However, in order to faithfully model the event selection cuts and facilitate a direct comparison between theory and experiment at the level of fiducial measurements, it is necessary to include the decay of unstable particles in the calculations.
The application of unfolding corrections back to, e.g. stable Higgs bosons, can potentially spoil the formal accuracy of such a comparison.
Moreover, the interplay between fiducial cuts and radiative corrections can also impact the perturbative stability of the prediction.
This has been observed for the case of \(H\to4\ell\) decays
\cite{Chen:2019wxf} in the different treatment of the lepton isolation
(rejecting leptons as opposed to jets), as well as in the Higgs Dalitz
decay \cite{Chen:2021ibm} in the almost degenerate low-mass region.
Fiducial cuts close to back-to-back kinematics were found to potentially induce large power corrections \cite{Salam:2021tbm} that were also identified in \cite{Chen:2021isd} through large higher-order corrections.
Finally, in case the decay products involve colored particles, a proper combination of higher-order corrections to the production and decay sub-processes should be taken into account.
Ambiguities in combining the individual corrections that arise in such calculations have been explored \cite{Gauld:2019yng} for the case of associated Higgs production with the subsequent decay of the Higgs into bottom quarks, revealing potential inadequacies in estimating missing higher orders by employing the customary re-scaling procedure.

\paragraph*{Photon identification and isolation} The measurement of scattering reactions involving final-state photons requires the application of isolation requirements in order to tame the overwhelming backgrounds from non-prompt photons.
Current calculations at NNLO rely on a convenient trick of imposing a smooth isolation prescription \cite{Frixione:1998jh} to eliminate the fragmentation component without jeopardizing infrared safety.
This isolation prescription however cannot be implemented exactly in
experimental measurements and the resulting mismatch between theory
and experiment has been the subject of many studies \cite{SM:2012sed,Andersen:2014efa,Andersen:2016qtm,Catani:2018krb,Cieri:2015wwa,Hall:2018jub}, with further refinements proposed such as the hybrid isolation prescription \cite{Siegert:2016bre,Amoroso:2020lgh,Chen:2019zmr}.
The intrinsic ambiguity in the theory predictions that stems from the removal of the collinear fragmentation contribution can have an impact at the percent level.
Precise predictions for processes with external photons thus demand a
treatment of the fragmentation contribution at NNLO.
The necessary subtraction terms and their integrated counterparts have been computed recently in the antenna formalism \cite{Gehrmann:2022cih}.
This allows the same fixed-cone isolation as in the experiments to be
used, entirely removing the aforementioned ambiguity.
Predictions based on these calculations further pave the way for a determination of the photon fragmentation function at NNLO.

\paragraph*{Jet flavor and identified hadrons} The identification of jet flavor within a calculation based on massless flavored quarks suffers from conceptual problems that are linked to infrared safety.
These issues can be mitigated by an appropriate adjustment of the jet definition \cite{Banfi:2006hf}, which however cannot be easily applied in the experimental measurements.
Accounting for the mismatch of the jet definition between theory and experiment through (un-)folding can induce sizable corrections \cite{Gauld:2020deh} that can spoil the formal accuracy of the prediction and introduces an additional source of uncertainty.
While a calculation based on massive flavored quarks does not suffer
from such issues and can use any jet definition, the calculation itself becomes substantially more challenging due to the additional scale in the problem.
What is more, the finite mass renders the calculation infrared safe through logarithmic terms of the form \(\log(m_q^2/Q^2)\).
Such logarithmic corrections might not yet be alarming in the case of
bottom quarks, but could jeopardize the fixed-order expansion for lighter flavors and should ideally be resummed to all orders through DGLAP evolution equations.
Such a calculation would require the extension of NNLO calculations with identified hadrons, and consequently the inclusion of the associated hadron fragmentation functions.
Important first steps have been taken within the antenna formalism and
the \NNLOJET framework to handle such identified heavy-flavor configurations.

\paragraph{Going beyond NNLO}
Recent years have seen the advent of differential N$^3$LO predictions for benchmark \(2\to1\) processes, unlocking the next level of precision phenomenology.
These calculations all employ subtraction methods that rely on an existing NNLO calculation for the same process but with an additional resolved jet.
In the context of an N$^3$LO calculation, this additional emission is allowed to become unresolved to probe the genuine N$^3$LO infrared limits.
This poses a formidable challenge on the stability and efficiency for the underlying NNLO calculation to obtain reliable results.

The \NNLOJET framework has been successfully used to perform differential calculations at this order both using the projection-to-Born (P2B) method \cite{Cacciari:2015jma} as well as the \(q_T\) subtraction method \cite{Catani:2007vq}.
The P2B method relies on an inclusive calculation that retains the differential information on the Born kinematics.
The DIS structure function \cite{Vermaseren:2005qc,Moch:2008fj} and the Higgs boson rapidity distribution \cite{Dulat:2018bfe} available at this order constitute such inclusive predictions that enabled the fully differential calculation for jet production in DIS \cite{Currie:2018fgr,Gehrmann:2018odt} and Higgs production \cite{Chen:2021isd}.
While the \(q_T\) subtraction method does not rely on any inclusive calculation as input, the introduction of a resolution variable as a slicing parameter makes this approach substantially more prone to numerical instabilities and thus also very challenging computationally.
Several optimizations in \NNLOJET that target the stability of both the
amplitudes and the phase space have enabled
the stable computation of the cross section and rapidity distributions for Higgs \cite{Cieri:2018oms} and the Drell-Yan process \cite{Chen:2021vtu}.

\subsubsection{\texorpdfstring{\nlox}{NLOX}}
\nlox is a one-loop provider that allows for the automated calculation
of one-loop QCD and electroweak (EW) corrections to Standard Model
(SM) scattering amplitudes. Based on Feynman-diagram techniques, it
has been developed to optimize the manipulation of symbolic and tensor
structures recurring in one-loop amplitudes.  The first release of the
code (v1.0)~\cite{Honeywell:2018fcl} consisted of the main
tensor-integral reduction library, \tred, and several
pre-generated processes. Further processes have been produced upon
demand for various projects and are publicly available from the \nlox library of processes.

Since v1.0 the code has seen a substantial number of important
additions aimed at improving the numerical stability of the one-loop
amplitudes and the interface to external codes. Among the most
important additions present in the current
v1.2~\cite{Figueroa:2021txg} are improved stability checks in the
tensor reduction performed by \tred and new routines that
check at runtime the IR pole structure of virtual amplitudes against
the IR pole structure of the corresponding real emission, order by
order in the QCD+EW couplings. These new IR checks are implemented
using color-correlated amplitudes that are now also independently
obtainable from the code.

\nlox utilizes \qgraf~\cite{Nogueira:1991ex},
\form~\cite{Vermaseren:2000nd,Kuipers:2012rf}, and \python to
algebraically generate \cpp code for the virtual QCD and EW one-loop
contributions to a certain process in terms of one-loop
tensor-integral coefficients.  The tensor-integral coefficients are
calculated recursively at runtime through standard reduction methods
by the \cpp library \tred, an integral part of \nlox.

\nlox uses dimensional regularization for UV and IR singularities.
For QCD renormalization, a modified on-shell scheme is used for the
wave-function and mass renormalization of massive quarks, while the
$\msbar$ scheme is used for massless quarks and gluons, where,
however, in the latter case heavy-quark-loop contributions are
decoupled by subtracting them at zero momentum. EW renormalization is
implemented in the on-shell renormalization scheme or, in the presence
of potential resonance channels, in the complex-mass
scheme. As EW input scheme choices \nlox provides
both the $\alpha(0)$ and the $G_\mu$ EW input schemes. All core functionalities of \nlox v1.0 have
been presented in detail in Ref.~\cite{Honeywell:2018fcl}.

NLOX has been successfully interfaced with NLO QCD Monte Carlo event
generators. In particular it has been used in the \powhegbox
implementation of $t\bar{t}W^\pm$~\cite{FebresCordero:2021kcc}, $t\bar{t}Z$, and
$t\bar{t}l^+l^-$~\cite{Ghezzi:2021rpc}, as well as for studies of
$W+b$-jets~\cite{ Reina:2011mb} and $Z+b$-jet~\cite{Figueroa:2018chn}
studies in both the 4- and 5-flavor schemes. In one of its earlier
stages it further partook in a technical comparison of tools for the
automation of NLO EW calculations \cite{Proceedings:2018jsb}.

Current and previous releases of the \nlox package, the process
archives, and a simple set of instructions on how to download and install the 
various components can be found on \url{http://www.hep.fsu.edu/~nlox/}.

\subsubsection{\texorpdfstring{\OpenLoops}{OpenLoops}}

\OpenLoops~\cite{Cascioli:2011va, Buccioni:2017yxi,Buccioni:2019sur} 
is a fully automated generator of scattering amplitudes at
tree and one-loop level.  It is based on a numerical
recursion that combines the building blocks of Feynman diagrams 
by means of process-independent operations, which depend only on the Feynman rules of
the theoretical model at hand~\cite{Cascioli:2011va,
  Buccioni:2017yxi}.
The \OpenLoops algorithm supports the generation of scattering amplitudes
for arbitrary NLO calculations within the full Standard Model, i.e.~for all
amplitudes that involve strong, electroweak, Yukawa and
scalar integrations at tree and one-loop level.
The code consists of a process-independent part and a collection of libraries
that contain automatically generated numerical code for specific
scattering processes.  
The public \OpenLoops process library covers a very wide range of
processes at hadron and lepton colliders.  To date it includes more than
2600 one-loop amplitudes for the QCD and electroweak corrections to
different (crossing-independent) partonic processes.  Tree amplitudes for
the associated real-emission processes are also available.  
When relevant, such amplitudes are available in different variants,
e.g.~with massless or massive $b$-quarks, including or excluding heavy-quark
loops, etc.  Dedicated process libraries that implement special
approximations, such as color expansions or specific diagrammatic filters, can
be generated in an automated way and made publicly available upon request.

The \OpenLoops\,2 code~\cite{Buccioni:2019sur} 
has native {\sc Fortran} and \textsc{C/C++} interfaces that
allow its functionalities to be accessed by other codes in a flexible way.
Besides the choice of processes and parameters, the interfaces support the
calculation of LO, NLO, and loop-induced matrix elements and building blocks
thereof, as well as various color and spin correlators relevant for the
subtraction of IR singularities at NLO and NNLO.  
Additional interface functions give access to the SU(3) color basis and the
color flow of tree amplitudes.
\OpenLoops can be used as a plug-in by
\sherpa~\cite{Gleisberg:2008ta,Bothmann:2019yzt},
\powheg~\cite{Alioli:2010xd}, \herwig~\cite{Bellm:2015jjp},
\geneva~\cite{Alioli:2012fc}, and \whizard~\cite{Kilian:2007gr}, which
possess built-in interfaces that control all relevant \OpenLoops
functionalities in a largely automated way, requiring only little user
intervention.
In combination with these Monte Carlo programs \OpenLoops has been used
to generate NLOPS samples for several processes, 
including nontrivial multi-particle and multi-scale
LHC production processes such as $t\bar t b \bar b$~\cite{Cascioli:2013era,Jezo:2018yaf}, 
$W^+W^- b\bar b$~\cite{Jezo:2016ujg} and $4\ell/2\ell2\nu$~\cite{Cascioli:2013gfa,Alioli:2021wpn} at 
NLO~QCD and $HZ/HW+$jet~\cite{Granata:2017iod} at NLO QCD+EW.

Technically, in \OpenLoops one-loop amplitudes are expressed as linear
combinations of tensor integrals, which can be reduced to scalar integrals
using external reduction programs like the \collier library~\cite{Denner:2016kdg} 
or, alternatively, a built-in algorithm based on the on-the-fly reduction 
method~\cite{Buccioni:2017yxi}.
The required scalar integrals are evaluated by means of 
\Collier~\cite{Denner:2016kdg} and
{\sc OneLoop}~\cite{vanHameren:2010cp}.

One of the attractive features of \OpenLoops---and of automated
amplitude generators in general---lies in the fact that many important
aspects of scattering amplitudes (such as the color and helicity management,
the reduction of tensor integrals, etc.) can be managed and systematically
improved in a process-independent way.
For this reason, many technical improvements or new user-requested features
can be addressed in a general manner and then deployed through the
full process library in a fully automated way.

The \OpenLoops algorithm is designed to guarantee
excellent CPU efficiency for challenging multi-leg amplitudes at one loop.  
In fact, the automation of one-loop calculations has enabled 
NLO calculations for a wide spectrum of $2\to N$ processes with
$N=4,5$ final-state partons and even beyond.
In \OpenLoops, the CPU cost of helicity- and color-summed one-loop
corrections scales linearly with the number of Feynman diagrams, 
and applications to processes with up to $\mathcal{O}(10^5)$ one-loop diagrams 
per partonic channels are possible.
For instance, \OpenLoops has been used to compute the NLO QCD
corrections to $pp\to t\bar t+3$jet~\cite{Hoche:2016elu} 
and $t\bar t b\bar b+$jet~\cite{Buccioni:2019plc}, 
which are among the most complex NLO calculations to date.

\OpenLoops is widely used also for low-multiplicity
$2\to 2$ and $2\to 3$ one-loop calculations, which are typically much
faster. On average the CPU cost of one-loop amplitudes grows by a
factor 20--30 for each additional external leg.
For this reason, the further optimization of CPU efficiency is not a priority for such
simple processes. In principle, when available, analytic one-loop
amplitudes for specific $2\to 2$ and $2\to 3$ processes can be significantly
faster.
However, the potential NLO speedup that can be achieved by replacing $2\to
2$ or $2\to 3$ \OpenLoops amplitudes by analytic ones is typically very
modest, since \OpenLoops amplitudes are already fast enough.  In particular
they are typically quite far from being the bottleneck of a full NLO
calculation.

\OpenLoops\,2~\cite{Buccioni:2019sur} implements the
on-the-fly reduction technique~\cite{Buccioni:2017yxi} to control
numerical instabilities.
A sophisticated, automated monitoring scheme is used to assess the
stability of calculations and trigger different calculational
techniques when they are beneficial.

In the case of NNLO applications, where one-loop amplitudes need to be evaluated
in the unresolved regions of phase space, numerical instabilities are
potentially very large, and the stability techniques play a key role in order to ensure the correct cancellation
of infrared singularities.
Thanks to the above mentioned techniques and further optimizations 
in the unresolved regions, \OpenLoops has been extensively and successfully
applied in $2\to 2$ NNLO calculations and also in some of the first $2\to 3$
NNLO calculations~\cite{Kallweit:2020gcp,Czakon:2021mjy}.
In particular, \OpenLoops is used as a building block of
\Matrix~\cite{Grazzini:2017mhc} for the calculation of a wide spectrum of
color-singlet and heavy-quark production processes at NNLO QCD.  
In this context, the automation of EW corrections in \OpenLoops opens
the door to ubiquitous NNLO\,QCD+NLO\,EW calculations in 
\Matrix~\cite{Grazzini:2019jkl}. Moreover, \OpenLoops+\sherpa 
support automated NLO\,QCD+NLO\,EW simulations
at fixed-order~\cite{Kallweit:2014xda,Kallweit:2017khh}, 
and matched to QCD+QED parton showers in an approximation where QED
radiation is integrated out at the level of the virtual 
EW amplitudes~\cite{Kallweit:2015dum,Gutschow:2018tuk}.

\OpenLoops implements various renormalization schemes, including 
the possibility of using the weak mixing angle as
input parameter~\cite{Chiesa:2019nqb}.  Unstable particles can be handled by means of a
flexible implementation of the complex-mass scheme~\cite{Denner:2005fg}.
Moreover, \OpenLoops\,2 implements a fully general mechanism for the correct
choice and renormalization of EW couplings  in processes with resolved,
unresolved or initial-state photons.  A thorough documentation of all
aspects of QCD and EW corrections can be found in~\cite{Buccioni:2019sur}.

The ongoing development activities include the preparation of a new version
of \OpenLoops that will feature a new implementation of the on-the-fly
reduction technique~\cite{Buccioni:2017yxi} at the level of tensor integrals.  As already
demonstrated in~\cite{Chen:2021azt}, this new reduction algorithm will drastically
improve the numerical stability of NLO calculations for loop-induced
processes, which require squared one-loop amplitudes with unresolved
radiation.
Moreover, it will lead to
additional significant speed and stability improvements for 
the interference of Born and one-loop amplitudes, 
especially in the context of NNLO and complex multi-leg NLO
applications.

Recent progress also includes the development of two-loop
rational terms~\cite{Pozzorini:2020hkx,Lang:2020nnl,Lang:2021hnw} 
and the automated construction of two-loop
integrands~\cite{Pozzorini:2022ohr}. These developments represent important 
steps towards the automation of NNLO calculations within {\sc OpenLoops}.

\subsubsection{\texorpdfstring{\recola}{Recola}}
\recola \cite{Actis:2012qn,Actis:2016mpe}  is an automated tree- and one-loop matrix element generator
for the Standard Model and Beyond the SM theories, originally
tailored to computations within the
SM in the 't~Hooft--Feynman gauge. The generalized version, \recola2
\cite{Denner:2017vms,Denner:2017wsf}, allows for computations in the
SM, in BSM theories, and in SMEFT, as far as corresponding model files
exist.   
\recola generates
all the needed ingredients for tree-level and one-loop computations for arbitrary
processes. The matrix elements are generated on-the-fly in memory, \ie
without generating process source code.  \recola uses the one-loop
library \collier \cite{Denner:2016kdg} for the calculation of one-loop
scalar and tensor integrals.

\recola1.4.0 provides
tree-level and one-loop amplitudes with helicities and colors of
external particles. Squared amplitudes are given either polarized or
unpolarized. In addition, color-correlated and spin-correlated squared
amplitudes are available for dipole subtraction in NLO calculations \cite{Catani:1996vz}.
\recola uses dimensional regularization for UV and IR singularities
and optionally mass regularization for collinear singularities. It
supports the $\alpha(0)$, $\alpha(M_Z)$, and $G_F$ schemes for the
renormalization of the electromagnetic coupling and fixed and variable
flavor schemes for the renormalization of the strong coupling.
Unstable particles can be treated with the complex-mass scheme
\cite{Denner:2005fg,Denner:2019vbn}. Alternatively, factorizable
contributions can be selected for use of a pole approximation
\cite{Denner:2019vbn}.

\recola 2.2.3 extends these capabilities by 
allowing for model files of SM extensions and
the use of the Background-Field
Method \cite{Denner:1994xt}. It implements memory optimizations for
processes related by crossing symmetry and offers tree-level and one-loop spin- and
color-correlated squared amplitudes.  A \Python interface is
also available.

\recola has been used together with the private Monte Carlo integration codes
\bbmc and \mocanlo for various cutting-edge fixed-order calculations,
including the electroweak corrections to vector-boson scattering
\cite{Biedermann:2016yds,Biedermann:2017bss,Denner:2019tmn, Denner:2020zit}, electroweak
corrections to top-pair production in association a $W$-boson that decays to leptons
\cite{Denner:2021hqi}, and QCD corrections to
top-pair production in association with bottom pairs
\cite{Denner:2020orv}.

Recently,
\recola 1.4.2 has been extended to enable the selection of helicity
states for intermediate electroweak bosons and top quarks in resonant
tree-level and one-loop SM amplitudes.  
This feature, in combination
with a double-pole approximation \cite{Denner:2000bj}, has been
employed for the calculation of QCD and electroweak corrections to
the production of polarized vector-boson pairs decaying to 
leptons \cite{Denner:2020bcz,Denner:2021csi}.

\recola has been interfaced to several public Monte Carlo
generators.  In \sherpa \cite{Sherpa:2019gpd}, the interface
\cite{Biedermann:2017yoi} is fully general for NLO QCD and EW
corrections and was used for several NLO computations
\cite{Schonherr:2018jva,Reyer:2019obz,Baberuxki:2019ifp,Brauer:2020kfv,Caletti:2021oor,Reichelt:2021svh,Bothmann:2021led}.
A public interface to \whizard2.4.1 exists
\cite{Brass:2018xbv}, while an interface to \herwig exists but is not yet
public \cite{Platzer:2022}.
Moreover, a basic interface between \recola and \powhegbox has been used
to construct an event generator for same-sign $W$-boson scattering
at the LHC including electroweak corrections \cite{Chiesa:2019ulk}.
This interface has been subsequently generalized for the computation of NLO QCD+EW corrections to di-boson production~\cite{Chiesa:2020ttl}.

Finally, it is worth mentioning that \recola has been benchmarked against
several other programs in
Refs.~\cite{Andersen:2016qtm,Proceedings:2018jsb} at the level of
phase-space points and differential distributions for various
processes.

\subsubsection{\stripper}

The \stripper code has been at the forefront of NNLO QCD calculations for many years.  It is a C++ implementation of the  four-dimensional formulation of the sector-improved residue subtraction scheme~\cite{Czakon:2010td,Czakon:2011ve,Czakon:2014oma}.  A detailed description of the implementation is given in Ref.~\cite{Czakon:2019tmo}.

This approach was initially employed in the pioneering calculation of top-quark pair production at NNLO~\cite{Czakon:2014xsa,Czakon:2015owf,Czakon:2016ckf} and has since been substantially extended.  It has now been used to provide LHC predictions for inclusive jet~\cite{Czakon:2019tmo}, three photon~\cite{Chawdhry:2019bji}, $W+c$~\cite{Czakon:2020coa}, identified $B$-hadron~\cite{Czakon:2021ohs}, polarized $W$-pair~\cite{Poncelet:2021jmj}, diphoton+jet~\cite{Chawdhry:2021hkp} and three-jet~\cite{Czakon:2021mjy} production.

\subsection{QCD parton showers and dipole showers}
\label{sec:hadron_collider_evolution}
QCD parton showers are of particular relevance to high-energy physics and play a crucial role in our understanding of physics at hadron colliders.
QED parton showers are similarly important for lepton collider physics
and are discussed in the section on lepton colliders, Sec.~\ref{sec:lepton_colliders}.

The main role of parton showers in event generators is to connect physics processes over (potentially) several orders of magnitude in momentum.
They relate the high scales of hard processes, treated with standard
perturbative methods, and the low momentum scales of hadronization and
beam fragmentation, which are usually described by phenomenological
methods that rely on a rather qualitative understanding of the relevant dynamics.
Parton showers can be understood as a numerical implementation of
evolution equations, where the evolution in momentum scales is driven
by the emissions of secondary quanta -- in QCD parton showers these
secondary quanta are quarks and gluons.
This has made parton showers an essential component of any realistic simulation of collider processes for several decades, reflected in ongoing development and refinement of their construction principles.
The earliest parton showers were the traditional virtuality- and angular--\-ordering-based approaches in \pythia~\cite{Sjostrand:1985xi,Bengtsson:1986et} and \herwig~\cite{Marchesini:1983bm}, and the dipole shower implementation in \ariadne~\cite{Gustafson:1987rq,Kharraziha:1997dn,Lonnblad:1992tz}.
Improvements of these algorithms and new approaches are now deployed within \herwig~\cite{Corcella:2000bw,Gieseke:2003rz,Platzer:2009jq}, \pythia~\cite{Sjostrand:2004ef,Cabouat:2017rzi}, and \sherpa~\cite{Kuhn:2000dk,Schumann:2007mg}, and the more recent additions \vincia~\cite{Giele:2007di,Giele:2011cb} and \dire~\cite{Hoche:2015sya}.

Recently, more attention has been paid to parton showers by the theory community.
This is because 
current limits in our understanding of the parton shower translate directly into limiting systematics in the interpretation of experimental measurements.
One example of such limitations concerns the calibration of the jet energy scale at the LHC.
Systematic differences between generators, notably an understanding of the difference in the parton showering between gluons and quarks~(see e.g.~\cite{CMS:2016lmd}), translate into the largest systematic uncertainty on the jet energy scale over a broad range of momenta (see e.g.~\cite{ATLAS:2020cli}), ultimately affecting hundreds of experimental publications from the LHC.
Differences between generators are also a limiting factor in the search for new physics and for measurements at the highest energy scales accessible in the laboratory, for example through uncertainties in tagging efficiencies for highly boosted objects (see e.g.~\cite{ATLAS:2021wkg}). 
This latter issue is likely to be exacerbated in the coming years,
since machine learning techniques could enable us to exploit the
information within each event's pattern of parton shower radiation, if
the parton shower can be understood sufficiently well.
Furthermore,  improving the physics of parton showers has the
potential for substantial physics gains across the broad spectrum of Higgs physics, BSM searches and SM measurements, as long as we can control the systematics associated with the quantitative understanding of their predictions.
We will address multiple paths for systematically improving parton showers in the following.

\subsubsection{Quantifying the Logarithmic Accuracy of Parton Showers}
The accuracy of a parton shower can be discussed through its
connection with resummation and logarithmic accuracy,
specifically control of the terms $\as^n L^m$ where $L$ is the logarithm of the ratio of two disparate momentum scales.
Analytic resummation techniques can achieve high logarithmic accuracy~(e.g.\ N$^3$LL for the Drell-Yan and Higgs-boson $p_T$ distribution~\cite{Bizon:2017rah,Bizon:2018foh}).
The advantage of parton showers over analytic resummations is that
they provide an effective resummation for any observable, as well as
the fully differential information that makes hadron-level predictions
possible.
This is crucial for experimental observables that cannot be obtained from analytic resummation.
However, so far, obtaining fully differential information at the event
level has come at the price of lower formal accuracy,
specifically leading-logarithmic (LL) accuracy, which, broadly speaking, accounts for terms $\as^n L^{n+1}$.
The next set of logarithmic terms would be NLL, $\as^n L^{n}$, and for $L \sim 1/\as$, NLL terms are nominally of order one.

Recent work has introduced a taxonomy for analyzing the logarithmic
accuracy of parton
showers~\cite{Dasgupta:2018nvj,Dasgupta:2020fwr,Forshaw:2020wrq,Nagy:2020dvz}
and
led to the development of final-state dipole shower schemes that demonstrably achieve NLL accuracy for a large class of observables. At the same time, the size of higher logarithmic corrections needs to be viewed in relation to major systematic uncertainties in parton showers~\cite{Hoche:2017kst}.

There are many classes of logarithmic terms that a shower can aim to reproduce: for example Sudakov double logarithms for a variety of event-shape type observable classes, collinear single logarithms, and large-angle soft single logarithms.
Several groups are investigating the question of logarithmic accuracy within the broad class of showers that use the large-$N_C$ concept of color dipoles to organize the radiation pattern within the parton shower, which can in principle address all of these classes of terms simultaneously.

An important class of shower algorithms are of the angular--\-ordered kind, which implement the coherent branching formalism, a technique that is known to reproduce full-color NLL accuracy for double logarithmic global observables~\cite{Catani:1990rr,Catani:1992ua}.
A careful treatment of kinematic is needed to maintain that accuracy in a complete shower, as discussed in~\cite{Bewick:2019rbu,Bewick:2021nhc} for both final and initial states.
The collinear-oriented formalism of the \herwig angular--\-ordered
shower easily includes collinear spin correlations~\cite{Knowles:1988hu,Marchesini:1991ch,Richardson:2018pvo}.
However, angular ordering provides an incomplete treatment of soft single (NLL) logarithms~\cite{Banfi:2006gy}, which affects non-global observables~\cite{Dasgupta:2001sh}, though the numerical impact is modest in many circumstances.

Building on well established principles of logarithmic resummation,
significant progress has been made in the past two years in developing
NLL accurate final-state shower schemes with a certain freedom in
choosing recoil and dipole-partitioning schemes.
The \panscales
group~\cite{Dasgupta:2020fwr,Hamilton:2020rcu,Karlberg:2021kwr,Hamilton:2021dyz},
has outlined the core principles required to achieve NLL accuracy
in a shower, developed several showers that satisfy those principles
and provided numerical demonstrations of NLL accuracy for many
different classes of observable, covering the double-logarithmic
(global) observables, non-global observables and multiplicity, across
a range of new showers, including full color and spin treatment for
the majority of them.
%
The Forshaw-Holguin-Pl\"atzer
approach~\cite{Forshaw:2020wrq,Holguin:2020joq} has used amplitude
level evolution~\cite{Forshaw:2019ver}
to study improved dipole shower algorithms that deliver NLL accuracy
for global observables
while retaining the accuracy of non-global observables inherent to a dipole shower algorithm. It is targeted at an implementation in \herwig's dipole shower, which already includes spin correlations \cite{Richardson:2018pvo}.
The \textsc{Deductor} group has studied the summation of large logarithms for the thrust distribution in electron-positron annihilation with special attention to color \cite{Nagy:2020rmk}.
To some extent, these issues have also been discussed in the context of initial-state radiation~\cite{Nagy:2009vg}.

The above body of work lays foundations and provides an essential proof-of-principle for work that can be anticipated over the coming decade.
One major milestone that should be achievable in the next few years is the development of multiple fully-fledged public shower codes that are NLL--\-accurate across a full set of observables, and that can be used  for mainstream experimental and phenomenological applications.
The specific tasks that remain depend to some extent on the specific
algorithm. However, broadly speaking they include:
a) the extension of NLL accurate dipole showers, and their associated logarithmic analysis, to processes with incoming hadrons (for both
hadron colliders and DIS);
b) the inclusion of quark-mass effects in the showers, including their study from the point of view of logarithmic accuracy;
c) developing an understanding of the interplay between hard-process matching and logarithmic shower accuracy; and
d) where not already done, interfacing and testing with public hadronization codes such as those in \herwig, \pythia and \sherpa.

Some parts of this work (notably accuracy of the treatment of logarithms related to soft physics) connect with the handling of spin and color degrees of freedom and their associated quantum mechanical interferences.
This includes subtle questions connected with super-leading logarithms~\cite{Forshaw:2006fk,Forshaw:2008cq} and recent progress on their understanding~\cite{Catani:2011st,Forshaw:2012bi,Nagy:2019rwb,Becher:2021zkk,Forshaw:2021fxs}, discussed below in Sec.~\ref{sec:spin-color}.

A second major milestone is to provide accuracy beyond NLL, at the very least for the leading-color terms.
One motivation is that NLL accuracy implies control only of terms of order $1$.
Despite it being called \emph{next}-to-leading logarithmic, an ``order
1'' control therefore means that an NLL shower can be considered
analogous to only having LO control over fixed-order matrix elements,
which is widely acknowledged to be inadequate for precision
calculations of hard scattering processes.
There are many facets to achieving accuracy beyond NLL, including:
a) handling the matching of fixed-order calculations with showers (already alluded to above), where the suitable inclusion of NLO matching coefficients is generally a technical requirement for NNLL accuracy --- in particular, existing schemes may need to be adapted in order to be suitable as a baseline for NNLL accuracy;
b) including higher-order splitting kernels within showers, with some work on this topic already having been started in recent years, as discussed in section~\ref{sec:fully-diff-HO-showers};
c) including the associated virtual corrections, both those directly connected with the next-order splitting kernels (see e.g.~\cite{Dasgupta:2021hbh}) and those that connect with the higher orders in the cusp anomalous dimension (see e.g.~\cite{Dulat:2018vuy}); and
d) performing analytic or semi-analytic calculations that provide references and/or validations of the many different kinds of logarithmic contributions that need to be included in the shower (e.g.\ recent advances in subleading-logarithmic calculations for non-global logarithms~\cite{Banfi:2021owj,Banfi:2021xzn,Becher:2021urs}).

\subsubsection{Spin and Color Correlations}
\label{sec:spin-color}
Most parton showers algorithms are essentially probabilistic and
ignore or approximate quantum interference effects.
Amplitude-level evolution  provides a new class of algorithms.
Several quantum mechanical effects like spin correlations, or color correlations to some extent, can already be implemented in existing algorithms of the angular ordered or dipole-shower type.
In particular, these algorithms can reproduce the exact color
contributions to global observables that measure the deviation from the two jet limit,
and a recent proposal suggests how to extend this beyond the two jet limit~\cite{Forshaw:2021mtj}.
Furthermore, spin correlations can be included using the
Collins-Knowles algorithm~\cite{Collins:1987cp,Knowles:1988hu},
as is done in 
the modern \herwig angular ordered shower~\cite{Richardson:2018pvo}.

In the following, we highlight recent developments 
to dipole-type showers.
The \herwig and \panscales groups have
extended the efficient Collins-Knowles
algorithm~\cite{Richardson:2018pvo,Karlberg:2021kwr}, connecting it with
proposals to measure quantum interference structures in jet
fragmentation~\cite{Chen:2020adz,Karlberg:2021kwr}.
A further extension can 
simultaneously handle spin correlations
for soft and collinear emissions in the large-$N_c$ limit~\cite{Hamilton:2021dyz}.
The interplay of soft spin
correlations with subleading color corrections is still unclear, and
only
an amplitude-level treatment can clearly address this.
The question remains unanswered of how best to 
measure spin correlations in QCD jets in order to
explore these effects experimentally (see~\cite{Chen:2020adz} for a recent proposal).

Color matrix element corrections are an intermediate step to including subleading color correlations in the emission contributions~\cite{Platzer:2018pmd}.
These approaches can be easily included in existing algorithms using weights~\cite{Platzer:2018pmd,Isaacson:2018zdi,Hoche:2020pxj}, although numerical efficiency presents a significant challenge.

It is also possible to design shower algorithms with color corrections
in any number of clusters of commensurate-angle
energy-ordered emissions~\cite{Hamilton:2020rcu}, an approach taken by
the \panscales group.
The effective systematic expansion parameter is the maximum number
of particles in a cluster up to which exact color is used.
An implementation with up to two particles in a
cluster has been found to be highly efficient.
In the simple $e^+e^- \to 2\,\text{jets}$ process where it has been
tested, it exactly reproduces full-color NLL terms for all global observables
(as does a 1-particle-in-a-cluster version, with which
Ref.~\cite{Forshaw:2021mtj} bears similarities).
For non-global observables, while full-color accurate only to order
$\alpha_s^2 L^2$, it has been found to be able to reproduce all-order
full-color single-logarithmic calculations
\cite{Hatta:2013iba,Hatta:2020wre} to within their statistical
accuracy of about $1\%$.
Open issues include the extension to hadron-collider processes,
whether and how it can be extended to clusters with more than two
particles, and the interplay with super-leading logarithms.

The above approaches are, however, not full--\-color evolution in the
sense of an evolution at amplitude level, which is necessary to
exactly resum all sources of logarithms stemming from multiple
exchanges and emissions in unresolved limits.
This is particularly relevant for non-global and super-leading logarithms and for logarithms of global observables in events with four or more hard legs.
\subsubsection{Amplitude Evolution}
An accurate theoretical formalism for including quantum effects should
be based on 
a simultaneous dressing of the amplitude and its conjugate.
This can be understood by considering the nature of non-global observables, their duality to evolution equations like the JIMWLK equation~\cite{Jalilian-Marian:1997ubg,Jalilian-Marian:1997jhx,Weigert:2000gi,Iancu:2001ad,Iancu:2000hn,Ferreiro:2001qy}, and their generalization to include hard-collinear contributions (see e.g.~\cite{Somogyi:2005xz}) or even their structure beyond leading order~\cite{Catani:1999ss,Catani:2000pi}. 
Two existing approaches are \textsc{Deductor}~\cite{Nagy:2007ty, Nagy:2012bt, Nagy:2014mqa, Nagy:2015hwa, Nagy:2017ggp, Nagy:2019pjp, Nagy:2019bsj} and \textsc{CVolver}~\cite{DeAngelis:2020rvq}. They provide a formulation of the evolution at the level of the cross section density operator.

\textsc{Deductor} formulates evolution in color as evolution of the density matrix.
The density matrix is a linear combination of  basis elements $\iket{\{c\}_m} \ibra{\{c'\}_m}$, where $m$ denotes the number of final-state partons and $\{c\}_m$ and $\{c'\}_m$ are color basis states. 
At each parton splitting, $m \to m+1$, and the size of the color space
increases, while 
a virtual parton exchange leaves $m$ unchanged but acts as a linear operator on $\iket{\{c\}_m}$ or $\ibra{\{c'\}_m}$. 
It is straightforward to express first order parton evolution using
the exact color matrices in the QCD Lagrangian.
The resulting expressions, however, are impractical to compute numerically.
\textsc{Deductor} uses the LC+ approximation~\cite{Nagy:2012bt}, which omits certain terms in the exact-color evolution operators. 
The LC+ approximation can be applied starting with any product of basis states, $\iket{\{c\}_m}\ibra{\{c'\}_m}$, and it is exact except for soft gluon emissions or exchanges. 
It is implemented as part of \textsc{Deductor} both for electron-positron annihilation and for hadron-hadron scattering and is computationally efficient~\cite{Nagy:2015hwa}.
\textsc{Deductor} also provides a numerical estimate of the accuracy of the LC+
approximation for 
a cross section of interest~\cite{Nagy:2019pjp, Nagy:2019bsj}.

The parton shower \textsc{CVolver} solves evolution equations in color space and resums a specific class of observables. 
\textsc{CVolver} samples over intermediate color structures in the color flow
basis, and systematically expands  them around the large-$N_c$
limit~\cite{DeAngelis:2020rvq}, for both real emissions, and the
virtual evolution. Cancellation patterns present in the sub-leading
color contributions~\cite{Forshaw:2021mtj} are combined with a sophisticated sampling of color structures and a systematic expansion of the virtual evolution. 
Recently, the full-color resummation of non-global observables has been achieved within this framework, reproducing results which have otherwise been obtained with complementary methods using an ``equivalent Langevin formulation''~\cite{Hatta:2020wre}.
The ability to compare dedicated resummation and parton shower
predictions is an important cross-check of the algorithm. 

\subsubsection{Mass effects}
A realistic description of data requires that parton showers
include finite quark masses (both
genuine heavy-quark masses, and in the case of light quarks effective
non-perturbative masses)~\cite{Marchesini:1989yk,Corcella:1998rs,Norrbin:2000uu}.
Finite quark masses impact the splitting functions, the evolution variables (and the integration limits), and
kinematic mapping used to define the recoil
strategy~\cite{Catani:2002hc,Gehrmann-DeRidder:2009lyc,StollThesis}.
These effects are manifest as power corrections to the mass parameter
over the scale of the parton shower and
are responsible for the dead cone in 
gluon radiation from a heavy quark~\cite{Frixione:1997ma}.
As the parton shower evolves from the hard scale of the process down to the hadronization scale, the impact of mass 
effects become relatively more important.   How important and where depends on the
value of the mass parameters. 
So far,
however, they have been considered to be beyond any claimed
accuracy by parton showers, and as such, including them fully or
partially, or not at all, has not been seen as making the shower any
more accurate, giving a certain freedom in how to exactly account for
mass effects, and at which level.

Despite the lack of a formal statement about the importance of finite
quark masses, there are sizable effects when a gluon splits into a
heavy quark pair \cite{ATLAS:2015jru}, a significant signal and
background in and beyond the Standard Model.
A number of research directions can be envisaged to address questions of shower accuracy involving
quark masses, for example as concerns the treatment of logarithms of
the ratio of the hard scale to the quark mass.
Of particular interest is to
determine the correct evolution variable to account for such
logarithms and its relation
to the evolution variable used for massless quarks.

\subsubsection{Fully differential higher-order corrections}
\label{sec:fully-diff-HO-showers}
Most current parton shower algorithms are based on leading-order splitting functions. 
The possibility of including next-to-leading order corrections was explored several decades ago~\cite{Kato:1986sg,Kato:1988ii,Kato:1990as,Kato:1991fs}, but did not receive widespread attention until recently. 
The implementation of a fully differential parton evolution that
resums logarithms at NNLL accuracy for different observables will require the faithful implementation of splitting functions to second order in the strong coupling expansion in a fully differential form. 
Various technical challenges need to be solved in order to achieve this goal.
One potential requirement is a proper treatment of negative weighted events, which can be achieved through the weighted Sudakov veto algorithm~\cite{Hoeche:2009xc,Lonnblad:2012hz,Platzer:2011dq}.
A second requirement is a fully differential computation of splitting functions, including virtual corrections and real emission corrections, during the parton-shower evolution.
First proposals for the corresponding Monte-Carlo techniques were made in~\cite{Hartgring:2013jma,Li:2016yez}. 
The connection to the known NLO collinear and soft anomalous dimensions was established in~\cite{Hoche:2017iem,Dulat:2018vuy}, which also provided the first resummed predictions with differential higher-order kernels, as well as first (approximate) perturbative uncertainty estimates for a parton shower.
A soft-collinear overlap removal technique was discussed in~\cite{Gellersen:2021eci} and implemented in both \sherpa and \pythia.
Computations of relevant splitting functions have been performed
in~\cite{Loschner:2021keu},
providing systematic power counting and highlighting the relation
between recoil scheme, partitioning of soft radiation, and the leading
power expansion.
These computations can be applied to both probabilistic and
amplitude-level shower algorithms.
Integrated combinations of a subset of the real and virtual 
final-state splitting functions have been presented
in~\cite{Dasgupta:2021hbh}.
Although a complete solution to the problem of higher-order evolution
kernels is not yet available, notable progress has been made in both
conceptual understanding and implementation of necessary pieces.

\subsubsection{The sector shower approach}

\vincia{}~\cite{Brooks:2020upa} provides an alternative to parton
splittings based on a dipole-antenna formalism similar to that pioneered by the LEP-era \ariadne shower~\cite{Gustafson:1987rq,Lonnblad:1992tz}. 
Final-state shower aspects in common with \ariadne include $2\mapsto 3$ splitting kernels each of which contains a full soft-eikonal factor plus additional terms to reproduce the  DGLAP kernels in collinear limits, and $2\mapsto 3$ recoil kinematics in which both parents share the transverse recoil. 
This differs from DGLAP, dipole-like, and related approaches which
partially fraction the eikonal terms  and assign all of the transverse recoil to a single parent for each branching; however it still amounts to a local recoil scheme in the parlance of~\cite{Nagy:2017ggp,Dasgupta:2020fwr}. 
\vincia{} also uses a Lorentz-invariant evolution variable which, for massless final-state partons, is the same as that of \ariadne{}.

For hadron collisions, \vincia extends interleaved backwards evolution~\cite{Sjostrand:1985xi,Sjostrand:2004ef} to antenna showers with initial-initial (II) and initial-final (IF) color connections~\cite{Ritzmann:2012ca,Fischer:2016vfv}, with the coherence properties of especially the latter validated, e.g., in a recent VBF study~\cite{Hoche:2021mkv}. 
Other characteristic features include accurate treatments of mass
effects~\cite{Gehrmann-DeRidder:2011gkt}, coherent ``resonance-final''
(RF) antenna functions with collective recoils for showers in decays
of colored resonances such as top quarks~\cite{Brooks:2019xso}, and
the insertion of resonance decays as $1\to n$ branchings interleaved with the overall perturbative evolution at decay-specific scales~\cite{Brooks:2021kji}.

The most significant difference between the present version of
\vincia{} and other shower models is in how the branching phase space is populated. 
In \vincia, it is partitioned into non-overlapping sectors~\cite{Lopez-Villarejo:2011pwr,Brooks:2020upa}, each of which is characterized by a specific gluon being the softest in the event, specifically the gluon that has the smallest resolution scale.
Each such sector only receives contributions from a single splitting kernel, which in turn contains the full soft pole for the given gluon as well as its collinear pole structure up to a kinematic sector boundary at $z=1/2$, beyond which the neighboring phase-space sector takes over. 
Since only a single kernel populates each sector, the resulting algorithm is both Markovian and has a unique inverse.
Including $g\mapsto q\bar{q}$ splittings introduces an ambiguity when multiple same-flavor quark pairs have to be summed over; \vincia's sector antenna shower can then be viewed as maximally bijective.

\subsection{Matching QCD fixed-order calculations to parton showers}
\label{sec:hadron_collider_matching}

Matching parton showers to fixed-order calculations, and merging cross
sections for multiple jet multiplicities into one inclusive sample
have been central aspects of event generator development in the last
two decades, and continue to be an active field of research.
It is an important topic since these methods combine the best aspects of
perturbative calculations with those of parton showers, allowing
precise calculations to be compared to data.
This section reviews the different paradigms and state-of-the-art
approaches.
Before going into more detail, we should highlight the difference
between matching and merging.
\textit{Matching} is a modification of a fixed-order calculation and a 
combination with a parton shower so as to produce the fixed-order cross section up to higher order corrections. 
This is a well defined procedure and amounts to subtracting the parton shower contribution from the fixed order calculation at the same order, and locally in phase space. 
All other ambiguities in carrying out this calculation will be discussed in the next section. 
The accuracy of the parton shower limits the applicability of the matching paradigm. 
An alternative to matching is multijet merging.
Multijet \textit{merging} combines parton showers and fixed-order
predictions for multijet production so that there is no overlap and the transition in between jet bins is smooth and provides the physical Sudakov suppression for exclusive event topologies. 
Recent development in multijet merging allows for a combination of fixed-order calculations beyond NLO for a certain class of processes.

\subsubsection{NLO matching}
\label{sec:nlo-matching}

NLO matching by means of the \mcatnlo~\cite{Frixione:2002ik}
method has become a standard for multi-purpose event generators either through built-in capabilities as \herwig \cite{Hamilton:2008pd,Platzer:2011bc} or \sherpa \cite{Hoeche:2011fd,Hoeche:2012fm} offer them, or through external tools like \madgraph \cite{Alwall:2014hca}, which can be used both in conjunction with \herwig, \pythia \cite{Corke:2010zj}, and \vincia \cite{Hoche:2021mkv}. 
The combination of the ${\cal O}(\alpha_s)$ expansion of the parton shower with the fixed-order NLO calculation requires so-called Monte-Carlo subtraction terms. 
In their differential form, these terms can be used to render real-emission events finite, and their integrated counterpart will render the virtual corrections, together with collinear mass-factorization counterterms, finite. 
One subtlety is that the subtraction terms must be constructed so that
the pole structure of the fixed-order calculation is fully reproduced
by the parton shower. A standard parton shower is typically
inadequate, because it does not reproduce the local, full-$N_c$ color
correlations upon a fixed order expansion, except in the simplest
processes, but this problem can be overcome through generic matrix-element
corrections, or by implementing color matrix element corrections.
Parton-level events generated in this manner can then be combined with
an event generator just as for LO processes.

The \powheg{} matching method combines a matrix-element corrected parton showers with Born-local NLO $K$ factors. 
In showers featuring a matrix element
correction, the full real emission matrix element is
employed in place of the approximate splitting function, admitting a fixed
order expansion in terms of the full real emission matrix element.
This can then be used to simplify the subtracted NLO calculation. 
This \powheg{} algorithm is implemented in the \powhegbox program~\cite{Alioli:2010xd}, as well as \matchbox~\cite{Platzer:2011bc},
\whizard~\cite{ChokoufeNejad:2015kpc} and some versions of \sherpa~\cite{Hoeche:2010pf}. 
Besides the simplification of the NLO matched calculation and the reduction in negative weights, this paradigm also offers in principle more flexibility to run different shower algorithms with the same matched events. 
However, care must be taken when the shower has a different ordering
variable and different shower phase space limits.  This case might
require more complicated evolution such as truncated showering.
Further complications arise specifically in processes with resonances for which the ratio of real-emission to Born matrix elements squared can yield significant, but unphysical, values due to the steep shape of a resonance peak. 
This issue has been addressed in so-called resonance aware algorithms~\cite{Jezo:2015aia} and work has only started to understand how these processes should be handled in a more general framework.

The latest addition to the NLO matching methods is \KrkNLO. 
The crucial advantage of this approach with respect to other techniques is its simplicity. 
The \KrkNLO method is implemented in \herwig 7 and owes its simplicity to the use of parton distribution functions in a novel factorization scheme that is tailored to match the Parton Shower event generator~\cite{Jadach:2015mza}. 
Extending the \KrkNLO method to higher precision and more processes  (currently it is only available for Higgs~\cite{Jadach:2016qti} and $Z$ boson production~\cite{Jadach:2015mza}) and its automation using the \matchbox machinery are important steps in the development of the method. 
It has also been proposed to combine the \KrkNLO and \mcatnlo{}
methods~\cite{Nason:2021xke} to obtain the best features of each,  namely:  
(a) applicability to general showers, as with the \mcatnlo{} and
\powheg{} methods; and 
(b) positive-weight events, as with the \KrkNLO and \powheg{} methods. 
It is planned to implement this combined method in \herwig 7 and investigate it in practice.  

Algorithms that implement matrix-element corrections (MECs) are an ideal starting point for NLO matching.
The systematic treatment of MECs has been a cornerstone of development
of the \vincia\ parton shower from the early
days~\cite{Giele:2007di,Giele:2011cb},
thus making it a prime candidate.
A framework for (semi-)automated calculation of Born-local $K$ factors is currently being developed in \vincia.

\subsubsection{(N)LO multijet merging}

Multijet final states can be described by both matrix element and
parton shower calculations.
While matrix element calculations give the best description of hard
jet configurations, parton showers give the best description of soft and collinear configurations. 
Multijet merging combines the strength of each approach by describing
hard emissions with matrix element calculations and incorporating softer jet evolution by parton showers. 
Typically, a jet resolution criterion, or cut, is used to separate the hard from the soft phase space regions. 
Due to computational limitations, only a relatively small number of hard jets in the hard region are described by matrix elements, with further emissions then approximated by parton shower emissions. 
The resummation of logarithms $\log(Q_\mathrm{hard}/Q_\mathrm{cut})$
between the hard scale $Q_\mathrm{hard}$ and the introduced cut
$Q_\mathrm{cut}$
is included through parton shower Sudakov form factors, either in their analytic form~\cite{Catani:2001cc,Krauss:2002up} or dynamically generated by the parton shower itself~\cite{Lonnblad:2001iq,Lonnblad:2011xx}, a procedure often dubbed CKKW and CKKW-L. 
Parton-shower histories are constructed recursively to accomplish this
and to determine the strong coupling scales.
For high parton multiplicities, the number of possible histories grows
factorially, and a "winner-take-all" strategy can be employed, see e.g.~\cite{Hoche:2019flt}. 
In cases where the shower evolution variable does not coincide with the merging cut, a truncated shower approach might be necessary~\cite{Hoeche:2009rj,Hamilton:2009ne}. 
These original methods and their variations have been implemented in a range of event generators and results have been systematically compared~\cite{Alwall:2007fs}, providing evidence for their robustness and predictive power in describing multi-jet final states.

Typically, the integrand of the parton-shower Sudakov does not reproduce the matrix element present in the fixed order calculation. 
Thus, including higher-multiplicity matrix elements typically changes
the inclusive cross section of the merged calculation, especially for
low merging scales.
The only possible way to avoid this problem is to correct the parton shower with the full matrix-elements, in which case no merging would be necessary.
A way to reinstate the fixed order inclusive cross section is to
introduce unitarized prescriptions like UMEPS~\cite{Lonnblad:2012ng}.
An alternative method is to iteratively include matrix-element corrections (MECs), as in the \vincia{} antenna shower~\cite{Giele:2011cb,Fischer:2016vfv}. 
When relying on the parton shower as a (Sudakov-weighted) phase-space
generator,
care has to be taken, as strongly-ordered showers are not \textit{a
  priori}  able to completely fill the phase space. To this end, direct higher-order emissions need to be taken into account~\cite{Li:2016yez,Hoche:2017iem}.

Multiple schemes have been proposed for combining next-to-leading order matrix elements of different jet multiplicities. They have been compared in a systematic fashion in a number of community
efforts~\cite{Andersen:2016qtm,Bellm:2019yyh,Buckley:2021gfw}.
The underlying idea is the same as for leading order multijet merging. 
Multiple samples are combined into one calculation while preserving both the fixed order accuracy of the matrix elements and the logarithmic accuracy of the shower, avoiding double counting. 
A method that can be seen as the NLO equivalent of CKKW-L has been presented in~\cite{Gehrmann:2012yg,Hoeche:2012yf}, and an alternative approach was presented in~\cite{Frederix:2012ps}. 
Next-to-leading order equivalents to the unitarized multijet merging approach are available in~\cite{Lonnblad:2012ix,Platzer:2012bs}. 
An alternative unitarization approach is again given by relying on matrix-element correction schemes, as has been demonstrated in \vincia~\cite{Hartgring:2013jma}.
With the improved accuracy of multijet merging prescriptions, a
systematic and efficient study of uncertainties becomes
relevant~\cite{Cormier:2018tog,Bothmann:2016nao,Gellersen:2020tdj}.

Sector showers such as \vincia reduce the complexity of matching, merging~\cite{Brooks:2020mab}, and matrix-element-correction schemes~\cite{Lopez-Villarejo:2011pwr},
since any given configuration with at most one quark pair has only a single unique (leading-color) shower history, independent of final-state multiplicity. 
The exception is if multiple same-flavor quark pairs are present, when
all quark permutations that yield viable quark-antiquark clusterings have to be taken into account. 
However, since the history is again deterministic for each such permutation (further gluon emissions still do not add to the complexity), this growth is still much milder than the original one; at most factorial with the number of same-flavor quark pairs. 
\vincia's sector-merging implementation~\cite{Brooks:2020mab} shows approximately constant scaling with the number of additional jets both for CPU event-generation time and for memory allocation. 
This eliminates all bottlenecks associated to the parton shower in multi-jet merging at any multiplicity, while retaining the accuracy of the CKKW-L method. 
As such, sector-based extensions to state-of-the-art NLO merging schemes should be straightforward and will be a high priority in \vincia in the near future, contributing to the kind of ``aggressive R\&D'' considered necessary by the LHC experiments in view of tight computing-budget projections for the high-luminosity phase~\cite{ATLAS:2020pnm} (cf.\ Sec.~\ref{sec:performance_hadron_colliders}).
An automated, sector-shower based MEC framework is currently being implemented in \vincia and will become available in the near future as well. 
In this context, the inclusion of on-the-fly NLO corrections via a generalization of the scheme in~\cite{Hartgring:2013jma} is being investigated as an alternative to existing NLO merging schemes.

The proper combination of QCD and QED effects in matching and multi-jet merging can be important for reaching the precision targets of current and future experiments. Naively, the interference between sub-dominant Standard-Model interactions and QCD can be of similar size as subleading QCD corrections. A first assessment of the impact of QCD/QED interference effects in parton showers has been presented in~\cite{Gellersen:2021caw}. This simulation includes QED, QCD at fixed color, and complete tree-level matrix element corrections for individual $N_c=3$ color configurations to embed interference. In this particular study it was found that QCD/QED interference effects are small.

\subsubsection{Combination with NNLO calculations}
\label{sec:he_collider_nnlo_matching}
Various techniques have been proposed for the combination of NNLO
accurate perturbative calculations with parton showers.
A good algorithm for the combination of NNLO calculations with parton showers
should attain NNLO accuracy for observables inclusive in the QCD radiation 
beyond the Born level, while preserving the logarithmic structure (and accuracy)
of the parton-shower simulation after matching.
In the following, we will discuss two main methods (in alphabetical order) currently applied by the experiments, as well
as a new proposal that may pave the way to fully differential matching.

\subsubsection*{The Geneva method}
The \geneva approach aims to incorporate the effects of higher logarithmic resummation in an appropriate jet resolution variable into event generators. By carrying out this resummation to an order consistent with the fixed-order accuracy, it effectively mediates between the fixed-order calculation and the parton shower. The result is a systematic combination of calculations at next-to-next-to-leading order in perturbation theory with the parton shower, which extends to arbitrary processes and is performed in the spirit of a traditional matching for analytically resummed calculations. \geneva also provides considerable flexibility in terms of how exactly the higher order resummation is achieved, and different formalisms can be used to obtain the resummed calculation.

The theoretical background and a first application of the method to the process of $e^+e^- \to jj$ were first presented in~\cite{Alioli:2012fc}. In this case, the resolution variable, which distinguishes between events at Born level and those with one or more additional emissions, was chosen to be the two-jettiness, a quantity closely related to thrust. 
The matching of resummed and fixed order (FO) pieces was achieved via a multiplicative approach.
The first application of the method to color-singlet production at a hadron collider was the study of the neutral Drell-Yan process~\cite{Alioli:2015toa}. This entailed an additive, rather than multiplicative, matching of resummed and FO contributions and a change to the shower interface. The zero-jettiness observable, $\Tau_0$~\cite{Stewart:2010tn} was used as a resolution variable, with the resummation again obtained via Soft-Collinear Effective Theory (SCET). Several subsequent applications of \geneva to other color-singlet processes at hadron colliders follow a similar approach.

SCET allows the cross section for color-singlet production, differential in the zero-jettiness, to be factorized as~\cite{Stewart:2009yx}
\begin{equation}
\label{eq:Tau0Factorization}
\frac{\df \sigma^{\rm SCET}}{\df \Phi_0 \df \Tau_0}=
\sum_{ij} \frac{\df\sigma_{ij}^B}{\df\Phi_0} H_{ij} (Q^2, \mu) \int\!\df t_a\, \df t_b \, B_i (t_a, x_a, \mu)\,
B_j (t_b, x_b, \mu) \, S\Bigl(\Tau_0 - \frac{t_a + t_b}{Q}, \mu \Bigr)
\,,\end{equation}
where $\df\sigma_{ij}^B/\df\Phi_0$ is the Born cross section for the
hard scattering.
The hard function $H_{ij}(Q^2, \mu)$ depends on the corresponding Born and
virtual squared matrix elements, and the sum runs over all possible
initial states. The $B_i(t, x)$ are inclusive  beam functions computed perturbatively in terms of standard PDFs $f_j$; schematically $B_i = \sum_j \mathcal{I}_{ij}\otimes f_j$. Finally, $S(k)$ is the hemisphere soft function for beam thrust.

Equation~\eqref{eq:Tau0Factorization} has a corresponding counterpart in full QCD, which, for $\Tau_0>0$ is given by the NLO cross section for color-singlet plus jet production.
An important feature of \geneva is that the cancellation of singularities between the approximate result in Eq.~\eqref{eq:Tau0Factorization}, and the exact result in the one-jet phase space is local in the resolution variable $\Tau_0$ by virtue of the use of a $\Tau_0$-preserving phase-space mapping. Although still non-local in the remaining phase space variables, this allows a lower $\Tau_0^{\cut}$ to be chosen than what would be possible in a strict slicing method~\cite{Gaunt:2015pea}. This cancellation requires  the complete set of singular terms of the NNLO calculation to be present in the SCET result, which are only fully captured beginning at NNLL$'$. 

Each of $H$, $B$, and $S$ depend only on a single characteristic
scale, so that the perturbative expansions of these constituent functions do not feature any large logarithms when a suitable choice of scale is made. 
In Eq.~\eqref{eq:Tau0Factorization}, however, all ingredients must be evaluated at an arbitrary common scale $\mu$, whose dependence exactly cancels between the different functions. We achieve this by using the renormalization group evolution in the effective theory to evolve each function from its own scale to $\mu$. We thus obtain the resummed $\Tau_0$ spectrum as:
\begin{align} \label{eq:resummedspectrum}
\frac{\df \sigma^{\rm NNLL'}}{\df \Phi_0 \df \Tau_0}
&= \sum_{ij} \frac{\df \sigma_{ij}^B}{\df \Phi_0} H_{ij}(Q^2, \mu_H)\, U_H(\mu_H, \mu)
\otimes \bigl[ B_i (x_a, \mu_B) \otimes U_B(\mu_B, \mu) \bigr] \nonumber \\
& \qquad \otimes \bigl[ B_j (x_b, \mu_B) \otimes U_B(\mu_B, \mu) \bigr]
\otimes \bigl[ S(\mu_S) \otimes U_S(\mu_S, \mu) \bigr]
\,\end{align} 
where the large logarithmic terms arising from the ratios of scales have been resummed by the RGE factors $U_X(\mu_X, \mu)$.

The final step of the \geneva method is the interface to the parton
shower. If the resolution variables were closely related to the
ordering variable of the shower, the shower matching would be
considerably simplified -- for example, using the transverse momenta
of the color singlet system and the hardest jet as resolution
variables, one can perform the matching to a transverse momentum
ordered shower such as \pythia following the same approach as
\powheg. When using $N$-jettiness, however, the mismatch between
resolution and ordering variable requires a more careful treatment. In
practice, a truncated shower of the kind described in
Ref.~\cite{Nason:2004rx}
is used with appropriate vetoing. 

\begin{figure}[t]
    \centering
    \includegraphics[width=0.44\textwidth]{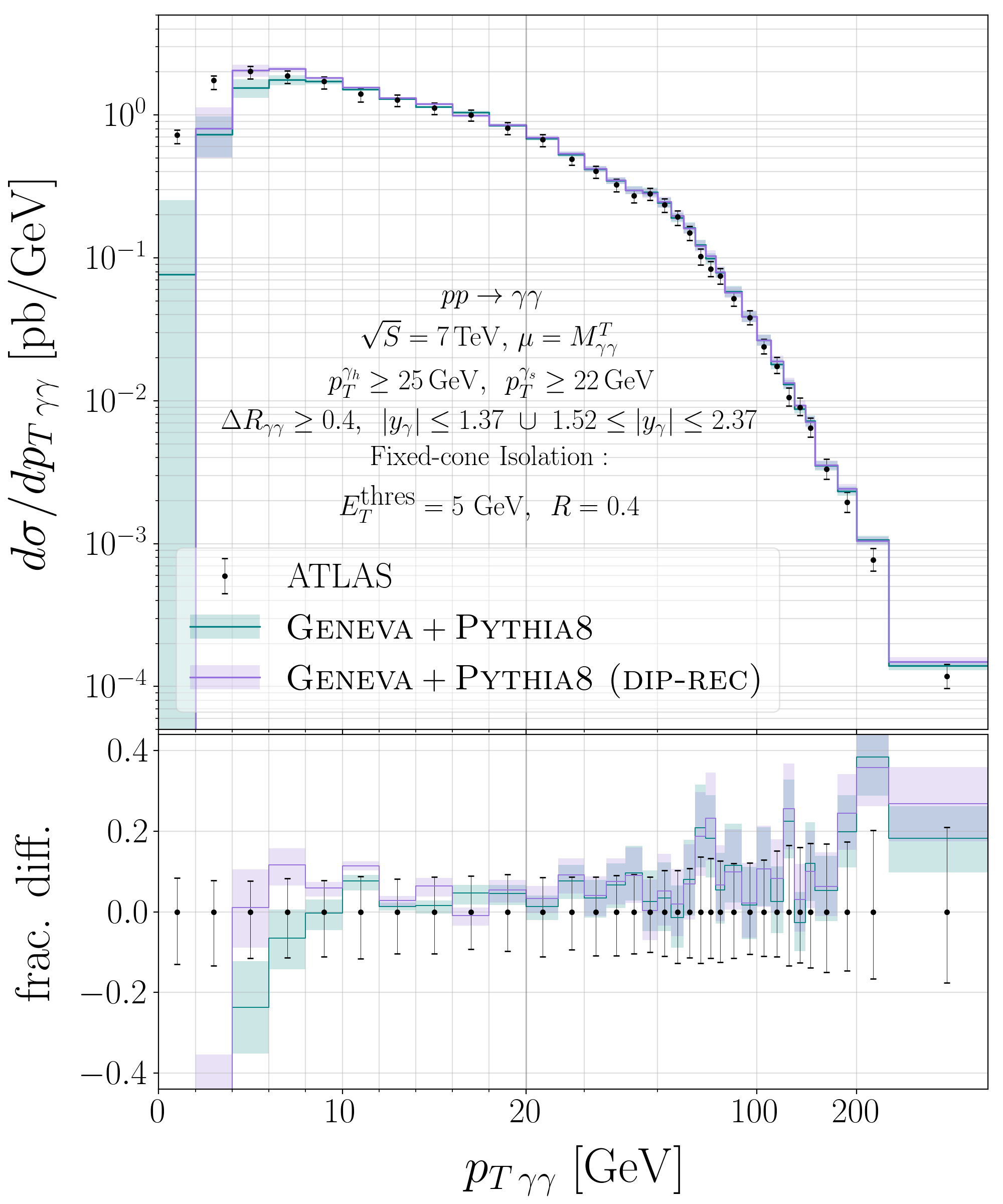}\hfill
    \raisebox{4.5mm}{\includegraphics[width=0.55\textwidth]{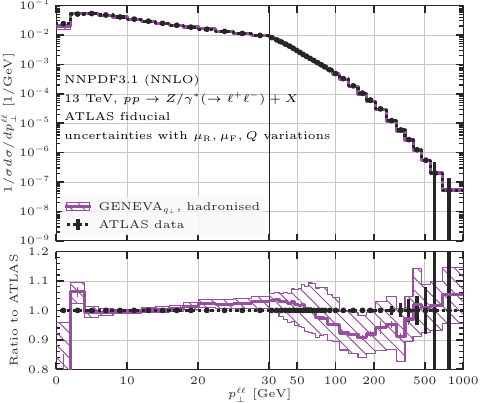}}
    \caption{Left: \textsc{Geneva} predictions for the diphoton transverse momentum in $pp\to \gamma\gamma$ in comparison to experimental data from ATLAS. 
    Right: \textsc{Geneva}$_{p_T}$ predictions for $pp\to \ell^+\ell^-$, achieved using $p_T$ resummation provided by \RadISH in comparison to 13 TeV data from ATLAS.
    \label{fig:geneva}}
\end{figure}

The \geneva approach has been applied to several processes of interest
at the LHC, in particular $VH$ production with different decay
modes~\cite{Alioli:2019qzz,Alioli:2020fzf}, diphoton
production~\cite{Alioli:2020qrd}, $ZZ$
production~\cite{Alioli:2021egp} and $W\gamma$
production~\cite{Cridge:2021hfr}. Figure~\ref{fig:geneva} (left) shows
predictions for di-photon production in comparison to ATLAS
experimental data. This process is of interest to the LHC experiments
due to the intricate QCD dynamics and because it is the irreducible background to Higgs production with decay to di-photon.

Although it has been convenient to exploit resummation in a SCET
framework for several applications of \geneva, this is not, in fact, a
necessity. Any formalism which is able to provide the resummed
spectrum in a valid resolution variable can be utilized. In
Ref.~\cite{Alioli:2021qbf}, the program \RadISH was used as
part of the \geneva implementation.
\RadISH resums the transverse momentum for color-singlet
processes at N$^3$LL accuracy via a direct QCD
approach~\cite{Monni:2016ktx}.
A calculation of the transverse momentum distribution of the $Z$ boson
with this method was in good numerical agreement with exact resummed result. It was also possible to compare for the first time two NNLO+PS generators which use the same framework but different resolution variables, in order to assess the size of the differences in their predictions for exclusive observables. A result from this work is shown in Fig.~\ref{fig:geneva} (right). This prediction has direct bearing on the $W$ mass measurement discussed in Sec.~\ref{sec:physics_goals_hadron_colliders}.

\geneva is not limited to color-singlet processes, and its application
to the production of colored particles is restricted only by the
availability of resummed calculations in a suitable resolution
variable at the required accuracy.
In Ref.~\cite{Alioli:2021ggd}, a resummation framework in SCET was established for the zero-jettiness in top-quark pair production. This will allow the \geneva approach to be extended to processes involving colored heavy particles, including not only the $t\bar{t}$ process but also, in principle, the $t\bar{t}X$ cases, should the appropriate two-loop amplitudes become available.

An important consideration when matching fixed order calculations to
parton showers is the relationship between the ordering variable of
the shower and the jet resolution variable used in the matching. The
flexibility that \geneva provides with respect to the choice of the
latter means that one could use  either numerical or analytic resummation to match to showers with NLL accuracy while use a non-standard evolution parameter. Once such a shower is publicly available, this would be an interesting avenue to explore.

\subsubsection*{The NNLOPS and \texorpdfstring{MiNNLO$_\text{PS}$}{MiNNLO} methods}
\label{sec:minnlo}

The \minnlo{}~\cite{Monni:2019whf,Monni:2020nks}  technique  was introduced 
as an extension of the \minlo{} procedure
of~\cite{Hamilton:2012np,Hamilton:2012rf}.
The latter method 
obtained NNLOPS accuracy using a reweighting procedure, and it was applied to some simple LHC processes, namely
Higgs-boson production~\cite{Hamilton:2013fea}, the
Drell-Yan
process~\cite{Karlberg:2014qua} and Higgs to bottom quark decays~\cite{Bizon:2019tfo}, and more
complicated LHC processes, such as the two Higgs-strahlung
reactions~\cite{Astill:2016hpa,Astill:2018ivh}, and the production of
two opposite-charge leptons and two neutrinos
($W^+W^-$)~\cite{Re:2018vac}. 

Compared to NNLOPS, the \minnlo{} procedure addresses more directly 
the requirement of NNLO accuracy and thus, besides not requiring any
reweighting,
can be more easily generalized to processes beyond
massive color-singlet production.  It has the following features:
\begin{itemize}
\item NNLO corrections are calculated directly during the generation
  of the events and without any additional \textit{a posteriori} reweighting. 
\item No merging scale is required to separate different multiplicities in the
  generated event samples.
\item The matching to the parton shower is performed according to the
  \POWHEG{} method~\cite{Nason:2004rx,Frixione:2007vw,Alioli:2010xd} and preserves the leading 
  logarithmic (LL) structure of transverse-momentum ordered 
  showers. For a different ordering variable, preserving the
    accuracy of the shower is more subtle. Not only does one need to veto shower radiation that has relative transverse momentum greater than the one generated by \powheg, but also one has to resort to truncated showers~\cite{Nason:2004rx,Bahr:2008pv} to compensate for missing collinear and soft radiation.
\end{itemize}
This method has applied to the production of
color singlet systems such as
$Z\gamma$~\cite{Lombardi:2020wju,Lombardi:2021wug},
$W^+W^-$~\cite{Lombardi:2021rvg}, $ZZ$~\cite{Buonocore:2021fnj}, $VH$
including the $H\to b\bar b$ decay at \nnlops{}~\cite{Zanoli:2021iyp},
and was extended to deal with the production of massive colored final
states~\cite{Mazzitelli:2020jio,Mazzitelli:2021mmm}, particularly
top-quark pair production. The corresponding generator also
handles the decay of the top quarks at leading order and 
approximates off-shell effects.

The \minnlo~method formulates a
NNLO calculation fully differential in the phase space $\PhiB$ of the
produced color singlet \FF{} with invariant mass $Q$. It starts from a
differential description of the production of the color singlet and a
jet (\FJ{}) by means of the \powheg{} matching technique
~\cite{Nason:2004rx,Frixione:2007vw,Alioli:2010xd} for
\FJ{} production, supplemented with a modified local $K$-factor 
${\bar B}(\PhiBJ)$. 
The function ${\bar B}(\PhiBJ)$ is the central ingredient of
\minnlo{}. It is derived using the observation that the NNLO
cross section differential in the transverse momentum of the 
color singlet ($\ptlocal$) and in the Born phase space
$\PhiB$ is given by
\begin{equation}
\label{eq:start}
  \frac{\mathd\sigma}{\mathd\PhiB\mathd \ptlocal} = 
  \exp[-\tilde{S}(\ptlocal)]
  \left\{D(\ptlocal)+\frac{R_f(\ptlocal)}{\exp[-\tilde{S}(\ptlocal)]}\right\}\,,
\end{equation}
where $R_f$ contains terms that are integrable in the $\ptlocal\rightarrow 0$ limit, and 
\begin{equation}
\label{eq:Dterms}
  D(\ptlocal)  \equiv -\frac{\mathd \tilde{S}(\ptlocal)}{\mathd \ptlocal} {\cal L}(\ptlocal)+\frac{\mathd {\cal L}(\ptlocal)}{\mathd \ptlocal}\,.
\end{equation}
$\tilde{S}(\ptlocal)$ represents the Sudakov form factor, while ${\cal L}(\ptlocal)$
contains the parton luminosities, the squared virtual matrix elements
for the underlying \FF{} production process up to two loops as well as
the NNLO collinear coefficient functions. Explicit expressions can be found in~\cite{Monni:2019whf,Monni:2020nks}.
A crucial feature of the \minnlo{} method is that the renormalization
and factorization scales are set to $\muR\sim\muF\sim \ptlocal$.
Eventually, one obtains the following expression for the
${\bar B}(\PhiBJ)$ function
\begin{align}
\label{eq:Bbar}
      {\bar B}(\PhiBJ)& \equiv \exp[-\tilde{S}(\ptlocal)]\Bigg[ B(\PhiBJ)\left(1+\abarmu{\ptlocal}\left[\tilde{S}(\ptlocal)\right]^{(1)}\right) + V(\PhiBJ)\nonumber \\
       &  + D(\ptlocal)^{\rm (\ge 3)} F^{\rm corr}(\PhiBJ)   \Bigg] + \int d\Phi_{\rm rad} R(\PhiBJ,\Phi_{\rm rad})\tilde{S}(\ptlocal)\,, 
\end{align}
where $D(\ptlocal)^{(\ge 3)}$ corresponds to the $\mathcal{O}(\ge\alphaS^3)$
contributions in Eq.~\eqref{eq:Dterms}, $B$ and $R$ are the squared tree-level matrix elements for FJ and FJJ
production, and $V$ is the virtual matrix element for FJ. The factor
$F^{\tmop{corr}}(\PhiBJ)$ encodes the dependence on the full
FJ phase space, as discussed in detail in Sec.~3 of
\cite{Monni:2019whf}.

In the case of the production of colored final states, such as
top-quark pair production, the starting point to derive the singular term in
Eq.~\eqref{eq:start} is the more complex expression
\begin{equation}
\sum_{c=q,\bar{q},g}
  \frac{|M^{(0)}_{c\bar{c}}|^2}{2 m_{t\bar{t}}^2}\int\frac{d^2\vec{b}}{(2\pi)^2} e^{i \vec{b}\cdot
  \ptvec } e^{-S_c \left(\frac{b_0}{b}\right)}
\times\sum_{i,j}\Tr({\mathbf H}_c{\mathbf \Delta})\,
 \,({C}_{ci}\otimes f_i) \,({C}_{\bar{c} j}\otimes f_j) \,,
\label{eq:bspace}
\end{equation}
which describes the production of a pair of heavy quarks at small
transverse momentum. Here $b_0=2\,e^{-\gamma_E}$, $b=|\vec{b}|$. $S_c$
is the same Sudakov radiator which also enters the description of the
production of a color singlet system at small transverse momentum.
The first sum in Eq.~\eqref{eq:bspace} runs over all possible flavor
configurations of the incoming partons $p_1$ of flavor $c$ and $p_2$
of flavor $\bar c$.
The collinear coefficient functions $C_{ij}$
describe the structure of constant terms related to the emission of
collinear radiation, and the parton densities are denoted by $f_i$ and are
evaluated at $b_0/b$. The operation $\otimes$ denotes the standard
convolution over the momentum fraction $z$ carried by initial-state
radiation.
The factor
$\Tr({\mathbf H}_c{\mathbf \Delta})\, \,({C}_{ci}\otimes f_i)
\,({C}_{\bar{c} j}\otimes f_j)$
has different expressions for the $q\bar q$ and $gg$ channels and has
here a symbolic meaning. In particular, it has a rich Lorentz
structure that we omit for simplicity, which
is a source of azimuthal correlations in the collinear
limit~\cite{Catani:2010pd,Catani:2014qha}.

All quantities in bold face denote operators in color space, and the
trace $\Tr({\mathbf H}_c{\mathbf \Delta} )$ in Eq.~\eqref{eq:bspace}
runs over the color indices.
The hard function ${\mathbf H}_c={\mathbf H}_c(\Phi_{\rm
  t\bar{t}};\alpha_s(m_{t\bar{t}}))$ is obtained from the subtracted
amplitudes and the ambiguity in its definition corresponds to using a
specific resummation scheme~\cite{Bozzi:2005wk}.
The operator ${\mathbf \Delta}$ encodes the structure of the quantum
interference due to the exchange of soft radiation at large angle
between the initial and final state, and within the final state. It is
given by ${\mathbf \Delta}={\mathbf V}^\dagger{\mathbf D}{\mathbf V}$,
where~\cite{Catani:2014qha}
\begin{align}
\label{eq:soft}
{\mathbf V} &= {\cal
  P}\exp\left\{-\int_{b_0^2/b^2}^{m_{t\bar{t}}^2}\frac{\mathd q^2}{q^2}{\mathbf
  \Gamma}_t(\Phi_{\rm t\bar{t}};\alpha_s(q))\right\}\,.
\end{align}
The symbol ${\cal P}$ denotes the path ordering (with increasing
scales from left to right) of the exponential matrix with respect to
the integration variable $q^2$. ${\mathbf \Gamma}_t$ is the anomalous
dimension accounting for the effect of real soft radiation at large
angles, and 
${\mathbf D}={\mathbf D}(\Phi_{\rm t\bar{t}},\vec{b};\alpha_s(b_0/b))$
encodes the azimuthal dependence of the corresponding constant terms,
and is defined such that $[{\mathbf D}]_\phi={\mathbf{1}}$, where
$[\cdots]_\phi$ denotes the average over the azimuthal angle $\phi$ of
$\ptvec$.

The strategy to arrive at a \minnlo{} improved $\bar B$ function is
the same as for the color-singlet case. We expand
Eq.~\eqref{eq:bspace} taking care of not spoiling the NNLO counting
and maintaining leading
logarithmic accuracy to arrive at an expression in transverse momentum
space that can be used to correct the $\bar B$ function in order to
achieve NNLO accuracy. All details are given in
Refs.~\cite{Mazzitelli:2020jio,Mazzitelli:2021mmm}.

The \minnlo{} method can currently be used to describe processes with
generic color structure that, at the Born level, do not include light
partons in the final state. The only requirement is the availability
of the resummation of a suitable kinematic variable (such as, for
instance, the transverse momentum of the heavy system) at the desired
perturbative accuracy as well as the NNLO computation of the hard
process.
The next milestone is likely to be the matching for a process with a
light parton in the final state, such as Higgs plus jet or Drell-Yan
plus jet. Similarly, NLO EW corrections may compete with NNLO QCD ones
and a joint inclusion of both is another future important goal.
A crucial observation is that the current research activity in methods
to improve the logarithmic accuracy of the shower to NLL for a variety of observables,
and eventually to NNLL level, is likely to add new requirements for matched
NLOPS and NNLOPS methods, which at the moment only preserve the 
accuracy of the shower at LL level.


\subsubsection*{Progress towards fully differential matching}
Despite the great success of the methods presented above, an NNLO matching technique that provides the analog of
\mcatnlo\ or \powheg\ is still lacking. Such a technique could be provided by the NLO matrix-element-corrected parton
shower in \vincia.
With the sectorization of the branching phase space, it becomes
exceptionally simple to identify regions associated to
strongly-ordered shower sequences and regions that would normally not
be captured with a strongly-ordered shower, such as direct $n\mapsto
n+2$ branchings \cite{Li:2016yez}.
Since such  higher-order branchings correspond to multiple,
simultaneously unresolved partons, the associated splitting kernels
have a close connection to higher-order subtraction terms in
fixed-order calculations. In particular, they share the same infrared
singularity structure. This feature is exploited in \vincia's approach
to NNLO matching, described in \cite{Campbell:2021svd}, which is the
first NNLO matching scheme to be based on a direct correspondence
between the parton shower and the hard matrix elements at NNLO. While
it has so-far been limited to a proof-of-concept implementation for
hadronic $Z$ decays, it is being actively developed.   Work is in
progress to generalize this result to processes of wider phenomenological interest.

\subsubsection{Matching at \texorpdfstring{N$^{\,3}$LO}{N3LO} precision}
Calculations at third order in perturbative QCD are the most precise predictions available for hadron collider observables to date. In the recent past, such N$^3$LO calculations have moved from analytic calculations of total cross sections to fully differential cross sections, partly obtained within numerical tools~\cite{Dulat:2017prg,Currie:2018fgr,Dreyer:2018qbw,Cieri:2018oms,Mondini:2019gid,Chen:2021isd,Billis:2021ecs,Camarda:2021ict,Re:2021con}. Only very recently, this progress has been supplemented with a method to match N$^3$LO calculations to parton showers~\cite{Prestel:2021vww,Bertone:2022hig}. The aim of such developments is to produce \emph{event generators} that yield physical results for inclusive observables, in phase space regions with well-separated partons, and, crucially, also in phase-space regions with multiple unresolved partons. The produced events can straightforwardly be combined with hadronization, and the modeling of multiple interactions between the constituents of composite hadronic initial states. Thus, N$^3$LO+PS calculations will allow for the prediction of the (collinear) dynamics of QCD at unprecedented precision, and could thus provide an important baseline for high-precision measurements at future colliders. 

The field of N$^3$LO+PS matching is still in its infancy, with only a single technique, the so-called \tomte method, available to date. However, given the rapid progress in combining N$^3$LO fixed-order results with analytic resummation~\cite{Billis:2021ecs,Camarda:2021ict,Re:2021con}, the emergence of alternative methods in the near future is plausible. The \tomte master formula (Eq.~$(24)$ of~\cite{Prestel:2021vww}) is process independent, and constructed with a simple procedure:
\begin{itemize}
\item[$a)$] Add a physically meaningful real-emission pattern for $m$-parton states, created by reweighting fixed-order results with appropriate all-order shower factors.
\item[$b)$] Subtract what you have added (\ie\ the real-emission pattern) from the next-lowest multiplicity $(m-1)$.
\item[$c)$] Add a physically meaningful, and appropriate, higher-order exclusive cross section for the $(m-1)$-parton states.
\item[$d)$] Complement the exclusive $(m-1)$-parton cross section with its $m$-parton real-emission pattern. 
This yields an $(m-1)$-parton matched calculation that may serve as a ``physically meaningful real-emission pattern" for the next-lower multiplicity $(m-2)$. Thus, repeat the procedure from point $b)$ on, until the minimal multiplicity (at N$^3$LO precision) is reached.
\end{itemize}
The usage of exclusive higher-order cross sections is by choice, since it allows for a convenient separation of the ``subtract" and ``complement" contributions, which helps by introducing non-parametric shower bias into fixed-order expansions. Such biases arise since the subtractions are based on the specific implementation of unitarity in the parton shower. No such notion exists at fixed order, such that biases may arise if the subtractions are directly involved in producing exclusive fixed-order results from inclusive fixed-order cross sections, as will \eg\ be the case for UNLOPS~\cite{Lonnblad:2012ix,Platzer:2012bs,Bellm:2017ktr}. A similar effect in turn also suggests that biases could arise in UN2LOPS calculations~\cite{Hoeche:2014aia,Hoche:2014dla} for processes with Born contributions that vary substantially over phase space, \eg\ vector-boson + jets production.

Since programs to produce fully differential N$^3$LO calculations are not public, the \tomte method has only been studied with ``toy calculations" for  $e^+e^-\rightarrow$ jets and Drell-Yan lepton-pair production in hadron-hadron collisions, which allows for rigorous tests of its theoretical consistency and numerical feasibility. Sample distributions can be found in~\cite{Prestel:2021vww,Bertone:2022hig}.

\subsection{Electroweak evolution}
\label{sec:hadron_collider_ewps}
At a high-energy lepton collider, new particles can be produced through $s$-channel $\mu^+\mu^-$ annihilation at beam collision energy, with a mild spread due to initial-state radiation.
Until such a threshold is discovered, the vector-boson-fusion processes dominate due to the collinear enhancement.
As emphasized in~\cite{Dawson:1984gx,Chiesa:2013yma,Chen:2016wkt,Bauer:2017isx,Bauer:2018arx,Han:2020uid,Han:2021kes,Buarque:2021dji}, the particles collinearly radiated from the collider beams should be treated as partons, with the large logarithm resummed in the distribution functions (PDFs). The (semi-)inclusive cross section can be factorized into the partonic cross sections convoluted with the corresponding PDFs, in analogy to the high-energy proton collider cases, cf.~Eq.~\eqref{eq:factorization_main}. The same effects will be important at very high energy hadron colliders, such as a potential FCC-pp.

\begin{figure}
    \centering
    \includegraphics[width=0.49\textwidth]{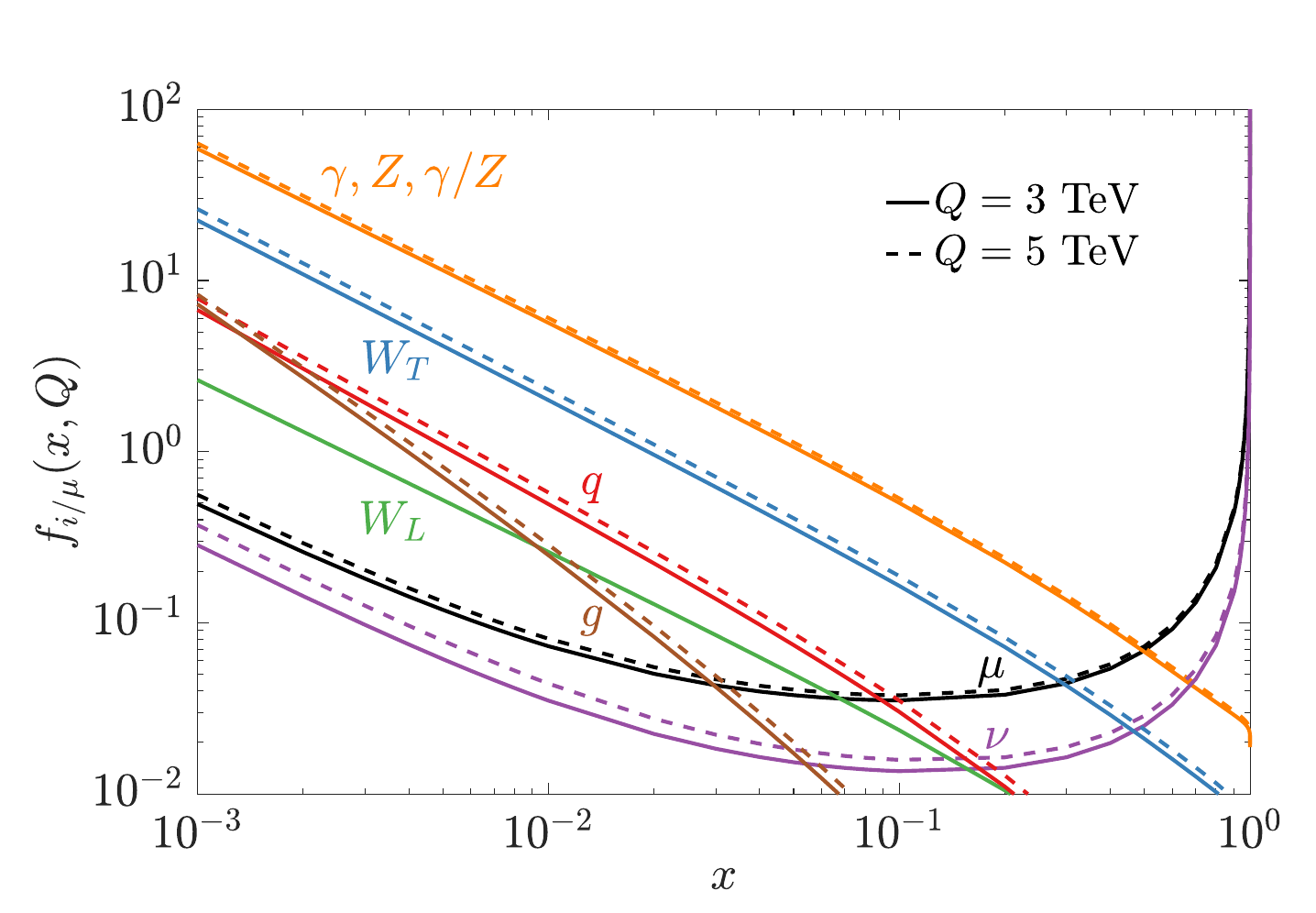}
    \includegraphics[width=0.49\textwidth]{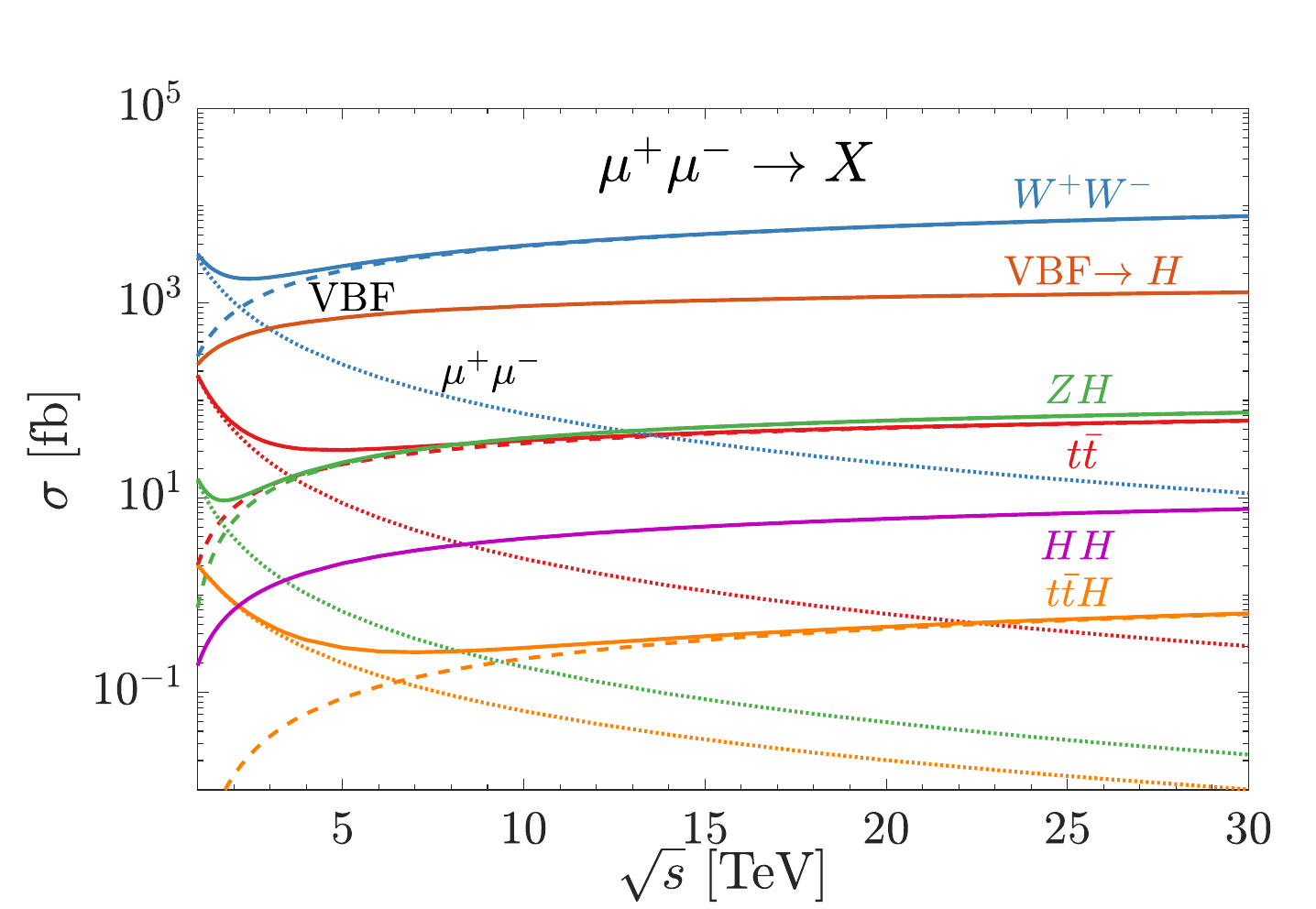}
    \caption{The distributions for the EW PDFs (left) and the corresponding predictions for a few representative SM processes at a high-energy muon collider (right). Adapted from Ref.~\cite{Han:2020uid}.}
    \label{fig:PDFvsXSec}
\end{figure}

Below the EW scale $(\mu_{\rm EW}\sim M_Z$), only light particles, \emph{i.e.}, quarks, leptons, the photon and the gluon, behave as active partons. The PDFs evolve in terms of the DGLAP equations~\cite{Dokshitzer:1977sg,Gribov:1972ri,Altarelli:1977zs} of QED and QCD groups. 
At an energy well above $\mu_{\rm EW}$, all the Standard Model particles become essentially massless. As a result, the heavy particles, $W/Z/H/t$, also act as partons, causing PDFs to run according to the full Standard Model gauge group $SU(3)_c\otimes SU(2)_L\otimes U(1)_Y$. Across the threshold $\mu_{\rm EW}$, a matching between these two energy regions is required, and one switches to the SM unbroken phase, with conditions
$f_{\gamma}(x,\mu_{\rm EW}^2)\neq0,~f_{Z}(x,\mu_{\rm EW}^2)=f_{\gamma Z}(x,\mu_{\rm EW}^2)=0$. By solving the DGLAP equations, we can obtain the complete EW PDFs for a high-energy muon beam, with scales $Q=3$ and 5 TeV shown in Fig.~\ref{fig:PDFvsXSec} (left). We see that the valence lepton sharply peaks at $x\approx1$, while others peak at $x\approx0$, reflecting the $1/(1-x)$ and $\log(x)$ behaviors, respectively. We also show the quark and gluon distributions, with $q=\sum_{i=d}^{t}(q_i+\bar{q}_i)$, which come from higher-order splittings. The longitudinal gauge boson $(W_L,Z_L)$ does not run with scale $Q$, as a residual effect of the electroweak symmetry breaking. For completeness, we also show the predictions of a few representative Standard Model cross sections in Fig.~\ref{fig:PDFvsXSec} (right). As expected, the annihilation cross sections dominate when the energy crosses the threshold, and then decrease following their $1/s$ behavior. Instead, vector-boson fusion processes start dominating at high energies.

Electroweak PDFs resum the large collinear logarithms in initial-state radiations. Similarly, the final-state logarithms can be factorized and resummed in fragmentation functions (FFs)~\cite{Bauer:2018xag}, or equivalently as Sudakov factors simulated with the electroweak parton showers (PS)~\cite{Chen:2016wkt}. This has been implemented in approximate form in a number of event generators~\cite{Christiansen:2014kba,Krauss:2014yaa}.
Recently, a complete angular ordered electroweak parton shower has become available in \herwig7~\cite{Masouminia:2021kne,Darvishi:2021het}. 
The \vincia parton shower also includes a full-fledged electroweak shower that models all possible collinear branchings in the Standard Model~\cite{Kleiss:2020rcg,Brooks:2021kji}. Due to the chiral nature of the electroweak theory, the shower evolution is based on the helicity shower formalism~\cite{Fischer:2017htu} in which intermediate particles are assigned definite helicities. 
Further features include full mass dependence in the kinematics and dynamics, a treatment of neutral boson interference effects and a veto procedure to avoid double-counting from different hard processes. \vincia also includes a sophisticated treatment of resonance decays, which are interleaved with the shower evolution. This is especially relevant in the context of an electroweak parton shower, which is able to produce highly-energetic resonances.
A systematic comparison among the different generators would be useful to assess the systematics of electroweak evolution.

\subsection{General-purpose and high-energy resummation and matching}
\label{sec:hadron_collider_resummation}

There is a rich interplay between (semi-)analytic resummed calculations and
Monte Carlo event generators. On the one hand, it is vital to have predictions  of known formal accuracy for specific observables, to benchmark
the more general, but limited in accuracy, event generators. On the other hand,
general purpose event generators aim to provide fully exclusive physical predictions including non-perturbative effects, while still being based on perturbative inputs, and can hence guide the transition from parton to hadron level states also for analytic calculations. In practice, the computation of analytic predictions quickly becomes very challenging, in particular in cases with many colored particles in an event. In such cases, the framework of an event generator is ideally suited to incorporate a largely automated implementation of such calculations.

\subsubsection{\texorpdfstring{\caesar}{Caesar} type resummation}
\label{sec:hadron_collider_caesar}
A formalism for soft gluon resummation, convenient for automation,
is provided by the \caesar approach \cite{Banfi:2004yd}. The necessary ingredients
for NLL resummation of suitable observables, and matching to reach
NLL$^\prime$+NLO accuracy, is implemented as a plugin \cite{Gerwick:2014gya,
  Baberuxki:2019ifp}  to the \sherpa \cite{Sherpa:2019gpd, Gleisberg:2008ta}
event generator. It allows the resummation of the recursive IRC safe observables
of the original \caesar formalism, as well as of event shapes including soft
drop grooming in various  settings \cite{Marzani:2019evv, Baron:2020xoi}. The
latest addition are (groomed) jet substructure observables, in particular jet angularities, in hadronic production processes \cite{Caletti:2021oor, Reichelt:2021svh}. The resummation of soft gluons around complicated
color structures is automated using an interface to the matrix element
generator \comix \cite{Gleisberg:2008fv}. Some components of \caesar like
resummation, in particular in some cases the multiple emission
effects \cite{Banfi:2001bz} and, for example for jet observables, the non-global contributions \cite{Dasgupta:2001sh}, have to be evaluated numerically. 
Since those can be done once and then applied in several settings, their evaluation is currently left to separate codes. An additional benefit of
using an event generator framework relies on the already automated capabilities of \sherpa to perform tasks like fixed order corrections, optimized phase space integration and the handling of interfaces
to external resources like loop integrals and parton distribution functions. In
a similar vein, aspects of resummation in SCET have been automated
in \cite{Farhi:2015jca, Balsiger:2018ezi} based on
the \madgraph \cite{Alwall:2014hca} framework. 

There are various matching schemes commonly used to combine fixed-order and
resummed calculations to obtain distributions that are accurate both, at a given
order in \alphaS and a certain (resummed) logarithmic accuracy that match or exceed the accuracy of the individual ingredients. A particular strength of the \sherpa implementation of \caesar resummation is the access to the fully differential, both in
kinematics and in particular flavor, matrix elements from \comix. This implies
that both initial and final-state flavors are known at event level. Hence,
flavor sensitive clustering algorithms like the one proposed in
\cite{Banfi:2006hf} (BSZ algorithm) can easily be implemented. With this at hand, the fixed order cross sections to be used in the matching can be separated into channels corresponding to the lowest order (Born) phase space, in an IR safe
way. This is an important ingredient in matching schemes that allow to
automatically achieve NLL$^\prime$ accuracy by matching NLL and LO (or NLO)
calculations \cite{Banfi:2010xy}. This type of clustering has been implemented
and used for matching in the \sherpa resummation plugin for observables in
color neutral- \cite{Baberuxki:2019ifp} and hadronic \cite{Baron:2020xoi}
initial states. A downside of this approach is that it defines a global flavor
structure of an event, but not necessarily that of an object obtained from a
different clustering strategy, like for example the leading anti-$k_t$ jet of
the event. Ref. \cite{Reichelt:2021svh} avoided this problem in the context of
calculations of jet substructure observables by iterative re-clustering the
constituents of an anti-$k_t$ jet with the BSZ algorithm, thus defining its
flavor in an IR safe manner. This can contribute to the topic of quark--gluon discrimination that is of more general interest, see for
example \cite{Larkoski:2014pca, Andersen:2016qtm, Gras:2017jty, Mo:2017gzp,
Reichelt:2017hts, Larkoski:2019nwj, Amoroso:2020lgh, Caletti:2021ysv}, beyond the technical developments of event generators and resummation tools.

\begin{figure}
  \begin{minipage}{.49\textwidth}
    \includegraphics[width=\textwidth]{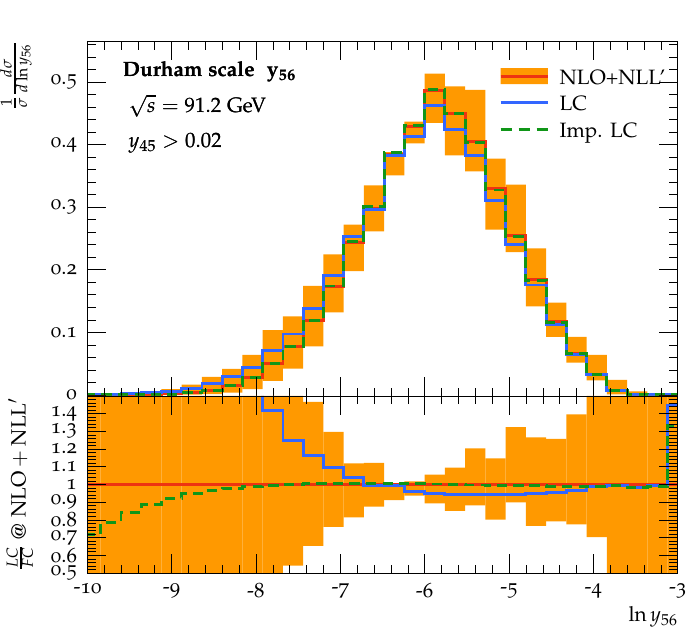}
  \end{minipage}\hfill
  \begin{minipage}{.49\textwidth}
    \includegraphics[width=\textwidth]{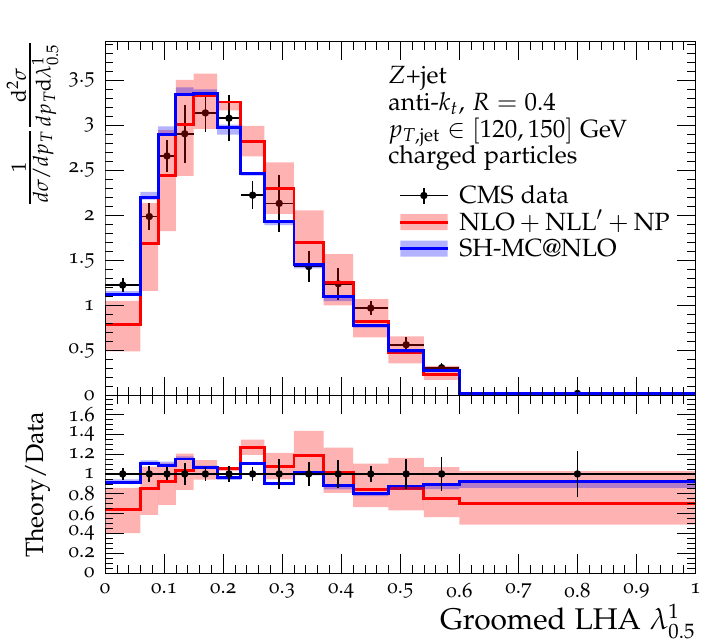}
  \end{minipage}
  \caption{Left: Durham jet resolution scale $y_{56}$ describing the $5\to6$ jet
    transition, at NLO+NLL$^\prime$ compared to various simplified color
    treatments. Right: Les Houches angularity $\lambda^1_{0.5}$ of the leading
    jet in Z+jet production, including the effect of soft drop grooming. Data
    from CMS compared to the NLO+NLL$^\prime$+NP prediction and a full \sherpa
    simulation.}\label{fig:sherpa_caesar}
\end{figure}

A concrete example for the need of benchmarks with quantifiable accuracy are the efforts to include subleading color effects in parton showers \cite{Platzer:2012np,
Nagy:2012bt, Platzer:2018pmd, AngelesMartinez:2018cfz, Forshaw:2019ver,
Nagy:2019pjp, DeAngelis:2020rvq, Holguin:2020joq, Platzer:2020lbr,
Hoche:2020pxj}, cf.\ Sec.~\ref{sec:spin-color}. In several of the mentioned studies, the prime example was the
Durham jet resolution scale $y_{n,n+1}$. While at low multiplicities, i.e. for
$y_{23}$ the color structure is simple, insights can be gained by considering
higher multiplicities, leading to more complicated color insertion operators in
the Born amplitudes. In the left of Fig.~\ref{fig:sherpa_caesar}, a resummed
prediction for $y_{56}$ obtained with the \sherpa framework is
shown \cite{Baberuxki:2019ifp}. The main plot illustrates the full
NLL$^\prime$+NLO result (red solid line) including the perturbative uncertainty
(orange band) obtained from variations of renormalization- and resummation
scales, as well as power corrections. This is compared to the calculation in the
strict $N_c\to\infty$ limit (LC, blue solid line). As an intermediate step, closer
to what traditionally is implemented in parton showers, the improved leading
color (Imp.~LC, green dashed line) scheme makes the necessary simplifications
in the soft functions but evaluates the $SU(N_c)$ Casimir invariants and anomalous dimensions at the physical value $N_c = 3$, see App.~A of \cite{Baberuxki:2019ifp} for details. Approximations at different accuracy
are further discussed in~\cite{Baberuxki:2019ifp}. These results will inform the construction
of improved parton showers, see Sec.~\ref{sec:hadron_collider_evolution}.

Resummed predictions can on the other hand largely benefit from event generator based studies of non-perturbative corrections. In \cite{Reichelt:2021svh}, such a study was used to derive transition matrices, describing non-perturbative shifts simultaneously in the transverse momentum of a jet and in a jet substructure observable.
For the Les Houches angularity \cite{Berger:2003iw, Almeida:2008yp, Andersen:2016qtm}, there is a major contribution in adding shifts in the angularity bins to the perturbative distribution, and a minor but significant effect of adding contributions from neighboring $p_T$ bins, which saturates when  adding contributions from regions further away in $p_T$.
Figure~\ref{fig:sherpa_caesar} (right) shows the final prediction obtained after folding the resummed, perturbative, predictions with these corrections (red solid line) in comparison to data obtained by the CMS experiment~\cite{CMS:2021iwu}. 
Additionally, the result of a full fledged \sherpa simulation at MC@NLO level is displayed. The wealth of data available, in this case just from a single experimental analysis, but covering multiple parameter choices in the observable- and jet definitions, in regards to soft drop grooming and final-state selection criteria also demonstrates the need for automation of resummed calculations.

The current automation in the \sherpa resummation framework is limited to NLL
resummation. This could be extended in the future however. The \caesar approach
has been generalized to cover NNLL accuracy in the \ares
formalism \cite{Banfi:2014sua} which has successfully been applied to two jet final states \cite{Banfi:2016zlc, Arpino:2019ozn}. These formulas could be automated in a similar way. Recently, first results for non-global logarithms at higher orders have become available \cite{Banfi:2021owj, Banfi:2021xzn}. The resummation of the transverse momentum of color singlets has also been formulated in a very similar language in \cite{Monni:2016ktx,Bizon:2017rah}. 
Based on the \sherpa framework a first resummed calculation for the $q_T$ and $\Delta\phi$ distributions in $W$ and $Z$ boson production at N$^3$LL$^\prime$ accuracy based on SCET has recently been presented \cite{Ju:2021lah}.

\subsubsection{High Energy Jets}
\label{sec:high-energy-jets}

Predictions based on a fixed-order evaluation of the hard matrix element 
give very precise results for cross sections and e.g.~distributions in transverse momenta.
However, there are regions of phase space where an expansion in $\alpha_s$ is
slowly convergent. For example, consider the cross section for 2$\to$2 scattering of
partons in the region where the Mandelstam variable $s$ is much larger than
$|t|$. At order $\alpha_s^{n+2}$ there are terms which are of the form
$\log^n(s/|t|)$. At energies accessible by the LHC the quantity
$\left( \alpha_s \log(s/|t|)\right)^n$ can be of order one. Logarithms of
this type are called high-energy logarithms. At high-energy colliders they
can be large and should in this case be summed to all orders
to ensure stability of the perturbative predictions. This motivates an
alternative classification of accuracy, where N$^m$LL refers to all terms in
the perturbative series of the form $\alpha_s^{n+2} \log(s/|t|)^{n-m}$.

The LL contribution to QCD amplitudes, to all orders in $\alpha_s$, takes a
simple factorized form
\cite{Fadin:1975cb,Lipatov:1976zz,Kuraev:1977fs}. Rather than describing the
hard matrix element of a collision at the LHC via fixed-order perturbation
theory we can instead take as our starting point matrix elements which are LL
accurate. High Energy Jets (\HEJ) is a framework that builds on this LL
accuracy. However, \HEJ utilizes Monte Carlo integration for phase space
integration, which means minimal approximations to the amplitude need to be
made, and no approximations to phase space. This in turn allows \HEJ to
make all-order predictions for not only asymptotically large energies, but
also at the scales of energy accessible by the LHC and for arbitrary cuts and
analyses.

To describe an inclusive cross section to LL accuracy it is required only to integrate additional real radiation over the so called Multi-Regge-Kinematic (MRK) region of phase space, where for momenta
$p_a,p_b\to p_1,\cdots, p_n$ the outgoing momenta are strictly ordered in
rapidity, $y$, while the transverse momenta $\mathbf{p}_i$ are of the same
finite magnitude. As an illustration, we give the squared
amplitude for the scattering of distinguishable quarks $q$ and $Q$ to $n$
partons:
\begin{align}
\label{eq:LLMRK}
\begin{split}
    \left|\mathcal{M}^{\mathrm{LL \ MRK}}_{qQ\to n}\right|^2&=\frac{4s^2}{N_C^2-1}\,\frac{g_s^2C_F}{|\mathbf{p}_1|^2}\,\frac{g_s^2C_F}{|\mathbf{p}_n|^2}
    \left(\prod_{i=2}^{n-1}\frac{4g_s^2C_A}{|\mathbf{p}_i|^2}\right)
    \left(\prod_{i=1}^{n-1}e^{2\alpha(\mathbf{q}_{i})\left(y_{j+1}-y_j\right)}\right).
\end{split}
\end{align}
The only real-emission corrections which are
relevant to LL order are emissions of gluons within the rapidity span of the
outgoing quarks. Virtual corrections exponentiate in this limit and are
described by the Regge trajectory of the associated $t$-channel momentum
$\mathbf{q}_i$,
\begin{align}
	\label{eq:trajectory}
	\begin{split}
	\alpha(\mathbf{q}_{i})=-g_s^2C_A\frac{\Gamma(1-\epsilon)}{(4\pi)^{2+\epsilon}}\frac{1}{\epsilon}\left(\frac{\mathbf{q}_{i}^2}{\mu^2}\right)^\epsilon.
	\end{split}
\end{align}
The IR singularity here can be regularized by the procedure described in \cite{Andersen:2011hs}.
\par
While expressions such as Eq.~\eqref{eq:LLMRK} are useful for probing the
asymptotic limit of QCD, the kinematic approximations are too severe to be
directly applied to the LHC. This motivates constructing a framework which
maintains the logarithmic accuracy of Eq.~\ref{eq:LLMRK} while relaxing many
of the kinematic approximations used to obtain it.

At leading order, the amplitude for $qQ\to qQ$ scattering
already has a factorized form without making any
kinematic approximations. It is given as the contraction of two
currents defined by the external quark lines.
Using them as building blocks one can construct a framework 
which is LO accurate for this process without loosing
LL accuracy \cite{Andersen:2009nu}.  The leading-order amplitude for
$qg\to qg$ contains $s$- and $u$-channel diagrams as well as a $t$-channel
diagram, but it can be shown that for the dominant helicity configurations,
the $qg\to qg$ amplitude can be written exactly in the same form as 
$qQ \to qQ$ with $C_F$ replaced by a momentum-dependent factor~\cite{Andersen:2009he}.
The final form of the LL squared amplitudes for pure-QCD processes is
an extension of Eq.~\eqref{eq:trajectory}, with real-emission corrections
described in terms of the Lipatov vertex \cite{Lipatov:1976zz,Andersen:2009nu}.
This framework can be extended to the production of a $W$, $Z$, 
or Higgs boson or same-sign $W$-pair production in association with jets~\cite{Andersen:2009nu,Andersen:2012gk,Andersen:2016vkp,Andersen:2018kjg,Andersen:2021vnf}. 

The \HEJ framework may be systematically improved by 
increasing the logarithmic accuracy of the all-orders amplitudes. Leading
powers in $s/|t|$ in matrix elements lead to leading logarithms after
integration.  Regge theory gives the scaling of matrix elements in terms of
the spin of the particles exchanged in the $t$-channels of the planar Feynman
diagrams that can be drawn for a given process when the legs are ordered in
rapidity. A flavor/momentum configuration contributes at LL if it permits an
exchange of a gluon in all $t$-channels. Beyond that, a flavor/momentum
configuration contributes at N$^m$LL if it permits an exchange of a gluon in
all but $m$ of the $t$-channels (see Ref.~\cite{Andersen:2020yax} for further
details).
\par
The first NLL component to be included in \HEJ was the description of a gluon emission where that
gluon was more extreme in rapidity than one of the outgoing
quarks~\cite{Andersen:2017kfc}.  In~\cite{Andersen:2020yax}, the necessary
pieces have been calculated to include all-order corrections to all configurations 
at 3 jets and above whose leading contribution is at NLL order. 
This also includes the potential emission of a $W$ boson from these pieces.
This is a gauge-invariant subset of the full NLL correction to
inclusive 2$j$ or $W+2j$ production.

The \HEJ framework may also be systematically improved by matching to
fixed-order calculations, in particular to NLO~\cite{Andersen:2020yax}.
Requiring that the fixed-order expansion of \HEJ{} agree with the fixed-order
result up to $\alpha_s^3$ without spoiling the logarithmic accuracy of \HEJ
leads to the following matching term:
\begin{align}
\frac{\sigma^{\text{NLO}}_{2j}}{\sigma^{\text{HEJ@NLO}}_{2j}} &=1+(f_{2j}^{(3)}-h_{2j}^{(3)}) \prod_{n=0}^{\infty}(-1)^n\alpha_s^{(n+1)}\frac{(f_{3j}^{(3)}+h_{2j}^{(3)})^n}{(f_{2j}^{(2)})^{(n+1)}}\;,
\end{align}
where $f$ are the fixed-order coefficient functions and $h$ are the corresponding \HEJ{} coefficients.
In the limit of $s\gg|t|$ this ratio tends to one, where in particular, $f_{2j}^{(3)} \to h_{2j}^{(3)}$.
Applying this term as a reweighting factor to the \HEJ{} resummed result gives the desired 
NLO accuracy fully differentially in the two- and three-parton phase space.

\subsection{Event generators for ultra-peripheral collisions}

Ultra-peripheral collisions (UPCs) are a special class of hadronic interactions.  They are photonuclear or two-photon interactions that occur when a nucleus emits a photon which interacts with the other nucleus, or with a photon emitted by the other nucleus \cite{Bertulani:2005ru,Baltz:2007kq}.  They typically require that the two nuclei do not interact hadronically (so the impact parameter $b>2R_A$, where $R_A$ is the nuclear radius.  Some reactions that have been studied have been vector meson photoproduction (including the $\rho$, $\rho'$, $J/\psi$, $\psi'$ and $\Upsilon(1S)$, photoproduction of dijets, and two-photon production of dilepton pairs and light-by-light scattering \cite{Klein:2020fmr}. Recently, photonuclear and two-photon interactions have also been observed in peripheral heavy-ion collisions. 

UPC Monte Carlos should model the photon flux carried by protons or heavier nuclei, accounting for the geometric non-hadronic-interaction requirement; the photon $p_T$ spectrum is also important.  The Monte-Carlo event generator STARlight~\cite{Klein:2016yzr} is commonly used for these simulations, to estimate cross-sections and kinematics, and to study detector acceptance.  It can simulate a wide variety of final states, including vector meson photoproduction, two-photon production of dilepton pairs, and more exotic reactions like axion production~\cite{Knapen:2016moh}.

For photoproduction, the photon flux is coupled with the cross-section for the specific process.  Coherent photoproduction of vector mesons is of particular interest because the cross-sections are large - comparable to the hadronic cross-section with lead beams at the LHC~\cite{Klein:1999qj}, and because the final state is distinctive - usually a two-prong decay with total transverse momentum less than 100~MeV/c.  Coherent photoproduction of vector mesons on heavy targets can modelled based on a Glauber or dipole calculation, using either calculations or data for photoproduction on proton targets.   One complication with coherent photoproduction in UPCs is that we cannot know which nucleus is the photon emitter and which is the target; the two possibilities interfere destructively.  The interference varies with the impact parameter, so calculations must carefully account for the geometry to correctly simulate the $p_T$ spectrum. This impossibility of uniquely identifying the photon source introduces a two-fold ambiguity regarding the photon energy \cite{Klein:1999gv}. It also complicates interpretation of the signal in terms of energy-dependent cross-sections. 

For two-photon interactions, the photon fluxes from the two nuclei can be convolved, and multiplied by the cross-section for the desired two-photon process.  In addition to STARlight, the UPCGen code may be used for this type of simulation~\cite{Burmasov:2021phy}.

Double-diffractive (double-Pomeron) interactions are also possible, and are dominant in proton-proton interactions.  They may be simulated using the SuperChic 3 Monte Carlo code~\cite{Harland-Lang:2018iur}, which also includes some photon-mediated channels. 

One additional complication for heavy ions is that $Z\alpha\approx 0.6$ ($Z$ is the nuclear charge), so many interactions involve multi-photon exchange.  Simulations must account for the possibility of these additional photons as well~\cite{Baltz:2002pp}.   STARlight can be used for this purpose, and the n$^0_0$n afterburner~\cite{Broz:2019kpl} can also be employed to simulate neutron emission from the additional interactions.

\subsection{Computing performance and portability}
\label{sec:performance_hadron_colliders}
The Physics Event Generator Working Group (WG) of the
HEP Software Foundation (HSF) has led the charge in studying
the performance of event generators for the LHC physics program.
A detailed description of their activities
and HSF-related developments can be found in the review document~\cite{HSFPhysicsEventGeneratorWG:2021xti},
produced by the WG in 2021 for the second phase of the LHCC review~\cite{bib:lhccreview,bib:lhcc145,bib:CERN-LHCC-2021-014} 
of computing for High-Luminosity LHC (HL-LHC),
and in the published WG paper
on the specific challenges in
Monte Carlo event generator software for
HL-LHC~\cite{HSFPhysicsEventGeneratorWG:2020gxw},
which had been prepared during the first phase of
the LHCC review in 2020 and 2021.
The previous documents from the HSF Generator WG have also been submitted as standalone Snowmass White Papers~\cite{snowmass_hsf_refs}.
In this section, we give an overview of this work and also briefly extend the discussion of event generators to other experimental facilities as well. 

\subsubsection{Profiling and benchmarking} 

Historically, event generators have been the fastest steps in the
simulated data workflow for LHC experiments and, therefore, have not
faced as much scrutiny as detector simulation and
reconstruction. However, this has changed in recent years with
experiments~\cite{ATLAS:2020pnm,cmshllhcprojections}, for some
scenarios, projecting computing needs to exceed the expected resource
budgets by several factors, with event generation comprising the
largest, single component of the workflow for some experiments.
Given this, CPU performance improvements can naturally have a large impact on the ability of the experiments to reach their physics goals.

Previous HSF Generator WG activities have led to
the generator groups 
performing more in-depth profiling of the 
CPU costs and possible computational inefficiencies 
of their software.  
By clearly identifying some performance bottlenecks, this has already
allowed major improvements in several generators.
Several concrete examples are:
\begin{itemize}
\item the latest version of the \sherpa event generator configuration
  used by ATLAS is expected to be three times faster than the previous
  one~\cite{ATLAS:2021yza} for one of the largest CPU-consuming
  processes, vector boson production in association with multiple jets, with no appreciable loss in the quality of the physics modeling
\item helicity recycling in \mgamc\ 
gives speed-ups of 
around a factor~$\sim$2 for many complex
processes~\cite{mattelaer-ostrolenk}
\item an optimized program flow for weight variation calculations in \sherpa,
leads to a factor of about~4 speed-up for the number of variation weights typically employed by ATLAS in representative processes~\cite{Bothmann:2022thx}
\item and a new interface to MCFM’s analytic one-loop amplitudes significantly speeds up the evaluation of one-loop amplitudes compared to automated tools~\cite{Campbell:2021vlt}. 
\end{itemize}
  
In parallel, a more detailed study of the CPU costs of MC generation campaigns in the experiments is progressing in order to better understand how the event generators perform where it really matters, e.g., on the Worldwide LHC Computing Grid for LHC simulations.
Establishing better mechanisms to collect CPU costs and generation efficiencies (e.g. due to sampling, merging or filtering) from the production systems of experiments would allow for an easier comparison between different experiments and different generators.
It is also important that experiments share this knowledge, especially
in published papers or public notes.

\subsubsection{Efficient phase-space sampling} 
The choice of a phase-space sampling is one of the most important
ingredients
in determining the efficiency of unweighted event generation.
Phase-space sampling may be especially inefficient for
high-multiplicity physics processes, where many kinematic combinations
can yield peaks.   Some methods have been proposed and studied to
improve the situation.   For example, neural networks have been
applied in \sherpa~\cite{Bothmann:2020ywa,Gao:2020zvv} to optimize
integration variables.
Other improvements are new, physics-based algorithms in \mgamc\, such as
the use of multi-weight channels and optimizing the order of integration of the virtuality of $t$-channel particles~\cite{mattelaer-ostrolenk},
which may lead to speed-ups by factors~$\sim$2 to~$\sim$5 for typical LHC processes (and to much larger speed-ups by three or four orders of magnitude for some VBF processes).
However, many of these improvements  have not yet been carefully validated
or used for large scale productions by the experiments.

\subsubsection{Negative weights} 
As described in Sec.~\ref{sec:xcuts_computing}, negative weights result in a drastic decrease in statistical power. Considering $t\overline{t}$ and $Hb\overline{b}$
production~\cite{Frederix:2020trv} as worst-case scenarios, 
the negative weight fractions are $r$=25\% and $r$=40\%, respectively.
This results in the need to generate 4 times and 25 times as many events to
achieve the same statistical precision on MC predictions.

Several strategies have emerged to mitigate this problem.
In particular,
significant progress 
on improved NLO matching prescriptions
has been achieved 
in both \mgamc\ and \sherpa.
The recent work on the MC@NLO-$\Delta$ 
technique~\cite{Frederix:2020trv} 
reduces the negative weight fractions from $\sim21$\% to $\sim7$\% for $W$-boson plus jet production, and provides even larger reductions for problematic processes like Higgs boson production in association with $b$-quarks, where the fraction reduces from $\sim40$\% to $\sim30$\%.
However, there is not yet a clear timeline 
for its use in a new production version of \mgamc,
partly because its implementation requires a strict coordination
between \mgamc\ and \pythia. 
Similar developments in \sherpa have led to a negative weight fraction reduction from $\sim18$\% to $\sim9$\% for $Z$-boson production in association with multiple jets~\cite{ATLAS:2021yza}.

By design, negative weights can almost be fully avoided in the \powheg NLO matching prescription~\cite{Frixione:2007vw}.  
Reference~\cite{Nason:2020lxx} describes an interface between the \powhegbox and \mgamc\ that allows generic event generation 
from \mgamc\ interfaced with \powheg to use \powheg NLO matching to avoid negative weights.
It is important, however, that the physics modeling of \powheg is different from that of \mcatnlo,
and the predictions which are at the same level of fixed-order accuracy may be different at particle
level and have different systematics~\cite{Frederix:2020trv}. As such, the elimination of negative weights
through \powheg is not always an option.
Another prescription that can avoid negative weights is \KrkNLO~\cite{Jadach:2015mza} which utilizes a different method than \mcatnlo or \powheg. However, it has been rarely used at the LHC, and it is available only for a small number of processes. 
As mentioned in section~\ref{sec:nlo-matching}, one can also
envisage relatively modest modifications of the underlying matching
procedures in order to eliminate otherwise irreducible sources of
negative weights using the MAcNLOPS method~\cite{Nason:2021xke}.

It is important to note, however, 
that improvements in precision calculations, 
such as NNLO in QCD matrix element calculations,
and 
new parton shower algorithms
(e.g., DIRE)
are expected to come 
with new sources
and a different level of negative weights depending on the process. 
Therefore, despite the recent progress, 
negative weights will likely remain an issue that needs close attention.

\subsubsection{Event weights for systematic uncertainties} 
Many analyses at the LHC depend on reweighted MC events for systematic uncertainty calculations. PDF variations and variations of the renormalization and factorization scale in matrix elements and parton showers can be handled by computing alternative weights for the same physical event, saving CPU cycles. However, this cannot be done in some cases, such as for parton shower starting scale and underlying event tune variations (including color reconnections and intrinsic transverse momentum). This results in extra load in MC production systems of the experiments because of the additional CPU cycles to produce each sample with different parameters. 
Better flexibility in (re-)deriving scale and PDF weights can be obtained by writing more information to the event record (at least in \sherpa), but the CPU and storage implications of this have not yet been assessed.

\subsubsection{Porting software to modern hardware: GPUs and vectorization on CPUs} 
Some aspects of event generators naturally offer an ideal workflow to exploit data parallelism on GPUs and vector CPUs, such as when the same calculations must be repeated in exactly the same way for different phase-space points (for instance, the matrix element calculation for different event kinematics).
This makes it possible to process different events in parallel in lockstep, using either 
single instruction, multiple threads (SIMT) processing with no thread divergence on GPUs, or
single instruction, multiple data (SIMD) vectorization on CPUs.
Among others, two recent activities in this area are especially interesting.

One promising development  
is the ongoing work~\cite{mg4gpu}
on reengineering 
the matrix element calculation in
\mgamc\ for GPUs and vector CPUs.
With respect to a single threaded non-vectorized CPU version on a modern Intel CPU, in double precision 
a throughput increase of a factor $\sim$4 was obtained on a modern Intel CPU through SIMD vectorization compared to a single-threaded non-vectorized implementation for C++ double precision calculations.  A throughput increase by two to three \textit{orders of magnitude} was observed using an Nvidia V100 GPU.   Both for CPU vectorization and V100 GPUs, an additional throughput increase by a factor $\sim$2 was obtained by moving from double precision to single precision (at the cost of some numerical instabilities).   These speed-ups were observed for the simple physics process $e^+e^-\rightarrow\mu^+\mu^-$, but seem to be confirmed by more recent preliminary studies for $gg\rightarrow t\bar{t}(g*)$ processes, with up to three gluons in the final state.
This work is making rapid progress
and should lead to a new production version 
of \mgamc\ in 2022,
also thanks to a collaboration
between theorists and software engineers
which was largely promoted
by the HSF generator WG. 

Another promising development is 
a study of the known variants of the Berends–Giele recursion~\cite{Berends:1987me} 
for gluon matrix elements, assessing their strengths and weaknesses on various computing 
platforms as well as providing a comparison to existing general-purpose matrix element
generators.
This was presented in~\cite{Bothmann:2021nch}. In a chip-to-chip comparison for 
similarly priced hardware, and focusing on the case of less than seven outgoing gluons, it was found that 
the best-performing GPU algorithm had evaluation times per event at least an order 
of magnitude smaller than the best CPU evaluation times. 
These results did not use any symmetries exclusive 
to the gluon-only case, and are representative of what can be expected from a 
full-fledged GPU matrix-element calculator for the Standard Model.

Other approaches to exploit GPUs for event generators have also been recently proposed, including some based on general purposed libraries such as \textsc{TensorFlow}~\cite{Carrazza:2020,Carrazza:2020qwu,Carrazza:2021zug}.
Further R\&D in porting and optimizing generators for heterogeneous computing architectures is essential to facilitate new and upcoming HPC resources and their unmatched parallelism. The GPU-based architectures of these resources will also be ideal for AI/ML approaches to accelerate and complement existing algorithms for MC event generation.

\section{Neutrino experiments}
\label{sec:neutrino_experiments}

This section is dedicated to event generators for neutrino-nucleus interactions, with a focus on accelerator-based experiments.
Understanding these interactions is the main challenge to constraining and measuring the
properties of neutrinos and exploring BSM signatures in neutrino experiments.

The extraction of fundamental neutrino parameters such as mixing angles, CP phase and mass ordering requires the knowledge of the incoming neutrino beam energy. While in all other high-energy and nuclear physics experiments the beam energy is known to a high precision, this is not the case for the neutrino beam energy, where only broad distributions are known. The beam energy has to be reconstructed event-by-event from the observation of the final state of the reaction; the latter is --- due to detector acceptance limitations --- also only partially known. This is where generators are needed, and without them the fundamental neutrino parameters cannot be accurately measured.

Another complication arises from the fact that all presently running or planned experiments use nuclear targets. This requires not just the generation of the fundamental events, but also --- in addition --- a reliable description of final-state interactions in the target. The latter in turn requires the knowledge and implementation of nuclear physics structure and reaction theory.

To understand the importance of properly modeling neutrino-nucleus interactions, we can give a few key examples of the role of cross section uncertainties in experimental measurements.
Regarding measurements of standard oscillation physics parameters, an interesting example of the importance of accurate cross section modeling can be found on how changes on the neutrino-nucleus interaction modeling has erased NOvA's preference for non-maximal mixing~\cite{NOvA:2017ohq, NOvA:2018gge}.
Looking to the future, we highlight that the DUNE collaboration has estimated that an uncertainty on signal interaction rates greater than 1\% will substantially degrade the sensitivity to CP violation and the neutrino mass hierarchy~\cite{DUNE:2015lol}.
Neutrino-nucleus modeling is also important for Hyper-Kamiokande ---a water Cherenkov detector cannot distinguish electrons from photons. Due to that, neutral current $\pi^0$ production is a significant background for the $\nu_e$ appearance channel, which is crucial for the measurement of the  CP phase, and therefore a proper modeling of the backgrounds will have direct impact on the sensitivity to CP violation.
If we turn our attention to new physics searches, we note that the recent MicroBooNE analyses on the $\nu_e$ interpretation of the MiniBooNE low energy excess are all dominated by cross section uncertainties~\cite{MicroBooNE:2021jwr, MicroBooNE:2021nxr, MicroBooNE:2021sne}, which range roughly between 15-25\%.
The future SBND experiment, as the near detector of the SBN program, will collect an enormous sample of neutrino events. 
This will make neutrino event generators even more important in bridging high statistics near detector measurements to predictions in other detectors, such as DUNE, and in searches for new physics.

Nevertheless, 10 to 20\% discrepancies between event generator predictions and data have been observed in double differential cross sections~\cite{MINERvA:2020zzv} and exclusive observables such as proton multiplicity~\cite{Palamara:2016uqu}.
Similar disagreements are found when using neutrino event generators to describe neutrino-electron scattering data~\cite{Ankowski:2020qbe, CLAS:2021neh}.
It becomes clear that improving the description of neutrino-nucleus interactions is crucial for the success of current and future neutrino experiments.
In the following, we review the experimental needs related to the description of neutrino-nucleus scattering, 
the current status of generators, generator tuning procedures. and future requirements of event generators to enable neutrino experiments to develop their full potential.

The programmatic aims and event generator requirements of current and
next-generation neutrino facilities is reviewed in Sec.~\ref{sec:nuexpt-needs}.
An overview  of the main physics
ingredients needed to simulate realistic neutrino-nucleus interactions
is presented in Sec.~\ref{sec:nu:key_components}.
Recent developments and priorities specific 
to commonly-used generator frameworks are summarized in Sec.~\ref{sec:nu:generators}.
Special attention is given  to simulations of and
constraints from charged-lepton scattering on nuclei in Sec.~\ref{sec:nu:ch-lep},
due
to its strong potential to constrain model ingredients that 
can otherwise be challenging to disentangle through neutrino data alone.
Section~\ref{sec:nu:needs} focuses on several issues that need to be addressed urgently
to realize the key physics goals, including upgrades
to tuning and uncertainty quantification, as well as development of geometry/flux
and theory interfaces.

\subsection{Experimental needs and opportunities}
\label{sec:nuexpt-needs}

Event generators need to make robust predictions over a wide range of neutrino beam energies, from hundreds of MeV to tens of GeV.
Until recently, accelerator-based neutrino oscillation experiments have placed significant emphasis on simulating neutrino interactions
for neutrino beam energies $E_\nu$ around 1 GeV~\cite{NuSTEC:2017hzk}.
This is an energy region where quasielastic (QE) interactions are dominant, and 2-body kinematics allow for the reconstruction of the neutrino energy~\cite{Katori:2016yel}.
The significant impact of nucleon correlations within nuclei has lead to additional theoretical interest to this energy region~\cite{NuSTEC:2017hzk,Katori:2016yel}.
Among future projects, DUNE uses a wide-band energy beam ($E_\nu \sim 1-6$) GeV, while HyperK uses a narrow-band low energy beam ($E_\nu \sim 600$~MeV).
DUNE, HyperK, and JUNO~\cite{JUNO:2015zny} also intend to measure atmospheric neutrinos with high-statistics; energies $\le 20$ GeV are key to determining the neutrino mass ordering. As these projects move on to higher energy regions, precise simulation of neutrino interactions above the QE region is important.

\subsection{Key physics components}\label{sec:nu:key_components}

\begin{figure}
    \centering
    \includegraphics[width=0.47\textwidth]{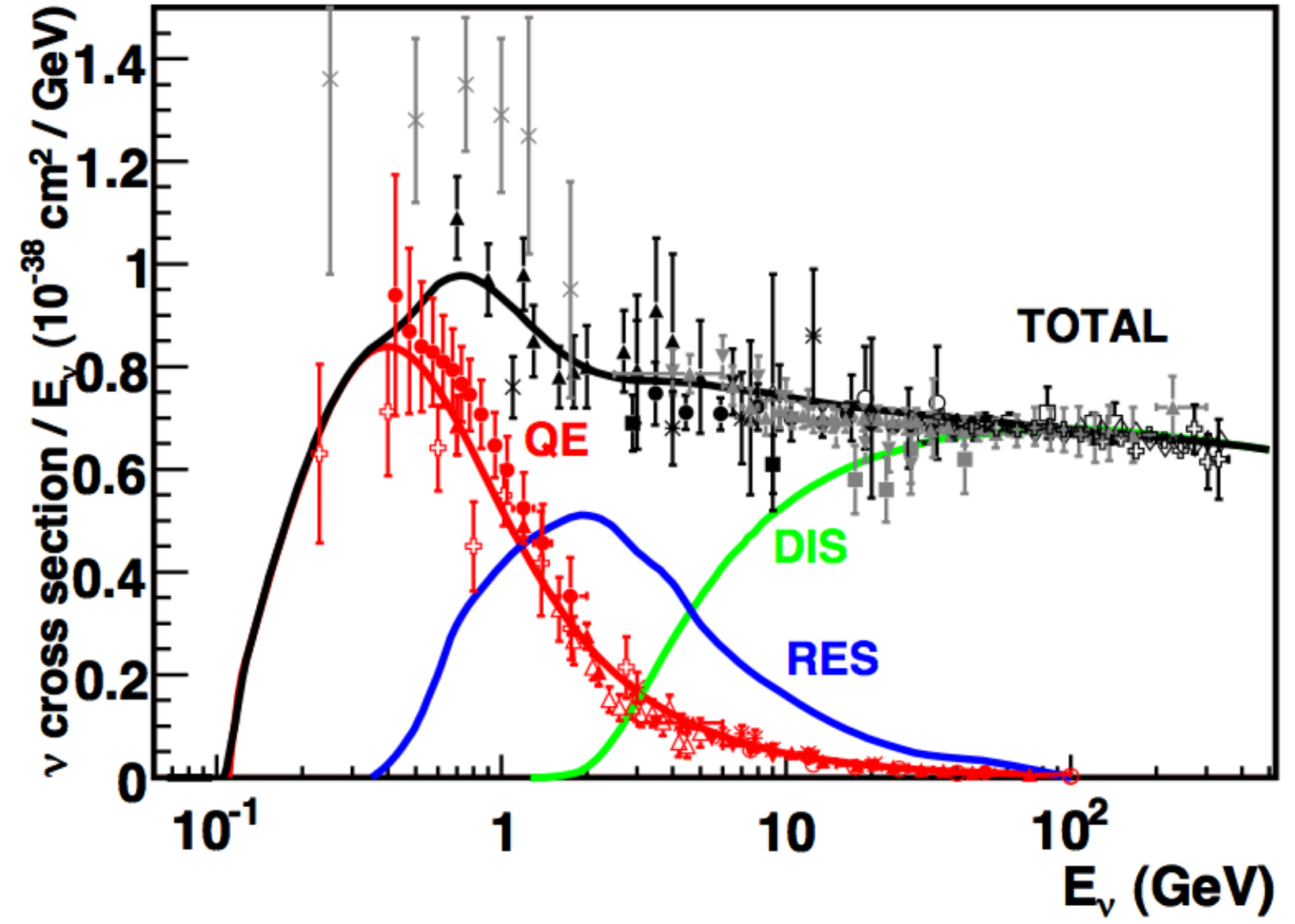}
    \hfill
    \includegraphics[width=0.47\textwidth]{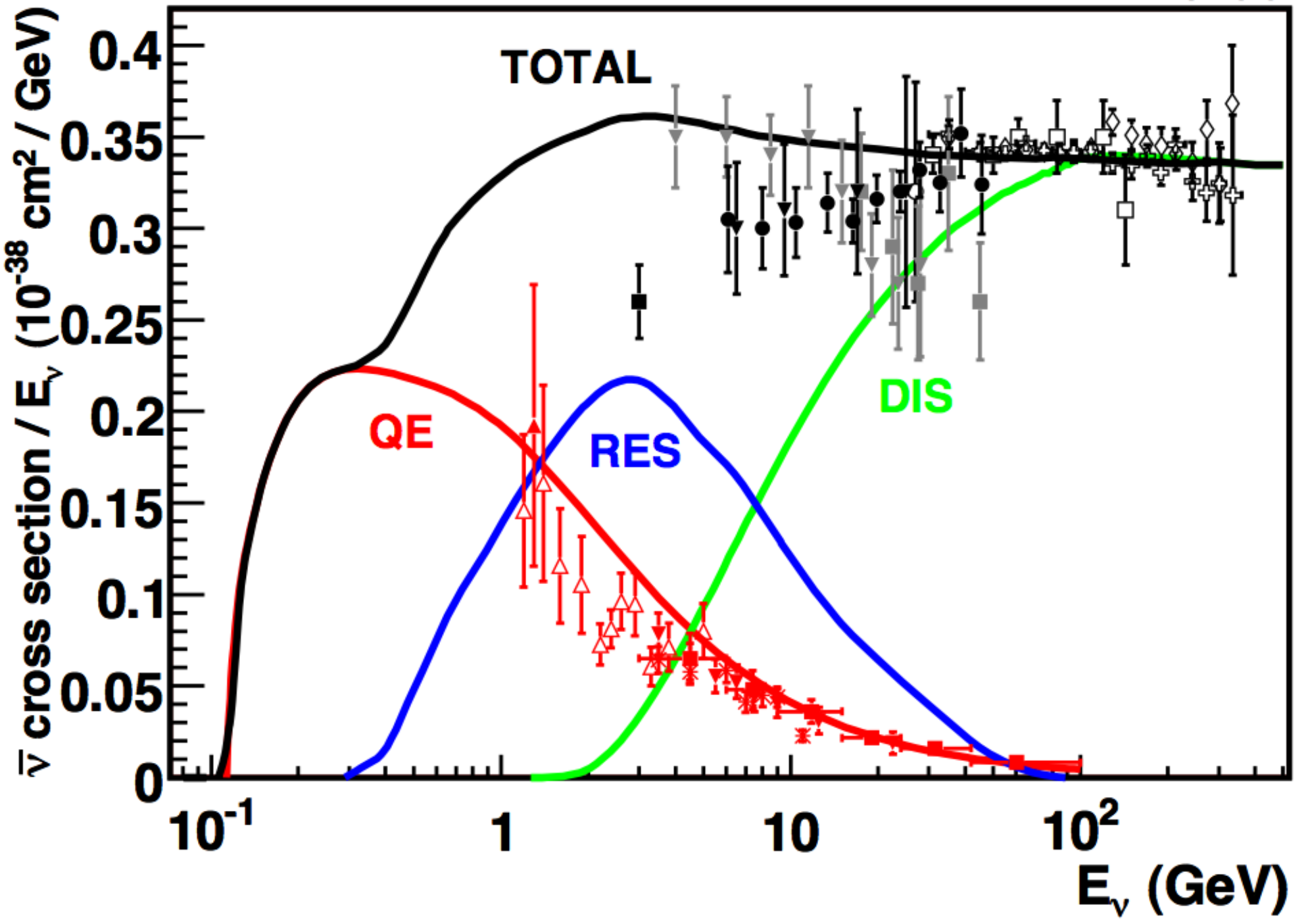}
    \caption{Comparison of neutrino (left) and anti-neutrino (right) cross sections. Data is from~\cite{Formaggio:2012cpf} and the theory calculations are from~\cite{Casper:2002sd}. The figure is reproduced from~\cite{Formaggio:2012cpf}.}
    \label{fig:nu_xsec}
\end{figure}

In different energy regimes, different interaction models become more or less important.  These models are quasielastic scattering (Sec.~\ref{sec:nu:QE}), meson exchange currents (Sec.~\ref{sec:nu:MEC}), hadron resonance production (Sec.~\ref{sec:nu:RES}), and deep inelastic scattering (Sec.~\ref{sec:nu:DIS}). 
The breakdown of the scattering cross section for the different processes versus the (anti-)neutrino energy is shown in Fig.~\ref{fig:nu_xsec}.
A discussion of the especially important, but poorly understood, matching between the resonance production channel and the deep inelastic scattering is
outlined in Sec.~\ref{sec:nu:RES-DIS}.
Details on the nucleon form factors applied to account for the non-point like nature of nuclei and nucleons can be found in Sec.~\ref{sec:nu:form_factors}.
Furthermore, the modeling of hadron interactions as they escape the nucleus using cascade algorithms is covered in Sec.~\ref{sec:nu:FSI}.
Finally, the current status and recent developments in simulating non-standard interactions,
which requires the simulation of leptonic final states at higher multiplicities, is  discussed in Sec.~\ref{sec:nu:BSM}. For additional details see the theoretical tools for neutrino scattering white paper~\cite{Ruso:2022qes}.

\subsubsection{Quasielastic Scattering}\label{sec:nu:QE}
Quasielastic (QE) neutrino scattering on nuclei is the dominant interaction process for incident neutrino energies $E_\nu$ on the order of hundreds of MeVs. 
In this kinematic region, the lepton interacts with individual bound nucleons via a one-body current operator. The current operator is a sum of a vector contribution, depending on a combination of the proton and neutron electric $G^{(n,p)}_E$ and magnetic $G^{(n,p)}_M$ form factors; and an axial term, written in terms of the axial $F_A$ and induced pseudoscalar $F_P$ form factors. 

QE electron scattering on nuclei can be used to constrain and validate neutrino scattering predictions, and
data from experiments on light $A<4$ nuclear targets have been used to extract the neutron charge and magnetic form factors as well as investigate 
possible medium modifications of the nucleon form factors.

Instead of the sharp peak in the excitation spectrum observed in true elastic scattering,
the nucleon motion inside the nuclear medium yields a broad peak.
The width of this peak provides a direct measure of the average momentum of nucleons in nuclei, and has been used to determine nuclear Fermi momenta.

Scaling of the inclusive cross section is another important feature of the QE region. This refers to the fact that the inclusive electromagnetic cross section divided by an appropriate function describing the elementary electron-nucleon elastic scattering process, no longer depends on two variables---the energy and momentum transfer $\omega$ and ${\bf q}$---but only upon a single variable $y(\omega,{\bf q})$.
Violations of scaling quantify the effectiveness of the Impulse Approximation (IA) picture of QE scattering.

Different approaches have been developed to account for quantum-mechanical effects in Final-State Interactions (FSI) in the quasielastic region. 
To correct the Impulse Approximation results, the real part of an optical nuclear potential is added to the free energy spectrum of the outgoing nucleon and the cross section is folded with a function accounting for rescattering effects. In the Relativistic Mean Field approach, FSI between the outgoing nucleon and the residual nucleus are accounted for by solving the associated Dirac equation using the same mean field as used for the bound nucleon. For more details on FSI see Sec.~\ref{sec:nu:FSI}.

\subsubsection{Meson Exchange Current}\label{sec:nu:MEC}

An important correction to QE scattering is the meson exchange current (MEC) that accounts for multi-nucleon
correlations in the initial state.
For electron QE scattering, this contribution is a direct consequence 
of current conservation.   The inclusion of a nucleon-nucleon interaction term in the Hamiltonian describing
the initial nuclear state requires the introduction of a two-body current operator to guarantee gauge invariance. The role of two body currents
in electron scattering has been studied in connection with scaling violations in the longitudinal and transverse electromagnetic scaling functions in the quasielastic region~\cite{Carlson:1994zz,Carlson:2001mp}. 

MiniBooNE measurements played a central role in directing theoretical interest towards the MEC by revealing the presence of additional nuclear processes in neutrino scattering.
The discrepancy observed between the CCQE data---measuring one lepton and no pion final state---and different theory predictions including only the one body current contribution, suggested the occurrence of events with two particle-two hole final states induced by two-body currents not included in the theory calculations~\cite{Katori:2009du,MiniBooNE:2010bsu,Martini:2009uj}.
According to the definition above, these events cannot be distinguished from those with one particle-one hole final states.
Therefore, they are often referred to as CCQE-like.
The two-body current contribution to neutrino-nucleus scattering has been studied by different groups~\cite{ValenciaModel,Susav2,Martini:2009uj},
and some differences are obtained by the different theoretical approaches.

Two of the most used models~\cite{ValenciaModel,Susav2} in generators
include processes involving nucleons and $\Delta$ resonances, and therefore have applicability to neutrino energies only up to a few GeV.  The effects of virtual pions are also included.  The key challenges to modeling the MEC are the proper inclusion of two-nucleon correlations, binding energy, and in-medium corrections.

\subsubsection{Resonance Production}
\label{sec:nu:RES}
In resonance production (RES), a neutrino probe interacts 
with a single bound nucleon, exciting it to any of a large number 
of baryonic resonances.   These resonances decay 
into a final-state nucleon and mesons — pions, predominantly.
Because of the beam energy, resonance production will be one of the dominant processes in the DUNE data~\cite{DUNE:2015lol}.
 
Most neutrino generators describe resonance production through implementations of 
the Rein-Sehgal (RS) model~\cite{Rein:1980wg}. The RS model is a generalization of the 
relativistic quark model of Feynman, Kislinger, and Ravndal~\cite{Feynman:1971wr}, originally 
adapted to predict pion photoproduction using matrix elements.
While the RS model quantifies the resonant process, there are also important non-resonant
contributions to the same final states.
Most neutrino generators 
combine the resonant RS piece with a smooth background function based on a 
rescaling of the Bodek-Yang DIS model~\cite{Bodek:2002vp,Bodek:2002ps,Bodek:2010km,Bodek:2021bde} discussed in more detail in 
Sec.~\ref{sec:nu:DIS}.
Recent improvements to generators include the treatment of low lying resonances~\cite{Berger:2007rq}.  In addition, there is work by Kabirnezhad~\cite{PhysRevD.102.053009} to implement the Rein model that includes resonant and non resonant processes and their  interference in a formalism that treats electron and neutrino processes equally.  

A major challenge is to include all the information on in-medium corrections from analysis of earlier hadron and photon experiments.  Another is to include all meson processes from non-resonant and resonant processes in a coherent way.  This has been done by the Sato-Lee group~\cite{Kamano:2019gtm} and that work is starting to move into the generators.  A NuSTEC workshop~\cite{NuSTEC:2020nsl} explored the situation in detail. 

\subsubsection{Deep Inelastic Scattering}
\label{sec:nu:DIS}
Neutrino Deep Inelastic Scattering (DIS) is described using the parton model and perturbative QCD, and thus has overlap with
the discussion in Sec.~\ref{sec:eic}. The primary interaction
is characterized by the scattered neutrino interacting 
predominantly with quark-gluon, ({\it i.e.}, {\it partonic})
degrees of freedom within the target nucleus. These interactions
occur incoherently via charge- or neutral-current exchange, and come into
play beginning in the few-GeV region, ultimately dominating the
neutrino-nuclear cross section at higher energies.  Traditionally,
QCD factorization theorems~\cite{Collins:1989gx} permit the neutrino-scattering cross section
to be decomposed into perturbatively-calculable Wilson coefficients and
characteristic parton distribution functions (PDFs) of the struck
nucleus. It should be stressed that this picture of neutrino
interactions with individual partons at higher energies --- a process
that is itself the subject of substantial theoretical and phenomenological
focus --- becomes still more complicated at lower energies, particularly
the few-GeV region of greatest relevance for DUNE. The reason for this is
that a number of nonperturbative QCD effects, including the insufficiency
of a purely leading-twist description~\cite{Ellis:1982cd} of the DIS interaction and the presence
of target-mass effects~\cite{Brady:2011uy,Schienbein:2007gr,Georgi:1976ve}, must be consistently and systematically incorporated.
In addition, there is a phenomenological ambiguity concerning the threshold and
onset of the DIS region as distinct from the resonance and shallow-inelastic
scattering (SIS) regimes. Different neutrino generators generally abide by
distinct kinematic definitions in terms of $W^2$ and $Q^2$ in their default
settings. We discuss the issues associated with this resonance-to-DIS transition
in greater detail in Sec.~\ref{sec:nu:RES-DIS}.

Currently, neutrino generators implement phenomenological
models to provide an effective description of the neutrino DIS cross section.
The prevailing model is that of Bodek and Yang (BY)~\cite{Bodek:2010km}, which
computes the DIS cross section in terms of leading-order PDFs (GRV98)~\cite{Gluck:1998xa} rescaled
via phenomenological factors to empirically describe available data at softer
$Q^2$ and $W^2$. This approach is generally taken with a combination of \pythia~\cite{Sjostrand:2006za}
and dedicated string models for hadronization in event generation, see Sec.~\ref{sec:fragmentation}.

\subsubsection{Form factors}
\label{sec:nu:form_factors}
In the interaction between a neutrino and a nucleus, we can distinguish three different levels: the quark, nucleon, and nuclear level.
At the quark level,  the neutrino interacts with the point-like quarks in the parton model. 
Calculations are carried out using the Standard Model Lagrangian or extensions including 
beyond-the-standard-model (BSM) physics.
At the nucleon level and at four-momentum transfers comparable to the nucleon mass, 
the composite nature of the nucleon is represented by form factors.
Since most current and future experiments require nuclear targets, 
we need to consider how the nucleons and the hadrons produced in the interaction behave inside the nucleus. This nuclear level requires input from nuclear physics. Thus, to extract a Lagrangian parameter, such as Standard Model or BSM coupling, from data, we need to ``unfold" the data from the nuclear to the nucleon level, and then from the 
nucleon to the quark level.

Unlike the case for lepton-{\it nucleon} interactions, there are no rigorous, QCD-based factorization theorems to separate among each of these three levels in lepton-{\it nuclear} scattering, but this approach seems to be the most phenomenologically sound.  It is sometimes not even explicitly stated when discussing the neutrino-nucleus interaction. We would like to isolate and control the effects coming from the quark, nucleon, and the nuclear level. Luckily, there has been considerable progress in the last 10 years in our theoretical and experimental knowledge of the nucleon form factors. To some extent these were motivated by experimental anomalies. For the proton electric form factor it was the ``proton radius puzzle" \cite{Pohl:2010zza}, related to different values of the form factor slope at zero $q^2$ from different experiments. For the nucleon axial form factor it was the value of the axial mass reported by MiniBooNE \cite{MiniBooNE:2007iti}. 

The A1@MAMI experiment~\cite{A1:2010nsl,A1:2013fsc} has provided high-precision data both for the electric and magnetic proton form factors at low momentum transfers of interest to the neutrino physics community. However, there are some tensions to the PRad@JLab experiment~\cite{Xiong:2019umf} at the level of the electric form factor and to the older dataset at the level of the magnetic form factor. These questions are going to be investigated with future measurements of electron- and muon-proton scattering. Now, modern fits for the nucleon vector form factors are largely determined by the measurements of the A1@MAMI.

Besides the experimental progress, an important theoretical improvement from the last 10 years was done regarding the parameterization of the form factors. Prior to 2010, several models were used ranging from the historical dipole model, to multi-parameter models based on assumed functional forms. In \cite{Hill:2010yb},  it was first suggested to use the $z$ expansion for baryonic form factors, and further studies on its use were detailed in \cite{Lee:2015jqa}. The $z$ expansion relies on the known analytical properties of the form factor. It is used extensively in flavor physics for meson form factors,  where it is the universally accepted parameterization and is a common instrument for fits of the nucleon form factors. 

These studies of $z$-expansion parameterization were combined with all available electron-nucleon scattering data~\cite{Lee:2015jqa}, extended by atomic physics and neutron scattering constraints, and further optimized for the kinematics of neutrino oscillation experiments~\cite{Borah:2020gte}. The resulting fit for all nucleon vector form factors is presented in the form of $z$-expansion coefficients with the corresponding covariance matrices and allows a rigorous error propagation to any experimental observables.

The nucleon axial form factor is extracted mainly from the deuterium bubble chamber data from the 1970s and 1980s with additional $\beta$-decay constraints on the normalization. Besides the partially lost expertise and unknown incoming neutrino fluxes, an updated model-independent $z$-expansion fit~\cite{Meyer:2016oeg}, which does not rely on flux predictions, was recently performed. This fit is also suitable for the rigorous propagation of errors to experimental observables.

Lattice QCD calculations can be used to predict nucleon form factors and other few-body observables that can be used as inputs to nuclear many-body calculations, as detailed in the lattice white paper~\cite{Ruso:2022qes}. Results are often complementary to those obtained from experiment; for example, different flavor combinations of vector and axial form factors can all be calculated with similar computational resource requirements. Vector-current form factors are precisely determined by electron scattering experiments and provide validation for lattice QCD results. State-of-the-art nucleon electromagnetic elastic form factor calculations using approximately physical values of the quark masses can achieve few-percent uncertainties for $Q^2 \lesssim 1$ GeV$^2$ and show good agreement with experimental determinations~\cite{Alexandrou:2017ypw,Ishikawa:2018rew,Shintani:2018ozy,Alexandrou:2018sjm,Park:2021ypf}. Nucleon axial-current elastic form factor calculations can achieve similar precision (and the range of $Q^2$ used in current calculations can be extended), and lattice QCD axial form factor results with fully controlled uncertainties and higher precision than experimental determinations will soon provide a valuable input to nuclear many-body calculations. Axial form factor calculations in particular face challenges from $N\pi$ excited states that are found to have large axial-current transition matrix elements in chiral EFT~\cite{Bar:2018xyi,Bar:2019gfx} and lattice QCD~\cite{Jang:2019vkm,RQCD:2019jai,Park:2021ypf} and therefore complicate extractions of nucleon elastic axial form factors. Calculations including careful treatment of these excited-state effects are found to pass consistency checks related to chiral Ward identities that were apparently violated in earlier calculations~\cite{Jang:2019vkm,RQCD:2019jai,Park:2021ypf}; however, there is tension between these lattice QCD results and phenomenological determinations of axial form factors from deuterium bubble-chamber data~\cite{Meyer:2022mix}. Further refined lattice QCD calculations including additional ensembles and critical investigations of the systematic uncertainties in both lattice QCD and experimental form factor determinations will be critical for obtaining reliable axial form factor results with few-percent uncertainties over the next 5-10 years.

Lattice QCD can also be used to predict inelastic nucleon transition form factors such as $N \rightarrow N\pi$, including the effects of the $\Delta(1232)$ and higher-energy nucleon resonances. Calculations of $N\pi$ scattering and $N\rightarrow N\pi$ transition form factors are at a much more exploratory stage than calculations of nucleon elastic form factors, but work in this direction is ongoing~\cite{Andersen:2017una,Silvi:2021uya,Barca:2021iak}. Further, variational methods that can be used to reliably include $N\pi$ states in transition form factor calculations can also be used to explicitly subtract their effects from nucleon elastic form factor calculations, and the development of resonant form factor calculations will proceed synergistically with that of elastic form factor calculations. Accurate and precise determinations of $\Delta(1232)$ and other resonance production amplitudes will be essential for achieving few-percent total cross-section uncertainties for experiments such as DUNE where the neutrino flux is peaked close to the resonance region. Lattice QCD results for nucleon resonant form factors would provide complementary information to experimental neutrino scattering data~\cite{Alexandrou:2006mc,Alexandrou:2007dt,Hernandez:2007qq,Alexandrou:2010uk,Hernandez:2010bx}, including for example detailed descriptions of the angular dependence of resonance production amplitudes, that could be used to inform nuclear many-body theories and event generators where nucleon-level resonant amplitudes enter as inputs~\cite{Rocco:2019gfb}. Although the computational costs of lattice QCD calculations increase exponentially with the number of nucleons included, it is also possible to compute nuclear matrix elements for few-nucleon systems directly from lattice QCD that can be used to inform models of meson exchange currents~\cite{Savage:2016kon,Davoudi:2020ngi,Parreno:2021ovq}. Exploratory calculations of resonant nucleon transition form factors, few-nucleon axial matrix elements and form factors, and also structure functions relevant to neutrino DIS are reviewed in a Snowmass paper~\cite{Constantinou:2022yye}, will mature over the next few years and it will become increasingly important to develop robust pipelines connecting theory results for few-nucleon systems to neutrino event generators.  

\subsubsection{Resonance and DIS matching}
\label{sec:nu:RES-DIS}

For $E_\nu \sim $ $1-10$ GeV,  different effects are important and must be combined into a
a sensible interaction model.   This is referred to as the the shallow-inelastic scattering (SIS)~\cite{SajjadAthar:2020nvy} regime,
and is characterized by $Q^2$ $\le 1$ GeV$^2$ (not deep, but ``shallow") and $W$ greater than the pion production threshold (``inelastic").
The theory behind SIS is interesting in itself, and includes inelastic processes with higher baryonic resonances, quark-hadron duality, and nuclear DIS. 

At this time, no neutrino event generator accounts for the interference effects from the overlap of all states that would be needed to simulate SIS correctly. Instead, neutrino interactions are treated at a nucleon level and added incoherently. Various corrections, most notably nuclear corrections, are applied subsequently.
The interaction models are also connected to avoid discontinuity in simulation. However, it is extremely challenging to simulate neutrino interactions smoothly in all kinematic variables.
First, the high-$W$ region of RES is dominated by higher baryonic resonances. However, none of these are identified by neutrino scattering experiments and models have not been validated in this region. Second, RES and DIS are treated incoherently, although they are related by quark-hadron duality~\cite{SajjadAthar:2020nvy}.
The Bodek-Yang model~\cite{Bodek:2002ps}, based on electron scattering data, attempts to merge these two processes, but, again, it is not fully validated by neutrino data. For nuclear targets, we also expect modification of the PDF which is again well understood for electron scattering but not for neutrino scattering. Consequently, these theoretical problems need to be addressed before we can implement the models in event generators. 

\subsubsection{Hadronization at Low Energies}
\label{sec:nu:had}

\begin{figure} [tbp]
\centering
\includegraphics[width=0.49\textwidth]{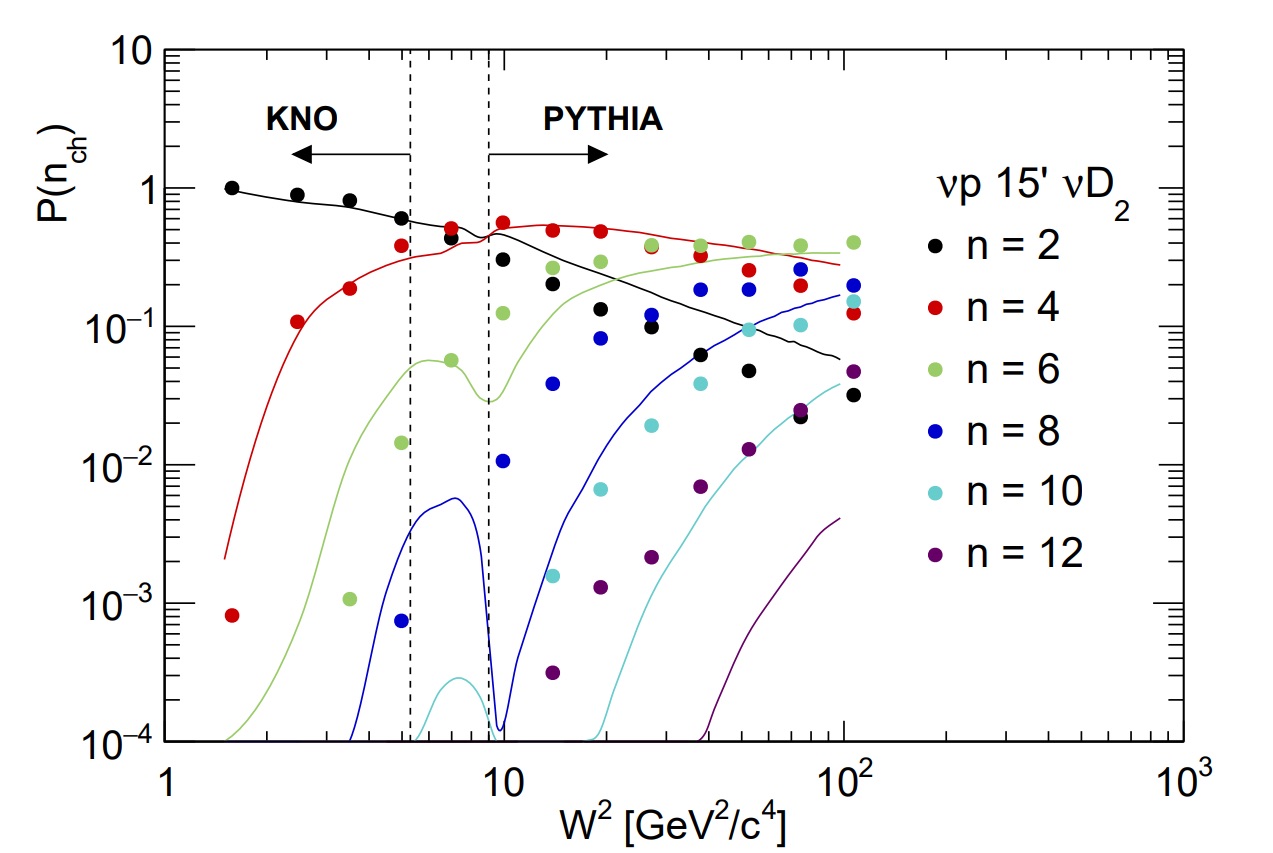}
\includegraphics[width=0.49\textwidth]{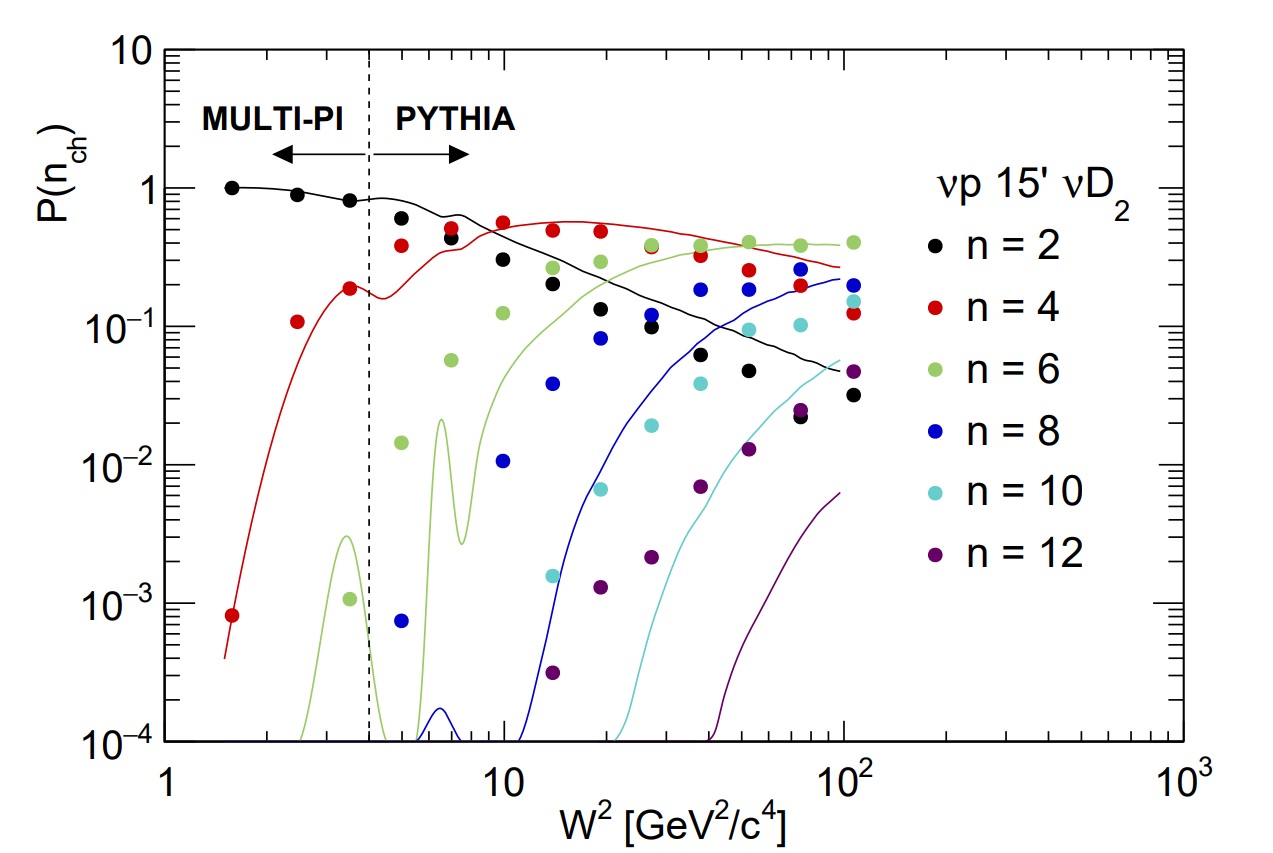}
\caption{\label{fig:SISTopo}
Data-MC comparisons for normalized topological cross-sections for muon neutrino - proton scatterings. The left is from \GENIE v3 (tune \texttt{G18\_02a\_02\_11a})~\cite{GENIE:2021wox} and the right is from the \textsc{NEUT} v5.4.0.1~\cite{Bronner:2016gmz}. Data are from the Fermilab 15' bubble chamber~\cite{Zieminska:1983bs}. MC simulations show large non-physical discontinuities of topological cross-sections and these simulations suggest event-by-event predictions of hadronic final states are extremely challenging using current methods.}
\end{figure}

Neutrino experiments have detectors that are fully active with 4$\pi$ coverage.
This allows the measurement of exclusive final state, and presents a challenge to event generators.
Hadron production is based on phenomenological models such as the Lund string model~\cite{Sjostrand:1982fn}.
However, the assumptions behind the models do not clearly match the typical invariant mass in the RES-DIS transition region ($<3$~GeV).
For this reason, the AGKY model was developed for \GENIE, based on the charged hadron multiplicity  extracted from data
and applying the KNO scaling law~\cite{Koba:1972ng}.
The predicted multiplicities are smoothly connected to \pythia6~\cite{Sjostrand:2006za} for the high-$W$ region.
In this way, continuous topological cross-sections are drawn in all of phase space. Figure~\ref{fig:SISTopo} shows the $\nu_\mu p$ normalized topological cross-sections simulated by \GENIE v3~\cite{GENIE:2021wox} and \textsc{NEUT} v5.4~\cite{Bronner:2016gmz} compared with bubble chamber data~\cite{Zieminska:1983bs}. There are visible discontinuities where the low-$W$ hadronization models based on the KNO scaling switches to \pythia (note, \textsc{NEUT} uses \pythia5). These discontinuities originate from using the string model at very low $W$ ($<3$ GeV).
The dispersion from the string model is  narrower than that predicted from KNO scaling based on low-energy bubble chamber data.
It may be possible to tune the string model for this low $W$ region~\cite{Katori:2014fxa}, but, again, this is likely beyond the expected applicability of the model.  Alternatively, this discontinuity can be avoided by moving this transition to a much higher $W$ region where the string model is well validated. In general, matching hadron final-state predictions from neutrino event generators to data is challenging. Efforts to predict high-multiplicity data from high-resolution neutrino experiments~\cite{Chukanov:2016lra,MicroBooNE:2018xad,NINJA:2020gbg,NINJA:2020bvx} has recently begun.

\subsubsection{Final-State Interaction Models}\label{sec:nu:FSI}
An important
aspect of neutrino interactions is that hadronic states, produced in the primary collision, must propagate
through the nuclear environment to be detected.
To model these interactions completely, the interference of all interaction mechanisms yielding the same final state must be included.
In practice, this is not computationally feasible. Two approaches to solve this problem have been developed: intranuclear cascades and hadronic transport equations. For a discussion of hadronic transport at higher energies see Sec.~\ref{sec:hadronic_transport}.

\paragraph{Intranuclear Cascade}
In intranuclear cascades, particles are assumed to propagate classically through the nucleus between consecutive scatterings upon background nucleons~\cite{Serber:1947zza,Metropolis:1958sb,Bertini:1963zzc,Cugnon:1980zz}. The interactions are traditionally modeled using free-space elementary cross sections~\cite{said} with modifications to account for Pauli blocking and non-relativistic in-medium effects~\cite{Pandharipande:1992zz,Salcedo:1987md}.  Although these methods are simple, they describe a wide variety of data with hadron and electromagnetic beams.

\begin{figure}
    \centering
    \includegraphics[width=0.47\textwidth]{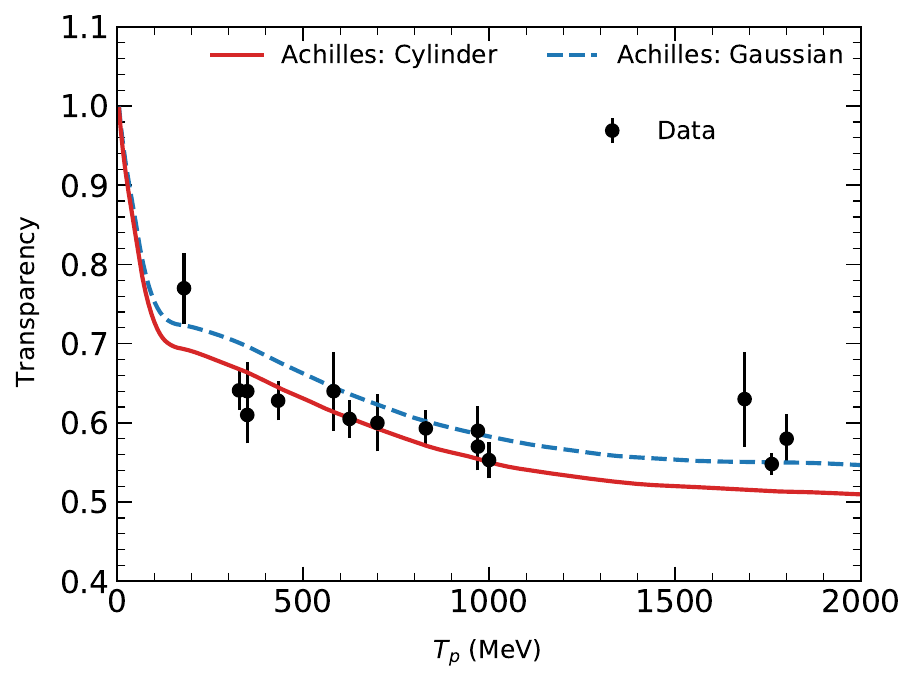}
    \hfill
    \includegraphics[width=0.47\textwidth]{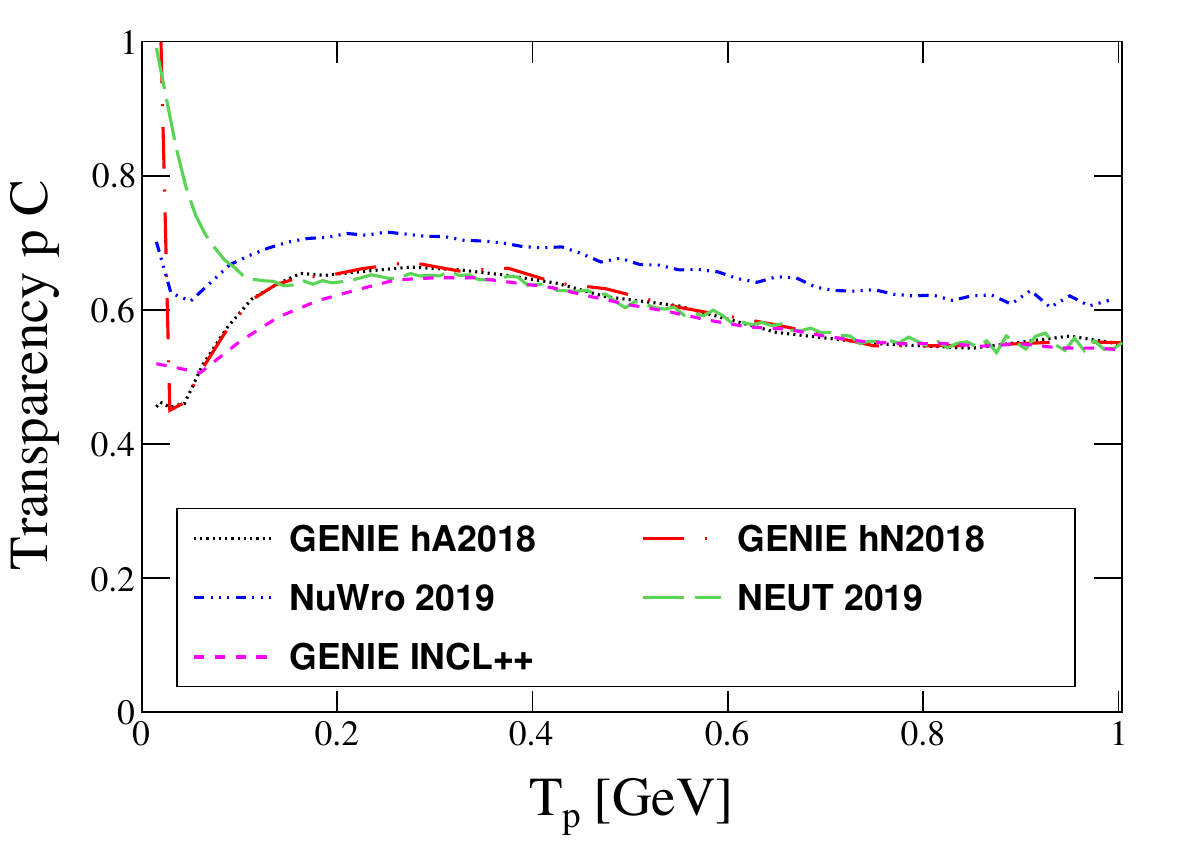}
    \caption{Carbon transparency as a function of the kicked proton kinetic energy. Left: Shows results for \achilles for the cylinder and Gaussian interaction models compared to data from~\cite{Garino:1992ca,ONeill:1994znv,Abbott:1997bc,Garrow:2001di,JLabE91013:2003gdp,E97-006:2005jlg}, figure is reproduced from~\cite{Isaacson:2020wlx}. Right: Shows results from \GENIE hA2018, \GENIE hN2018, \neut 2019, \nuwro 2019, \GENIE INCL++, figure is reproduced from~\cite{Dytman:2021ohr}.}
    \label{fig:cascade}
\end{figure}

An IAEA study~\cite{iaea} examined inclusive cross sections for neutron and proton scattering off nuclei at a variety of energies, and more recent studies can be found in~\cite{Niewczas:2019fro,Isaacson:2020wlx, Dytman:2021ohr}. In addition to nucleon-nucleus scattering, the papers also investigated the proton transparency. This is a new and interesting observable for traditional neutrino event generators to study. A comparison between available algorithms in neutrino event generators for transparency can be found in Fig.~\ref{fig:cascade}.
Both studies found that current FSI methods based on hadron-nucleon scattering are able to model both total reaction cross section and transparency data for nuclear targets sufficiently for existing data.
The authors of~\cite{Dytman:2021ohr} found that, in their intranuclear cascades, the transparency is sensitive to a variety of factors such as nucleon-nucleon correlations and formation zones. 
Nevertheless, a different approach developed in~\cite{Isaacson:2020wlx} found that the transparency is strongly affected by the modeling of formation zones and only mildly sensitive to correlations. To investigate the different conclusions, a detailed study is required.
Further comparison to data, particularly for exclusive and differential observables, is needed for this to become a standard validation tool. However, these datasets are currently unavailable or not precise enough to distinguish the differences among models.
For additional details on how each generator implements the intranuclear cascade, see Sec.~\ref{sec:nu:generators}.

\paragraph{Hadronic Transport Models}

Hadronic transport models were originally developed to simulate the dynamical evolution of nucleus-nucleus collisions in heavy-ion experiments~\cite{Bertsch:1984gb,Stoecker:1986ci,Bauer:1986zz,Bertsch:1988ik,Danielewicz:1991dh}.
These transport models attempt to solve the Kadanoff-Baym equations~\cite{Kadanoff,Botermans:1990qi}.
Because these equations are exceedingly complex, approximations are made, 
such as gradient expansion, which leads to the Boltzmann-Uehling-Uhlenbeck equations. This set of truncated semi-classical kinetic equations is used to describe the dynamics of hadronic systems explicitly in phase space and time in the GiBUU event generator~\cite{Cassing:1990dr,Teis:1996kx,Buss:2011mx,Mosel:2019vhx}.
Hadronic transport algorithms have been benchmarked against nucleus-nucleus and hadron-nucleus scattering data in~\cite{Cassing:1990dr,Gallmeister:2009ht}.

\subsubsection{BSM and Higher multiplicity SM processes}\label{sec:nu:BSM}
Neutrino experiments have traditionally focused on only charged ($\nu_l A \rightarrow l X$) and neutral ($\nu_l A \rightarrow \nu_l X$) current final states in their analyses. However, with the statistics available at current and future short-baseline experiments~\cite{MicroBooNE:2015bmn,Machado:2019oxb} and future long baseline experiments~\cite{DUNE:2020jqi,DUNE:2021tad}, the ability to study other leptonic final states are of extreme interest. Additionally, the simplest interpretation of the low energy excess from the MiniBooNE experiment~\cite{MiniBooNE:2007uho,MiniBooNE:2008yuf,MiniBooNE:2010idf,MiniBooNE:2013uba,MiniBooNE:2018esg,MiniBooNE:2020pnu} are either disfavored by global analyses~\cite{Dentler:2018sju, Diaz:2019fwt, Boser:2019rta} or by recent  constraints from the MicroBooNE experiment~\cite{MicroBooNE:2021zai, MicroBooNE:2021jwr,MicroBooNE:2021nxr,MicroBooNE:2021rmx, Arguelles:2021meu}. 
Many models have been proposed to explain the excess~\cite{Gninenko:2009ks,McKeen:2010rx,Gninenko:2010pr,Dib:2011jh,Gninenko:2012rw,Masip:2012ke,Ballett:2016opr,Magill:2018jla,Fischer:2019fbw,Bertuzzo:2018ftf,Bertuzzo:2018itn,Ballett:2018ynz,Arguelles:2018mtc,Ballett:2019cqp,Ballett:2019pyw,Abdullahi:2020nyr,Abdallah:2020vgg,Abdallah:2020biq,deGouvea:2019qre,Dentler:2019dhz,Chang:2021myh}, including some that lead to a higher multiplicity of final-state leptons. 
Nevertheless, none are implemented into current event generators. Furthermore, next generation experiments will require an efficient pipeline to go from a new physics theory to simulated events to enable a robust new physics search program similar to that at the LHC.

Regarding higher multiplicity final-state leptons, within the Standard Model, neutrino trident production has only been observed to be nonzero at the $3-4\sigma$ level~\cite{CHARM-II:1990dvf, CCFR:1991lpl, NuTeV:1998khj}. 
The statistics obtained at next generation experiments should be sufficient to have a 5$\sigma$ discovery of
neutrino tridents~\cite{Ballett:2018uuc,Magill:2016hgc}. Furthermore, the trident process is sensitive to new neutral current processes~\cite{Ballett:2019xoj,Altmannshofer:2014pba}.
However, event generators do not currently have this process implemented. This prevents
an accurate simulation of the Standard Model prediction, especially after including experimental cuts. 
It is therefore necessary that generators begin to expand beyond simulating only single lepton final states.

Currently, neutrino event generators generally do not implement BSM physics models (\nuwro, \neut, GiBUU), with only limited exceptions ({\it e.g.}, a recent \GENIE implementation of boosted dark matter and dark neutrinos~\cite{GENIE:2021npt}). In this latter example, the events are generated as a $2 \rightarrow 2$ process, with subsequent on-shell decays afterwards. When taking such an approach, it is important to ensure that all spin-correlations are taken into account. For example, these spin correlations are important for separating Dirac from Majorana heavy neutral leptons~\cite{Formaggio:1998zn,Balantekin:2018ukw,deGouvea:2021ual,deGouvea:2021rpa}, and for studying $\tau$ neutrino charged current interactions~\cite{Hagiwara:2003di,Graczyk:2004uy,Sobczyk:2019urm,Machado:2020yxl,Hernandez:2022nmp}. 
At this moment, the effort involved in incorporating a BSM model into an neutrino generator is a critical
limiting factor in detailed searches at current experiments; as such, there is strong interest in formulating
more flexible and generalizable approaches.

A first prototype generator (\achilles) was developed in~\cite{Isaacson:2021xty} using the UFO interface~\cite{Degrande:2011ua} within the \comix event generator~\cite{Gleisberg:2008fv} (for additional details on the UFO file format see Sec.~\ref{sec:new_physics}).
This framework leverages the fact that new physics first appearing at neutrino experiments will most likely not modify the nuclear physics significantly and that the matrix element can be factorized into a leptonic current and a hadronic current. This allows us to use the tools developed for arbitrary tree level process calculation at the LHC and the most advanced nuclear physics models, without having to drastically modify either component. Additional details can be found in the discussion of the \achilles generator (Sec.~\ref{sec:nu:achilles}).

\subsection{Neutrino generator frameworks}
\label{sec:nu:generators}
In this subsection we describe active projects for neutrino event generators.
We list the generators that cover the full phase space in alphabetical order, followed by developing generators.

\subsubsection{\texorpdfstring{\GENIE}{GENIE}}
\label{sec:genie}

Significant recent activity in \GENIE has been devoted to implementing improved models of quasielastic and 2p2h interactions.
The historical and default QE model is based on a calculation by Llewellyn-Smith~\cite{LlewellynSmith1972}.  Two more modern treatments have been implemented, each usable for a wide variety of target nuclei.
The Valencia QE model is based on a local Fermi gas representation of the nuclear ground state and adds corrections for Coulomb effects and long-range nucleon-nucleon correlations, the latter using a Random Phase Approximation (RPA) approach~\cite{ValenciaModel}. In a second new QE model, SuSAv2, the nuclear responses are computed using scaling functions based on Relativistic Mean Field (RMF) theory~\cite{GENIESuSAv2,SuSAv2CC}. Both of these new QE models have a corresponding implementation of charged-current 2p2h scattering based upon similar physical assumptions. In the case of SuSAv2, the QE and 2p2h channels may also be simulated in a consistent way for electromagnetic scattering~\cite{PhysRevD.103.113003,e4vNature21}

In addition,
several other theoretical descriptions of QE scattering on specific nuclei are in advanced stages of development.
First is the Short-Time Approximation (STA), an \textit{ab initio} calculation that uses quantum Monte Carlo methods to evaluate lepton-nucleus scattering cross sections~\cite{STA}. The STA approach has significant strengths, including the ability to predict kinematic distributions for a two-nucleon final state and to treat interference between one- and two-nucleon contributions to the cross section correctly. However, it is computationally expensive, requiring many CPU hours to achieve precise results. A first STA implementation in \GENIE, able to reproduce the model predictions for inclusive electron scattering on $^{4}$He, relies upon interpolation of tables of precomputed nuclear responses~\cite{GENIESTA}. Efforts to extend the technique to simulate the hadronic final state (while preserving proper correlations with the sampled lepton kinematics) are underway.
Second is an implementation of the Hartree-Fock Continuum Random Phase Approximation (HF-CRPA) calculations developed primarily by a group at Ghent University~\cite{GENIECRPA,PhysRevC.65.025501,PhysRevC.92.024606}. These provide a more realistic description of nuclear structure compared to standard \GENIE models, leading to sizable differences in the theoretical predictions for small energy and momentum transfers.

For high-energy
neutrino scattering (of particular interest for the upcoming FASER$\nu$
facility or neutrino telescopes), the High-Energy DIS
(HEDIS) package was recently interfaced and incorporated,
thereby extending \GENIE DIS capability beyond the default Bodek-Yang
model noted in Sec.~\ref{sec:nu:DIS}. The HEDIS framework
is based on charge-current DIS structure functions computed
from arbitrary PDFs at NLO QCD accuracy.   These QCD-based
approaches to DIS need to be extended to softer kinematics near and in the
RES-DIS transition region relevant for DUNE, discussed in Sec.~\ref{sec:nu:RES-DIS}.

\GENIE v3.2~\cite{GENIE:2021npt} features 4 FSI models: hA, hN, INCL++~\cite{Mancusi:2014eia} and Geant4.
The hA model, which is the long-standing default, is an effective data-based effort that provides a good description of a wide variety of hadron-nucleus data.
The hN model is very similar to what is standard in \nuwro and \neut.
The hN, INCL++ and Geant4 models have stronger nuclear medium corrections and additional FSI particles with better theoretical basis,
and provide a better description of kinematic observables involving low energy hadrons~\cite{GENIE:2021npt}.
Comparisons of \GENIE, \nuwro, and \neut predictions to hadron-nucleus scattering data and hadron transparency
reveal interesting deviations~\cite{Dytman:2021ohr} for low energy protons and pions with energy near the $\Delta (P_{33}(1232))$ resonance,
a region that is very prominent in neutrino interactions.
There is a strong need for pion transparency data which is only studied for high energy pions of energies well above the $\Delta$ resonance.

Continued model exploration and development will be
an ongoing priority within \GENIE and the larger \GENIE
user base in the coming years. These development
efforts will extend along the areas discussed above,
and can also be expected to involve a growing
concentration on the few-GeV region in preparation
for LBNF/DUNE. This involves particular focus on
DIS and shallow-inelastic scattering, which require
a sustained program of model development and refinements
to the underlying theory.
In a broader model-development context, the input of charged-lepton
data as discussed in Sec.~\ref{sec:nu:e4nu} is expected to play a significant role in guiding model
selection and subsequent tuning.

The neutrino flux and geometry interfaces and the wealth of mature and extensively tested drivers implementing these interfaces constitute one of the most well-known and desirable \GENIE features.
There is a strong community desire to reuse the \GENIE experimental interfaces to test alternative physics generators. This drove the implementation of an ``Event Library'' interface~\cite{GENIEEventLibrary} and the development of a generic {\em EvtLib} \GENIE generator.
The purpose of the {\em EvtLib} generator is to read from an external library of cross sections and pre-computed final particle kinematics (most likely computed using an alternative neutrino generator). 
For each interacted neutrino selected by \GENIE, the generator will use the appropriate cross section from the file, and then use the kinematics from the library entry with the closest-matching energy. 
Within the limits of the library statistics, this will then reproduce the physics of the  external generator, but making use of the flux and geometry handling of \GENIE.
The details of the event library file structure are described in the code and in the manual~\cite{Andreopoulos:2015wxa}.

%

\subsubsection{GiBUU}
The Giessen model is an extensive general theory framework to describe nuclear collisions, from relativistic heavy-ion collisions to neutrino-induced reactions on nuclear targets.
The treatment of final-state interactions (FSI) is built on quantum-kinetic transport theory \cite{Kad-Baym:1962} and as such can be used as a generator, called GiBUU, for the full final state of a reaction. The underlying theory is described in detail in \cite{Buss:2011mx}; this article also contains the details of the actual implementation of theory and its parameters in the code. The code is being updated from year to year and its source-code is freely available \cite{gibuu}.

GiBUU is not only a neutrino generator, but has been used for an analysis of a wide class of nuclear reactions. Its origins go back to the description of relativistic heavy-ion reactions and to $p+A$ and $\pi + A$ reactions. The same handling of final-state interactions was then used for investigations of particle production in $\gamma + A$ and $e+A$ reactions. The treatment of final-state interactions has thus been checked in wide kinematic regimes. The code also allows for off-shell transport and is thus suited to deal with nucleon spectral functions not only in the initial state but also in the final state. GiBUU's treatment of collisions  relies on a local interaction rate \cite{Lang:1993}; it does not use mean-free-paths for the handling of collisions.

\subsubsection{\texorpdfstring{\neut}{NEUT}}

The following section is largely reproduced from \cite{Hayato:2021heg}, and the interested reader is directed there for more details of the model and implementation choices made in \neut.

\neut is primarily a neutrino--nucleus scattering simulation program library and provides a complete model capable of predicting the observations for a wide range of neutrino scattering experiments. \neut is capable of simulating neutrino--nucleon and coherent neutrino--nucleus interactions in a number of reaction channels over a neutrino energy range from 100 MeV to a few TeV. Additionally, \neut incorporates initial- and final-state nuclear effects for interactions with nuclei from boron to lead. One of the most important nuclear effects is the re-scattering of hadrons, which are produced in the primary neutrino--nucleon interaction, as they propagate out of the nuclear medium. This re-scattering can result in hadron absorption, extra hadron production or knock-out, or distortion of the nuclear-leaving particle kinematic spectra. The \neut hadron propagation model has also been used to simulate low-energy pion--nucleus scattering, both to tune the model to the experimental data~\cite{PinzonGuerra:2018rju} and to simulate pion propagation in neutrino-scattering experimental simulations. Finally, \neut can also simulate various nucleon decay channels to support experimental searches for the process.

Current modeling improvements include the implementation of the state-of-the-art CCQE and multi-nucleon model by Amaro~\emph{et al.}~\cite{amaro_neutrino-nucleus_2021}, and single pion production models by Kabirnezhad~\cite{PhysRevD.102.053009} and by Sato~\emph{et al.}~\cite{PhysRevD.98.073001}. The implementation of a rudimentary QE-only electron-scattering simulation in \neut is underway. This will enable improved and extended validations of the implemented physics that is most critical to T2K analyses.

\neut currently depends on a version of \cernlib that has not been maintained since 2005.
Building and distributing the library for modern compilers and operating systems takes time away from more important development efforts. \neut relies on \cernlib for reading configuration files, random number generation and some common mathematical operations, and \pythia v5.72~\cite{Sjostrand:1993yb} for simulating SIS and DIS interactions.
\pythia versions 5 and 6, written in \fortran, are also not maintained by their authors.
Resolving the dependency on \pythia v5 is not trivial, due to the removal of the ability to generate final-state lepton kinematics in later versions.
Removing the dependency on \cernlib is a high priority for near-future maintenance.

Because of the in-house nature of \neut development and analysis usage, it is not yet open source.
This is undesirable. Exposure to more users and use cases will result in code, interface, and physics improvements. However, the lack of human resources render it difficult to support \neut as a more general tool. Work has begun, in collaboration with other neutrino interaction simulation stake-holders, to define, test, and implement a new community-designed event format and event generation API~\cite{Barrow:2020gzb}. These critical future developments will be implemented in \neut as they become defined and mature.

\subsubsection{\nuwro}

\nuwro~\cite{Golan:2012wx} is a Monte Carlo neutrino event generator developed by the theory group at the
University of Wrocław in 2005. \nuwro is focused on the needs of accelerator-based neutrino experiments, i.e., the $\sim$1~GeV energy region, but can be used in the energy range from $\sim$100~MeV to $\sim$100~GeV.
\nuwro models neutrino interactions in the {\it impulse approximation}.
In this scheme,  any simulated event is a two-step process with an initial (hard) neutrino scattering off a bound nucleon (or a pair of nucleons) followed by final-state interactions (FSI) of knocked-out nucleons, hyperons and produced pions. Together with various supplemental nuclear effects, this simple picture can reproduce many experimental results for both inclusive~\cite{Juszczak:2010ve, Golan:2013jtj, Bonus:2020yrd} and exclusive~\cite{Gonzalez-Jimenez:2017fea, Nikolakopoulos:2018gtf, Niewczas:2019fro} cross sections measurements.

\paragraph{Initial nuclear state}

Charged-current quasielastic scattering (CCQE) is the dominant interaction channel within the scope of NuWro development interests. Prominently, the hole spectral function (SF) approach described in papers of Benhar~\cite{Benhar:1994hw, Benhar:2005dj} is available for carbon, oxygen, calcium, and iron. Additionally, in the same framework, one can also perform simulations using an approximate solution for the argon nucleus. The most profound feature of the SF approach is that it contains a contribution from nucleon short-range correlated (SRC) pairs giving rise to a characteristic large momentum tail in the nucleon momentum distribution. In that case, interactions only occur on one target nucleon, with the other being merely a spectator (but also affected by FSI effects). Within this picture, following the recipe from Ref.~\cite{Ankowski:2014yfa}, it is possible to go beyond the plane-wave impulse approximation and effectively include the distortion of the final nucleon wave function by an optical potential. The aforementioned effective spectral function, proposed in Ref.~\cite{Ankowski:2005wi}, is a convenient approximation to include the essential features of a hole spectral function, like realistic target nucleon momentum distributions.

Finally, NuWro contains a few other solutions that address different aspects of nuclear modeling. An alternative approach that attempts to account for the SRC effects on top of a Fermi gas is the Bodek-Ritchie model~\cite{Bodek:1981wr}. It is also possible to describe target nucleon using density- and nucleon momentum-dependent potentials~\cite{Juszczak:2005wk}. In the case of LFG and CCQE interactions, long-range correlations calculated with the RPA technique can be optionally included~\cite{Graczyk:2003ru}.

\paragraph{Hard interaction models}

The quasielastic interactions are described using the Llewellyn-Smith~\cite{LlewellynSmith:1971uhs} formalism with several options for vector and dipole axial vector form factors. For antineutrinos, there is a possibility to produce hyperons with a mechanism similar to CCQE.

The NuWro approach to single-pion production uses a model optimized for the $\Delta(1232)$ resonance peak region. $\Delta$ is the only resonance explicitly included, with nucleon-$\Delta$ form factors taken from Ref.~\cite{Graczyk:2009qm}. The additional non-resonant background contribution is modeled as a fraction of the DIS dynamics, extrapolated down to the pion production threshold. This contribution is added incoherently to the $\Delta$-excitation model, as described in Ref.~\cite{Juszczak:2005zs}. In the region of $W\in (1.3, 1.6)$~GeV, NuWro employs a linear interpolation between the RES and DIS pion production cross sections. Additionally, for nuclear target reactions, the $\Delta$ self-energy is included as an approximation based on Ref.~\cite{Sobczyk:2012zj}, and the effect of the finite $\Delta$ lifetime is included, as discussed in Ref.~\cite{Golan:2012wx}. Finally, the angular distribution of pions resulting from $\Delta$ decays follows the values of density matrix elements reported by ANL and BNL experimental studies~\cite{Radecky:1981fn, Barish:1978pj}.

In the DIS region, NuWro uses the Bodek-Yang prescription~\cite{Bodek:2002vp} for the inclusive cross section. Hadronic final states are generated using \pythia6~\cite{Sjostrand:2006za} fragmentation routines~\cite{Sjostrand:2006za}, with modifications described in Ref.~\cite{Nowak:2006sx}. NuWro performance has been optimized to reproduce experimental results for charged hadron multiplicities, see Ref.~\cite{Nowak:2006xv}. For DIS events, formation zone effects are included~\cite{Golan:2012wx}.

Simulation of $2p2h$ events can be done using several models for the overall contribution to the cross section and kinematics of final-state leptons. For the charged current reaction, the default is the Valencia model~\cite{nieves_2011, Gran:2013kda}. Other options include the transverse enhancement model~\cite{Bodek:2011ps}, Marteau-Martini model~\cite{Marteau:1999jp, Sobczyk:2003nx, Martini:2009} and SuSAv2 model~\cite{Megias:2016fjk}.  A novelty is an option of using the phenomenological $2p2h$ model of Ref.~\cite{Bonus:2020yrd}, which is tuned to reproduce the inclusive T2K and MINERvA CC$0\pi$ cross section data. As for the hadronic part, NuWro uses a model proposed in Ref.~\cite{Sobczyk:2012ms}. The Valencia and SuSAv2 $2p2h$ models were implemented using five tabulated nuclear response functions for carbon and oxygen nuclei. Extrapolation to heavier targets is possible using methods similar to those proposed in Ref.~\cite{Schwehr:2016pvn}.

Besides the most significant reaction channels, NuWro also models purely leptonic neutrino-electron interactions~\cite{Zhuridov:2020hqu} with a much smaller cross section than neutrino-nucleon scatterings but non-negligible in specific kinematic regions.

\paragraph{Final-state interactions}

NuWro cascade model employs the algorithm proposed by Metropolis \textit{et al.}~\cite{Metropolis:1958wvo, Metropolis:1958sb}, with several new physical effects and updated hadronic cross sections.
Hadrons in the cascade are treated in a semi-classical approach propagating in straight-line steps using the standard non-interaction probability formula. Pauli blocking is implemented locally using information about nucleon density. The cascade terminates when all the moving hadrons leave the nucleus or do not have enough kinetic energy and are stuck in nuclear potential.
The remnant nucleus is generally left in an excited state, but its de-excitation is not modeled.

In Ref.~\cite{Niewczas:2019fro}, the nucleon part of the NuWro cascade has been studied and benchmarked  on the nuclear transparency data collected from $(e,e'p)$ scattering experiments.
The essence of the model lies in nucleon-nucleon cross sections, which replicate the PDG dataset~\cite{ParticleDataGroup:2016lqr}, the fraction of single-pion production adjusted to the fits found in Ref.~\cite{Bystricky}, and the center-of-momentum frame angular distributions of Ref.~\cite{Cugnon:1996kh}. For nucleon-nucleon elastic interactions, in-medium modifications are taken from Ref.~\cite{Pandharipande:1992zz}, while for inelastic reactions the model from Ref.~\cite{Klakow:1993dj} is adopted. Short-range correlation effects are included by reducing the nuclear density near every hadron-nucleon interaction point~\cite{ANLdensity, Carlson:2014vla} with a compensating factor applied at larger distances~\cite{Niewczas:2019fro}.

For pions in the $\Delta(1232)$ region, pion-nucleon cross sections are described with the Salcedo-Oset~\cite{Salcedo:1987md} model. At larger pion energies, free pion-nucleon total cross sections are taken from experimental data, and differential cross sections are provided by the SAID model~\cite{said}. 

\paragraph{Electron scattering}

The electron scattering module (eWro) is available for QE interactions and nuclear models described above. Computations are done exactly as for neutrinos with interaction vertex and intermediate boson propagator replaced by what follows from the Standard Model for electromagnetic interactions. Axial form factors are set to zero, and vector form factors are accordingly modified. The default option is that outgoing electrons are selected from a piece of spherical angle.
The SF model in NuWro 21.09 performs remarkably well for certain quasielastic-dominated kinematics setups in inclusive electron scattering experiments compared to Fermi gas models.

There are plans to extend this procedure to pion production, but it requires a replacement of the existing model, which is constructed for neutrino interactions only.

\paragraph{Flux and detector geometry}

While performing experimental neutrino interaction analyses, one can incorporate NuWro output into detector simulations by using realistic neutrino fluxes and detector geometry described with \geant4. The framework has proven successful in handling the experimental environment of detectors such as ND280 (T2K) and MicroBooNE.

\subsubsection{\texorpdfstring{\achilles}{Achilles}}
\label{sec:nu:achilles}
\achilles is a lepton-nucleus event generator focused on modularity. The ultimate goal of the generator is to provide a flexible framework allowing for quick implementation of new nuclear models and leptonic processes. The primary interactions are then interfaced to a novel impact parameter based cascade to create fully exclusive events. The details of these two components are briefly discussed below, along with the future directions for the generator.

\paragraph*{Primary Interaction Modeling}
In the \achilles event generator, the differential cross section is expressed in terms of the contraction of a leptonic current with a hadronic current, instead of using tensors.
A major advantage of expressing the cross section in terms of currents instead of tensors is in handling interference effects. 
These effects may be important in BSM scenarios in which the dominant contribution arises from the interference term between the Standard Model process and the new physics.
Furthermore, this factorization provides a flexible interface to quickly change either the nuclear model or the leptonic process independently. 
By separating these two components, \achilles is able to interface sophisticated nuclear models currently available in neutrino event generators with robust BSM generation tools developed for efficient computation of arbitrary physics models at the LHC. The BSM generation is handled via an interface to the \comix event generator~\cite{Gleisberg:2008fv} and the UFO file format~\cite{Degrande:2011ua}.
Additionally, \achilles provides a complete $n$-body phase space with appropriate energy conservation modifications to correctly account for the nuclear medium.
One major complication in interfacing with the UFO file format is in the handling of the nuclear form factors. 
To simplify the requirements on the user, \achilles assumes that all physics models can be expressed as a linear combination of the photon vector form factors ($F_1$ and $F_2$) and a universal axial form factor ($F_A$). Additional details on how this interface is implemented can be found in~\cite{Isaacson:2021xty}.

\paragraph*{Impact Parameter Cascade}
Traditional cascade algorithms rely on propagating particles according to the mean free path based on an estimate of the local nuclear density and nucleon-nucleon cross sections. This approach neglects
the correlations between nucleons within a nucleus. There have been some attempts to address this through modifications of the nuclear density. \achilles takes a different approach to handling
the cascade propagation via the generation of nuclear configurations. These configurations can either neglect the correlations by using a mean field approach, or capture the correlations through the use of calculations
in a quantum Monte-Carlo~\cite{Bogner:2009bt,Barrett:2013nh,Hagen:2013nca,Carlson:2014vla}.

Since all the nucleons are given a fixed starting position, using the density to calculate a mean-free path is not possible. Therefore, a new method of propagation is used, which is inspired from the multiple parton interaction models used at the LHC~\cite{Sjostrand:1987su}. At each time step, a probability for the propagating particle to interact is calculated, and an accept-reject test is used to determine if such an interaction occurs.
Two proposed implementations of the probability distribution have so far been investigated. These are the \textit{cylinder} and \textit{Gaussian} interaction probabilities.
The \textit{cylinder} probability mimics interactions of billiard ball like particles with a radius proportional to the nucleon-nucleon cross section, while the \textit{Gaussian} probability is directly inspired from the multiple particle interaction models used for LHC generators~\cite{Sjostrand:1987su}. Additionally, Pauli blocking is used with a local Fermi momentum to ensure
Fermi-Dirac statistics are obeyed. Furthermore, to account for multiple coherent scatterings, a formation zone is used to define a period of time in which a particle may not re-interact~\cite{Landau:1953gr}. For additional details see~\cite{Isaacson:2020wlx}.

\paragraph*{Ongoing Developments}

Currently, only quasielastic scattering and coherent scattering are implemented in the generator. The quasielastic scattering is based on the spectral function formalism of~\cite{Benhar:2006wy,Rocco:2018mwt}, and the coherent scattering has two form factor options~\cite{Helm:1956zz,Lovato:2013cua}. The \achilles generator is currently in the process of implementing the spectral function approach to all remaining interaction regions. This will allow the generator to simulate the full flux range required by the current and next generation experiments. Additionally, pions are being implemented into the cascade to account for proper propagation of nucleons at higher energies.


\subsection{The importance of electron-nucleus scattering data}
\label{sec:nu:ch-lep}

The abundance of electron scattering data, and the similarities between electron-nucleus and neutrino-nucleus cross sections, provides a rich testing ground for validating neutrino event generators. Here we discuss at a high level the impacts that electron-nucleus scattering has on improving neutrino event generators. For additional details, the reader is referred to the dedicated Snowmass white paper titled: ``Electron Scattering and Neutrino Physics''~\cite{Ankowski:2022thw}.

\subsubsection{Neutrino-nucleus event generators and electron scattering data}

The primary goal of the accelerator-based neutrino program is to extract the oscillation parameters from the reconstructed energy spectra of collected events. The process heavily relies on neutrino cross sections implemented in Monte Carlo generators, especially for experiments employing the calorimetric reconstruction method, such as DUNE. Specifically, the generators are invoked to model the contributions of sub-threshold particles, particles for which identification could not be performed, and especially neutrons, which are a major missing energy channel~\cite{Friedland:2018vry,Friedland:2020cdp}. 

While detailed near-detector studies play an essential role in tuning the cross sections, they by themselves may not be sufficient to resolve all physical ambiguities. In fact, the intricacies of the energy reconstruction problem have already impacted past oscillation results, even in the presence of a near detector~\cite{NOvA:2016vij,NOvA:2017ohq,NOvA:2018gge}. In its quest to robustly measure the subtle effects of the CP-violating phase and the mass hierarchy from the difference between the oscillations of neutrinos and antineutrinos, DUNE will need to face even more challenging accuracy requirements for the cross sections underlying the oscillation analysis.

At the kinematics similar to that of DUNE, the MINERvA Collaboration performed dedicated studies of neutrino cross sections. The collaboration demonstrated that existing Monte Carlo generators were not able to consistently explain all of the collected data, no matter which generator or which tune was used~\cite{MINERvA:2014rdw,MINERvA:2019kfr,MINERvA:2021wjs}. Particularly dire problems were observed for pion production,  
where measurements for different channels within the same experiment could not be reconciled. The origin of this issue is difficult to pinpoint~\cite{MINERvA:2019kfr}. This is a consequence of the complexity of the problem~\cite{Ankowski:2020qbe}: flux-averaging greatly diminishes differences between different interaction channels at the inclusive level, different interaction mechanisms yield the same final states, only nuclear cross sections are available with the desired precision, and scattering involves multicomponent vector and axial currents.

In Ref.~\cite{Ankowski:2020qbe}, it has been proposed to diagnose Monte Carlo generators by taking advantage of the similarities between electron and neutrino interactions~\cite{Ankowski:2019mfd}, which differ predominantly at the primary vertex. Electron-scattering measurements feature monoenergetic and adjustable beams, cross sections that are higher by many orders of magnitude, and feasible targets ranging from complex nuclei to deuterium and hydrogen. As long as the cross sections for neutrinos and electrons are calculated consistently in the generators, and they are (over)constrained by precise electron data, the uncertainties of Monte Carlo simulations employed in neutrino physics can be reduced to those stemming from the axial contributions, to be constrained by the near-detector measurements. 

\begin{figure}
    \begin{center}
\phantomsubfloat{\label{fig:GENIE&data_ArH_a}}
    \phantomsubfloat{\label{fig:GENIE&data_ArH_b}}
        \includegraphics[width=0.95\textwidth]{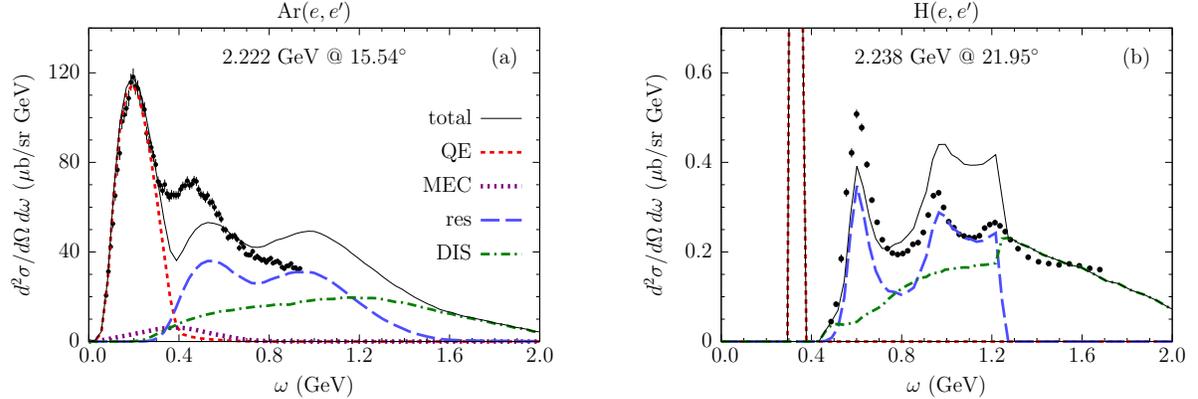}
       \caption{Comparison of the predictions of \genie 2.12 for the double differential cross sections for electron scattering on (a) argon and (b) hydrogen with the data~\cite{Dai:2018gch,Liang:2003,Liang:2004tj}, as reported in Ref.~\cite{Ankowski:2020qbe}.}
        \label{fig:GENIE&data_ArH}
            \centering
    \end{center}
\end{figure}

To illustrate the advantage of the precisely controlled kinematics, quoting the results of Ref.~\cite{Ankowski:2020qbe} in Fig.~\ref{fig:GENIE&data_ArH} we compare the predictions of the Monte Carlo generator \genie---employed by MINERvA---with electron scattering data. The beam energies, $\sim$2.2 GeV, are selected to correspond to the peak of the DUNE flux. Consistently with the findings of MINERvA~\cite{MINERvA:2019kfr}, various issues are observed for pion production. The $\Delta$ peak is heavily underestimated. The contribution of higher resonances is overpredicted. Additionally, the calculated position of the $\Delta$ peak is shifted with respect to the argon data. As neutrino results mix contributions of different energy transfers, due to flux integration, such issues are difficult to diagnose.

Surveying the broad kinematics of DUNE for different targets, the analysis~\cite{Ankowski:2020qbe} identified the origins of the issues in \genie and offered suggestions for possible improvements. Among others, 
\begin{itemize}
    \item the issue of the $\Delta$ strength can be traced back to the cross sections for individual nucleons, calling for an improved pion production model, as can be seen by comparing Figs.~\ref{fig:GENIE&data_ArH_a} and~\ref{fig:GENIE&data_ArH_b},
    \item the double-counting of the cross section in the region of higher resonances occurs as a consequence of adding the cross sections for resonant and deep-inelastic pion production, as shown in Fig.~\ref{fig:GENIE&data_ArH_b},
    \item the implementation of the deep-inelastic scattering requires a revision, as \genie overestimates the H data that the underlying model describes by construction by as much as 60\%--110\%,
    \item to correct the position of the $\Delta$ peak, one should reexamine its implementation in the nuclear framework, as well as take into account the effect of final-state interactions.
\end{itemize}

The problem of modeling the transition between resonant pion production and deep-inelastic scattering is universal between generators, and common for all transitions between different interaction mechanisms that lead to the same final states (see Sec.~\ref{sec:nu:RES-DIS}).
Due to the lack of a consistent theoretical approach on the market, in such situations generator developers need to resort to {\it ad hoc} prescriptions, which lead to discontinuities or regions of double-counting, such as those presented here for pion production in \genie. To meet the stringent accuracy requirements of the cross sections for DUNE, it is, therefore, essential to stimulate theoretical developments aiming to consistently treat all relevant interaction mechanisms.

To guide and validate these developments, the neutrino community has an urgent need for more electron-scattering data at the kinematics relevant for DUNE, especially for argon and for exclusive cross sections, the accurate knowledge of which is of fundamental importance for the calorimetric reconstruction of neutrino energy~\cite{Friedland:2018vry,Friedland:2020cdp}. One of such proposed experiments is LDMX at SLAC~\cite{Ankowski:2019mfd}, the other is e4nu, as described below.

\subsubsection{\texorpdfstring{e4$\nu$}{e4v}}
\label{sec:nu:e4nu}
The extraction of oscillation parameters by neutrino experiments~\cite{PhysRevLett.123.151803,PhysRevLett.121.171802,DUNE} relies on the comparison between the event rates for a given neutrino flavor relatively close to the neutrino production point to those at a significant distance away as a function of the reconstructed neutrino energy.
The yield at each point of the neutrino energy spectrum is obtained using the measured neutrino-nucleus interactions in the corresponding detectors.
This approach demands very small uncertainties on the underlying neutrino-nucleus interactions that are implemented in neutrino event generators.
Unfortunately, neutrino-nucleus cross sections are extremely small and accelerator-produced neutrino beams cover wide ranges of energies.
A fairly small number of data sets is available~\cite{NuSTEC:2017hzk} for validation, which is further limited due to the low statistics and the wide spread of the relevant neutrino fluxes.

The reported uncertainties on the oscillation parameters are dominated by cross-section related sources~\cite{PhysRevLett.123.151803,PhysRevLett.121.171802}.
Those uncertainties can be significantly constrained by validating electron-nucleus interaction predictions on data.
Since electrons and neutrinos are both leptons, their interactions with the atomic nuclei are very similar. Neutrinos interact via vector and axial-vector currents, while electron interactions take place via purely vector currents.

Electron scattering has two major advantages over neutrino scattering.
First, electron beams are monoenergetic, while neutrino beams cover a broad energy range, allowing tighter constraints on the kinematics.
Second, electron-nucleus cross sections are orders of magnitude greater than the equivalent neutrino counterparts, leading to much smaller  statistical uncertainties.

Therefore, we can use electron-nucleus scattering to identify regions of inaccurate modeling in neutrino event generators by comparing data and event generator  predictions for different targets, beam energies, topologies and kinematics. Here we will focus on the results from the $e4\nu$~\cite{e4vNature21} collaboration, and in particular the comparison between the data they obtain and \GENIE. 

\begin{figure} [tbp]
\centering
\includegraphics[width=0.49\textwidth]{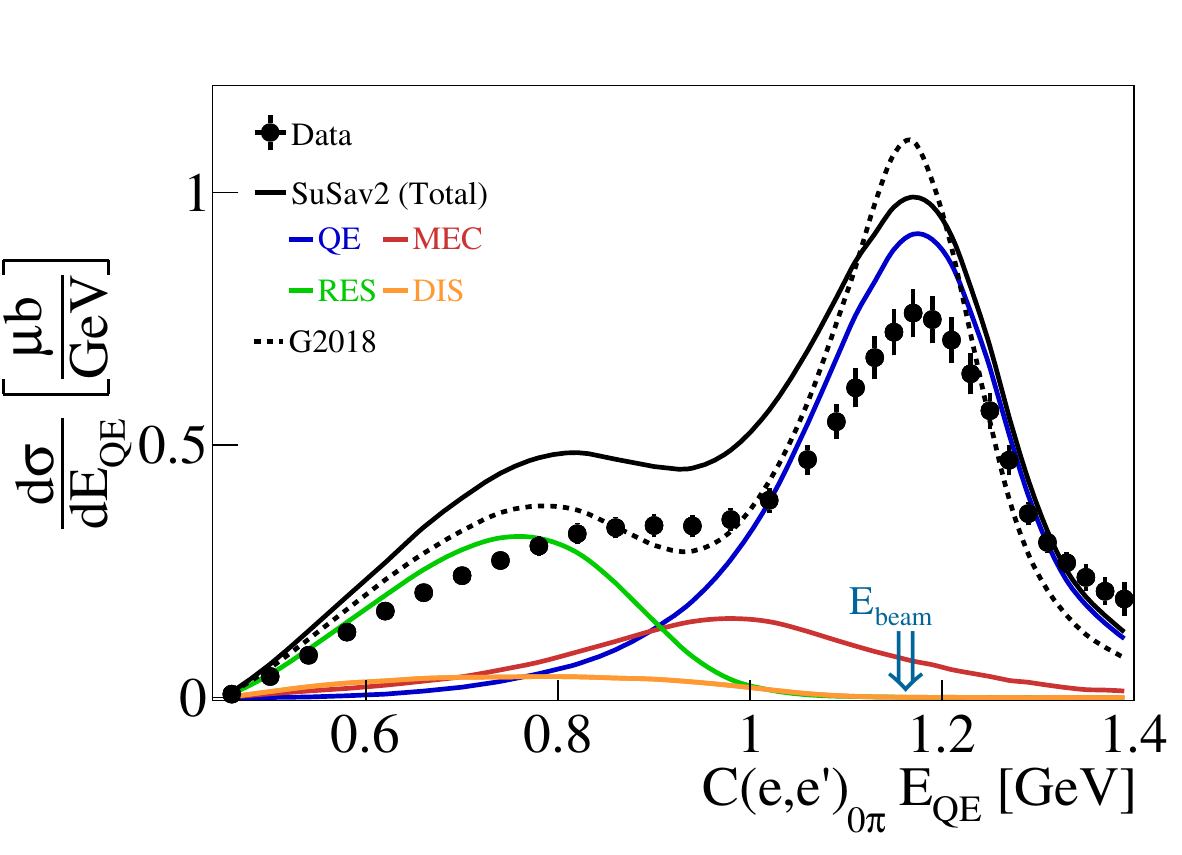}
\includegraphics[width=0.49\textwidth]{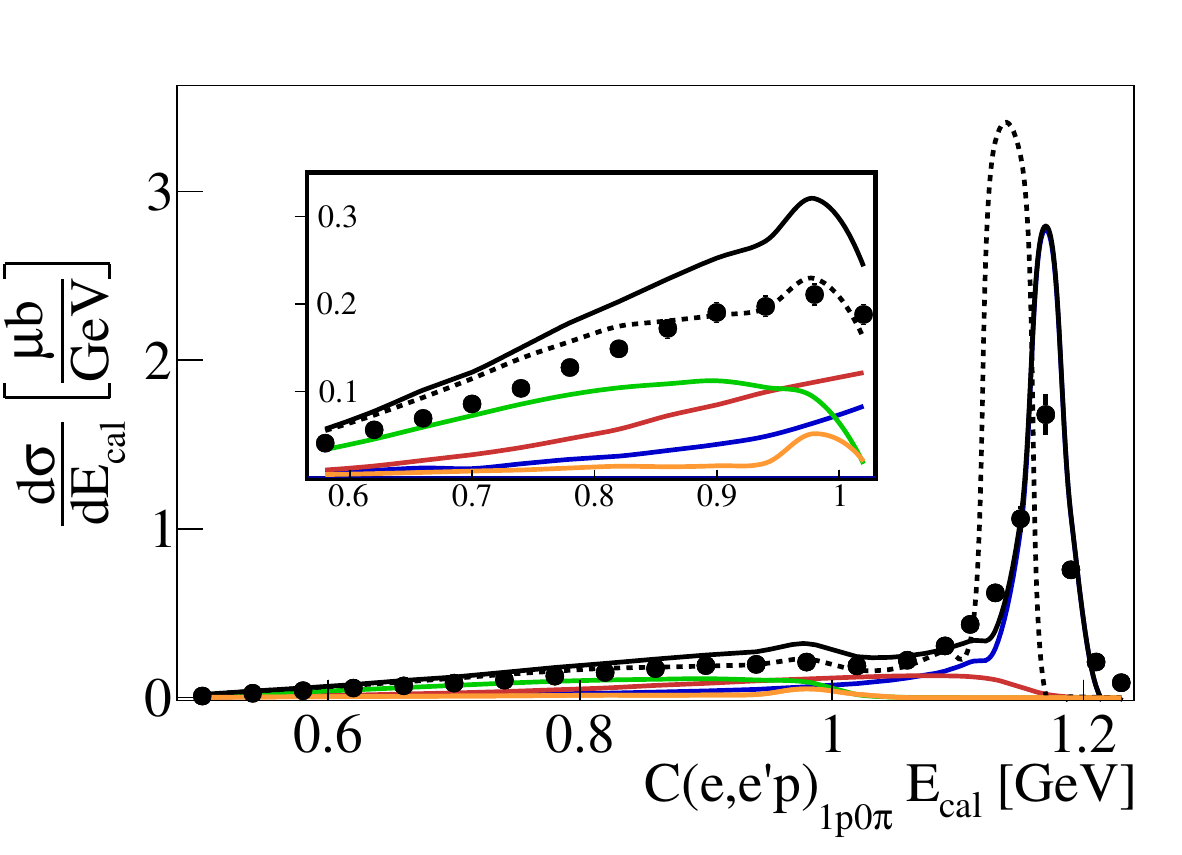}
\caption{\label{e_reco} (left) The 1.159\,GeV C(e,e')$_{0\pi}$ cross-section plotted as a function of the reconstructed energy $E_{QE}$ and (right) the 1.159\,GeV C(e,e'p)$_{1p0\pi}$ cross-section plotted as a function of the reconstructed energy $E_{cal}$, for data (black points), \GENIE SuSAv2 (solid black curve) and \GENIE G2018 (dotted black curve).
The colored lines show the contributions of different processes to the \GENIE SuSAv2 cross-section: QE (blue), MEC (red), RES (green) and DIS (orange). Error bars show the 68\% (1$\sigma$) confidence limits for the statistical and point-to-point systematic uncertainties added in quadrature. Error bars are not shown when they are smaller than the size of the data point. Normalization uncertainty of 3\% not shown. Figure adapted from~\cite{e4vNature21}.}
\end{figure}

While \GENIE does a reasonable job describing inclusive electron-nucleus quasielastic scattering, it appears to dramatically overpredict inclusive electron scattering cross sections in the resonance and DIS regions for both electron-nucleus and electron-deuteron scattering~\cite{PhysRevD.103.113003}.  This points to errors in the \GENIE description of the vector part of the elementary electron-nucleon cross section.
\GENIE reproduces 1.16 GeV  zero-pion $(e,e')$ and $(e,e'p)$ electron scattering moderately well (see Fig.~\ref{e_reco}).  Different models in \GENIE (e.g., SuSAv2 and G2018) make different predictions.  The G2018 model introduces a 20 MeV error into the reconstructed $E_{cal}$ beam energy.  

However,  this agreement is significantly worse at higher electron energies, where \GENIE dramatically overpredicts the cross section for events reconstructing to energies much less than the incident energy. The disagreement appears to be driven by \GENIE overpredicting resonance and DIS cross sections.  This is particularly important because only 25-50\% of events reconstruct to the correct incident energy.  See Ref.~\cite{e4vNature21} for more details.

We can also use monoenergetic electron beams to test the accuracy of the incident energy reconstruction algorithms commonly used across neutrino experiments.
The two main algorithms are for inclusive scattering, assuming a QE interaction:
\begin{equation}
\centering
E_{QE}=\frac{2M_N\epsilon+2M_N E_l-m_l^2}{2(M_N-E_l+k_l\cos\theta_l)}  
\end{equation}
where $E_l$ and $\theta_l$ are the energy and angle of the scattered lepton and $\epsilon$ is a binding energy term, and for semi-exclusive scattering
using the outgoing proton (and other hadron) and lepton energies: 

\begin{equation}
\centering
E_{\text{cal}}=E_l+T_p+E_{\text{binding}}.
\end{equation}

We can also use electron scattering to test the usefulness of single transverse variables (STVs). The perpendicular missing momentum is
$P_T=P^e_T+P^p_T$
(where $P_T^e$ and $P_T^p$ are the three-momenta of the detected lepton and proton perpendicular to the direction of the incident lepton).
At  $P_T\le 200$ MeV/c almost all events reconstruct to the correct cross section and \GENIE describes those events well. At intermediate $P_T$ ($200 \le P_T\le 400$ MeV/c) only a small fraction of events reconstruct to the correct beam energy and \GENIE estimates the cross section of events that reconstruct to lower energies. At large $P_T$ ($P_T>400$ MeV/c), none of the events reconstruct to the correct beam energy and \GENIE significantly overestimates the  cross section of events that reconstruct to lower energies. This shows that neutrino experiments can use $P_T$ to select events that reconstruct to the correct energy and that \GENIE needs further modeling improvements for large $P_T$.

As more electron-nucleus scattering data becomes available and is published, more specific areas of improvement in the event generators will be identified.  The next $e4\nu$ analysis will be primarily examining the $1p1\pi$ channel, in order to focus on the RES and DIS aspects of event generators.  Far more data is currently being taken with the CLAS12 detector at Jefferson Lab to help improve the underlying modeling and tune free parameters in neutrino event generators~\cite{eforvproposal}.


\subsection{Future Generator Requirements}
\label{sec:nu:needs}

Currently, the neutrino event generator community has started conversations on a plan to reach the needed precision for the current and future experiments.
To meet these goals, it is necessary to develop complete error budgets for theory calculations.
The systematic uncertainties should include estimates from all approximations taken, uncertainties on non-tunable parameters, and an appropriate range for tunable parameters.
Providing this information should complement the work on improving global tunes of neutrino interaction models to contain a robust uncertainty from this procedure.
There have been two workshops discussing the future of neutrino event generators, and culminated in a white paper summarizing the importance of developing community tools~\cite{Barrow:2020gzb}.
The community has focused on a number of important tools that would be of significant impact. These include a geometry and flux driver, common output formats, and a theory-generator interface.
 While this is a good start in addressing the needs of the community, there is still much work to be done. This will require continued workshops and meetings similar to the Les Houches meetings held bianually by the collider community.
 
Furthermore, while the issue of confronting voluminous generator code is a substantial problem by
itself, it is further complicated by the diverse base of developers and users; these include,
in the former case, arrays of Monte-Carlo experts who lead coding efforts and theorists who produce models for implementation.
Unlike the LHC community, these two groups tend to be mutually exclusive.
Bringing these two
distinct groups together can lead to inhomogeneities that must be managed on implementation. At the human level, these features require better organization of efforts. Beyond
enhanced coordination, there is also a need to further diversify the neutrino-generator developer community. This entails expanding the theoretical model-building community
with HEP theorists who carry insights
related to certain BSM scenarios or cross section methods typical of event generation for
hadron colliders. Analogously, core generator development would greatly benefit from input
and involvement of computer science and HPC specialists.

\subsubsection{Neutrino generator uncertainty quantification and tuning}\label{sec:nu:uncert}
Precise theoretical predictions of neutrino scattering cross sections on Ar are essential to achieve the maximum sensitivity of DUNE and maximize its potential to discover new physics. Total cross-section uncertainties of 2\% are assumed in sensitivity studies described in the DUNE Conceptual Design Report, and a decrease from 3\% to 1\% cross-section uncertainty is estimated to lead to a factor of two decrease in the total exposure required to achieve 5$\sigma$ discovery of CP violation~\cite{DUNE:2015lol}. 
Neutrino-oscillation parameters are extracted from the energy distribution of the oscillated neutrinos, which has to be reconstructed from the hadronic final states observed in the detector and, in the case of charged-current transitions, from the kinematics of the outgoing lepton and hadrons. Current oscillation experiments report significant systematic uncertainties associated with the assumptions made in the neutrino-nucleus interaction models. For example, the NOvA experiment reports a cross-section and FSI uncertainty of 7.7(8.6)\% and a total systematic uncertainty of 9.2(11)\% for signal (background) events~\cite{NOvA:2018gge}.
Propagating the uncertainty of the nuclear physics models employed in event generators would at present require generation of multiple sets of events. To reduce the computational cost of the simulation, existing Monte Carlo events are typically reweighted.  This procedure only allows a subset of model uncertainties to be gauged, for example because events in unpopulated regions of phase space cannot be re-weighted into existence. Precise estimates of the uncertainty associated with the theoretical calculations are not available at the moment. Promising strategies to pursue in the future include using different interactions obtained within Chiral Effective Field Theory to describe the nuclear wave functions and including in the nuclear calculations elementary matrix elements obtained from Lattice QCD (LQCD), when they become available.%

It is vital for the future success of neutrino event generators, that all nuclear models implemented contain a complete error budget. If physics parameters are not fully determined from theory, these need to be exposed for tuning along with an appropriate range of validity~\cite{uBooNEGENIETune}.
 
Next-generation uncertainty quantification will permit a better understanding of how to tune models to experiment, and will be another significant focus.
As discussed above, many model types must be grafted together --- an aspect which introduces a variety of potential parameterizations from both nucleon-level
and nuclear models; understanding both model and parametric uncertainties (including potential correlations) in this context is highly nontrivial, and requires dedicated tools. We expect further development of frameworks like the Professor package or involving
Bayesian techniques aimed at disentangling uncertainties for future global analyses and comprehensive tunes.
In addition to next-generation computational tools, the charged-lepton
data and associated analyses discussed in Sec.~\ref{sec:nu:ch-lep} will play
a valuable role in error quantification by providing important guidance to
generator tunes.

Problems with tuning arise when different, unrelated theoretical models for different elementary processes are being implemented. Examples are the treatment of QE scattering and of MEC contributions that are obtained from different assumptions on the nuclear ground state. Another example is pion production and absorption which should be linked by time-reversal invariance but are often tuned independently from each other~\cite{Mosel:2019vhx}.
As a result, significant coordination between theoretical model-builders and
generator developers will be necessary to guarantee consistent implementations
of models and their subsequent parametric optimization during comprehensive
tunes.

\subsubsection{Flux and Geometry Driver}
Neutrino experiments are unique in the field of high energy physics, in that the hard interaction occurs inside the detection material. These detectors are made of a collection of different elements. In order to accurately simulate the events, the interaction vertex has to be placed correctly and the nucleus has to be selected based on the material. Furthermore, the neutrino flux is unique in that it is both broad in size and energy distribution (relative to collider beams for example). This leads to complications in properly modeling the incoming energy and direction of the initial neutrino.

Currently, each generator implements their own method of handling the flux and geometry requirements of neutrino experiments.
While including effects from the flux are straight-forward, the plethora of different file formats provided as official fluxes from the experiments add additional complications. When discussing the flux driver, we focus on a tool to provide a common interface to take in all official flux files and pass the needed information on to the generators.
Here, the geometry driver is responsible for properly placing the hard interaction vertex within the detector, but not responsible for modeling geometry considerations pertaining to acceptance of final-state particles. 
This raises two major issues for current and future experiments.
First, this is an unnecessary duplication of effort that consumes person power.
Second, this creates a large barrier for theorists (who are traditionally not involved with event generator development) to quickly compare their results to experimental data. This drastically increases the time for model improvements to be compared to experimental data, due to the time scales involved with implementations in generators.
The idea behind developing a common tool for these components would be similar in spirit to how the LHC handles PDFs through the use of \lhapdf~\cite{Buckley:2014ana}.
The ultimate goal would be to relegate the common experimental setups into a community maintained program instead of all generators reimplementing identical handling of this input data.

Within the community white paper~\cite{Barrow:2020gzb}, two options to address this issue emerged:
\begin{itemize}
    \item \GENIE provides a model in which user hooks are provided to the flux and geometry drivers within the \GENIE build system. The other generators are then responsible for the development
    of the interface to \GENIE. This would be considered part of the \GENIE incubator process, in which each new addition is reviewed and validated by the \GENIE collaboration before being made publicly available.
    Additionally, this would require an MOU between \GENIE and any other generator using the interface defining responsibilities and licensing.  An initial implementation of a general interface of event libraries from alternate generators to \GENIE flux and geometry tools has recently been included.  It is described in Sect.~\ref{sec:genie} and Ref.~\cite{Andreopoulos:2015wxa}.
    \item The community develops an independent flux and geometry driver. This would duplicate work that is already available in current generators, but would be made completely public for all current and
    future generators going forward. Some work has already been done on developing prototypes for this tool, but nothing has been made public yet. However, due to the interplay between geometry and cross section calculations, decisions need to be made on the scope and interface for the tool. A major issue with this approach boils down to determining the party responsible for maintenance of the tool.
\end{itemize}

For further discussions on requirements for the flux and geometry driver see Sec.~2.2 of~\cite{Barrow:2020gzb}.

\subsubsection{Common Output Format}
\label{sec:nu:output}
Rather than observing a common standard, each generator currently has its own custom event format.  \nuisance\ can convert all popular event formats to a consistent structure, but this is
fragile and requires close coordination with the event generator authors.
A common event generator output format would allow for a single point of change and maintenance.
Developing a common output format was also a major focus of~\cite{Barrow:2020gzb}. The advantages of developing a common output format were given as: enabling the sharing of experimental tunes and uncertainties more straightforward, facilitating hand-offs between generators to study the impact of different components in more detail, and creating a straightforward method of reading in the events to external tools (such as detector simulation and analysis tools). 

It was proposed to investigate if the \hepmc3~\cite{Buckley:2019xhk} format would be able to handle the needs of the community. A major advantage to this approach would be that maintenance of the tool would be a shared burden across the entire HEP field and not only the responsibility of the neutrino generators. A few complications specific to neutrino generators in the development of a common output format are: 
\begin{itemize}
    \item Appropriately delineating between different primary process definitions (i.e., the definition of the DIS region in each generator).
    \item Preserving the nuclear options used in calculating the primary process
    \item The frame in which the momentum are defined
    \item Methods of recording the status of the event at boundaries of different models (i.e., hadronization, FSI, unstable particle decays).
\end{itemize}

The other option proposed was to leverage the work done by the NUISANCE collaboration to develop a method of outputting their internal representation. The advantage of this approach is that NUISANCE has already developed the conversions for the neutrino generators, and would require only minimal work. The major disadvantage is that this would not be a great long term solution due to the maintenance involved for each minor change of an event generator output format.

For a more detailed discussion on these issues see Sec.~2.3 of~\cite{Barrow:2020gzb}. Also, for more details on the HepMC3
file format and how this relates to generators for other experiments see Sec.~\ref{sec:xcuts_interfaces}.

\subsubsection{Theory-Generator interfaces}
\label{sec:nu:thy-interface}
Precise theoretical calculations of neutrino scattering cross sections on target nuclei are essential to the success of the experimental program at short- and long-baseline neutrino oscillation facilities. In the last decade there has been tremendous progress in the field of nuclear methods made possible by the increasing availability of computing resources and the development of new algorithms.
Despite this remarkable progress in the development of new theoretical models, they are not routinely used in neutrino experiments. This is primarily due to a software development bottleneck: typically several years are required for a completed theory calculation to be implemented and fully vetted in a neutrino event generator. As a consequence of this long developmental bottleneck, nuclear models in event generators tend to be outdated or over-simplified. While there has been much work on shortening the implementation timeline through the use of interpolation tables~\cite{GENIESTA} and separating the leptonic and hadronic currents from each other~\cite{Isaacson:2021xty}, a general procedure for interfacing event generators with state-of-the-art calculations needs significant developments. Possible additional avenues to address this interface include factorizing the events at each stage of the simulation and developing a universal
interface~\cite{Barrow:2020gzb}. 

A general interface should address a few key requirements in the development process. Firstly, the interface should be general enough to handle all current and future interaction models. This involves defining a method
to handle arbitrary multiplicity for incoming and outgoing particles. For the incoming particles, currently predictions either have a two particle (QE, RES, DIS) or a three particle (MEC) initial state. 
In terms of final-state particles, the multiplicity needs to be flexible to not only handle the hadronic current, but also to accommodate arbitrary leptonic currents for BSM processes. Secondly, the interface should appropriately handle the propagation of uncertainties through the calculation (additional details on uncertainty quantification can be found in Sec.~\ref{sec:nu:uncert}). 

To facilitate faster implementation of theoretical models, there is a need to produce flexible and
generalizable interfaces; these would substantially
defray the cost of incorporating new models into the
generator, a process which typically entails implementing individual models in an {\it ad hoc} fashion. The burden represented by this current need
involves not only a significant encumbrance in human
effort, it also imposes a significant limitation on the number and types of models that might be included for event generation; this in turn impacts our ability
to explore potential model
uncertainties and biases due to the restriction to
a relatively smaller model collection.

Surrogate models for computationally expensive, but state-of-the-art, nuclear calculations are necessary for in including them in event generators. Nuclear calculations that fall into this category are the QMC~\cite{Lovato:2017cux} and the Short-Time approximation~\cite{STA}. One such approach was to tabulate the calculations in some standardized format to be read in by event generators~\cite{GENIESTA}.
Event generators would then interpolate these tables to generate events. The choice of variables used for tabulation depends on the calculation under consideration, and can lead to complications if flexible variable combinations are not appropriately standardized. However, there are a few major disadvantages to this approach. Firstly, a table will be needed for each nucleus, neutrino scattering process, and neutrino energy. Additionally, many of the tables are fully inclusive in the hadronic final state, limiting their predictive power at neutrino experiments. Finally, extending the tables to include a sufficient number of variables to completely describe the hadronic final state will result in memory management issues.

Similar to what was developed for the collider-oriented theory community~\cite{Alwall:2006yp},
a common output format for intermediate states for each stage of the simulation can be provided. This would enable a theorist to calculate a series of events for
their new model and supply them to an event generator for final-state interactions, etc. However, there are many issues that are unique to the nuclear environment in which these interactions take place that need to be addressed.

Developing a common interface to theory calculations would also allow for theorists to directly attach their model to the event generators. Although this seems ideal, there are many complications preventing a simple implementation. One of the major complications in directly using code provided by theorists is that an agreed upon standard for passing physics parameters across the interface needs to be developed. This also will involve choosing which party is responsible for handling physics components, such as form factors, and maintaining consistency with other parts of the code. Additionally, a method of ensuring that all interactions for a given neutrino flux are covered without double counting cross sections needs to be developed. 
Furthermore, a method to ensure that all nuclear models used in a calculation are consistent with the others, this can be complicated if a user is mixing and matching theory calculations from different authors. 
Traditionally, the nuclear codes are developed in \fortran, while the generators are developed in \cxx. Thus, any interface should provide a general use \fortran wrapper instead of forcing a rewrite of the nuclear code. Finally, in developing this interface, it will be important to keep the leptonic current and hadronic current separate. This would enable theorists to focus either on the nuclear components or investigate modifications to the leptonic process without having to be an expert in the opposite field. A beginning implementation of what such an interface could look like has been developed in~\cite{Isaacson:2021xty}.

For a more detailed discussion on these issues see Sec.~2.5 of~\cite{Barrow:2020gzb}.

\section{Electron-Ion Collider}
\label{sec:eic}

Although the building blocks of the nucleon have been known for decades, a comprehensive theoretical and experimental understanding of how the quarks and gluons form nucleons and nuclei, and how their strong dynamics determines the properties of nucleons and nuclei, has been elusive.
Most of the information about the nucleon’s inner structure has emerged from the study of deep-inelastic scattering (DIS) in lepton-hadron scattering experiments~\cite{Breidenbach:1969kd,Bloom:1969kc,Miller:1971qb}, activity which has established QCD as the theory of the strong interaction. In DIS, a high-energy lepton scatters off a hadron and excites that hadron to a final state with much higher mass. Information on the quark momentum density can be determined by detecting the scattering electron and the additional hadrons produced in the reaction. Correspondingly, information on the gluon density is derived from logarithmic scaling-violations when analyzing DIS data at a range of virtualities~\cite{Gluck:1980cp}, or through the photon-gluon fusion process~\cite{Ellis:1988sb}. Information on structure and dynamics beyond a picture of hadrons as collections of fast-moving partons can be obtained by measuring correlations of the struck quark and the further remnants of the hadron. In some cases, the high-energy lepton diffractively scatters $(ep \to epX)$, leaving the hadron intact, with no further signature of hadronic products~\cite{Collins:1997sr,H1:2006zyl}. Such processes offer another context to examine QCD,
especially at low $x$.

Dual advances in perturbative QCD and computation have laid the foundation to imaging quarks and gluons and their dynamics in nucleons and nuclei. The theoretical accuracy of modern perturbative QCD calculations has recently been advanced to NNLO and beyond (see Sec.~\ref{sec:hadron_collider_hard_process}),
including implementations of heavy-quark mass dependence and thresholds~\cite{Witten:1975bh,Aivazis:1993pi,Collins:1998rz} in general-mass schemes~\cite{Guzzi:2011ew,Gao:2021fle}; these advances enable lepton-hadron scattering as a discovery tool via precision measurements and the observation of new particles, both on its own or in strong synergy with hadron-hadron facilities. 

This section describes the science case of the U.S.-based Electron-Ion Collider (EIC) and its connection to the HEP science program; it also summarizes the status of R\&D on MC event simulation tools for the EIC.
The EIC targets the exploration of QCD to high precision, with a particular focus on unraveling the quark-gluon substructure of the nucleon and of nuclei. It will be designed and constructed in the 2020s, with an extensive science case as detailed in the EIC White Paper~\cite{Accardi:2012qut}, the 2015 Nuclear Physics Long Range Plan~\cite{Geesaman:2015fha}, an assessment by the National Academies of Science~\cite{NAS:2018eic}, and the EIC Yellow Report~\cite{AbdulKhalek:2021gbh}. 
The Yellow Report has been important input to the DOE CD-1 review and decision. It describes the physics case, the resulting detector requirements, and the evolving detector concepts for the experimental program at the EIC; these aspects of the EIC program were developed within the Yellow Report alongside a wide array of community-driven impact studies as we note below.
In 2021, the host labs for the EIC, Brookhaven National Laboratory and Jefferson Lab, issued a call for detector collaboration proposals. Three proto-collaborations, ATHENA, CORE, and ECCE, followed the call and developed a detector concept that addresses the White Paper and NAS Report science case and meets the requirements from the Yellow Report. 

The status report and plans are also relevant for other envisioned high-energy colliders such as the Large Hadron electron Collider (LHeC) and the Future Circular Collider in electron-hadron mode (FCC-eh)~\cite{LHeC:2020van, FCC:2018byv}. 

\subsection{Importance of the EIC physics program}
The physics extracted from the EIC is applicable to the important systematic uncertainties at the LHC and neutrino experiments.
These measurements will provide precise data on the hadronic and nuclear bound states involved in hadron-hadron collisions and neutrino-nucleus interactions.
In particular, understanding the validity of factorization theorems and probing their kinematic boundaries~\cite{Moffat:2017sha,Guerrero:2015wha} amounts to crucial tests of QCD and is essential to improving the interface between high-accuracy perturbative QCD calculations and non-perturbative inputs. 

For $pp$ collisions at the LHC, this systematic limitation most clearly appears in the form of the PDF uncertainty on many standard-candle observables of particular relevance for BSM searches. For instance, the inclusive Higgs-production process proceeds predominantly (at the level of 90\%) through gluon fusion, $gg \to H$, such that uncertainties in the gluonic content of the proton can impede the realization of higher precision in detailed studies of Higgs production and decay. This message extends to a number of other high-interest HEP quantities, including those in the electroweak sector such as gauge-boson masses, {\it e.g.}, $M_W$, or the Weinberg angle, $\sin^2 \theta_W$~\cite{EICWP,Hobbs:2022jhm}.

Analogously, on the side of neutrino scattering phenomenology as discussed in greater detail in Sec.~\ref{sec:neutrino_experiments}, uncertainties come from an array of sources, including incomplete knowledge of the PDFs, especially for the few-GeV region wherein there can be a nontrivial interplay of power-suppressed corrections due to high-twist or target-mass corrections. Beyond this, it will be necessary to achieve a better understanding than presently available for nuclear PDFs (for instance, relevant to the DUNE program, which will conduct $\nu ^{40}$Ar scattering) as well as elastic quantities like the axial form factor of the nucleon, which is crucial to the quasi-elastic cross section simulated in the primary interaction of neutrino generators.

The EIC will undertake a systematic program to investigate each of these inputs, and, in so doing, can be expected to profoundly influence precision in HEP. These points are reviewed in greater detail in another Snowmass contribution devoted to intersections between the EIC program and HEP phenomenology~\cite{EICWP}.

The kinematic coverage of the EIC follows from the wide range in GeV-scale energies. It will span, $20 \le \sqrt{s} \le 140$ GeV, which will permit explorations of the transition region between interactions dominated by perturbative and non-perturbative QCD dynamics~\cite{AbdulKhalek:2021gbh,Hobbs:2022jhm}. As such, the EIC will play an important role as a machine for precision QCD, with a wide-reaching program with implications for both Energy and Intensity Frontier activities. 

The EIC science program entails the comprehensive {\it tomography} of the nucleon and of nuclei. It will furnish high-precision data capable of deepening our understanding of the quark-gluon structure of nucleonic matter and its bulk properties that emerge from the underlying QCD dynamics~\cite{Boer:1997nt,Mulders:1995dh,Mulders:2000sh,Meissner:2007rx,Diehl:2003ny}. This will provide extensive information on the collinear PDFs of the nucleon and nuclei as well as a variety of other distributions and form factors, including transverse-momentum dependent (TMD) distributions~\cite{Mueller:2012sma,Ji:2004wu,Collins:2004nx,Echevarria:2011epo,Collins:1350496} and generalized parton distributions (GPDs)~\cite{Diehl:2003ny,Belitsky:2005qn,Goeke:2001tz,Ji:2004gf,Boffi:2007yc,Guidal:2013rya,Chouika:2017dhe}. For the collinear PDFs, the EIC will be sensitive to the unpolarized PDFs of the nucleon~\cite{AbdulKhalek:2021gbh,Khalek:2021ulf,Hobbs:2019gob} as well as of both light~\cite{Jentsch:2021qdp} and heavy nuclei~\cite{AbdulKhalek:2019mzd}; information on the spin-dependent PDFs will be accessible~\cite{Zhou:2021llj} for the nucleon and lighter ions, such as d and $^3$He. It should be stressed that these objectives of the EIC program are of a strongly cross-cutting nature with respect to many of the other generators and physics programs discussed in this white paper; this owes mainly to the fact that non-perturbative physics inputs are a recurring element of event generator frameworks, as discussed in Secs.~\ref{sec:factorization} and~\ref{sec:fragmentation}. 

The EIC will extract this information through a diverse set of DIS measurements as discussed at greater length in a dedicated Snowmass white paper~\cite{EICWP}. These consist of measurements of the DIS reduced cross section over a wide range in $(x,Q^2)$, carried out using both neutral- and charge-current interactions. By merit of their large kinematic coverage, these inclusive cross section measurements can act as an empirical lever arm to better constrain scaling violations and, with them, the gluon content of the nucleon and nuclei. Similarly, access to a variety of processes in terms of charge- and neutral-current interactions can serve as a powerful discriminant to aid flavor separation.

In addition to the inclusive measurements, an extensive battery of EIC phenomenology related to final-state tagging, semi-inclusive hadron production~\cite{Aschenauer:2019kzf}, deeply-virtual Compton scattering (DVCS)~\cite{Kumericki:2016ehc}, DIS jet production~\cite{Arratia:2020azl}, and other basic QCD processes can provide still further access to PDFs and related tomographic information. In addition to constraints to hadronic and nuclear structure, the EIC measurements discussed above will allow incisive tests of the QCD factorization theorems that underlie the separation of short- and long-distance physics that is essential to modern event generation, as discussed in Sec.~\ref{sec:factorization}.

\subsection{Status of Event Generators}

The EIC User Group collected information on the community’s specific software tools and practices in a survey after the submission of the Yellow Report and another survey after the submission of the detector collaboration proposals. Everyone involved in the simulation efforts and related physics and detector studies was invited to participate in the surveys. Part of each survey was a question about the event generation tools being used for EIC simulations. The answers show three main categories of event generator usage: 
\begin{enumerate}
    \item For detector studies, the \textsc{Geant}4 particle gun is used to create primary particles that then undergo physics interactions and produce secondary particles, all for a given detector geometry.
    \item For general detector and physics studies, the general-purpose event generators \herwig, \sherpa, and in particular \pythia are being widely used. These generators are currently not well adapted to simulate $eA$ collisions and polarization effects and are mainly used for $ep$ studies. The EIC community has started to validate \herwig, \pythia, and \sherpa for the DIS process based on a selection of the results from the HERA experiments H1, HERMES, and ZEUS~\cite{Gawas:2021}. Part of these results have been added as \rivet analyses~\cite{Bierlich:2019rhm,Buckley:2010ar}, see Sec.~\ref{sec:xcuts_analysis}. The EIC-specific development on general-purpose event generators is summarized in Sec.~\ref{sec:eic:dev}. 
    \item For specific physics studies and detector studies, the EIC community has developed a large number of specialized event generators. According to the survey, prominent examples are BeaGLE (cf.~Sec.~\ref{sec:eic:dev}), eSTARlight~\cite{Lomnitz:2018juf}, lAGER~\cite{Gryniuk:2020mlh}, and Sartre (cf.~Sec.~\ref{sec:eic:dev}). The event generators from the HERA community, \textit{e.g.}, \ariadne{}~\cite{Lonnblad:2006pt,Lonnblad:1992tz} or RAPGAP~\cite{Jung:1993gf}, are challenging to install and to support on modern computing systems and not widely used even though they are able to describe the HERA measurements well. The only exception is DJANGOH~\cite{Spiesberger:2005} as the only available generator with a merging of higher order QED and QCD effects that has been well studied. The development of some of the specialized event generators is described in Sec.~\ref{sec:eic:dev}. 
\end{enumerate}

\subsection{Ongoing Research and Development}
\label{sec:eic:dev}

For the science program of the EIC, event generators must be able to simulate the collision of spin-polarized electrons with spin-polarized light ions ($p$, $d$, \(^3\mathrm{He}\)) and unpolarized heavy ions (up to uranium) at high precision and
combine higher order QED effects, {\it e.g.}, radiative effects, and higher-order QCD effects. Ideally, these would be general purpose event generators that can be validated and tuned for various processes and experiments. This section summarizes the R\&D of the event generator community to fulfill these requirements. It also describes the construction of two domain-specific event generators by the EIC community, BeAGLE for the simulation of collisions of electrons with light ions including the nuclear response and Sartre for saturation studies. Worth adding is an established R\&D project from the heavy-ion community. JETSCAPE is an event-generator for ultra-relativistic heavy-ion collisions that is developed by a collaboration of physicists, computer scientists, and statisticians from 13 institutions~\cite{Putschke:2019yrg, JETSCAPE:2019udz, JETSCAPE:2020mzn, JETSCAPE:2021ehl}. For more details on JETSCAPE see Sec.~\ref{sec:HImonolithicMCEG}. XSCAPE will be an extension of JETSCAPE that will allow the simulation of $eA$ based on CGC and parton propagation models. 

\subsubsection{General-Purpose Event Generators}

The use of general purpose event generators for the EIC is important in order to take advantage of physics models that are able to describe various processes beyond DIS and to use the wealth of data collected at other experiments for global validation and tuning. The three main generators, \herwig, \pythia, and \sherpa, are currently only able to simulate $ep$ but not $eA$ collisions in general. Many improvements are therefore needed for the EIC. 

The transition region between DIS, photo-production and Vector Meson Dominance (VMD)~\cite{Bauer:1977iq,Melnitchouk:1992eu} has to be modeled. This is especially problematic for $eA$, as any Glauber calculation will need to consider the quark degrees of freedom of the (virtual) photon when modeling the number of wounded nucleons in the \(\gamma-N\) process. An example implementation is Angantyr, an extension of \pythia for multiparton interactions to heavy ion collisions and eventually lepton-hadron collisions, as described in detail in Sec.~\ref{sec:hi-initial-state}. The Glauber calculation, as implemented in, {\it e.g.}, \pythia 8/Angantyr \cite{Bierlich:2018xfw}, also suffers from currently only relying on a standard Woods-Saxon geometry for nuclei. Recent developments in \textit{ab initio} calculations based on Nuclear Lattice Effective field theories~\cite{Freer:2017gip} ought to be utilized for the small geometries foreseen at the EIC.

Also, experience from the description of $pA$ collisions at the LHC~\cite{ATLAS:2015hkr} shows that the fluctuations in the nucleon wave functions are important to consider in a Glauber calculation, and especially for DIS large effects from the fluctuations in the photon are expected~\cite{Bierlich:2019wld}, cf.\ Sec.~\ref{sec:xcuts_photoproduction}.

Another lesson learned from the LHC is that there are collective effects even in pp collisions, {\it e.g.}, flow and strangeness enhancement, and these are so far not very well modeled by the general purpose event generators~ \cite{Greif:2020rhi}, cf.~Sec.~\ref{sec:fragmentation} and~\ref{sec:ion_colliders}.
Whether these will be important also at the EIC remains to be seen, but it is necessary to have good models that can show the effect.

Finally, spin-dependent effects are only partially included in parton showers (see Sec.~\ref{sec:spin-color} for the current status) and hadronization models (see Sec.~\ref{sec:fragmentation}). For the latter only a first implementation based on the string and \(^3P_0\) model exists that is able to describe the Collins effect and the production of vector mesons in the fragmentation of polarized quarks~\cite{Kerbizi:2021pzn, Kerbizi:2021gos}. 

\subsubsection{BeAGLE}

The \textbf{B}enchmark \textbf{eA} \textbf{G}enerator for \textbf{LE}ptoproduction (BeAGLE) Monte Carlo model simulates inelastic scattering in $eA$ collisions, including the nuclear response~\cite{Beagle}. The current version is BeAGLE 1.1.4. It has been used extensively in recent years to enable the design and optimization of the integrated interaction and detector region, in particular the far-forward detection for $eA$ physics at the EIC~\cite{AbdulKhalek:2021gbh}. The future development of this code is uncertain primarily due to lack of manpower and reliable funding.

BeAGLE is a hybrid model that uses the \textsc{DPMJet}~\cite{Roesler:2000he}, \pythia\,6~\cite{Sjostrand:2006za}, PyQM~\cite{Dupre:2011afa}, FLUKA~\cite{Bohlen:2014buj,Ferrari:2005zk} and \lhapdf\,5~\cite{Whalley:2005nh} codes to describe high-energy lepton-nuclear scattering. Overall, steering and optional multi-nucleon scattering (shadowing) is provided in BeAGLE as well as an improved description of Fermi momentum distributions of nucleons in the nuclei (compared to DPMJet). DPMJet is not designed for light nuclei, so substantial changes were made for the case when the nucleus is a deuteron~\cite{Tu:2020ymk}. The geometric density distribution of nucleons in the nucleus is provided primarily by PyQM, while the quark distributions within that geometry are taken from the EPS09 nPDF~\cite{Eskola:2009uj}. BeAGLE allows the user to provide ``Woods-Saxon'' parameters, including non-spherical terms, to override the default geometric density description. The parton-level interactions and subsequent fragmentation is carried out by \pythia\,6. The optional PyQM module implements the Salgado-Wiedemann quenching weights to describe partonic energy loss~\cite{SW:2003}.  Hadronic formation and interactions with the nucleus through an intranuclear cascade is described by DPMJet. The decay of the excited nuclear remnant is described by FLUKA, including neutron and proton evaporation, nuclear fission, Fermi breakup of the decay fragments and finally de-excitation by photon emission. 

The ability to model the nuclear response is a particularly unique feature of this code, especially the modeling of the excited nuclear remnant and it’s decay, including photonic de-excitation. The authors intend to preserve the best features of this code/model into the future, but the exact path forward is unclear due to the fact that the code is in \fortran and links to other \fortran codes ({\it e.g.}, \pythia 6 rather than \pythia 8~\cite{Sjostrand:2014zea}). 

Medium-term, BeAGLE can be used as an afterburner to another primary model, \textit{e.g.},  Angantyr, to apply the nuclear breakup model to the spectator nucleons. A similar approach has already been used to study the physics of tagged Short-Range Correlations~\cite{Hauenstein:2021zql}.  The Generalized-Contact-Formalism (GCF) generator~\cite{Pybus:2020itv} simulates the hard interaction between the electron and a pair of nucleons and leaves the rest to BeAGLE as is. Longer-term we need to migrate to C${++}$ and \pythia8 and probably move in one of two directions: 1) engage more completely with FLUKA (PEANUT) using it for the intra-nuclear cascade as well as the nuclear response or 2) model the nuclear remnant decay with an open-source code such as ABLA~\cite{KAITANIEMI:2011yzc} or GEMINI$^{++}$~\cite{Mancusi:2010tg}. The open-source approach may require further development, for instance implementing photonic de-excitation. As discussed above, funding/manpower issues are currently limiting our ability to implement many of these improvements.

\subsubsection{Sartre}

Sartre is a specialized event generator~\cite{Toll:2012mb,Toll:2013gda} for exclusive diffractive vector meson production ($\Upsilon$, $J/\psi$, $\phi$, $\rho$) and DVCS in $e$+$p$ and $e$+$A$ collisions based on the dipole model \cite{Lappi:2010dd}. It is able to generate coherent and incoherent events  treating longitudinally and transversely polarized photons separately. It can also describe Ultra Peripheral Collisions (UPC) in $p$+$p$, $p$+$A$, and $A$+$A$. The most common nuclei are included. The heart of Sartre is an implementation of the impact parameter dependent dipole model bSat (also known as IPSat)~\cite{Kowalski:2006hc,Watt:2007nr,Kowalski:2003hm}. Saturation is introduced in the bSat model through an exponential term in the scattering amplitude. In order to study the impact of saturation on the production cross-section a non-saturated version of the bSat model, called bNonSat,is realized in addition by linearizing the dipole cross-section. Sartre is thus the only event generator that allows for the comparison of saturated and non-saturated scenarios. The model bSat and bNonSat parameters are determined through fits to HERA data \cite{Mantysaari:2018nng}. 

To calculate and generate the total cross-sections including the essential fluctuations, rather complex integrations have to be performed for each phase-space point (in $e$+$A$ around 1000 4D integrals each) rendering an on-the-fly calculation while generating events impractical. Therefore, the moments of the amplitudes  are calculated  and stored in large lookup tables. Sartre thus consist of two parts: an event generator part and a table generator part. Most users will only use the former. The availability of lookup tables is limited to few vector mesons and nuclei. 

Sartre is not a stand-alone program but a set of C++ classes and C functions that form the API. A set of example main programs is provided to allow for a rapid start.  Work is currently in progress to implement parton-level fluctuations that will allow to generate incoherent events also in $e$+$p$. In the long term it is planned to expand Sartre  to also generate inclusive diffractive events.

\section{Forward Physics Facility}
\label{sec:forward_facilities}
The main LHC experiments are principally focused on physics associated with large momentum transfer $Q^2$. However, the majority of interactions occur with momentum transfer at the GeV-scale, producing a large flux of mesons in the far-forward direction. To give an example, the HL-LHC will produce about $10^{18}$ pions, about $10^{17}$ kaons and about $10^{15}$ D-mesons within 2~mrad of the beam axis~\cite{Kling:2021fwx}. Due to these large rates, even light and weakly coupled particles can be produced abundantly. Indeed, the LHC produces an intense, strongly collimated and highly energetic beam of neutrinos of all three flavors in the far forward direction. In addition, yet undiscovered new light states, such as the dark photon, could be produced in large numbers as well~\cite{Feng:2017uoz}. 

To exploit this physics opportunity, several new experiments will start their operation in the upcoming Run~3 of the LHC, starting in 2022: FASER$\nu$~\cite{FASER:2019dxq, FASER:2020gpr} and SND@LHC~\cite{SHiP:2020sos, Ahdida:2750060} will detect thousands of LHC neutrinos with TeV energies, while FASER~\cite{FASER:2018ceo, FASER:2018eoc, FASER:2018bac} will search for light long-lived particles predicted by models of new physics. For the high luminosity era of the LHC, several larger experiments are proposed. These would be housed in the Forward Physics Facility (FPF) which is a new dedicated cavern placed about 620\,m downstream from the ATLAS interaction point~\cite{Anchordoqui:2021ghd}. Both the existing experiments operating during LHC Run~3 and the experiments housed in the FPF will significantly extend the LHC's physics potential by i) searching for signatures of BSM particles, including the decay of long-lived particles, the scattering of dark matter and millicharged particles; ii) studying the interactions of neutrinos of all three flavors in the TeV energy range; iii) using the neutrino flux as a probe of forward particle production with many implications for QCD and astro-particle physics. 

MC event generators will be an indispensable part of the forward physics program, since they are used to describe particle production, propagation and interactions. In the following, we describe the current status, ongoing efforts and requirements on MC development for the FPF. For additional details on the FPF program, see the dedicated white paper~\cite{Feng:2022inv}.

\subsection{Hadronic Interaction Models} 

Cosmic ray event generators are an ideal tool to simulate hadronic interactions at the FPF. Cosmic rays are charged particles that enter the atmosphere with energies up to $10^{11}\,\mathrm{GeV}$ and beyond where they produce large cascades of high-energy particles (see also the contribution to Snowmass 2021 on ultra-high energy cosmic rays~\cite{UHECRWhitepaper}). The development of these so-called \emph{extensive air showers} is driven by hadron-ion collisions under low momentum transfer in the non-perturbative regime of QCD where hadron production can not be derived from first principles. Instead, effective theories and phenomenological models must be used in order to describe the air shower development, most importantly \emph{hadronic interaction models}, which can introduce large uncertainties in simulations~\cite{Ulrich:2010rg, Albrecht:2021cxw}.

These models need to describe interactions at energies and toward forward rapidities beyond those of current colliders, including various types of projectiles (\emph{e.g.} nuclei, protons, mesons) and targets (\emph{e.g.} nitrogen, oxygen). Commonly used hadronic interaction models in air shower physics are \textsc{Sibyll} \cite{Ahn:2009wx, Riehn:2015oba, Riehn:2017mfm}, \textsc{QGSJet} \cite{Ostapchenko:2005nj, Ostapchenko:2006vr, Ostapchenko:2010vb, Ostapchenko:2019few}, \textsc{EPOS} \cite{Werner:2005jf, Pierog:2009zt, Pierog:2017awp}, and \textsc{DPMJet} \cite{Ranft:1994fd, Ranft:1999fy, Ranft:2002rj, Roesler:2000he, Fedynitch:2015kcn}.  
All of them are accessible via the interfaces \texttt{CRMC} \cite{CRMC} or \texttt{impy} \cite{impy}, for example. Due to their focus on multi-particle production in the far-forward region, hadronic interaction models are ideal to simulate particle fluxes at the FPF. In turn, the FPF is the perfect facility to perform dedicated tests of these models, which will help to improve the modeling of high-energy particle interactions in extensive air showers.

All hadronic interaction models for air showers are based on \emph{Gribov-Regge field theory} (GRFT) \cite{Gribov:1967vfb} which uses \emph{Pomerons}, color-neutral objects that can be exchanged between partons, to relate the inelastic cross-section to the particle production using a unique amplitude for the Pomeron exchange. The Pomeron amplitude defines the evolution of the model in terms of the energy and impact parameter and it can generally be derived using two different approaches (for details see \emph{e.g.} \cite{Albrecht:2021cxw}): \emph{(i)} in the \emph{soft+hard approach}, the amplitude is the sum of a purely soft and a purely hard Pomeron based on external parton distribution functions; \emph{(ii)} in the \emph{semi-hard approach}, the Pomeron is the convolution of a soft and a hard component based on DGLAP equations \cite{Gribov:1972ri, Lipatov:1974qm, Altarelli:1977zs, Dokshitzer:1977sg}. In both cases, the Pomeron model also determines the treatment of non-linear effects of the cross-section at high energies, such as screening or saturation.  However, GRFT typically only conserves energy at particle production but not for the calculation of the cross-section \cite{Drescher:2000ha}. An exception is \textsc{EPOS}, where energy sharing is consistently used, leading to a multiplicity distribution which is in better agreement with data than classical GRFT \cite{Pierog:2017awp}. Because GRFT models are based on the Pomeron amplitude, parton distribution functions (PDFs) are not an important element in hadronic interaction models. The leading-order PDFs can be interpreted as the momentum distributions of the partons inside the nucleon. Pomerons can be connected to partons either based on external PDFs and mini-jet calculations of hard scatterings (soft+hard approach), or the PDFs are given by the Pomeron amplitude (semi-hard approach). While \textsc{EPOS} and \textsc{QGSJet} use custom PDFs in the semi-hard approach, \textsc{Sibyll} uses the GRV parameterization \cite{Gluck:1991ee} and \textsc{DPMJet} uses \textsc{CT14} \cite{Dulat:2015mca} as external PDFs.

Due to the nature of extensive air showers, hadronic interaction models need to support nuclear collisions. \textsc{DPMJet} uses the classical Glauber approach \cite{Glauber:1970jm, Miller:2007ri}, while \textsc{Sibyll} uses a semi-superposition model where a collision of a nucleus with $A$ nucleons on a nucleus with mass $B$ nucleons is approximated by $A\times p-B$ collisions. Here, each nucleon carries an equal fraction of the energy and the $p-B$ cross-section is based on the Glauber model \cite{Engel:1992vf}. \textsc{EPOS} and \textsc{QGSJet} use an extended GRFT, where the reduction of the nucleon-nucleon cross-section with respect to $p-p$ collisions is modeled with higher-order Pomeron interactions \cite{Drescher:2000ha, Werner:2005jf, Ostapchenko:2010vb}.

Diffractive collisions contribute up to about $25\%$ to the inelastic cross-section at the TeV scale and they are thus important for air shower simulations and physics at the FPF. Interactions where only a few GeV are transferred, \emph{i.e. low-mass diffraction}, are handled with \emph{Good and Walker theory} \cite{Good:1960ba} and the particles are simply excited and hadronized. However, if the momentum transfer is large enough to produce new particles or a jet, \emph{i.e. high-mass diffraction}, the resulting elasticity is much lower. In \textsc{EPOS}, all diffraction types are computed with a special diffractive Pomeron which is added to the semi-hard Pomeron amplitude. Diffractive beam remnants are usually hadronized through resonance decay or via string fragmentation.

The final hadronization of the excited system is crucial in order to reproduce the correct multiplicities in hadron collisions. All GRFT models are based on a cylindrical topology of the Pomeron which produces two strings carrying the color field between the beam remnant and the partons. The strings are hadronized based on different schemes, such as Lund \cite{Bengtsson:1987kr} (\textsc{DPMJet}, \textsc{Sibyll}) or Area Law \cite{Artru:1974hr, Drescher:2000ha} (\textsc{EPOS}), which are tuned to different data. \textsc{QGSJet} uses a custom hadronization scheme \cite{Ostapchenko:2019few}. However, the parameter tuning to data is more important than the details of the underlying hadronization scheme that is used. For an in-depth overview and further discussions on hadronic interaction models, see \emph{e.g.} Refs.~\cite{Ulrich:2010rg, Albrecht:2021cxw}. 

At the FPF, hadronic interaction models are the primary tools to simulate the production of light hadrons due to their focus on the forward region. In addition, the models, \textsc{DPMJet} and \textsc{Sibyll}, include charm production and they can be used to simulate the production of charmed hadrons, although those could also be described perturbatively, as discussed below. Hadronic interaction models are therefore a necessary ingredient both to predict the fluxes of neutrino, but also new BSM particles, that are produced in their decays.
   
As described in the discussion above, there are currently sizable uncertainties associated with the predictions of hadronic interaction models. On the one hand, this makes measurements at the FPF, in particular those of neutrino fluxes, an interesting physics objective by itself that can help us better understand forward particle production. On the other hand, this also induces systematic uncertainties for many other measurements, such as that of the neutrino interaction cross section. For these applications, it is essential to not only have a reliable prediction for the neutrino flux itself, but also to quantify the associated uncertainties.
    
While hadronic interaction models constitute a sophisticated modeling of microscopic physics, they also contain a sizable number of phenomenological parameters that have been tuned to data. One approach, that is used in both LHC and astroparticle physics, is to use the spread of generator predictions to obtain an estimate of the associated uncertainties.  While this approach captures some differences due to both tuning and underlying modeling, it is unclear on how to interpret the results statistically. An alternative approach that has been proposed to addresses this problem is to use tuning uncertainties~\cite{FASER:2020gpr,Buckley:2018wdv}. Here, not dissimilar to the way PDF uncertainties are defined, multiple additional tunes are obtained which deviate from the central tune in such a way that they are representative of a given confidence level. Similar to the ATLAS A14 tune~\cite{TheATLAScollaboration:2014rfk}, a dedicated forward physics tune including tuning uncertainties based on forward LHC data, for example from LHCf, could be used to quantify the neutrino flux uncertainties.

\subsection{Charm Production} 

Predictions for neutrino fluxes at the FPF rely sensitively on the rates  of forward light hadron and $D$-meson production. The latter can be evaluated in perturbative QCD, though its modeling is affected by various challenges such as large missing higher order
uncertainties and the treatment  of charm quark fragmentation. The production of charm quarks can be evaluated at NLO in the QCD expansion, either at fixed-order in the three-flavor-number scheme (3FNS) or in a general-mass variable-flavor-number scheme (GM-VFN)~\cite{Eskola:2019bgf, Cacciari:2012ny} such as FONLL or ACOT which accounts for potentially large mass logarithms in the collinear limit. Recently, the NNLO calculation for $B$-meson production at the LHC has been presented~\cite{Czakon:2021ohs}, though its applicability to the charm case requires further work. Given that charm production in the forward region is sensitive to the small-$x$ region down to $x\simeq 10^{-7}$ for the FPF acceptance, accurate calculations may require accounting for BFKL small-$x$ resummation effects~\cite{Altarelli:2008aj}, already relevant for the HERA kinematics~\cite{Ball:2017otu, xFitterDevelopersTeam:2018hym}, or for eventual non-linear QCD dynamics such as those describing gluon saturation.
 
Furthermore, a precise modeling of charm quark fragmentation into $D$-mesons is key for the neutrino flux predictions. One can treat charm hadronization analytically, via semi-perturbative heavy quark fragmentation functions~\cite{Nason:1999zj}, or by means of Monte Carlo hadronization models, such as those employed in the frequently-used \powheg matched to \pythia8 simulations. In addition, robust predictions for charm production in the forward region require state-of-the art determinations of the proton PDFs: given that HERA data stops at $x\simeq 3\cdot 10^{-5}$, without additional constraints in the small-$x$ region PDF uncertainties become huge. It has been shown that by tuning PDFs, and in particular the gluon, to reproduce the LHCb charm production measurements one can obtain reasonably precise PDFs~\cite{Gauld:2016kpd, PROSA:2015yid, Gauld:2015yia} that are required for FPF neutrino flux predictions.
    
An important caveat is that the lack of available data in the FPF kinematics forces us to  rely on the extrapolation of existing models for $D$-meson  production predictions, which will need to be carefully validated once data becomes available. In addition, fully exploring the FPF physics potential would benefit from the availability of an integrated event generator that implements the various options described above for the modeling of charm production, hadronization, and decay which could be directly  interfaced with the experimental software  describing the FPF instrumentation and detectors.
    
\subsection{Particle Propagation}
    
A large fraction of muons and neutrinos at the FPF come from the decay and interaction of particles far from the interaction point (IP)~\cite{Kling:2021gos}. To obtain accurate muon and neutrino fluxes at the FPF, we must propagate particles from the IP and include their interaction with the intervening material and magnetic fields. Between the IP and the FPF (which is located 617m downstream from the IP), there is approximately 400m of accelerator complex in line-of-sight. The accelerator contains a variety of very strong magnets used to circulate the two proton beams that is not simply in one direction, but has return fields in the yokes of the magnets. To accurately predict the fluxes, a radiation transport model is required.
    
BDSIM~\cite{Nevay:2018zhp} (which is based on \geant4~\cite{Allison:2016lfl}) and FLUKA~\cite{FLUKA:old} are two codes that are commonly used for this purpose. \geant4 and FLUKA are capable of propagating many subatomic particle species through matter and are conventionally used for detector and radio-protection simulations. BDSIM is a specialized code to create accelerator models in \geant4 based on `optical' descriptions of accelerators (a set of magnetic strengths). The 3D \geant4 model it creates can piece together the detailed geometry of tunnels, magnets and infrastructure required for an accurate model of the FPF. \geant4 is regularly developed and validated by a wide array of specialists and represents the state-of-the-art radiation transport simulations for simulating physics experiments.
    
Such 3D radiation transport models are crucial to understanding the environment of the FPF. They are fundamentally required to quantify systematic uncertainties such as the effect of magnetic fields, varied accelerator optics throughout the LHC Run, as well as those of the generators and particle decayers. \geant4 provides decay channels for many particles, but interfaces are available to pre-assign decay products at the generator level. With the far-forward nature of the FPF, such tools as BDSIM, \geant4 and their interfaces to standard MC generators (e.g. via \hepmc3~\cite{Verbytskyi:2020sus}) will be key for the FPF. 
    
Radiation transport simulations can be computationally expensive but fantastic infrastructure exists at CERN and worldwide to accommodate this. The highly detailed model created of the LHC~\cite{Walker:2020uqg, Walker:2019slk} and HL-LHC for FASER and the FPF can also act as input to much more computationally efficient tools such as the one introduced in Ref.~\cite{Kling:2021gos}.
    
\subsection{Neutrino Event Generators}
\label{sec:fpf:neutrino}
    
Monte-Carlo event generators modeling neutrino interactions will play a key role in the planning and interpretation of the FPF experiments. They will be needed at each step of the experiments, in predicting neutrino interaction event rates, the kinematics of final-state particles, and their subsequent interaction with target material in the detector (tungsten, argon, etc.) for all three neutrino flavors over a broad spectrum of neutrino energies. Furthermore, event generators will help in estimating systematic uncertainties associated with neutrino interactions that will be vital in disentangling eventual new physics signals. For a more detailed discussion on neutrino event generators, see Sec.~\ref{sec:neutrino_experiments}.
    
There exist several general-purpose neutrino MC event generators - \GENIE~\cite{GENIE:2021npt}, GiBUU~\cite{Buss:2011mx}, \neut~\cite{Hayato:2021heg}, NuWro~\cite{Juszczak:2005zs}, and FLUKA~\cite{Battistoni:2009zzb} - most of them are widely used by the accelerator neutrino community, neutrino facilities at Fermilab in the US and J-PARC in Japan, and are therefore optimized for accelerator neutrino energies (100s MeV to a few GeV). This said, some of these tools include functionalities relevant for the description of high (GeV to TeV energies) and ultra-high (PeV and beyond) neutrino energies, such as \GENIE via its HEDIS module.
    
Neutrinos at the FPF carry energies well above those of accelerator neutrinos and their interactions are hence dominated by deep-inelastic scattering, unlike for accelerator neutrinos where DIS is typically a subleading channel. Therefore, neutrino event generators for the FPF would need to account for underlying physical mechanisms relevant for the description of DIS interactions: i) proton and nuclear parton distributions and their associated uncertainties~\cite{Gao:2017yyd}, ii) higher-order QCD corrections (to the PDFs and DIS coefficient functions), iii) heavy quark mass effects (charm, bottom, and top), iv) contribution from low-$Q$ region where pQCD is not applicable, v) subleading scattering channels such as coherent scattering, resonant DIS, and quasielastic scattering. For instance, state-of-the-art pQCD calculations with modern proton and nuclear~\cite{Khalek:2022zqe,Eskola:2021nhw} PDFs and heavy quark mass effects~\cite{Bertone:2018dse} as well as the contribution of subleading channels are already implemented in the \GENIE/HEDIS framework~\cite{Garcia:2020jwr}. In this respect, neutrino generators for the FPF need to combine low-energy nuclear physics with non-perturbative models and perturbative QCD calculations in order to fully describe neutrino interactions with the target material at the FPF experiments. One key challenge is to carefully account for transition regions, keeping consistency between different regions and avoiding double-counting. For instance, a smooth modeling of the transition between the low-$Q$ (non-perturbative) and high-$Q$ (perturbative) regions of the neutrino DIS structure function is essential for the FPF program, see Sec.~\ref{sec:nu:RES-DIS} for additional details.

Another relevant aspect are hadronic and nuclear final-state effects. The FPF neutrino detectors, in particular an emulsion detector with its high spatial resolution, will have the ability to resolve the kinematics of the final state, for example the multiplicity and energies of charged particles. This in turn is sensitive to the modeling of the parton shower, hadronization and interaction of both partons and hadrons when passing through the dense matter of the target nucleus. As discussed in Ref.~\cite{Mosel:2022tqc} and Ref.~\cite{FPFWhitepaper}, there are a variety of open questions associated with the latter, including i) the modeling hadronization inside cold nuclear matter; ii) the description of formation time and formation zones; iii) the validity of color transparency; iv) and the importance of final-state interactions (FSI) and the baryon avalanche effect.

Finally, uncertainties associated with the modeling of neutrino interactions need to be well controlled for both precision SM and BSM measurements. Therefore, ideally, FPF measurements would adopt multiple event generators and different implementations of the underlying physics assumptions described above to systematically analyze their impact on FPF measurements, quantify systematic uncertainties and when possible compare the predictions to data. To this end, the data collected  by FASER$\nu$ and SND@LHC during LHC Run 3 will also be valuable to help constrain the interaction physics in generators. Furthermore, these generators need to have a common interface and output format, ideally an established one like \hepmc or LHEF, to allow them to be easily incorporated into the experimental analysis frameworks.

\subsection{Tools for BSM} 

In addition to SM measurements with neutrinos, the FPF will perform a variety of searches for new particles predicted by models of physics beyond the SM.  This includes searches for i) long-lived particles which have a microscopic lifetime and decay a few hundred meters away from the collision point, ii) stable particles, such as dark matter, which can scatter in the FPF neutrino detectors, and iii) milli-charged particles which are characterized by anomalously low energy deposits. On the one hand, if these particles are sufficiently light, they can be efficiently produced in the decay of light and heavy hadrons. Similar to the case of neutrino production, this requires a reliable modeling of forward hadron production as described by hadronic interaction models for light hadrons and perturbative calculations for heavy hadrons. On the other hand, particles with heavier masses can be produced directly in the hard interaction, which can be modeled with established LHC tools such as \pythia or \madgraph~\cite{Alwall:2014hca}. In this case, an accurate description of the particle kinematics, such as the new particles transverse momentum distribution, would be an important input. 

An additional application for MC generators in the context of BSM physics searches at the FPF is the modeling of long-lived particle decays. Since searches at the FPF mainly aim at light new particles, their decays can typically not be described by perturbative QCD. Instead, different approaches are required to obtain a realistic description of the particles lifetimes, decay branching fractions and decay kinematics. For example, it has been shown that fits to low energy scattering data can be used to predict the dark photon decay widths into various hadronic channels~\cite{Foguel:2022ppx}. An implementation of these results in common MC tools would be a valuable input for related experimental efforts. 

In order to facilitate BSM physics studies in the far-forward region of the LHC, the FORESEE~\cite{Kling:2021fwx} package was developed as a dedicated tool for phenomenological simulations. It relies on the hadron fluxes provided by above mentioned MC generators, and allows to obtain the forward fluxes of new particles from their decays.  The package can also be used to obtain the expected sensitivity reach associated with new physics models for different FPF experiments and signatures. In addition, the \achilles event generator~\cite{Isaacson:2021xty} provides a method to simulate arbitrary non-colored vector BSM interactions upon nuclei. This will provide methods of simulating the detection of both new (semi-)stable particles interacting within the detector as well as new physics coming from the neutrino sector.
    
\section{Lepton colliders}
\label{sec:lepton_colliders}
This section focuses on questions that relate to the simulation of events specifically at lepton colliders.  Historically, the DORIS/PEP/PETRA, SLC, TRISTAN and LEP experiments provided testing grounds for some of the earliest development of Monte Carlo event generators, and this experience was crucial in demonstrating the importance of such work.  Although many of the methods developed during that period remain applicable to future lepton colliders, not all of the solutions are capable of meeting the experimental demands of new machines.  These include exquisite tests of the Standard Model, at 10ppm precision for the TeraZ collider~\cite{Blondel:2018mad} or at future Higgs factories. 
The physics program of future Higgs/EW/top factories contains the measurement of Higgs couplings in a model-independent way to the mostly per-mil level, the $W$ mass at 1-2 MeV (while the experimental precision could be much better) and the top mass to a precision of 20-70 MeV~\cite{Baer:2013cma,Behnke:2013lya,ilc_snowmass,Linssen:2012hp,Abramowicz:2016zbo,FCC:2018evy,Heinemeyer:2021rgq}. Furthermore, these machines will be able to search for dark matter candidates and other invisible particles with masses essentially up to the kinematic limit of the collider.
To meet this challenge it will be necessary to model QED and photon radiation as precisely as possible. Also at high-energy (linear) lepton colliders most of the effects described in Sec.~\ref{sec:he_colliders} are highly relevant as well. In addition, the production at lepton colliders is dominated by a multitude of different, interfering electroweak channels. Hence, signal and background samples for lepton colliders are very inclusive in the final states, determined by the number of elementary fermions in the MC sample compared to stacks of $X+$ jets at hadron colliders. This section discusses some of the special aspects that do not apply to all high-energy colliders, but are of particular importance to lepton colliders. QED initial-state radiation is also addressed in a separate Snowmass white paper~\cite{Frixione:2022ofv}.

The importance of Monte Carlo event generators for data analysis
has been established in previous lepton collider experiments.
Here we will briefly review the main reasons for that and also compare the role
of the MC programs to semianalytic (SAN) calculations in which phase space
integration is performed partly analytically and partly numerically.
The most widely known SAN codes were ZFITTER~\cite{Bardin:1999yd} 
and TOPAZ0~\cite{Montagna:1998kp}.
All programs based on strictly collinear PDFs for the ISR also fall into
this category because the photon transverse momenta are integrated over,
and analytical formulas for the total longitudinal momentum are used.
Semianalytic calculations can be quite sophisticated and encapsulate
a lot of higher order leading and subleading effects,
see for instance~\cite{Jadach:1992aa,Ablinger:2020qvo,Bertone:2019hks}.
A serious limitation though is that they cannot deliver 
sub-percent precision for many important observables like cross sections
of the low angle Bhabha process or $W$-pair production.
Even for the muon-pair production process there is no SAN code which could provide
predictions for the most important event selection, i.e.\ a collinearity cut-off.
The hybrid solution used in LEP data analyses was to use ZFITTER results
and correct them using Monte Carlo predictions.
However, such a solution is not precise enough at future lepton colliders,
especially at the 10ppm precision level targeted by TeraZ~\cite{Blondel:2018mad}.

The standardization of tools is likely important in view of a future lepton collider.
For Higgs factories, a common software framework, EDM4HEP is being developed~\cite{Gaede:2021izq}, that is based on a common event data model from the generator over the detector level simulation towards reconstruction and analysis. This event data model is encoded in the event format LCIO~\cite{Gaede:2003ip}.

\subsection{Collective beam-beam interactions and beam transport}
\label{sec:beamstrahlung}

In hadron collisions, the energies and longitudinal momenta of the
scattered partons in any given event are not known and cannot be
reconstructed fully from the observable components of the final-state
particles' four momenta.
Lepton colliders could overcome this problem, if the energy of each
colliding lepton in a bunch was known precisely.  Unfortunately, the
physics of colliding beams causes a spread of the energy of
colliding leptons that cannot be ignored in precision measurements.

This spread is either caused by collective effects involving a
macroscopic number of particles in a bunch (beamstrahlung) or by
distortions of the bunches as they pass through the beam optics.
Ab-initio calculations of this spread are thus outside
of the scope of even the most
comprehensive of Monte Carlo event generators for high energy
collisions.  Fortunately, all these effects are independent of the
scale of individual hard scattering events and can therefore be
parameterized without introducing a dependence on a factorization
scale.  On the other hand, initial-state radiation (ISR)
depends on the hard scattering scale~$Q^2$ and
must be described by the event generator for the hard scattering,
even if it can be approximated in a
first step by a universal lepton energy distribution
function~$D(x,Q^2)$.  Only in this region
can the description be improved by hard matrix elements,
resummation and parton showers without inducing a double counting of
contributions (cf.~Secs.~\ref{sec:QEDcollfac} and~\ref{sec:lepton_colliders_ceex}).

Beamstrahlung~\cite{Blankenbecler:1987rg,Bell:1987rw,Jacob:1987ua,Chen:1988ec}
rose to prominence with the advent
of $e^+e^-$-linear colliders (LC).  Since the bunches meet only once, a
very high luminosity per bunch crossing is needed.  This requires the
use of dense bunches with very high space charges. In this case, the
leptons in one bunch are deflected by the strong electromagnetic field
generated by the opposing bunch.  While this deflection focuses the
bunch and increases the luminosity for oppositely charged bunches
significantly, it also leads to electromagnetic radiation, called
beamstrahlung, that reduces and spreads out the beam energy.

In addition, the energy distribution of particles in the bunch before
it enters the interaction region can often not be described with
sufficient accuracy by a simple Gaussian.  Instead, the transport of
the bunches through the acceleration structures and, in the case of
circular colliders, the result of repeated passages through the
interaction regions must be taken into account.

\subsubsection{Input to Monte Carlo Event Generators}
For a given collider proposal, with specific
beam optics and bunch shapes at the interaction point,
bunch crossings can be simulated using particle-in-cell methods from
plasma physics, \eg~with the programs \cain~\cite{Chen:1994jt} or \guineapig~\cite{Schulte:1998au}.
The results of such simulations
appear to be sufficiently reliable to extract polarization dependent joint
probability densities~$D(x_1,x_2)$ as a function of energy
fractions~$x_i=E_i/E_{\text{design}}$. The joint densities are in
general correlated and can not be factorized into a product of 
densities for each beam.  

The densities~$D(x_1,x_2)$ can then serve as input for the particle physics
simulations that are 
used for gauging the physics potential of future lepton colliders and
detectors.
For maximum flexibility, this input should be provided both as
functions~$D(x_1,x_2)$ and as efficient generators of pairs of
non-uniform random numbers~$(x_1,x_2)$, that are distributed according
to~$D(x_1,x_2)$.

The output of \guineapig is a finite set of $(x_1,x_2)$-pairs.  It is
not guaranteed that an arbitrary subset will have the same
distribution as the full set.  In addition, high precision simulations
might require a larger number of events than there are pairs in a given set.
For both reasons, it is preferable to model the distribution of the
$(x_1,x_2)$-pairs instead of using the set of pairs as a stream of
random numbers.  Also, well constructed parameterizations save
considerable disk space compared to huge sets of pairs
for a large number of different designs.

In the analysis of experimental data, these
simulation results will be replaced by measurements of reference cross
sections that depend on the beam energy spectra.

\subsubsection{Existing Programs}

\paragraph{\textmd{\circe}}
\label{sec:circe}
In the case of early designs for TeV-scale linear colliders, a
factorized ansatz
\label{eq:circe1-parametrization}
\begin{equation}
  D_{p_1p_2}(x_1,x_2) = d_{p_1} (x_1)  d_{p_2} (x_2)\,,
\end{equation}
where the~$p_i$ denote the flavor and polarization of the colliding
particles, was sufficient.  Motivated by the theoretical
description~\cite{Blankenbecler:1987rg,Bell:1987rw,Jacob:1987ua,Chen:1988ec}, the program
\circe~\cite{Ohl:1996fi} used as a sum of an unperturbed
$\delta$-peak and a $\beta$-distribution with integrable singularities, i.e.
\begin{align}
  d_{e^\pm} (x) &= a_0 \delta(1-x) + a_1 x^{a_2} (1-x)^{a_3}\,, \\
  d_\gamma (x) &= a_4 x^{a_5} (1-x)^{a_6}\,.
\end{align}
In the absence of polarization dependence, each collider design was
therefore parameterized by seven real numbers~$a_{0,\ldots,6}$. These
numbers were derived with a least-squares fit from histogrammed
results of \guineapig.  They were distributed as \fortran common
blocks and as tables for event generators written in C++. This
simple parameterization was extensively used in the simulations that
formed the basis of the TESLA TDR~\cite{Badelek:2001xb,ECFADESYLCPhysicsWorkingGroup:2001igx}
and competing designs.

\paragraph{\textmd{\circetwo}}
\label{sec:circe2}
The parameterization of Eq.~\eqref{eq:circe1-parametrization} was however
inadequate for the energy distributions at a laser
backscattering~$\gamma\gamma$ and $\gamma e$ collider that was studied
as an option for TESLA~\cite{ECFADESYPhotonColliderWorkingGroup:2001ikq}.  
In addition, the wake field acceleration at CLIC~\cite{Aicheler:2012bya,Linssen:2012hp}
produces an energy spread that can not be described by a convolution 
of Eq.~\eqref{eq:circe1-parametrization} with a Gaussian.

In order to address these shortcomings, a successor~\circetwo
was developed.  \circetwo gives up the factorization,
$\delta$-peaks and integrable singularities of
Eq.~\eqref{eq:circe1-parametrization} in favor of two-dimensional step
functions~$D(x_1,x_2)$ that can describe arbitrary shapes and
correlations using a technique similar to the \vegas
algorithm~\cite{Lepage:1977sw,Ohl:1998jn}
to spread out the bin contents as evenly as possible.  In addition,
the distributions can optionally be smoothed by a Gaussian filter.
There is also support for mapping integrable singularities before
histogramming the output of \guineapig, though this has not been
necessary during extensive applications for CLIC.
\circetwo is available both as part of \whizard (cf.~Sec.~\ref{sec:whizard}) as well as a separate \fortran
package, which can also be linked with \cxx programs.

\paragraph{\textmd{\madgraphee}}

In order to be able to combine beamstrahlung and ISR analytically,
the Monte Carlo event generator \madgraphee~\cite{Frixione:2021zdp}
uses a parameterization that
generalizes and improves on~\eqref{eq:circe1-parametrization}.
This reproduces the results of \guineapig simulations of single bunch
crossings well. For details, see~\cite{Frixione:2021zdp}.

\paragraph{\textmd{\kkmcee}}

The Monte Carlo Event generator \kkmcee~\cite{Arbuzov:2020coe} uses
the \foam algorithm~\cite{Jadach:1999sf,Jadach:2002kn} to describe the
output of \guineapig and combines this with a Gaussian beam energy
spread.

\paragraph{\textmd{\babayaga}}

The Monte-Carlo event generator \babayaga \cite{CarloniCalame:2000pz,
  CarloniCalame:2001ny,CarloniCalame:2003yt,Balossini:2006wc} provides 
an NLO accurate description of Bhabha scattering, matched to a QED 
parton shower. 
\babayaga is thus well suited for precision studies with permille-level
accuracy.

\paragraph{\textmd{\bhlumi/\bhwide}}

\bhlumi \cite{Janot:2019oyi,Jadach:1996is} and \bhwide \cite{Jadach:1995nk} 
belong to a family of event generators for high-precision calculations of 
small and wide angle Bhabha scattering. 
Primarily used for luminosity determination at LEP, they are well suited 
for precision predictions with sub-percent accuracy. 
Both tools use the YFS \cite{Yennie:1961ad} formalism to resum 
QED radiation and include dedicated fixed-order corrections.

\subsubsection{Benchmark Beam Distributions}

As described above, the preparation of beam descriptions requires
intimate knowledge of the design of the entire collider: acceleration,
beam transport and interaction region.  Codes for the simulation of
bunch crossings, \eg~\guineapig, are public.  However, in addition to a
few beam optics parameters, they also need the output of simulations
of acceleration and beam transport.  For this, expert input from
accelerator physicists is necessary.

In the case of the CLIC TDR, accelerator physicists provided 
quality controlled \guineapig output files for 24 different design options.
For each, \circetwo beam descriptions were produced, cross checked and made available 
for the particle physicists performing simulations.  A similar approach is
under way for the ECFA Higgs Factory studies.

This effort should be extended to the other lepton collider projects
and to maintain a curated set of beam descriptions on a public
server.  There is no need to track all minute optimizations of
collider parameters, but major changes should result in the production
of a new set, while keeping the old sets for reference.



\subsection{Collinear factorization in \texorpdfstring{$\epem$}{epem} collisions}
\label{sec:QEDcollfac}

The computation of $\epem$ cross sections in perturbative QED leads
to the emergence of logarithms of the ratio $m^2/E^2$, where $m$ is
the electron mass and $E$ a scale of the order of the hardness
of the process, such as the collider energy. Owing to the smallness
of $m^2/E^2$, these logarithms must be resummed. Different techniques
can be used to this end, relevant to different classes of logarithms.
Here we focus on effects that are due to the collinear and/or soft radiation
off incoming particles, whereby resummation is achieved by means of
collinear factorization.
The cross section for a generic production process in $\epem$ collisions
is written as follows:
\begin{equation}
\mathd\Sigma_{\epem}(P_{\lp},P_{\lm})=\sum_{kl}\int \mathd\yp \mathd\ym\,
{\cal B}_{kl}(\yp,\ym)\,\mathd\sigma_{kl}(\yp P_{\lp},\ym P_{\lm})\,.
\label{beamstr}
\end{equation}
The function ${\cal B}_{kl}$ collects all effects due to beam dynamics 
(cf. Sec.~\ref{sec:beamstrahlung}),
and is determined heuristically by means of MC simulations. The indices
$k$ and $l$ identify the particles that emerge from said dynamics; 
predominantly, they are equal to electrons/positrons and photons, with
the dominant contributions resulting from those particles whose identity
is the same as that of the corresponding beam. Finally, the short-distance 
cross section $d\sigma_{kl}$ is written in a collinear-factorized form:
\begin{equation}
\mathd\sigma_{kl}(p_k,p_l)=\sum_{ij}\int \mathd\zp \mathd\zm\,
\PDF{i}{k}(\zp,\mu^2,m^2)\,\PDF{j}{l}(\zm,\mu^2,m^2)
\times
\mathd\hsig_{ij}(\zp p_k,\zm p_l,\mu^2)\,.
\label{master0}
\end{equation}
Here, $i$ and $j$ are {\em parton} indices, denoting the objects that 
emerge from multiple-emission phenomena, dominated by collinear and soft
dynamics, perturbative in nature (in QED), and that initiate the
actual hard scattering. The cross section associated with the latter is
denoted by $d\hsig_{ij}$, and is assumed to be computed with massless
electrons. All electron-mass effects are accounted for by the quantities
$\PDF{\alpha}{\beta}$ (the PDFs), that describe the collinear dynamics 
mentioned above and that are responsible for resumming the large logarithms 
of $m^2/E^2$ -- the resummation is technically achieved by solving the RGE
for the PDFs; these are nothing but the DGLAP equations~\cite{Gribov:1972ri,
Lipatov:1974qm, Altarelli:1977zs,Dokshitzer:1977sg}. Importantly, the
role of these PDFs is quite similar to that played by their QCD 
counterparts in hadronic collisions, cf.\ Sec.~\ref{sec:factorization}.
However, at variance with the latter the QED PDFs are perturbatively computable.

Until recently, all simulations adopting the collinear factorization framework 
of Eq.~(\ref{master0}) employed PDFs accurate to LL+LO~\cite{Skrzypek:1990qs,
Skrzypek:1992vk,Cacciari:1992pz}. In view of the precision targets of
foreseen $\epem$ colliders, this constitutes a problem. In particular,
a LL+LO description renders it formally impossible to 
define an uncertainty associated with the description of the collinear
dynamics. This includes, but is not limited to, the fact that in this
context the coupling constant $\aem$ is arbitrary: one can make more or 
less educated choices for it, but cannot quantify the systematics that
such choices entail. Furthermore, a LL+LO description is not adequate to
properly take the running of $\aem$ into account; this is important,
in view of the fact that as one moves away from c.m.~energies of the
order of the weak-boson masses fixed-$\aem$ schemes such as the $\aem(\mZ)$
or the $G_\mu$ ones become increasingly problematic.

In a series of recent papers~\cite{Frixione:2019lga,Bertone:2019hks,
Frixione:2021wzh} the issues mentioned above have been addressed and
solved, thanks to the extension to the NLL+NLO accuracy of the PDFs.
These results, as well as those relevant to the description of the 
beam dynamics, have also started to be included~\cite{Frixione:2021zdp}
in the automated machinery of \aNLO~\cite{Alwall:2014hca}, which is 
thus being upgraded to a full-fledged $\epem$ simulation tool.
Electron PDFs feature an integrable singularity at $z\to 1$:
it is therefore essential to have a detailed analytical understanding
of this region, which is provided in~\cite{Bertone:2019hks,Frixione:2021wzh}.
Among the various systematics that can be reliably assessed 
based on these works, the choice of a collinear factorization scheme
is particularly important. The $\MSb$ and the $\Delta$ schemes
were used in~\cite{Bertone:2019hks} and~\cite{Frixione:2021wzh}, 
respectively (the latter being physically analogous to the more 
familiar DIS). The solutions relevant to the $z\to 1$ region are:

\begin{eqnarray}
\Gamma_{\lm}^{(\MSb)}(z,\mu)&\stackrel{z\to 1}{\longrightarrow}&
\frac{e^{-\gE\xi_1}e^{\hat{\xi}_1}}{\Gamma(1+\xi_1)}\,
\xi_1(1-z)^{-1+\xi_1}
\label{NLLsol3run}
\\*&&\phantom{aaa}\times
\Bigg\{1+\frac{\aem(\mu_0)}{\pi}\Bigg[\left(L_0-1\right)
\left(A(\xi_1)+\frac{3}{4}\right)-2B(\xi_1)+\frac{7}{4}
\nonumber\\*&&\phantom{\times aaa1+\frac{\aem}{\pi}\Bigg[}\;
+\left(L_0-1-2A(\xi_1)\right)\log(1-z)
-\log^2(1-z)\Bigg]\Bigg\}\,,
\nonumber
\\*
\Gamma_{\lm}^{(\Delta)}(z,\mu)\!&\stackrel{z\to 1}{\longrightarrow}&\!
\frac{e^{-\gE\xi_1}e^{\hat{\xi}_1}}{\Gamma(1+\xi_1)}\,
\xi_1(1-z)^{-1+\xi_1}
\nonumber\\*&&\phantom{aa}\times
\left[\frac{\aem(\mu)}{\aem(\mu_0)}+
\frac{\aem(\mu)}{\pi}L_0
\left(A(\xi_1)+\log(1-z)+\frac{3}{4}\right)\right],\phantom{aaaa}
\label{NLLsol8Delasy}
\end{eqnarray}
where $\mu_0$ is a scale of the order of the electron mass, assumed to be 
the starting point of the RGE evolution, and \mbox{$L_0=\log\mu_0^2/m^2$} --
thus, $L_0$ is a very small number. The difference between
Eqs.~(\ref{NLLsol3run}) and~(\ref{NLLsol8Delasy}) is striking: apart from
a common prefactor (which, incidentally, has the same functional form as
the LL+LO PDF), the $\MSb$ solution has $\log^k(1-z)$ terms which are
absent in the $\Delta$ scheme. While this may seem surprising at first,
one needs to bear in mind that PDFs are unphysical quantities. The fact
above suggests the presence of significant numerical cancellations between
the PDFs and the short-distance cross sections in the context of 
$\MSb$-based computations, which simply do not occur in the $\Delta$ scheme. 
Ultimately, from the physics viewpoint this must not matter, because 
differences between the two schemes will affect observables only at the
NNLL+NNLO level, and are thus expected to be at the $0.1\%$ level.
Still, the more physical nature of the $\Delta$-scheme result is 
appealing because it should entail a better numerical stability for
any given statistics, and because it could pave the way to incorporating,
at least partially, NLL effects into shower-based MC approaches.
Investigations of these matters, as well as detailed phenomenological
predictions for key production processes, are being considered.

A photon radiated from a beam lepton may interact with a photon from the other beam resulting in an effective $\gamma\gamma$ collision. The collision energy and the virtualities of the photons will vary and, especially due to latter, these events provide a rich spectrum of different kinds of processes. Only in the case of sufficiently high virtualities can these photons be considered as purely point-like particles. In this case, it is relatively straightforward to model the collisions of two photons using the (collinear) photon spectra from the EPA. High-virtuality photons will, however, have non-negligible transverse momentum and, in the case of two photon interactions, the relative angle of this $p_\mathrm{T}$ will vary and alter the invariant mass of the two-photon system with respect to the purely collinear momenta from the EPA. For accurate simulations, it is therefore necessary to construct the full kinematics of the photons and correct for this effect.

When the virtuality of the photon becomes small, $Q^2 \lesssim 1~\text{GeV}^2$, the (quasi-)real photon may fluctuate into a hadronic state. Even though such fluctuations are suppressed by a factor of $\alpha$, the large hadronic cross sections can make the processes with such resolved photons even the dominant ones especially for observables sensitive to lower values of $p_{\mathrm{T}}$. Similarly, as in the case of photoproduction in lepton-hadron collisions, the resolved photons contain a hadron-like part, often modeled with VMD \cite{Sakurai:1960ju, Gell-Mann:1961jim, Stodolsky:1966am}, and a perturbatively calculable part from a point-like splitting to quark-antiquark pairs. The VMD part models the photon as a linear combination of different vector mesons-states (usually up to $\phi$ or  $J/\Psi$) but can be extended also to higher-mass (continuum of) states in generalized VMD \cite{Sakurai:1972wk}. All these are usually included in the DGLAP-evolved photon PDFs but often only the sum of all contributions is provided. Thus, to generate collisions of two resolved photons require all the same machinery as for any hadron-hadron collisions including initial- and final-state showers and multiparton interactions (MPIs) adapted for the different structure of the DGLAP equations \cite{Witten:1977ju, DeWitt:1978wn, Bardeen:1978hg}.

The tricky parts in modeling collisions between two photons are properly combining all the different possible contributions and to avoid double counting them. Here the virtuality of the photon plays the main role: the hadron-like part of the PDFs should vanish rather rapidly when the virtuality reaches the vector-meson masses but the point-like part may survive to somewhat higher virtualities. Only well above these scales can the photon be considered purely point-like. Even in this case, there is a risk of double-counting when the target photon is resolved as such processes can be considered as DIS-type, where the lepton scatters off from a quark by exchanging a virtual photon, or direct-like, where photons act as initiators for the hard scattering. To some extent, the different contributions can be classified by ordering the different scales present in the scattering process. An example of such a model was implemented in \pythia~6~\cite{Schuler:1996en, Sjostrand:2006za} consisting of six different components separated by scales and with parameterized suppression factors to minimize the remaining double counting. The model worked reasonably well but the predictive power was somewhat limited to energies at which the parameters were derived.

\subsection{Soft photon resummation and matching to \texorpdfstring{N$^x$LO}{NxLO}}
\label{sec:lepton_colliders_ceex}
It is generally recognized that the work of Yennie-Frautschi-Suura (YFS)
of Ref.~\cite{Yennie:1961ad} was the first complete analysis of the 
infrared structure (IR) of QED and showed the 
cancellation of all soft (infrared) singularities 
between the real and virtual contributions of photons to all perturbative orders,
for any scattering process with an arbitrary number of external particles.
More discussion on the gauge invariance in the YFS analysis was added in~\cite{Grammer:1973db}.
Let us underline two important points in the YFS analysis.
First, in the original YFS paper the Lorentz invariant phase space,
without any approximation for many real photons, 
is present in the early stage of the analysis.
However, it is masked by the fact that it was quickly translated to an abstract Mellin space.
Moreover, most examples of real photon phase space integration were done using some approximations.
At the time (1961), it was unthinkable that the multiparticle phase space integration
could be performed numerically without any approximations.
Second, in the YFS paper it was proven not only how IR singularities cancel,
but also how to calculate in a systematic way
non-IR contributions perturbatively order-by-order,
and how to recombine them with the basic distributions resummed to an infinite order.

Altogether,  what was quoted for a long time as a main YFS result,
was the semi-inclusive result of the Born cross section 
multiplied by a simple exponential factor representing
resummation over an infinite number of real and virtual photons, 
with the total energy of all real photons limited by some small cut-off 
-- without higher order non-soft contributions and without real hard photons.
The real breakthrough in the mid 1980's in exploiting the full advantage 
of the hidden treasure of the YFS resummation scheme was with the advent of
the Monte Carlo technique, which allowed for an inclusion of an arbitrary number
of hard real photons without any approximation.
The first MC implementation was provided for initial-state radiation (ISR) 
near the $Z$ resonance~\cite{Jadach:1988rr} and was extended and generalized
beyond the original YFS scheme over the next decade in many other programs.
Even in the relatively simple case of the final version 
of the {\sc YFS2} MC~\cite{Jadach:1988gb} for ISR
(used as part of {\sc KoralZ} \cite{Jadach:1991ws})
it was necessary to go beyond the original YFS scheme,
because of the presence of the narrow $Z$ resonance.
The next important implementation of the YFS scheme was the
BHLUMI\cite{Jadach:1988ec} program for the low-angle Bhabha process.
Here the main problem to be solved was the new
MC algorithm of generating the multiple hard
photon phase in the presence of a strong $t$-channel peak.
A further milestone was the inclusion of the final-state radiation (FSR)
in the {\sc YFS3} MC, 
which was incorporated into {\sc KoralZ}~\cite{Jadach:1993yv}.
Other, technically similar MC programs implementing YFS resummation
were {\sc KoralW}~\cite{Jadach:1998gi}, {\sc YFSWW}~\cite{Jadach:1996hi,Jadach:2001mp}
for $W$-pair production and \bhwide~\cite{Jadach:1995nk} for the wide angle Bhabha process.
They are reviewed in detail in~\cite{Kobel:2000aw,Grunewald:2000ju}.

The above MC programs implement the so-called exclusive exponentiation (EEX)
scheme, which is quite close to the original YFS work 
(albeit with some improvements for the narrow resonances),
in which separation of the  divergent infrared virtual and real contributions
and order by order calculation of the non-soft parts
is done at the level of the matrix element squared.
In the MC implementations of EEX
the non-soft QED perturbative corrections were always complete up to first order.
Dominant second and third order QED effects were also included.
The EEX scheme was also adopted in other MC projects (\sherpa, {\sc Winhac})
and is described in Sec.~\ref{sec:cross_cuts_soft_photons}.

In the EEX scheme it was quite difficult to resum consistently 
and efficiently to infinite order the contributions from the IR-divergent interferences
between ISR and FSR, especially in the presence of a narrow resonance.
The inclusion of the complete (transverse) spin effects 
in the EEX scheme was also practically infeasible.
These problems were addressed in an extension of the original YFS
scheme, dubbed the coherent exclusive exponentiation (CEEX).
In CEEX, the separation of the IR parts and order-by-order calculations
of the non-soft contributions is performed at the amplitude level
instead of spin-summed amplitude squared, while IR cancellations 
are still performed at the amplitude squared level.
The CEEX scheme was proposed in Refs.~\cite{Jadach:1998jb,Jadach:2000ir} 
and is implemented in the {\sc KKMC} program~\cite{Jadach:1999vf}.

Electroweak \order{\alpha^1} corrections can be added to the CEEX spin amplitudes 
as multiplicative form factors multiplying coupling constants in the Born-like 
amplitudes according to~\cite{Jadach:1999vf}. They have been implemented in
{\sc KKMC} with the help of the DIZET code~\cite{Arbuzov:2020coe}.
The Born-like amplitudes $\hbeta^{(2)}_0$ (see e.g.\ Sec.~\ref{sec:cross_cuts_soft_photons})
in fact include non-soft \order{\alpha^2} QED corrections
and are obtained according to Eq.~(\ref{eq:ceex-beta0})
in Sec.~\ref{sec:lepton_colliders_ceex}.
The only place in the SM corrections for $e^+e^-\to \mu^+\mu^-$ where pure EW
1-loop corrections coexists with IR divergent QED corrections
are two $\gamma-Z$ boxes.
The IR part of these boxes is removed in the above
Eq.~(\ref{eq:ceex-beta0}) of Sect.~\ref{sec:lepton_colliders_ceex}.
More details can be found in~\cite{Jadach:1998jb}.
The pure EW part including, non-soft QED corrections is cleanly
separated from the leading IR-divergent QED part (initial-final state interference),
which is resummed to infinite order, already at the amplitude level.%
An additional subtraction/resummation of the IR finite but numerically
sizable $\sim \ln^n(\Gamma_Z/M_Z)$ corrections is performed at the same time~\cite{Jadach:1998jb}.

In section 3.4 of~\cite{Blondel:2018mad}, it is sketched how the above
technique will work in disentangling the QED IR divergent part from two loop
and three-loop diagrams, which involve heavy EW bosons and photons simultaneously.
In particular this discussion covers the case of the two-loop example of the $WW$ box diagram,
with the $\mathrm{W}^+$ and $\mathrm{W}^-$ lines connecting the incoming electron line 
and the outgoing muon line, and with the insertion of an additional photon in all 
possible combinations between the charged particles.

The CEEX scheme of soft photon resummation has a number of important advantages
over the more common EEX scheme described in Sec.~\ref{sec:cross_cuts_soft_photons}:
(i) thanks to the use of spin amplitudes~\cite{Jadach:1998wp}
for massive fermions it was possible to implement full spin density matrix
(longitudinal and transverse) for the incoming $e^\pm$ beams and outgoing $\tau^\pm$.
This is important because tau pairs from the $Z$ decay are polarized 
and the decays of tau leptons are sensitive to spin polarization effects.
(ii) {\sc KKMC} implements complete%
\footnote{Except numerically small IFI pentaboxes, see \cite{Jadach:2000ir}.}
\order{\alpha^2} QED photonic corrections (no lepton pairs and multiperipheral diagrams). 
These are much more compact at the amplitude level than at the amplitude squared level,
due to proliferation of the interference terms in the latter case.
(iii) In the presence of the narrow $Z$ resonance, initial-final interference
corrections in the soft limit require summation over all photon partitions 
between the initial and final-state emitters, at the amplitude level. 
This additional soft resummation for narrow resonances 
was discussed in~\cite{Greco:1975rm}, with the real photon 
phase space integrals performed in an approximate way.
In the CEEX framework of {\sc KKMC} the same additional resummation 
could be implemented at the amplitude level, and the multiphoton 
phase space integration is exact.
(iv) Electroweak corrections are included at \order{\alpha^1},
while QED non-soft corrections are included up to \order{\alpha^2}. 
The flexibility to add pure EW corrections at one or two orders lower 
than the QED corrections is a feature which persists also at higher orders.
It is very valuable in high precision calculations for future 
electron colliders, because QED corrections are generally much larger
than electroweak ones.
(v) A recent proposal addresses the incorporation of unstable charged particles
like $W^\pm$ into the CEEX method~\cite{Jadach:2019yhw,Jadach:2019wol}.
This includes photon emission from the unstable resonance, the so called
non-factorizable interferences between production and decay.

Using the notation of~\cite{Jadach:2000ir}, the CEEX total cross section 
for the fermion pair production process at an electron collider,
$
 \mathrm{e}^{-}(p_a)+\mathrm{e}^{+}(p_b)\to \mathrm{f}(p_c)+\bar{\mathrm{f}}(p_d)+\gamma(k_1),\dots,\gamma(k_n)
$
reads as follows
\begin{equation}
  \label{eq:master-bis}
  \begin{split}
  \sigma^{(r)} =&\; 
  \sum_{n=0}^\infty \frac{1}{n!}
  \int \mathrm{d}\tau_{n} ( p_1+p_2 ;\; p_3,p_4,\; k_1,\dots,k_n)\;\\
  &\qquad\times\mathrm{e}^{2\alpha\Re B_4(p_a,\dots,p_d)}
  \frac{1}{4}\sum_{\rm spin} \left| \Mmf^{(r)}_n \left(p, k_1, k_2, \dots k_n \right) \right|^2,
  \end{split}
\end{equation}
where $\Mmf^{(r)}_n$ are the CEEX spin amplitudes,
$d\tau_n$ is the standard LIPS,
the virtual form factor $B_4$ is factorized (exponentiated)
and the real emission spin independent soft factors $\sfac$ are also factorized out.
The momenta $p_1,\ldots$ of the fermions are denoted collectively as $p$.
The spin amplitudes read
%
\begin{equation}
  \label{eq:beta-trunc}
  \begin{split}
  & \Mmf_n^{(r)}(p,k_1,k_2,k_3,\dots ,k_n) = 
  \prod_{s=1}^n \sfac(k_s) \Big \{ \hbeta^{(r)}_0(p)
    +\sum_{j=1}^n      \frac{\hbeta^{(r)}_1(p,k_j)}{\sfac(k_j) }
    +\sum_{j_1<j_2}    \frac{\hbeta^{(r)}_2(p,k_{j_1},k_{j_2}) }{\sfac(k_{j_1})\sfac(k_{j_2}) }
\\&~~~~~
    +\sum_{j_1<j_2<j_3}\frac{\hbeta^{(r)}_3(k_{j_1},k_{j_2},k_{j_3}) }
                   {\sfac(k_{j_1})\sfac(k_{j_2})\sfac(k_{j_3}) }
    +\sum_{j_1<j_2<...<j_r}\frac{\hbeta^{(r)}_r(k_{j_1},k_{j_2},...,k_{j_r}) }
              {\sfac(k_{j_1})\sfac(k_{j_2})...\sfac(k_{j_r}) }
    + \cdots
   \Big \},
  \end{split}
\end{equation}
such that the subtracted amplitudes $\hbeta^{(r)}_j$ are IR-finite.
In the \order{\alpha^2} ($r=2$) implementation of {\sc KKMC} we define
\begin{equation}
  \hbeta^{(2)}_0(p)= \Mmf^{(2)}_0(p) =
  \left[ \mathrm{e}^{-\alpha B_4(p)} \Meu^{(2)}_0(p) \right]\Big|_{{\cal O}(\alpha^2) },
  \label{eq:ceex-beta0}
\end{equation}
which includes QED and EW virtual corrections.
In the future implementation of the \order{\alpha^2} EW corrections,
they would also enter into to the $2\to 3$ non-soft components:
\begin{equation}
\begin{split}
  \hbeta^{(2)}_1(p,k_1)     
    = &\; \Mmf^{(2)}_1(p,k_1) -\hbeta^{(1)}_0(p)  \sfac(p,k_1), \;\;
\Mmf^{(2)}_1(p,k_1) = 
  \mathrm{e}^{-\alpha B_4(p)} \Meu^{(2)}_1(p,k_1) 
  \Big|_{{\cal O}(\alpha^2) }.
\end{split}
\label{eq:ceex-beta1}
\end{equation}
The implementation of the CEEX method is rather laborious in practice, and
porting it to other processes is cumbersome. The development of a new version,
in which one could generate matrix elements for many processes using semi-automated
procedures would be a desirable improvement.
Another yet unsolved problem is the inclusion of relatively simple and well known
\order{\alpha^3L^3},  $L=\ln(s/m_e^2)$, corrections without complicating 
the overall structure of the CEEX matrix element as much as in the present framework.
Such corrections are already included in the EEX matrix element in the {\sc KKMC}
program, but are still incomplete in CEEX.


\subsection{Event generators for lepton colliders}
\label{sec:lepton_collider_generators}
Various simulation frameworks are available for lepton colliders. 
In the following, we list a few which have specialized on this type
of simulation (in alphabetical order).

\subsubsection{\texorpdfstring{\mcmule}{McMule}}
\label{sec:mcmule}

In its current version, \mcmule~\cite{Banerjee:2020rww,Ulrich:2022man}
is a Monte Carlo integrator for fully differential fixed-order
calculations of QED processes. It builds on methods developed for
higher-order QCD calculations and adapts them to QED with massive
fermions. In many cases this results in a simplification. In
particular, in the absence of collinear singularities the simple
YFS~\cite{Yennie:1961ad} infrared structure of QED was exploited to
adapt the FKS method~\cite{Frixione:1995ms, Frederix:2009yq} and
develop an efficient subtraction method
FKS$^\ell$~\cite{Engel:2019nfw} to deal with infrared singularities in
phase-space integrations, in principle at all orders in a perturbative
expansion in the electromagnetic coupling $\alpha$.  However, keeping
non-vanishing fermions masses introduces additional (typically small)
scales to the problem. This results in serious complications, in
particular regarding the evaluation of loop amplitudes and the
numerical stability of phase-space integrations.

So far, several QED processes have been implemented in \mcmule at NNLO
in $\alpha$, most notably M{\o}ller scattering~\cite{Banerjee:2021qvi}
and photonic corrections to Bhabha scattering~\cite{Banerjee:2021mty}.
Diphoton ($e^+e^-\to \gamma\gamma$) and dilepton
($e^+e^-\to\ell^+\ell^-$ with $\ell$ being either $\mu$ or $\tau$) production are in
preparation.  Since Bhabha scattering is particularly relevant for
lepton colliders, we will use it to describe the \mcmule approach.

It should be stressed that as it stands, the \mcmule result is not
useful for high-energy lepton colliders. At least two extensions are
required. First, electroweak effects need to be included. Second,
dominant corrections beyond NNLO have to be added, in particular
initial-state radiation. While first steps to implement these in
\mcmule are being taken, these points will be discussed
elsewhere. Here we focus on the impact and calculation of the
\textit{complete} NNLO QED corrections of order $\alpha^2$ relative to
the Born term.  Such a calculation is possible with
today's techniques. It will at least substantiate error estimates of
current Monte Carlo tools, or -- depending on the required accuracy --
lead to additional corrections to be taken into account.

The fully differential photonic NNLO corrections are split into
double-real, real-virtual and double-virtual. For the double-real
corrections, the complete (eikonal subtracted according to the
FKS$^\ell$ scheme) tree-level amplitude squared for $e^+e^-\to
e^+e^-\gamma\gamma$ is integrated over the phase space. All
interference terms are kept and no soft-photon or massless-fermion
approximation is made.

The full one-loop amplitude for $e^+e^-\to e^+e^-\gamma$ that is
needed for the real-virtual corrections has been obtained with
\OpenLoops~\cite{Buccioni:2017yxi,Buccioni:2019sur}. This involves box
and pentagon one-loop diagrams with non-vanishing but typically small
fermion masses. The delicate numerical issues are resolved thanks to
the stability of \OpenLoops combined with next-to-soft
stabilization~\cite{Banerjee:2021mty}. For the latter, the analytic
result in the soft limit at subleading power in the photon
energy~\cite{Engel:2021ccn} is used to ensure a numerically stable
evaluation of the matrix element over the full phase space.

The two-loop amplitudes for $e^+e^-\to e^+e^-$ with massive fermions
required for the double-virtual part are not yet known. Starting from
the massless amplitudes~\cite{Bern:2000ie} the massive amplitudes can
be reconstructed \cite{Penin:2005eh,Becher:2007cu,Engel:2018fsb} up to
terms that are polynomially suppressed by the (small) mass. This
induces an error that parametrically is of the order
$(\alpha/\pi)^2\,m^2/Q^2$ where $m$ is the electron mass and $Q$
another scale of the process. Of course, the one-loop amplitude
squared is also required and treated the same way.

In order to get a first impression on the importance of the full NNLO
corrections we have compared to an approximate result obtained by
truncating the parton shower implemented in
B\myscalefont{ABA}Y\myscalefont{AGA}~\cite{Balossini:2006wc} to one additional emission beyond the
full NLO result. For a particular scenario with a center-of-mass
energy of 1020~MeV the full NNLO coefficient is about 15\% larger than
the parton shower. This results in a difference of 0.07\% for the
total cross section, well within the error estimate of
\cite{Balossini:2006wc}. As stated above, these numbers cannot simply
be taken over for the high-energy scenario. However, they indicate
that it is worthwhile to make use of the developments in the
calculation of higher-order corrections and include \textit{complete}
NNLO corrections whenever high accuracy is required.

\subsubsection{\texorpdfstring{\sherpa}{Sherpa}}
\label{sec:sherpa}

The \sherpa Monte Carlo~\cite{Sherpa:2019gpd,Gleisberg:2008ta} is a fully equipped 
hadron-level event generator capable of simulating collisions in 
many different collider environments, starting with the incident 
particle beams up until the final stable leptons, photons and hadrons 
that hit the detectors.
To this end, \sherpa factorizes the individual physics processes 
according to their characteristic scales, starting with the beam 
spectra resolving the incoming particle bunches into the colliding 
particle species, followed by resolving their substructure through 
PDFs and initial-state parton showers, culminating in a description 
of the hard interaction at fixed order (LO, NLO, or NNLO in some cases) 
in perturbation theory. 
The scattering products are then further evolved with final-state 
parton showers. 
If the interacting particles possess a substructure, multi-parton 
interactions are modeled in a similar fashion. 
The final partonic ensemble is then hadronized~\cite{Winter:2003tt}, and the initial 
primordial hadrons are decayed into (meta-)stable hadrons that 
reach the detector to be measured.

Of special importance at lepton-colliders are the beam spectra, 
describing the distribution of momenta and particle species within 
the incident bunches. In addition to a standard monochromatic beam, 
\sherpa offers descriptions of photon momentum distributions in 
electron bunches through either laser backscattering, 
where the initial lepton beam sources highly energetic photons through Compton scattering~\cite{Badelek:2001xb,Zarnecki:2002qr,Archibald:2008zzb}, or 
the equivalent photon approximation (Weizs\"acker--Williams method),
where the beam particles act as quasi-classical sources of collinear quasi-real 
photons \cite{vonWeizsacker:1934sx,Williams:1934ad,Budnev:1974de}.
An interface to \circe~\cite{Ohl:1996fi}, which parameterizes the $e^\pm$, and 
$\gamma$ beam-spectra based on the collider geometry, is anticipated
(cf.~Secs.~\ref{sec:circe} and~\ref{sec:circe2}). 
The thus generated electrons or photons can then either directly enter the hard interaction or be further resolved. 

To model the momentum distribution of incident electrons, 
\sherpa offers two different approaches. 
The first approach uses LL+LO QED PDFs for the electrons, 
that are solutions of the DGLAP evolution equations, cf. Sec.~\ref{sec:QEDcollfac}.
It can be combined with a traditional QED parton shower \cite{Hoeche:2009xc} 
to generate exclusive kinematic distributions for the collinear photons. 
In the second approach, the emission of soft photons is resummed to all 
orders using the Yennie-Frautschi-Suura (YFS) formalism~\cite{Yennie:1961ad}. 
In this method, photon emissions are considered in fully differential
form, get generated explicitly, and the treatment of their phase space is exact.
At the same time, the substructure of photons can be resolved using, e.g., 
the dedicated photon PDFs of \cite{Gluck:1991ee,Gluck:1991jc}, upon which 
the produced partons enter a traditional QCD and QED evolution through a 
parton shower.

The hard scattering interaction is then described at LO, NLO and, in some 
cases, at NNLO in QCD \cite{Hoeche:2011fd,Hoeche:2012yf,Gehrmann:2012yg,Hoche:2014uhw} accuracy, 
matched to the parton shower \cite{Schumann:2007mg,Hoche:2015sya}. 
NLO EW will be offered in the near future within the YFS soft-photon 
resummation.

\subsubsection{\texorpdfstring{\whizard}{Wizard}}
\label{sec:whizard}

\whizard is a multi-purpose event generator framework for very general high-energy physics simulations for all types of colliders and fixed-target experiments, that allows a high level of flexibility~\cite{Kilian:2007gr}. 

For hard-scattering processes, \whizard provides its own
(tree-level) matrix-element generator, \omegaMEG~\cite{Moretti:2001zz}, which is based on recursion relations and the
color-flow formalism~\cite{Kilian:2012pz} for QCD. 
This framework supports as hard-coded models the SM in different variants and the most studied BSM models. It has a general
interface to \feynrules~\cite{Christensen:2010wz} and
UFO~\cite{Degrande:2011ua}, cf. Sec.~\ref{sec:new_physics}
and supports Les Houches accord input files~\cite{Allanach:2008qq}. 
Loop-matrix elements for next-to-leading (NLO) QCD and EW processes
 can be accessed through
dedicated interfaces to one-loop providers (OLP) like \OpenLoops, \GoSam
and \recola, cf. Sec.~\ref{sec:hadron_collider_hard_process}. 

For all processes within all models, Feynman diagram selection is
possible, however should be treated with care 
due to possible gauge-invariance violations. Processes can be
factorized into production and decays by using the decay features of
\whizard, also consecutively in decay chains. All decays of a
resonance at a given final-state multiplicity can be auto-generated.
In resonant processes, intermediate resonances can be specified as
polarized, e.g. for the study of specific longitudinal or transversal EW gauge bosons like in multi-EW boson
production and vector-boson scattering~\cite{Brass:2018hfw,Alboteanu:2008my,Kilian:2014zja,Fleper:2016frz}.

\whizard ships with its 
own scripting language, {\sc Sindarin}, which allows arbitrary cuts to be
defined on the hard matrix elements, as well as selections for the
event generation. The hard matrix elements can be convoluted with a
large number of parton distribution functions, structure functions or
beam spectra: these include hadron-collider proton PDFs via \lhapdf
externally or from a selected set of internally shipped PDFs,
effective $W/Z$ approximation for the content of EW bosons inside
quarks or leptons, the effective-photon approximation (EPA), also
known as Weizs\"acker-Williams approximation. The latter is provided
in different variants, depending on the application for low-energy
$\gamma\gamma \to$ hadron background simulations or for high-energy
particle production.

For lepton collisions, electron/muon PDFs are provided at LL accuracy 
and are currently being extended to NLL. \whizard supports several options for structured
lepton beams: Gaussian beam spreads, individually adjustable for each
beam, beam spectrum files read in as tables of pairs of energy values
and an approximation to \guineapig spectra via its beamstrahlung
generators \circe~\cite{Ohl:1996fi} (cf.~Sec.~\ref{sec:circe}) and~\circetwo (cf.~Sec.~\ref{sec:circe2}).
\whizard allows initial-state particle of all
kinds to be polarized, both for scattering as well as  for decay
processes. For scattering processes, the polarization of both beams
can be arbitrarily correlated by specifying a spin density matrix. In
addition, polarization fractions can be given. Initial-state
beams can be asymmetric (like in flavor factories or HERA), and
crossing angles can be defined.

As the program has always had a strong focus on weak production
processes of multi-fermion final states, at lepton colliders it
contains a phase-space parameterization that is flexible enough to
adapt to many different (interfering) resonant production
channels. This builds upon an adaptive multi-channel Monte Carlo
integration encoded in the \vamp
subpackage~\cite{Ohl:1998jn}. In recent years, there have been
several attempts to make the MC integration much more efficient by
using highly parallelized computing architectures: using
Multi-Processing Interface (MPI), speed-ups of several tens up to
roughly a hundred can be achieved~\cite{Brass:2018xbv}. Also, \omegaMEG can produce very small expressions for the matrix elements in
the form of a so-called virtual machine, which are as efficient as
compiled code~\cite{ChokoufeNejad:2014skp}.

Turning to high-precision calculations at next-to-leading order in the
SM, \whizard has a complete implementation for NLO QCD and
electroweak (as well as mixed) corrections for hadron and lepton
colliders, based on its implementation of the FKS subtraction
scheme, standard and resonance-aware. In the same setup as for scattering processes, NLO decays can be calculated. The automated implementation has been validated for hadron and lepton colliders  and allows to calculate processes at NLO (QCD) for rather high
final-state multiplicities, e.g. $e^+e^- \to \ell\ell\nu\nu b\bar{b} H$
or $e^+e^- \to jjjjjj$~\cite{Rothe:2021sml,Brass:xxx}.  The
program provides a default framework to produce arbitrary differential distributions, using the concept of event groups with an analysis like e.g. \rivet~\cite{Bierlich:2019rhm,Buckley:2010ar}. Electroweak corrections for lepton colliders can be calculated already for massive leptons, while the infrastructure for
corrections for massless leptons is being validated right now. The NLL
electron PDFs are being implemented, and first results for this are
expected within 2022. To define infrared-safe quantities,
\whizard provides jet clustering with a FastJet interface as well
as photon recombination. To define fixed-order NLO cross sections in a 4- or 3-flavor scheme, the user can specify $b$-jet, $c$-jet and light
jet clustering.

In order to generate complete events, \whizard provides its own
parton shower, specifically a $k_T$-ordered parton shower, as well as, an
analytic, virtuality-order (QCD) shower~\cite{Kilian:2011ka}. In addition, \whizard
ships with the final \fortran version of \pythia
(6.427)~\cite{Sjostrand:2006za} which is specifically tuned to LEP2
hadron data, and with a dedicated interface to \pythia
8~\cite{Sjostrand:2014zea}; the latter allows to directly communicate
between the event records of \whizard and \pythia.

For inclusive jet samples at LO, \whizard can apply the MLM
merging algorithm, while for NLO matching between fixed-order and the
parton shower, a general, process-indepen\-dent version of \POWHEG{}
matching is available for QCD~\cite{ChokoufeNejad:2015kpc}.
\POWHEG{} matching for EW corrections is in preparation, while for mid-term planning also
alternative matching and merging schemes are envisioned. In order to
reconcile multi-fermion final states, e.g. $e^+e^- \to jjjj$ with
parton showering and hadronization, \whizard is able to find and
quantify contributions from underlying resonant subprocesses, and then
provide pseudo-shower histories to the final state weighted by the
relative cross sections of these underlying processes. This is
important to correctly model hadronic correlations in the final states, and
also inclusive observables like the total number of final-state
neutral and charged hadrons as well as photons.

One of the most important tasks for precision physics at future Higgs
factories will be a modeling of QED and photon radiation as precise
as possible. For the normalization of QED cross section, a resummation
based on collinear factorization gives very precise results. As
mentioned above, this is possible in \whizard at LL level and
will be available at NLL level soon. Specifically for the assessment
of experimental/systematic uncertainties, an exclusive simulation of
photon radiation is necessary. \whizard can of course produce
explicit photons from matrix elements; on the other hand, the energy
loss from LL electron PDFs in the collinear approximation is collected
in a single photon per beam. A heuristic approximation is possible to
generate a $p_T$ distribution for these photons, which can also be
applied to recoiling charged leptons from
Weizs\"acker-Williams/EPA~\cite{Fermi:1932xva,vonWeizsacker:1934sx,Williams:1934ad}. The photon distribution is generated using
a logarithmic scaling, while the event of the hard scattering (with
subsequent parton shower and hadronization) is boosted accordingly. It
has been shown that for both signal and background in monophoton dark
sector searches (where this distribution plays a key role), this heuristic
approach agrees very well with exact matrix-element
calculations~\cite{Kalinowski:2020lhp,1859727}. While collinear
resummation results in very accurate predictions for cross sections of
(inclusive) processes with initial-state radiation, combining the
highest possible precision of resummation with exclusive photon
radiation is achieved using the YFS
formalism~\cite{Yennie:1961ad}. The \whizard team is working on 
the implementation of this formalism for general processes. Furthermore,
\whizard will also get its own QED parton shower, which will be
available as a relatively slim stand-alone shower for pure QED, and
potentially also as an interleaved shower together with the QCD
shower.

\whizard has special support for the top threshold to model the
process $e^+e^- \to t\bar{t} \to bW^-\bar{b}W^+$. This process allows
for the determination of the top mass with the highest possible theoretical
precision. Analytic calculations using an NRQCD EFT approach can
include NNNLO corrections, which reduces the error to 30-70
MeV. This calculation, however, is completely inclusive. To study
experimental efficiencies and systematic uncertainties, Monte Carlo
samples resembling event weights close to the true top threshold cross 
sections are needed. \whizard includes a special treatment of a
matched calculation between the continuum NLO off-shell fixed-order
calculation and the NLL-threshold resummed calculation avoiding double
counting~\cite{Bach:2017ggt}, fully off-shell and exclusive in the
top decay products. This allows to study fully exclusive distributions.

In a similar framework, work has started to match resummed
QED-Coulomb corrections to the QED NLO/NNLO fixed-order calculation
for the $WW$ threshold~\cite{Beneke:2007zg,Actis:2008rb}. Like the top
threshold, this will be available as a dedicated process within a
specialized model. These simulations  allow experimental studies with
the desired accuracy to support measurements of the $W$ mass from a
threshold scan with a precision of 1-2 MeV or below.

\subsubsection{\texorpdfstring{\babayaga{}@NLO}{BabaYagaNLO}}
\label{sec:babayaga}
\babayaga{}@NLO~\cite{Balossini:2006wc,Balossini:2008xr} is a Monte Carlo
event generator that provides precise predictions for the processes
$e^+ e^- \to e^+ e^-$, $e^+ e^- \to \mu^+ \mu^-$ and
$e^+ e^- \to \gamma \gamma$ at NLOPS accuracy in QED.
It is a reference tool for luminosity measurements and other physics studies
at $e^+ e^-$ colliders with center of mass energy at the GeV scale
(flavour factories)~\cite{WorkingGrouponRadiativeCorrections:2010bjp}.
We will first discuss the QED fixed-order corrections implemented
in the code that are necessary to describe processes at low energies,
and then mention recent and future upgrades of the generator
that are necessary to make it suitable for simulations at
high-energy colliders.

In a nutshell, the theoretical formulation inherent to \babayaga{}@NLO~is
based on the matching of next-to-leading order (NLO) corrections with a
QED Parton Shower (PS) algorithm modeling multiple photon emission
exclusively~\cite{CarloniCalame:2000pz,CarloniCalame:2001ny}. 

The master formula for the cross section calculation reads as follows\footnote{
For the sake of simplicity, we limit the notation for our matching procedure
to the case of one radiating particle. The extension of Eq.~(\ref{eq:babayaga1}) to the 
realistic case, where every charged particle radiates photons, is almost straightforward.}:
\begin{equation}
  \di\sigma = F_{\textrm{SV}}\,\Pi \left(Q^{2},\epsilon\right)
  \sum_{n=0}^{\infty}\dfrac{1}{n!}\left(\prod_{i=0}^{n}F_{\textrm{H},i}\right)
  \modulo{\M_{n,\textrm{LL}}}^{2}\di \Phi_{n}
\label{eq:babayaga1}
\end{equation}
In Eq.~(\ref{eq:babayaga1}), $\Pi\left(Q^{2},\epsilon\right)$ is the Sudakov
form factor, which accounts for the exponentiation of leading logarithmic
(LL) contributions due to soft and virtual corrections, $Q^2$ and $\epsilon$ 
being the hard scale of the process\footnote{For charged fermion pair production,
we set in the code $Q^2 = s \, t / u$ in order to exponentiate also the dominant
contribution due to box diagrams and initial-final state interference, in addition
to the leading logarithms from initial and final state radiation. For 
$e^+ e^- \to \gamma \gamma$, only the choice $Q^2 = s$ describing initial state
radiation is physically meaningful.} and a soft-hard photon separator, respectively.  
$\sqmodulo{\M_{n,\textrm{LL}}}$ is the squared matrix element describing the emission
in LL approximation of $n$ hard photons, i.e. with energy larger than $\epsilon$. The 
factor $\di\Phi_{n}$ is the exact phase space element of the underlying process
accompanied by photon radiation. 

In Eq.~(\ref{eq:babayaga1}) the matching of the above PS ingredients with the NLO
QED corrections is realized by the fixed-order factors $F_{\textrm{SV}}$ and
$F_{\textrm{H}}$, which are process dependent and whose definitions can be found
in~\cite{Balossini:2006wc}. They are infrared/collinear safe correction factors
that account for those $O(\alpha)$ non-logarithmic terms entering the NLO calculation
and absent in the universal PS approach. The factor $F_{\textrm{SV}}$ takes
into account soft+virtual corrections, while $F_{\textrm{H}}$ is a hard 
Bremsstrahlung contribution including the exact matrix element of the $2 \to 3$
radiative process of the underlying $2 \to 2$ process under consideration. 
It is worth stressing that both  $F_{\textrm{SV}}$ and $F_{\textrm{H}}$ are applied 
at differential level on an event-by-event basis.

By construction, the matching procedure as in Eq.~(\ref{eq:babayaga1}) is such
that its $O(\alpha)$ expansion reproduces the NLO cross section and exponentiation
of LL contribution is preserved as in a pure PS approach. Moreover, as a by-product
of its factorized structure, the bulk of the photonic sub-leading contributions at NNLO,
i.e those of the order of $\alpha^2 L$, is automatically included by means of terms
of the type $F_{{\textrm{SV}}~|~{\textrm{H}}}~\otimes$~LL corrections~\cite{Montagna:1996gw}.

Concerning the treatment of the angular degrees of freedom in the PS approach,
the generation of the transverse momentum of fermions and photons at each branching
is achieved in \babayaga{}@NLO~according to the following function
\begin{equation}
I(k)~=~
\sum_{i,j=1}^N~
\eta_i \eta_j~
\frac{p_i\cdot p_j}{(p_i\cdot k)(p_j\cdot k)}~
E_\gamma^2
\label{idik}
\end{equation}
where $N$ is the number of involved charged particles, $p_\ell$ is the momentum
of the external fermion $\ell$, $\eta_\ell$ is a charge factor, see e.g.\
\cite{CarloniCalame:2001ny}, $k$ is the photon momentum, $E_\gamma$ is its energy
and the sum runs over all the external fermions. It is worth noting that Eq.~(\ref{idik}) 
is inspired to the expression of the differential cross section for a generic process
with emission of additional soft photons and accounts for the interference of radiation
coming from different particles.

For QED processes at flavour factories, the theoretical accuracy of  \babayaga{}@NLO~is 
at the 0.1\% level. This conclusion follows from a number of detailed tuned comparisons
with the results of independent codes, such as \bhwide~\cite{Jadach:1995nk} and
MCGPJ~\cite{Arbuzov:2005pt}, as well as with NNLO predictions and the estimate
of $O(\alpha^2 L)$ contributions according to realistic event selection criteria.

For the high-precision requirements of high-energy $e^+ e^-$ colliders, possible
improvements of \babayaga{}@NLO~are given by
\begin{enumerate}
\item the inclusion of the full set of NLO electroweak corrections.
  A first step towards this goal has been taken in Ref.~\cite{CarloniCalame:2019dom}
  for the study of the process $e^+e^-\to\gamma\gamma$ at FCC-ee energies;
\item the implementation into a QED PS algorithm of the parton evolution at 
  next-to-leading-logarithmic (NLL) accuracy, in order to improve
  the description of higher-order corrections;
\item the generalization of the matching algorithm to include also exact NNLO in QED.
\end{enumerate}

\subsubsection{\texorpdfstring{\bhlumi/\bhwide}{BHLumi/BHWide}}
\label{sec:bhlumi/bhwide}
\bhwide~\cite{Jadach:1995nk} is a Monte Carlo (MC) event generator for large-angle Bhabha  scattering 
with multiphoton emission.
It is based on the Yennie--Frautschi--Suura exclusive exponentiation~\cite{Yennie:1961ad}
in which all the infrared singularities are summed-up to the infinite order
and cancelled properly in the YFS form factor. 
The remaining non-IR residuals, $\bar{\beta}_n^{(l)}$, corresponding to 
the emission of $n$-real photons, are calculated perturbatively up to a given
order $l$, where $l\geq n$, and $(l-n)$ corresponds to a number of loops
accounted for in $\bar{\beta}_n^{(l)}$. 
In \bhwide, an arbitrary number $n$ of real photons with non-zero transverse momenta
and energies above some soft-photon cut-off are generated according to the YFS MC algorithm \cite{Jadach:1991by}.   
This cut-off can be set to an arbitrary low value without any danger of negative probability distributions, 
as it can happen in standard fixed-order calculations.
The non-IR residuals are calculated up to NLO: $\bar{\beta}_0^{(1)}$, corresponding no real-photon emission, includes one-loop 
electroweak  (EW) radiative corrections, and 
$\bar{\beta}_1^{(1)}$, corresponding to one-real photon emission, includes a tree-level hard-bremsstrahlung matrix element.
In $\bar{\beta}_0^{(1)}$, two libraries of the ${\cal O}(\alpha)$
virtual EW corrections are used: 
(1) the older one of Refs.~\cite{Bohm:1986fg,Berends:1987jm}, and 
(2) the newer of Ref.~\cite{Beenakker:1990mb} (recommended). 
In $\bar{\beta}_0^{(1)}$,  two independent matrix elements for single hard-photon radiation are implemented: 
(1) the one expressed in terms of helicity amplitudes~\cite{Jadach:1995nk}  (recommended), and 
(2) the formula of CALKUL~\cite{Berends:1981uq} for the matrix element squared. 
In both cases, finite electron/positron mass effects are taken into account.
The above two implementations agree numerically up to at least $6$ digits event-by-event.

The MC algorithm of  \bhwide is based on that of  
the event generator \bhlumi for small-angle Bhabha scattering~\cite{Jadach:1991by}
with three important extensions: 
(1) QED interferences between the electron and positron lines 
(``up-down'' interferences) are added,
(2) the full YFS form factor for the $e^+e^-\rightarrow e^+e^-$ process is implemented, and
(3) the exact NLO matrix element for the full Bhabha process (all channels and contributions) is included.
At a low-level MC stage, the multiphoton radiation is generated
for a $t$-channel scattering process, while an $s$-channel as well as all
interferences are reintroduced through appropriate MC weights.
This means that the program is more efficient when the $t$-channel contribution is dominant,
although it proved to work well also at the $Z$-peak. 

\bhwide offers two modes of event generation: 
(1) variable-weight events (faster, several MC weights can be computed in a single run),
and (2) unit-weight events (useful for experimental analyses where events
need to be processed through time-consuming detector simulations). 
Various input parameter options, to be set 
by the user, allow to choose between different contributions/corrections
to the cross section, such as EW corrections (two libraries),
vacuum-polarization parameterization, etc. Other input 
parameters allow to specify the necessary ingredients for 
event generation and cross-section calculations, such as collision
energy, physical parameters (masses, widths, etc.), the soft-photon cut-off, phase-space cuts, etc.    
For each generated event, information on four-momenta of the final-state electron,
positron, and all emitted photons is provided through a dedicated event record. 
In the variable-weight mode, it is supplemented with a value of the nominal
event weight and a vector of weight values corresponding to various
models or approximations. In the final stage of a program run, values of the total cross 
section and its statistical error corresponding to the generated event sample are computed
and provided together with other useful information. 

Several cross-checks and comparisons of \bhwide with other calculations were done
for the LEP experiments \cite{Jadach:1996gu,Placzek:1999xc} which were its first main users.
Based on these tests,  its overall theoretical
precision was estimated at the level of $\sim 1\%$ for the LEP energies.
\bhwide has also been used by some experiments at low-energy $e^+e^-$ colliders \cite{WorkingGrouponRadiativeCorrections:2010bjp}.  
Recently, it was helpful in studies of theoretical  precision of luminosity measurements at future lepton colliders
(FCC-ee, ILC, CLIC)  \cite{Jadach:2018jjo,Jadach:2021ayv},
where it can be complementary to \bhlumi in providing predictions for contributions to the small-angle Bhabha process 
other than a pure $t$-channel $\gamma$-exchange.
In order to match the expected experimental precision of the future lepton colliders, \bhwide needs further improvements,  
such as interfaces to modern NLO EW libraries, inclusion of the NNLO QED radiative corrections, and for FCC-ee, 
possibly also the NNLO EW corrections. 

\subsubsection{\texorpdfstring{\photos}{PHOTOS}}
\label{sec:photos}

In contrast to the above examples, \photos \cite{Barberio:1990ms,Barberio:1993qi,
  Golonka:2005pn,Davidson:2010ew} is not a complete stand-alone 
event generator for $e^+e^-$ collisions. It is, however, a widely used 
tool to approximate higher-order QED final state radiation corrections 
to an otherwise LO QED simulation. 
Its FSR resummation properties can be summarised by examining \photos' 
approximation of the all-orders decay rate $\mathrm{d}\Gamma$ 
in terms of a given LO decay rate $\mathrm{d}\Gamma_0$,
\begin{equation}
  \begin{split}
    \mathrm{d}\Gamma^\text{\photos}
    =\;&
      \mathrm{d}\Gamma_0
      \left\{
        1+
        \sum\limits_{c=1}^{n_\text{ch}}\sum\limits_{n_\gamma}
        \frac{\left(\alpha\,L_c\right)^{n_\gamma}}{n_\gamma!}
        \left[\prod\limits_{i=1}^{n_\gamma}\mathrm{d} x_c^i\right]
        \left(P_{\epsilon_{\text{cut}}}(x_c^1)\!\otimes\!.\;\!\!.\;\!\!.\!\otimes\! P_{\epsilon_{\text{cut}}}(x_c^{n_\gamma})\right)
      \right\}\;.
  \end{split}
\end{equation}
Therein, the emission kernels are summed over all charged particles 
$n_\text{ch}$ involved and $L_c$ is the logarithm of the ratio of the 
decaying particle's mass over the mass of the charged particle $c$. 
Finally, $x_c=\prod x_c^i$ is the energy energy fraction retained 
after the radiation process, with $n_\gamma$ photons being emitted.
The photons are distributed in phase space according to the 
Altarelli-Parisi splitting functions $P_{\epsilon_{\text{cut}}}(x)$, 
regulated with an infrared cut-off $\epsilon_{\text{cut}}$. 
These emission kernels can be modified by suitable 
weights to recover the correct coherent soft-photon limit, or 
exact higher-order corrections. 
Their precise definitions can be found in \cite{Barberio:1993qi}.

\subsection{Heavy flavor production and decay}
\label{sec:heavy_hadron_decays}

The production of excited heavy mesons and baryons, and their subsequent decay,
have always been a particular focus of the $e^+e^-$ collider physics program, especially
at flavor factories. Heavy-quark fragmentation differs from its light-quark counterpart,
and to some extent is more straightforward due to the appearance of a high mass scale.

Decay cascades of unstable hadrons down to the stable ground states are modeled in several event generators~\cite{Lange:2001uf,Bellm:2015jjp,Sherpa:2019gpd} using the same two basic ingredients. First, branching ratios are taken from experimental measurements for known decay channels or simulated inclusively through parton shower and hadronization models for the unmeasured ones. Second, decay kinematics are generated according to matrix elements including form factors, which are typically fitted from effective field theories or measurements. Dedicated implementations for particular phenomenological effects are available in some of the generators, two of which are exemplified in the following.

\subsubsection{Polarization effects}
\begin{figure}
\centering
\includegraphics[width=0.45\textwidth,clip,trim=80mm 0mm 0mm 0mm]{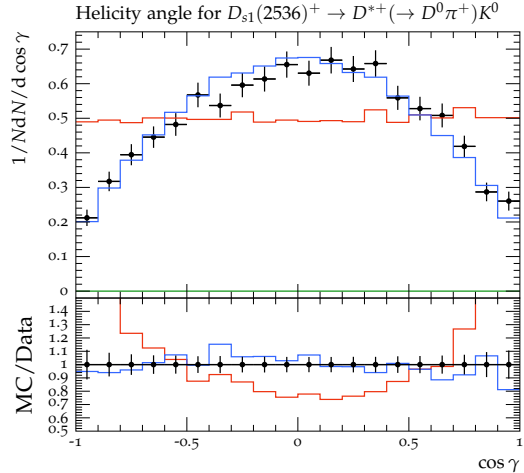}
\caption{Efficiency-corrected normalized signals as functions of $\cos \gamma$
in comparison to experimental data from BELLE~\cite{Belle:2007kff}.
Green Histograms show default predictions from \herwig{}, while red and blue histograms
are respectively predictions with only the implementation of HQEFT and with HQEFT effects
in the presence of improved hadronization and decay models.
\label{fig:heavy_flavor}}
\end{figure}
An interesting question is whether the polarization of the initial heavy quark may be detected
after hadronization, either in a polarization of the final ground state or in anisotropies 
in the decay products of the excited hadron. The result should hinge, in part, on a non-perturbative
parameter that can measure the net transverse alignment of the light degrees of freedom 
in the fragmentation process~\cite{Falk:1993rf}.
The effect can be simulated in the \herwig{} framework through the combination of various
physics models~\cite{LE-paper}:
\begin{itemize}
\item[(a)] Passing through the polarization of heavy hadrons at the end of parton shower by employing a prescription of the heavy-quark effective field theory (HQEFT) that allows for the detection of a net polarization of the initial heavy-quark, either in a polarization of the final ground state or in the decay products of the excited heavy mesons and heavy baryons~\cite{Falk:1993rf,LE-paper},
\item[(b)] improving the strong and radiative decay modes of the excited heavy mesons, i.e charm and bottom mesons, 
\item[(c)] re-tuning \herwig's cluster hadronization model by the use of {\it all} existing production rates of heavy hadrons, and
\item[(d)] improvement of the kinematic behavior of the cluster hadronization model, especially with respect to the treatment of light high-energy clusters.
\end{itemize}
Figure~\ref{fig:heavy_flavor} shows first results, demonstrating the immediate effects
of these modifications in the predictions of the experimental data from BELLE~\cite{Belle:2007kff}.

\subsubsection{Mixing}
\begin{figure}
\centering
\includegraphics[width=0.7\textwidth]{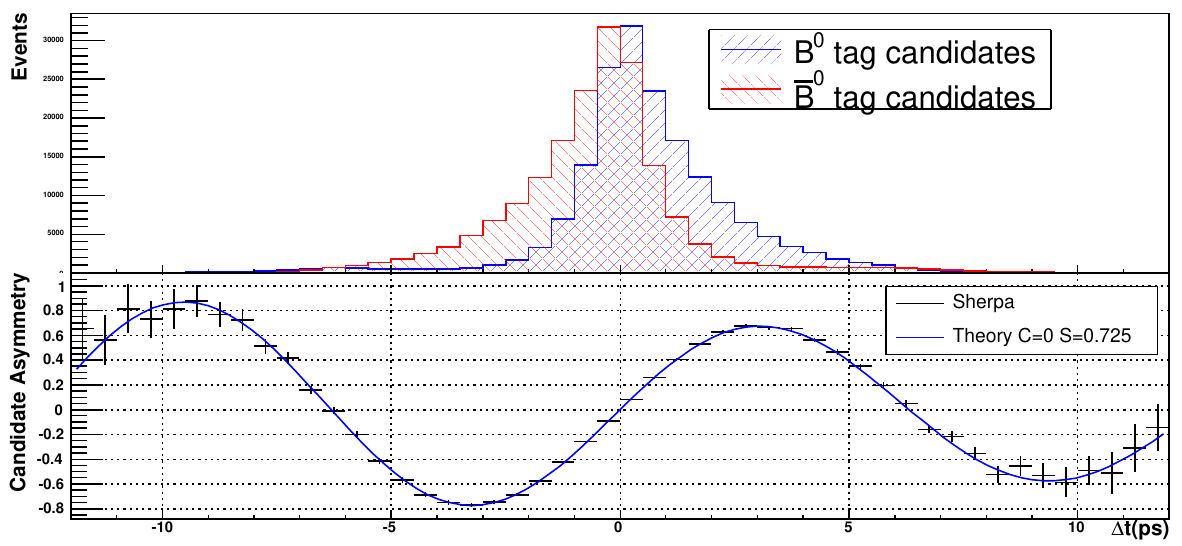}
\caption{CP violation in the interference between $B^0 \to J/\Psi K_S$ and $\bar B^0 \to J/\Psi K_S$ with $y=-0.05$.
\label{fig:bmixing}}
\end{figure}
Another interesting phenomenological aspect is the mixing of neutral mesons (e.g.\ $D^0$, $B^0$, $B_s$) and in particular its connection to CP violation, which has been studied in detail at flavor factories. The \sherpa{} framework allows a modeling of these effects in several ways:
\begin{itemize}
\item[(a)] Explicit mixing of neutral mesons in the event record, optionally with CP violation in the mixing, i.e.\ $|q/p|^2\neq 1$,
\item[(b)] Direct CP violation in decays, i.e.\ rate differences between the decay $i \to f$ and $\bar{i} \to \bar{f}$, implemented through separate branching ratio tables for particles and anti-particles,
\item[(c)] CP violation in the interference between mixing and decay amplitudes, for decays into final states common to both particle and anti-particle.
\end{itemize}
An example of the latter can be seen in Fig.~\ref{fig:bmixing} for the rate asymmetry between $B^0 \to J/\Psi K_S$ and $\overline{B^0} \to J/\Psi K_S$ decays in $\Upsilon(4S)\to B^0 \overline{B^0}$ events assuming $y\equiv \Delta\Gamma/(2\Gamma) = -0.05$.

\clearpage
\newpage
\section*{Acknowledgments}
The work of S. Alioli, A. Broggio, A. Gavardi, S. Kallweit, G. Marinelli, D. Napoletano and R. Nagar is supported by the ERC Starting Grant REINVENT 714788.
S. Alioli also acknowledges funding from Fondazione Cariplo and Regione Lombardia, grant 2017-2070 and from MIUR through the FARE grant R18ZRBEAFC.
The work by M.~D.~Baker was supported by Jefferson Science Associates, LLC under Contract No. DE-AC05-06OR23177 with the DOE, Office of Science, Office of Nuclear Physics and by the U.S. Department of Energy under Contract No. DE-SC0012704. 
This work is supported by a Royal Society Research Professorship
(RP$\backslash$R1$\backslash$180112: M.~van~Beekveld, G.~P~.Salam, L.~Scyboz),
by the European Research Council (ERC) under the European Union’s
Horizon 2020 research and innovation programme (grant agreement No.\
788223, PanScales: M.~Dasgupta, S.~Ferrario~Ravasio, K.~Hamilton, A.~Karlberg, R.~Medves, G.~P~.Salam, A.~Soto-Ontoso, G.~Soyez and R.~Verheyen), 
and by the Science and Technology Facilities Council (STFC) under
grants ST/T000864/1 (M.~van~Beekveld, G.~P.~Salam), ST/T001038/1 (M.~Dasgupta, M.~H.~Seymour), ST/T000856/1 (K.~Hamilton) and ST/T000694/1 (B.~R.~Webber).
The work of J.~M.~Campbell, S.~Gardiner, T.~Hobbs, S.~Hoeche, J.~Isaacson, S.~W.~Li, P.~Machado, S.~Mrenna, N.~Rocco and M.~Wagman was supported by the Fermi National Accelerator Laboratory (Fermilab),
a U.S. Department of Energy, Office of Science, HEP User Facility.
Fermilab is managed by Fermi Research Alliance, LLC (FRA), acting under Contract No. DE--AC02--07CH11359. 
The work of N. Darvishi is supported by the National Natural Science Foundation of China (NSFC) under grants No. 12022514, No. 11875003 and No. 12047503, and CAS Project for Young Scientists in Basic Research YSBR-006, by the Development Program of China under Grant No. 2020YFC2201501 (2021/12/28) and by the CAS President’s International Fellowship Initiative (PIFI) grant.
A.~Denner, T.~Ohl and G.~Pelliccioli are supported by the Federal Ministry of Education and Research, Germany (BMBF, grant 05H21WWCAA).
The work of M. Diefenthaler was supported by Jefferson Science Associates, LLC under Contract No. DE-AC05-06OR23177 with the DOE, Office of Science, Office of Nuclear Physics.
T.~Engel is supported by the Swiss National Science Foundation (SNF) under contract 200021\_178967.
The work of J.~R.~G is supported by the Royal Society through Grant URF\textbackslash{}R1\textbackslash{}201500.
I.~Helenius acknowledges support from the Academy of Finland, project number 331545 and has been funded as a part of the CoE in Quark Matter of the Academy of Finland.
S. Jadach  partly supported by the funding from the European Union’s Horizon 2020 research 
and innovation programme under Grant agreement no. 951754 
and from the National Science Centre, Poland, Grant no. 2019/34/E/ST2/00457.
The work of A.~Jentsch and Z.~Tu was supported by the U.S. Department of Energy under Contract No. DE-SC0012704, and A. Jentsch was also supported by the Program Development program at Brookhaven National Laboratory.
S.~Klein, B.~Nachman and C.~Wilkinson are supported by the U.S. Department of Energy (DOE), Office of Science under contract DE-AC02-05CH11231.
F.~Kling and J.R.~Reuter are supported by the Deutsche Forschungsgemeinschaft (DFG, German Research Association) under Germany’s Excellence Strategy-EXC2121 “Quantum Universe”-39083330. 
C.~Krause is supported by DOE grant DOE-SC0010008. 
M.R. Masouminia is supported by the UK Science and Technology Facilities Council (grant numbers ST/P001246/1).
The work of J. McFayden was supported by the Royal Society Fellowship Grant URF\textbackslash R1\textbackslash201519.
L.~Pickering is supported by a Royal Society University Research Fellowship (URF\textbackslash{}R1\textbackslash{}211661).
V. Pandey acknowledges the support from US DOE under grant DE-SC0009824.
C.T.~Preuss is supported by the Swiss National Science Foundation (SNF) under contract 200021-197130. 
A.~C.~Price is supported by the European Union’s Horizon 2020 research and innovation programme under the Marie Skłodowska-Curie grant agreement No 945422.
L.~Rottoli is supported by the SNF grant PZ00P2\_201878.
This work is supported by the Royal Society through a University Research Fellowship (URF\textbackslash{}R1\textbackslash{}180549: M.~Sch\"onherr) and an Enhancement Award
(RGF\textbackslash{}EA\textbackslash{}181033 and CEC19\textbackslash{}100349: W.~Ju, M.~Sch\"onherr)
as well as the STFC under grant agreement ST/P001246/1 (L.\ Flower, M.~Sch\"onherr).
S.~Schumann acknowledges support from the German Federal Ministry of Education and Research (BMBF, grant 05H21MGCAB) and the German Research Foundation (DFG, project number 456104544).
The work of A. Siodmok and J. Whitehead is supported also by the National Science Centre,
Poland, (Grant no. 2019/34/E/ST2/00457)
The work of A. Siodmok was also funded by the Priority Research Area Digiworld under
the program Excellence Initiative – Research University at the Jagiellonian University in Cracow.
T. Sj\"ostrand is supported by the Swedish Research Council, contract number 2016-05996.
P.~Skands is supported by ARC grant DP170100708. 
J.T.~Sobczyk was supported by NCN grant UMO-2021/41/B/ST2/02778.
D.~Soldin acknowledges support from the NSF Grant PHY-1913607.
The work of D.~Soper was supported by the United States Department of Energy under grant DE-SC0011640.
S.~Trojanowski is supported by the grant ``AstroCeNT: Particle Astrophysics Science and Technology Centre'' carried out within the International Research Agendas programme of the Foundation for Polish Science financed by the European Union under the European Regional Development Fund and by the Polish Ministry of Science and Higher Education through its scholarship for young and outstanding scientists (decision no 1190/E-78/STYP/14/2019).
R.~Winterhalder is supported by FRS-FNRS (Belgian National Scientific Research Fund) IISN projects 4.4503.16. The work by K.~Zapp has received funding from the European 
Research Council (ERC) under the European Union’s Horizon 2020 research and 
innovation programme (Grant agreement No. 803183, collectiveQCD). 
This research used resources of the Argonne Leadership Computing Facility, which is a DOE Office of Science User Facility supported under Contract DE-AC02-06CH11357.
The work is funded by LANL’s Laboratory Directed Research and Development (LDRD/PRD) program under project number 20210968PRD4.
Los Alamos National Laboratory is operated by Triad National Security, LLC, for the National Nuclear Security Administration of U.S. Department of Energy (Contract No. 89233218CNA000001).
This research is supported in parts by the Deutsche Forschungsgemeinschaft (DFG, German Research Foundation) under grant 396021762-TRR 257.
Much of the event generator development described above was performed in the context of the MCnet collaboration, which received funding from the European Union's Horizon 2020 research and innovation programme as part of the Marie Skłodowska-Curie Innovative Training Network MCnetITN3 (grant agreement no. 722104).
\bibliographystyle{apsrev4-1}
\typeout{}
\bibliography{main}

\end{document}